%% file: ms-qft-I.tex
\documentclass[12pt]{article} 
\input defs

\usepackage{epsfig}
\usepackage{amsbsy}
\begin{document}

\include{front0}
\include{ch000}

\include{ch001}

\include{ch002}

\include{ch003}

\include{ch004}

\include{ch005}

\include{ch006}

\include{ch007}

\include{ch008}

\include{ch009}

\include{ch010}

\include{ch011}

\include{ch012}
\include{ch013}
\include{ch014}

\include{ch015}
\include{ch016}

\include{ch017}

\include{ch018}

\include{ch019}
\include{ch020}

\include{ch021}
\include{ch022}

\include{ch023}
\include{ch024}

\include{ch025}

\include{ch026}

\include{ch027}
\include{ch028}
\include{ch029}
\include{ch030}
\include{ch031}

\include{ch032}

\end{document}

%% file: defs.tex
\def\square{\kern1pt\vbox{\hrule height 1.2pt\hbox{\vrule width 1.2pt\hskip 3pt
   \vbox{\vskip 6pt}\hskip 3pt\vrule width 0.6pt}\hrule height 0.6pt}\kern1pt}
\def\ts{\textstyle}

\def\sss{\scriptscriptstyle}
\def\gsim{\mathrel{\raise.3ex\hbox{$>$\kern-.75em\lower1ex\hbox{$\sim$}}}} 
\def\lsim{\mathrel{\raise.3ex\hbox{$<$\kern-.75em\lower1ex\hbox{$\sim$}}}} 
\def\eq#1{eq.$\,$(\ref{#1})} 
\def\eqs#1{eqs.$\,$(\ref{#1})} 
\def\Eq#1{Eq.$\,$(\ref{#1})} 
 
\def\fig#1{fig.$\,$(\ref{#1})} 
\def\figs#1{figs.$\,$(\ref{#1})}

\def\e{{\ts\varepsilon}}

\def\eps{\epsilon}

\def\lamph{\lambda_{\rm ph}}
\def\mph{m_{\rm ph}}
\def\msb{\overline{\rm MS}}

\def\os{{\rm OS}}
\def\pimsb{\Pi_{\sss\msb}}
\def\pios{\Pi_{\sss\os}}
\def\bfd{{\bf\Delta}}
\def\tbfd{{\bf\tilde\Delta}}
\def\td{{\tilde\Delta}}

\def\bfdmsb{{\bf\Delta}_{\sss\msb}}
\def\V{{\bf V}}
\def\vtmsb{\V_{3,{\sss\msb}}}
\def\tph{{\widetilde{\varphi}}}
\def\ra{\rangle}
\def\la{\langle} 

\def\d{\partial} 
\def\ph{\varphi} 
\def\Lam{\Lambda} 
\def\lam{\lambda}

\def\sign{\mathop{\hbox{sign}}}

\def\D{{\cal D}}
\def\H{{\cal H}}
\def\L{{\cal L}}
\def\O{{\cal O}}
\def\P{{\cal P}}
\def\T{{\cal T}}
\def\tr{{\rm Tr}}

\def\Tfi{{\cal T}_{f\leftarrow i}}

\def\p{{\bf p}} 
\def\x{{\bf x}} 
\def\y{{\bf y}} 
\def\k{{\bf k}}
\def\ph{\varphi}
\def\phd{\varphi^\dagger}

\def\ad{a^\dagger}
\def\bd{b^\dagger}

\def\adt{{\widetilde a}^\dagger}
\def\at{{\widetilde a}}
\def\dk{\widetilde{dk}}

\def\dtx{d^{3\!}x}

\def\dtk{d^{3\!}k}
\def\dtp{d^{3\!}p}

\def\dfx{d^{4\!}x}
\def\dfy{d^{4\!}y}
\def\dfz{d^{4\!}z}

\def\dfk{d^{4\!}k}

\def\dsx{d^{6\!}x}
\def\dsy{d^{6\!}y}
\def\dsk{d^{6\!}k}

\def\dsq{d^{6\!}q}
\def\dsxl{d^{6\!}\ell}
\def\ddx{d^{d\!}x}

\def\ddq{d^{d\!}q}
\def\ddl{d^{d\!}\ell}
\def\qbar{{\bar q}}

\def\ddqb{d^{d\!}{\bar q}}
\def\half{{\textstyle{1\over2}}}
\def\w{\omega}
\def\mut{{\tilde\mu}}
\def\psl{{\rlap{\kern1pt /}{p}}}
\def\pslp{{\rlap{\kern1pt /}{p}}\kern1pt{}'}

\def\Asl{{\rlap{\kern2.25pt /}{A}}}

\def\Dsl{{\rlap{\kern2.25pt /}{D}}}
\def\Asl{{\rlap{\kern2.25pt /}{A}}}
\def\dsl{{\rlap{\kern0.5pt /}{\partial}}}
\def\dslx{{\rlap{\kern0.5pt /}{\partial_x}}}
\def\dsly{{\rlap{\kern0.5pt /}{\partial_y}}}
\def\ksl{{\rlap{\kern0.5pt /}{k}}}
\def\kslp{{\rlap{\kern0.5pt /}{k}}{}'}
\def\lsl{{\rlap{\kern-0.5pt /}{\ell}}}

\def\eslp{{\rlap{/}{\varepsilon}}\kern1pt{}'}

\def\cd{\!\cdot\!}

\def\ubarp{\overline{u}\kern1pt{}'}

\def\vbarp{\overline{v}\kern1pt{}'}

\def\Psibar{\overline{\Psi}}
\def\Psibart{\Psibar^{\kern1pt\lower1pt\hbox{$\rm\sss T$}}}

\def\LK{Lehmann-K\"all\'en\ }
\def\Im{\mathop{\rm Im}}
\def\Re{\mathop{\rm Re}}

%% file: front0.tex
\rightline{$\phantom{X}$}

\vskip0.5in

\begin{center}
\large
Quantum Field Theory
\vskip0.25in
Part I: Spin Zero
\vskip0.5in
Mark Srednicki
\end{center}
\vskip0.1in
\begin{center}
Department of Physics

University of California

Santa Barbara, CA 93106
\vskip0.1in
mark@physics.ucsb.edu
\end{center}

\vskip0.5in

\baselineskip=16pt

\noindent
This is a draft version of Part I of a three-part 
introductory textbook on quantum field theory. 

\vskip0.2in

\vfill\eject

\begin{center}
\large{Part I: Spin Zero}
\end{center}

0) Preface

1) Attempts at Relativistic Quantum Mechanics 
(prerequisite: none)

2) Lorentz Invariance
(prerequisite: 1)

3) Relativistic Quantum Fields and Canonical Quantization
(2)

4) The Spin-Statistics Theorem 
(3)

5) The LSZ Reduction Formula
(3)

6) Path Integrals in Quantum Mechanics
(none)

7) The Path Integral for the Harmonic Oscillator
(6)

8) The Path Integral for Free Field Theory
(3, 7)

9) The Path Integral for Interacting Field Theory
(8)

10) Scattering Amplitudes and the Feynman Rules
(5, 9)

11) Cross Sections and Decay Rates
(10)

12) The \LK Form of the Exact Propagator
(9)

13) Dimensional Analysis with $\hbar=c=1$
(3)

14) Loop Corrections to the Propagator
(10, 12, 13)

15) The One-Loop Correction in \LK Form
(14)

16) Loop Corrections to the Vertex
(14)

17) Other 1PI Vertices
(16)

18) Higher-Order Corrections and Renormalizability
(17)

19) Perturbation Theory to All Orders: the Skeleton Expansion
(18)

20) Two-Particle Elastic Scattering at One Loop
(10, 19)

21) The Quantum Action
(19)

22) Continuous Symmetries and Conserved Currents  
(8)

23) Discrete Symmetries: $P$, $C$, $T$, and $Z$ 
(22)

24) Unstable Particles and Resonances 
(10, 14)

25) Infrared Divergences
(20)

26) Other Renormalization Schemes
(25)

27) Formal Development of the Renormalization Group
(26)

28) Nonrenormalizable Theories and Effective Field Theory
(26)

29) Spontaneous Symmetry Breaking 
(3)

30) Spontaneous Symmetry Breaking and Loop Corrections
(19, 29)

31) Spontaneous Breakdown of Continuous Symmetries 
(29)

32) Nonabelian Symmetries
(31)

\vfill\eject

%% file: ch000.tex

\noindent Quantum Field Theory  \hfill   Mark Srednicki

\vskip0.1in

\begin{center}
\large{0: Preface}
\end{center}

\vskip0.1in

This is a draft version of Part I of a three-part introductory textbook
on quantum field theory.  The book is based on a one-year course that
I have taught off-and-on for the past 20 years.  My goal is to present the
basic concepts and formalism of QFT as simply and straightforwardly as
possible, emphasizing its logical structure.  To this end,
I have tried to take the
simplest possible example of each phenomenon, and 
then to work through it in great detail.    

In this part, Spin Zero, the primary example is $\ph^3$
theory in six spacetime dimensions.  While this theory
has no apparent relevance to the real world, it has many advantages 
as a pedagogical case study.   For example, 
unlike $\ph^4$ theory, $\ph^3$ theory
has nontrivial renormalization of the propagator at one loop.
In six dimensions, the $\ph^3$ coupling is dimensionless,
and asymptotically free.
The theory is simple enough to allow a complete calculation of the
one-loop $\ph\ph\to\ph\ph$ scattering amplitude,
renormalized in an on-shell scheme.
This amplitude is singular in the limit of zero particle mass,
and this leads inexorably to an examination of infrared
divergences, alternative renormalization schemes, and the renormalization
group.

I have tried to make this book user friendly,
both for students and instructors.
Each of the three main parts is divided into numerous sections; each section
treats a single topic or idea, and is as self-contained as possible.
An equation is rarely referenced outside
of the section in which it appears; if an earlier formula is needed, 
it is repeated.
The book's table of contents includes, for each section,
a list of the sections that serve as the immediate prerequisites.
This allows instructors to reorder the presentation to fit individual
preference, and students to access topics of interest quickly.

Eventually, there will be more problems, and solutions to most of them.

Any comments, suggestions, corections, etc, would be very welcome.
Please send them to mark@physics.ucsb.edu.

Please note: {\it there are no references}.  
For thorough guides to the
literature, see, for example, Weinberg's {\it The Quantum Theory of Fields},
Peskin \& Schroeder's {\it Introduction to Quantum Field Theory},
or Siegel's {\it Fields}.

\vfill\eject

%% file: ch001.tex
\noindent Quantum Field Theory  \hfill   Mark Srednicki

\vskip0.5in

\begin{center}
\large{1: Attempts at relativistic quantum mechanics}
\end{center}

\vskip0.5in

In order to combine quantum mechanics and relativity, we must first
understand what we mean by ``quantum mechanics'' and ``relativity''.
Let us begin with quantum mechanics.

Somewhere in most textbooks on the subject, one can find a list of the ``axioms
of quantum mechanics''.  These include statements along the lines of

{\it The state of the system is represented by a vector in Hilbert space.

Observables are represented by hermitian operators.

The measurement of an observable always yields one of its eigenvalues as 
the result.} 

And so on.  We do not need to review these closely here.
The axiom we need to focus on is the one that says that the time evolution
of the state of the system is governed by the Schr\"odinger equation,
\begin{equation}
i\hbar{\d\over\d t}|\psi,t\ra = H|\psi,t\ra\; ,
\label{schro}
\end{equation}
where $H$ is the hamiltonian operator, representing the total energy.

Let us consider a very simple system: a spinless, nonrelativistic particle
with no forces acting on it.  In this case, the hamiltonian is
\begin{equation}
H={1\over 2m}{\bf P}^2\;,
\label{h00}
\end{equation}
where $m$ is the particle's mass, and $\bf P$ is the momentum operator.
In the position basis, \eq{schro} becomes
\begin{equation}
i\hbar{\d\over\d t}\psi({\bf x},t)=-{\hbar^2\over2m}\nabla^2\psi({\bf x},t) \;,
\label{schro2}
\end{equation}
where $\psi({\bf x},t)=\la{\bf x}|\psi,t\ra$ 
is the position-space wave function.
We would like to generalize this to relativistic motion.

The obvious way to proceed is to take
\begin{equation}
H=+\sqrt{{\bf P}^2 c^2 + m^2 c^4}\;,
\label{h2}
\end{equation}
which gives the correct energy-momentum relation.  If we formally expand
this hamiltonian in inverse powers of the speed of light $c$, we get
\begin{equation}
H=mc^2 + {1\over 2m}{\bf P}^2 + \ldots\;.
\label{h3}
\end{equation}
This is simply a constant (the rest energy), plus the usual nonrelativistic
hamiltonian, \eq{h00}, plus higher-order corrections.  With the hamiltonian
given by \eq{h2}, the Schr\"odinger equation becomes
\begin{equation}
i\hbar{\d\over\d t}\psi({\bf x},t)
 = +\sqrt{-\hbar^2c^2\nabla^2+m^2c^4}\;\psi({\bf x},t) \;.
\label{schro3}
\end{equation}
The square root of the differential operator looks nasty,
but it is perfectly well defined in momentum space.  Have we, then,
succeeded in constructing a relativistic form of quantum mechanics?

To answer this, we must first examine what we mean by ``relativity''.
Special relativity tells us that physics {\it looks the same\/} in all
{\it inertial frames}.  To explain what {\it this\/} means, let us first suppose 
that a certain spacetime coordinate system $(ct,{\bf x})$ represents (by fiat) 
an inertial frame.  Let us define $x^0=ct$, and write $x^\mu$, where $\mu=0,1,2,3$,
in place of $(ct,{\bf x})$.  It is also convenient (for reasons not at all
obvious at this point) to define $x_0=-x^0$ and $x_i=x^i$, where $i=1,2,3$.
More succinctly, we can introduce the {\it metric of flat spacetime},
\begin{equation}
g_{\mu\nu}=\pmatrix{ -1 & & & \cr
                     & +1 & & \cr
                     & & +1 & \cr
                     & & & +1 \cr} .
\label{gmunu}
\end{equation}
We then have $x_\mu = g_{\mu\nu}x^\nu$, where a repeated index is summed.
To invert this formula, we need
to introduce the inverse of $g$, which is confusingly also called $g$, except
with both indices up:
\begin{equation}
g^{\mu\nu}=\pmatrix{ -1 & & & \cr
                     & +1 & & \cr
                     & & +1 & \cr
                     & & & +1 \cr} .
\label{ginv}
\end{equation}
We then have $g^{\mu\nu}g_{\nu\rho}= \delta^\mu{}_\rho$, where
$\delta^\mu{}_\rho$ is the Kronecker delta (equal to one if its two indices
take on the same value, zero otherwise).  Now we can also write
$x^\mu = g^{\mu\nu}x_\nu$.  It is a general rule that, for
any pair of repeated indices, one must be a superscript, and one
must be a subscript.  Also, unsummed indices must match (in both
name and height) on the left and right sides of any valid equation.  

Now we are ready to specify what we mean by an inertial frame.  
If the coordinates
$x^\mu$ represent an inertial frame (which they do, by assumption), then so
do any other coordinates $\bar x^\mu$ that are related by
\begin{equation}
\bar x^\mu = \Lam^\mu{}_\nu x^\nu + a^\mu\;,
\label{lam}
\end{equation}
where $\Lam^\mu{}_\nu$ is a {\it Lorentz transformation matrix\/} 
and $a^\mu$ is a {\it translation vector}.  Both 
$\Lam^\mu{}_\nu$ and $a^\mu$ are constant (that is,
independent of $x^\mu$).  Furthermore, $\Lam^\mu{}_\nu$ must obey
\begin{equation}
g_{\mu\nu}\Lam^\mu{}_\rho \Lam^\nu{}_\sigma  = g_{\rho\sigma} \; .
\label{lam2}
\end{equation}
\Eq{lam2} ensures that the {\it invariant squared distance\/} between two
different spacetime points that are 
labeled by $x^\mu$ and $x^{\prime\mu}$ in one inertial frame, and by 
$\bar x^\mu$ and $\bar x^{\prime\mu}$ in another, is the same.  
This squared distance is defined to be
\begin{equation}
(x-x')^2 \equiv g_{\mu\nu}(x-x')^\mu (x-x')^\nu
= (\x-\x')^2-c^2 (t-t')^2 \;.
\label{x2}
\end{equation}
In the second frame, we have
\begin{eqnarray}
(\bar x-\bar x')^2 
&=& g_{\mu\nu}(\bar x-\bar x')^\mu (\bar x-\bar x')^\nu 
\nonumber \\
&=& g_{\mu\nu}\Lam^\mu{}_\rho \Lam^\nu{}_\sigma (x-x')^\rho (x-x')^\sigma
\nonumber \\
&=& g_{\rho\sigma} (x-x')^\rho (x-x')^\sigma
\nonumber \\
&=& (x-x')^2 \;, 
\label{x3}
\end{eqnarray}
as desired.

When we say that {\it physics looks the same}, we mean that two observers
(Alice and Bob, say)
using two different sets of coordinates (representing two different inertial frames)
should agree on the predicted results of all possible experiments.  In the case
of quantum mechanics, this requires Alice and Bob to agree on the value of the
wave function at a particular spacetime point, a point which is called $x$ by 
Alice and $\bar x$ by Bob.  
Thus if Alice's predicted wave function is $\psi(x)$, and Bob's
is $\bar\psi(\bar x)$, then $\psi(x)=\bar\psi(\bar x)$.
Therefore, the two functions $\psi(x)$ and $\bar\psi(x)$ are related by
\begin{equation}
\bar\psi(x)=\psi(\Lam^{-1}(x-a)) \;,
\label{psibar}
\end{equation}
where $(\Lam^{-1})^\mu{}_\nu = \Lam_\nu\vphantom{\Lam}^\mu$.
[This formula for $\Lam^{-1}$ follows from \eq{lam2}; see section 2.]
Furthermore,
in order to maintain $\psi(x)=\bar\psi(\bar x)$ throughout spacetime,
$\psi(x)$ and $\bar\psi(\bar x)$ must obey the {\it same\/} equation of motion.

Let us see if this is true of our candidate Schr\"odinger
equation for relativistic quantum mechanics, \eq{schro3}.
First let us define some useful notation for spacetime derivatives:
\begin{equation}
\d_\mu \equiv {\d\over\d x^\mu} = \left(+{1\over c}{\d\over\d t},\nabla\right) ,
\label{d}
\end{equation}
\begin{equation}
\d^\mu \equiv {\d\over\d x_\mu} = \left(-{1\over c}{\d\over\d t},\nabla\right).
\label{d1}
\end{equation}
Note that $\d^\mu x^\nu=g^{\mu\nu}$, so that our matching-index-height rule
is satisfied.  Also,
$\bar\d{}^\mu=\Lam^\mu{}_\nu\d^\nu$, which means that $\d^\mu$ transforms
in the same way as $x^\mu$.  To verify this, note that
\begin{equation}
g^{\rho\sigma}
= \bar\d{}^\rho \bar x{}^\sigma
= (\Lam^\rho{}_\mu\d^\mu)(\Lam^\sigma{}_\nu x^\nu + a^\mu)
= \Lam^\rho{}_\mu\Lam^\sigma{}_\nu(\d^\mu x^\nu)
= \Lam^\rho{}_\mu\Lam^\sigma{}_\nu g^{\mu\nu} \;,
\label{d2}
\end{equation}
and that $g^{\rho\sigma} = \Lam^\rho{}_\mu\Lam^\sigma{}_\nu g^{\mu\nu}$
is just another form of \eq{lam2} [again, see section 2].

Rewriting \eq{schro3} in this notation yields
\begin{equation}
i\hbar c\,\d_0\psi(x)
 = +\sqrt{-\hbar^2c^2{\textstyle\sum_{i=1}^3}\d_i^2+m^2c^4}\;\psi(x) \;.
\label{schro4}
\end{equation}
We should also have
\begin{equation}
i\hbar c\,\bar\d_0\bar\psi(\bar x)
 = +\sqrt{-\hbar^2c^2{\textstyle\sum_{i=1}^3}\bar\d_i^2
          +m^2c^4}\;\bar\psi(\bar x) \;.
\label{schro5}
\end{equation}
Now replace $\bar\psi(\bar x)$ with $\psi(x)$,
and use $\bar\d_\mu=\Lam_\mu{}^\nu\d_\nu$ to get
\begin{equation}
i\hbar c\,\Lam_0{}^\nu\d_\nu\psi(x)
 = +\sqrt{-\hbar^2c^2{\textstyle\sum_{i=1}^3}(\Lam_i{}^\nu\d_\nu)^2
          +m^2c^4}\;\psi(x) \;.
\label{schro6}
\end{equation}
It is obvious that \eq{schro6} does not have the same form as \eq{schro4}.
There is a {\it preferred frame\/} where the derivative that is not under
the square-root sign is purely a time derivative.  In any other
inertial frame, it is a linear combination of time and space derivatives. 
Thus, our candidate Schr\"odinger equation is {\it not\/} 
consistent with relativity.  

So, let's try something else: square the differential operators
on each side of \eq{schro4} before applying them to the wave function.  
Then we get
\begin{equation}
-\hbar^2c^2 \d^2_0\psi(x)
 = (-\hbar^2c^2\nabla^2+m^2c^4)\psi(x) \;.
\label{kg1}
\end{equation}
After rearranging and identifying $\d^2\equiv\d^\mu\d_\mu=-\d_0^2+\nabla^2$,
\begin{equation}
(-\d^2+m^2c^2/\hbar^2)\psi(x)=0 \;.
\label{kg2}
\end{equation}
This is the {\it Klein-Gordon equation}.  

To see if it is consistent with relativity,
start with Bob's version of the equation,
$(-\bar\d^2 + m^2 c^2/\hbar^2)\bar\psi(\bar x)$.
Replace $\bar\psi(\bar x)$ with $\psi(x)$, and note that
\begin{equation}
\bar\d^2=g_{\mu\nu}\bar\d^\mu\bar\d^\nu 
=g_{\mu\nu}\Lam^\mu{}_\rho\Lam^\mu{}_\sigma\d^\rho\d^\sigma=\d^2\;.
\label{dsq}
\end{equation}
Thus, we get \eq{kg2} back again!
This means that the Klein-Gordon equation {\it is\/} 
consistent with relativity:
it takes the same form in every inertial frame.

This is the good news.  The bad news is that we have violated one of the axioms
of quantum mechanics: 
\eq{schro}, the Schr\"odinger equation in its abstract form.
The abstract Schr\"odinger equation has the fundamental property of 
being first order in the time 
derivative, whereas the Klein-Gordon equation is second order.  
This may not seem too important, but in fact it has drastic 
consequences.  One of these is that the norm of a state,
\begin{equation}
\la\psi,t|\psi,t\ra 
= \int \dtx\, \la\psi,t|{\bf x}\ra\la{\bf x}|\psi,t\ra 
= \int \dtx\,\psi^*(x)\psi(x), 
\label{norm0}
\end{equation}
is not in general time independent.  
Thus probability is not conserved.
The Klein-Gordon equation obeys relativity, but not quantum mechanics.

Dirac attempted to solve this problem (for spin-one-half particles)
by introducing an extra discrete label
on the wave function, to account for spin: $\psi_a(x)$, $a=1,2$.  
He then tried a Schr\"odinger equation
of the form
\begin{equation}
i\hbar{\d\over\d t}\psi_a(x) 
=\Bigl(-i\hbar c(\alpha^j)_{ab}\d_j + mc^2(\beta)_{ab}\Bigr)\psi_b(x) \;,
\label{dirac}
\end{equation}
where all repeated indices are summed, and $\alpha^j$ and $\beta$ are
matrices in spin-space.  This equation, the {\it Dirac equation},
is linear in both time {\it and\/} space derivatives, and so has
a chance to obey the rules of both quantum mechanics and 
relativity.  In fact, it manifestly obeys the rules of quantum mechanics,
and so the norm $\sum_{a=1,2}\int \dtx\,|\psi_a(x)|^2$ is time independent,
as it should be.  To gain some understanding of its properties under
Lorentz transformations, let us consider the hamiltonian it specifies,  
\begin{equation}
H_{ab} = c P_j (\alpha^j)_{ab} + m c^2(\beta)_{ab} \; ,
\label{dirach}
\end{equation}
where $P_j$ is a component of the momentum operator.
If we square this hamiltonian, we get
\begin{equation}
(H^2)_{ab} = c^2 P_j P_k (\alpha^j\alpha^k)_{ab}  
+ m c^3 P_j (\alpha^j\beta+\beta\alpha^j)_{ab} + (m c^2)^2 (\beta^2)_{ab}\;.
\label{dirach2}
\end{equation}
Since $P_j P_k$ is symmetric on exchange of $j$ and $k$, we can replace
$\alpha^j\alpha^k$ by its symmetric part, $\half\{\alpha^j,\alpha^k\}$,
where $\{A,B\}=AB+BA$ is the anticommutator.  Then, if we choose matrices 
such that
\begin{equation}
\{\alpha^j,\alpha^k\}_{ab}=2\delta^{jk}\delta_{ab}\;,\quad
\{\alpha^j,\beta\}_{ab}=0\;,\quad
(\beta^2)_{ab}=\delta_{ab}\;,
\label{commab}
\end{equation}
we will get
\begin{equation}
(H^2)_{ab} =  ({\bf P}^2 c^2 + m^2 c^4)\delta_{ab} \;.
\label{dirach3}
\end{equation}
Thus, the eigenstates of $H^2$ are momentum eigenstates, with
$H^2$ eigenvalue ${\bf p}^2c^2 + m^2 c^4$.  This is, of course,
the correct relativistic energy-momentum relation.
While it is outside the scope of this section to demonstrate it,
it turns out that the Dirac equation is fully consistent with 
relativity provided the Dirac matrices obey \eq{commab}.
So we have succeeded in constructing a quantum mechanical, 
relativistic theory!

There are, however, some problems.  First, the Dirac matrices
must be at least $4\times4$, and not $2\times2$ as we would like
(in order to account for electron spin).  To see this, note that the
$2\times2$ Pauli matrices obey $\{\sigma^i,\sigma^j\}=2\delta^{ij}$, and
are thus candidates for the Dirac $\alpha^i$ matrices.  However,
there is no fourth matrix that anticommutes with these three
(easily proven by writing down the most general $2\times2$ matrix
and working out the three anticommutators explicitly).
Also, the Dirac
matrices must be even dimensional.  To see this, first define the matrix 
$\gamma \equiv \beta\alpha^1\alpha^2\alpha^3$.  This matrix obeys $\gamma^2=1$
and also $\{\gamma,\alpha^i\}=\{\gamma,\beta\}=0$.
Hence, using the cyclic property of matrix traces on $\gamma\beta\gamma$, we have
$\mathop{\rm Tr}\gamma\beta\gamma 
=\mathop{\rm Tr}\gamma^2\beta 
=\mathop{\rm Tr}\beta$.  
On the other hand, using $\beta\gamma=-\gamma\beta$, we also have
$\mathop{\rm Tr}\gamma\beta\gamma
=-\mathop{\rm Tr}\gamma^2\beta
=-\mathop{\rm Tr}\beta$.  
Thus, $\mathop{\rm Tr}\beta$ is equal to minus itself, and hence
must be zero.  (Similarly, we can show $\mathop{\rm Tr}\alpha^i=0$.)
Also, $\beta^2=1$ implies that the eigenvalues of $\beta$
are all $\pm 1$.  
Because $\beta$ has zero trace, these eigenvalues must sum to zero,
and hence the dimension of the matrix must be even.
Thus the Dirac matrices must be at least $4\times4$, and
it remains for us to interpret the two extra possible ``spin'' states.

However, these extra states cause a more severe problem than a mere overcounting.
Acting on a momentum eigenstate, $H$ becomes the
matrix $c\,\alpha\!\cdot\!\p + mc^2\beta$.  The trace
of this matrix is zero.  Thus the four eigenvalues
must be $+E(\p),+E(\p),-E(\p),-E(\p)$, where
$E(\p)=+(\p^2c^2+m^2c^4)^{1/2}$.
The negative eigenvalues are the problem:
they indicate that there is no ground state.
In a more elaborate theory that included interactions
with photons, there seems to be no reason why a
positive energy electron could not emit a photon
and drop down into a negative energy state.
This downward cascade could continue forever.
(The same problem also arises in attempts to interpret
the Klein-Gordon equation as a modified form of quantum mechanics.)

Dirac made a wildly brilliant attempt to fix this problem of
negative energy states.  His solution is based on an empirical
fact about electrons: they obey the Pauli exclusion principle.  
It is impossible to put more than one of them in
the same quantum state.  What if, Dirac speculated, all the negative
energy states were {\it already occupied\/}?  In this case, a 
positive energy electron could not drop into one of these states,
by Pauli exclusion!

Many questions immediately arise.  Why don't we see the negative
electric charge of this {\it Dirac sea\/} of electrons?  
Dirac's answer: because we're used to it.  
(More precisely, the physical effects of a uniform
charge density depend on the boundary conditions at infinity that we
impose on Maxwell's equations, and there is a choice that renders such a 
uniform charge density invisible.)  However, Dirac noted, if one of 
these negative energy electrons were excited into a positive energy state 
(by, say, a sufficiently energetic photon), it would leave
behind a {\it hole\/} in the sea of negative energy electrons.
This hole would appear to have positive charge, and positive energy.
Dirac therefore predicted (in 1927) the existence of the {\it positron},
a particle with the same mass as the electron, but opposite charge.  
The positron was found experimentally five years later.
  
However, we have now jumped from an attempt at a quantum description
of a {\it single\/} relativistic particle to a theory that apparently
requires an {\it infinite\/} number of particles.  Even if we accept this,
we still have not solved the problem of how to describe particles like
photons or pions or alpha nuclei that do {\it not\/} obey Pauli exclusion.

At this point, it is worthwhile to stop and reflect on why it has proven to
be so hard to find an acceptable relativistic wave equation for a 
single quantum particle.  Perhaps there is something wrong with our basic approach.

And there is.  Recall the axiom of quantum mechanics that says that
``Observables are represented by hermitian operators.''  This is not
entirely true.  There is one observable in quantum mechanics that is
{\it not\/} represented by a hermitian operator: time.
Time enters into quantum mechanics only when we announce that the ``state of the
system'' depends on an extra parameter $t$.  This parameter
is not the eigenvalue of any operator.  This is in sharp contrast
to the particle's position $\bf x$, which {\it is\/} the eigenvalue
of an operator.  Thus, space and time
are treated very differently, a fact that is obscured
by writing the Schr\"odinger equation in terms of the 
position-space wave function $\psi({\bf x},t)$.  Since space and time
are treated asymmetrically, it is not surprising
that we are having trouble incorporating a symmetry
which mixes them up.  

So, what are we to do?

In principle, the problem could be an intractable one:
it might be {\it impossible\/} to combine quantum mechanics
and relativity.  In this case, there would have to be
some meta-theory, one that reduces in the nonrelativistic limit
to quantum mechanics, and in the classical limit to relativistic
particle dynamics, but is actually neither.  

This, however, turns out not to be the case.  We can solve our
problem, but we must put space and time on an equal footing at the outset.
There are two ways to do this.  One is to demote position from its
status as an operator, and render it as an extra label, like time.  
The other is to promote time to an operator.

Let us discuss the second option first.  If time becomes an operator,
what replaces the Schr\"odinger equation?  Luckily, in relativistic
theories, there are plenty of times lying around.  We can use the {\it proper
time\/} $\tau$ of the particle (or, more technically, any affine parameter along
its worldline) as the ``extra'' parameter, while promoting coordinate time $T$
to an operator.  In the Heisenberg picture (where the state of the system
is fixed, but the operators are functions of time that obey the classical
equations of motion), we would have operators $X^\mu(\tau)$, where
$X^0=T$.  Relativistic quantum mechanics can indeed be developed along these lines,
but it is surprisingly complicated to do so.  (The many times are the problem;
any monotonic function of $\tau$ is just as good a candidate as $\tau$ itself
for the proper time, and this infinite redundancy of descriptions must be
understood and accounted for.)

One of the advantages of considering different formalisms is that they may
suggest different directions for generalizations.  For example, once we
have $X^\mu(\tau)$, why not consider adding some more parameters?  Then we would
have, for example, $X^\mu(\sigma,\tau)$.  Classically, 
this would give us a continuous
family of worldlines, what we might call a {\it worldsheet}, and so 
$X^\mu(\sigma,\tau)$ would describe a propagating {\it string}.  
This is indeed the starting point for string theory.

Thus, promoting time to an operator is a viable option, but is 
complicated in practice.
Let us then turn to the other option, demoting position to a label.  The first
question is, label on what?  The answer is, {\it on operators}.  Thus,
consider assigning an operator to each point $\bf x$ in space; 
call these operators $\ph(\x)$.  
A set of operators like this is called a {\it quantum field}.  In the
Heisenberg picture, the operators are also time dependent:
\begin{equation}
\ph(\x,t) = e^{iHt/\hbar} \ph(\x,0) e^{-iHt/\hbar}\,.
\label{hpic}
\end{equation}
Thus, both position and (in the Heisenberg picture) time are now labels on
operators; neither is itself the eigenvalue of an operator. 

So, now we have two different approaches to relativistic quantum theory,
approaches that might, in principle, yield different results.
This, however, is not the case: it turns out that
any relativistic quantum physics
that can be treated in one formalism can also be treated in the other.
Which we use is a matter of convenience and taste.  And, 
{\it quantum field theory}, 
the formalism in which position and time are both labels on operators,
is by far the more convenient and efficient.  
(For particles, anyway; for strings,
the opposite seems to be true, at least as of this writing.)

There is another useful equivalence: ordinary nonrelativistic quantum mechanics,
for a fixed number of particles, can be rewritten as a quantum field theory.
This is an informative exercise, since the corresponding physics is already
familiar.  Let us carry it out.

Begin with the position-basis Schr\"odinger equation for $n$ particles,
all with the same mass $m$, moving in an external potential $U(\x)$, 
and interacting with each other via an interparticle potential $V(\x_1-\x_2)$:
\begin{equation}
i\hbar{\d\over\d t}\psi
=\Biggl[\;\sum_{j=1}^n\left(-{\hbar^2\over 2m}\nabla^2_j + U(\x_j)\right)
+\sum_{j=1}^n\sum_{k=1}^{j-1}V(\x_j-\x_k)\Biggr]\psi\;,
\label{mschro}
\end{equation}
where $\psi=\psi(\x_1,\ldots,\x_n;t)$ is the position-space wave function.
The quantum mechanics of this system 
can be rewritten in the abstract form of \eq{schro} by first introducing
(in, for now, the Schr\"odinger picture) a quantum field $a(\x)$ and its
hermitian conjugate $\ad(\x)$.  We take these operators to have the
commutation relations
\begin{equation}
[a(\x),a(\x')]=0 \;, \quad 
[\ad(\x),\ad(\x')]=0 \;, \quad 
[a(\x),\ad(\x')]=\delta^3(\x-\x')\;, 
\label{comm}
\end{equation}
where $\delta^3(\x)$ is the three-dimensional Dirac delta function.
Thus, $\ad(\x)$ and $a(\x)$ behave like harmonic-oscillator
creation and annihilation operators that are labeled by a continuous index.
In terms of them, we introduce the hamiltonian operator of our quantum field theory,
\begin{eqnarray}
H &=& \int \dtx\;\ad(\x)\!\left(-{\textstyle{\hbar^2\over2m}\nabla^2}
                                  +U(\x)\right)\!a(\x) 
\nonumber \\
&&{} + \half\int \dtx\,d^3y\;V(\x-\y) \ad(\x)\ad(\y)a(\y)a(\x)\;. 
\label{ham} 
\end{eqnarray}
Now consider a time-dependent state of the form
\begin{equation}
|\psi,t\rangle = \int \dtx_1\ldots \dtx_n\;\psi(\x_1,\ldots,\x_n;t)
\ad(\x_1)\ldots\ad(\x_n)|0\rangle\;,
\label{state}
\end{equation}
where $\psi(\x_1,\ldots,\x_n;t)$ is some function of the $n$ particle positions
and time, and $|0\ra$ is the {\it vacuum state\/}, the state that is annihilated
by all the $a$'s: $a(\x)|0\ra=0$.  It is now straightforward (though tedious)
to verify that the abstract Schr\"odinger equation, \eq{schro}, is obeyed
if and only if the function $\psi$ satisfies \eq{mschro}.

Thus we can interpret the state $|0\ra$ as a state of ``no particles'', the
state $\ad(\x_1)|0\ra$ as a state with one particle at position $\x_1$, the
state $\ad(\x_1)\ad(\x_2)|0\ra$ as a state with one particle at position $\x_1$
and another at position $\x_2$, and so on.  The operator
\begin{equation}
N = \int \dtx\;\ad(\x)a(\x)
\label{num}
\end{equation}
counts the total number of particles.  It commutes with the hamiltonian, as
is easily checked; thus, if we start with a state of $n$ particles, we remain
with a state of $n$ particles at all times.

However, we can imagine generalizations of this version of the theory
(generalizations that would not be possible without the field formalism)
in which the number of particles is {\it not\/} conserved.  For example,
we could try adding to $H$ a term like
\begin{equation}
\Delta H \propto \int \dtx\left[\ad(\x)a^2(\x) + {\rm h.c.}\right]\;.
\label{dh}
\end{equation}
This term does {\it not\/} commute with $N$, and so the number of particles
would not be conserved with this addition to $H$.

Theories in which the number of particles can change as time evolves are a good
thing: they are needed for correct phenomenology.  We are already familiar with
the notion that atoms and molecules can emit and absorb photons,
and so we had better have a formalism that can incorporate this
phenomenon.  We are less familiar with emission and absorption (that is to say,
creation and annihilation) of electrons, but this process also occurs in
nature; it is less common because it must be accompanied by the emission or
absorption of a positron, antiparticle to the electron.  There are not a lot
of positrons around to facilitate electron annihilation, while $e^+ e^-$ pair
creation requires us to have on hand at least $2mc^2$ of energy available
for the rest-mass energy of these two particles.  The photon, on the
other hand, is its own antiparticle, and has zero rest mass; thus photons are
easily and copiously produced and destroyed.

There is another important aspect of the quantum theory specified by
\eqs{ham} and (\ref{state}).  Because the creation operators commute
with each other, only the completely symmetric part of $\psi$ survives
the integration in \eq{state}.  Therefore, without loss of generality,
we can restrict our attention to $\psi$'s of this type:
\begin{equation}
\psi(\ldots \x_i \ldots \x_j \ldots;t)
= +\psi(\ldots \x_j \ldots \x_i \ldots;t) \;.
\label{bose}
\end{equation}
This means that we have a theory of {\it bosons}, particles that (like
photons or pions or alpha nuclei) obey Bose-Einstein statistics.  
If we want Fermi-Dirac statistics instead, we must replace \eq{comm} with
\begin{equation}
\{a(\x),a(\x')\}=0 \;, \quad 
\{\ad(\x),\ad(\x')\}=0 \;, \quad 
\{a(\x),\ad(\x')\}=\delta^3(\x-\x')\;, 
\label{acomm}
\end{equation}
where again $\{A,B\}=AB+BA$ is the anticommutator.  Now only the fully
antisymmetric part of $\psi$ survives the integration in \eq{state}, and so
we can restrict our attention to 
\begin{equation}
\psi(\ldots \x_i \ldots \x_j \ldots;t)
= -\psi(\ldots \x_j \ldots \x_i \ldots;t) \;.
\label{fermi}
\end{equation}
Thus we have a theory of {\it fermions}.
It is straightforward to check that the abstract Schr\"odinger equation,
\eq{schro}, still implies that $\psi$ obeys the 
differential equation (\ref{mschro}).
[Now, however, the ordering of the last two $a$ operators in the last term of $H$,
\eq{ham}, becomes important, and it must be as written.]
Interestingly, there is no simple way to write down a quantum field theory
with particles that obey Boltzmann statistics, corresponding to a wave function
with no particular symmetry.  This is a hint of the {\it spin-statistics\/}
theorem, which applies to {\it relativistic\/} quantum field theory.  It
says that {\it interacting\/} particles with integer spin
must be bosons, and {\it interacting\/} particles with half-integer spin
must be fermions.  
In our {\it nonrelativistic\/} example, the interacting particles clearly
have spin zero (because their creation operators carry no labels that could
be interpreted as corresponding to different spin states), but can be either
bosons or fermions, as we have seen.

Now that we have seen how to rewrite the nonrelativistic quantum mechanics
of multiple bosons or fermions as a quantum field theory, it is time to try
to construct a relativistic version.

\vskip0.5in

\begin{center}
Problems
\end{center}

\vskip0.25in

1.1) Show that the state defined in \eq{state} obeys the abstract Schr\"odinger
equation, \eq{schro}, with the hamiltonian of \eq{ham}, if and only if
the wave function obeys \eq{mschro}.  Your demonstration should apply both
to the case of bosons, where the particle creation and annihilation operators
obey the commutation relations of \eq{comm}, and to fermions,
where the particle creation and annihilation operators
obey the anticommutation relations of \eq{acomm}.

1.2) Show explicitly that $[N,H]=0$, where $H$ is given by \eq{ham}
and $N$ by \eq{num}.

\vfill\eject

%% file: ch002.tex
\noindent Quantum Field Theory  \hfill   Mark Srednicki

\vskip0.5in

\begin{center}
\large{2: Lorentz Invariance}
\end{center}
\begin{center}
Prerequisite: 1
\end{center}

\vskip0.5in

A {\it Lorentz transformation\/} is a linear, homogeneous change of coordinates
from $x^\mu$ to $\bar x^\mu$, 
\begin{equation}
\bar x^\mu = \Lam^\mu{}_\nu x^\nu \;,
\label{lor}
\end{equation}
that preserves the {\it invariant squared distance\/} from the origin, 
$x^2=x^\mu x_\mu=g_{\mu\nu}x^\mu x^\nu=\x^2-t^2$;
this means that the matrix $\Lam^\mu{}_\nu$ must obey
\begin{equation}
g_{\mu\nu} \Lam^\mu{}_\rho \Lam^\nu{}_\sigma =  g_{\rho\sigma} \;,
\label{prod}
\end{equation}
where
\begin{equation}
g_{\mu\nu}=\pmatrix{ -1 & & & \cr
                     & +1 & & \cr
                     & & +1 & \cr
                     & & & +1 \cr} .
\label{g2}
\end{equation}
is the flat-space metric.

Note that this set of transformations includes ordinary spatial rotations:
take $\Lam^0{}_0=1$,
$\Lam^0{}_i=\Lam^i{}_0=0$, and
$\Lam^i{}_j=R_{ij}$, where $R$ is an orthogonal rotation matrix.

The set of all Lorentz transformations [that is, matrices obeying \eq{prod}] 
forms a {\it group\/}:  the product of any two Lorentz transformations
is another Lorentz transformation, the product is associative,
and every Lorentz transformation has an
inverse.  It is easy to demonstrate these statements explicitly.  For example,
to find the inverse transformation $(\Lam^{-1})^\mu{}_\nu$, note that
the left-hand side of \eq{prod} can be written as
$\Lam_{\nu\rho}\Lam^\nu{}_\sigma$, and that 
we can raise the $\rho$ index on both sides to get
$\Lam_\nu{}^\rho\Lam^\nu{}_\sigma=\delta^\rho{}_\sigma$.
On the other hand, by definition, 
$(\Lam^{-1})^\rho{}_\nu\Lam^\nu{}_\sigma=\delta^\rho{}_\sigma$.  Therefore
\begin{equation}
(\Lam^{-1})^\rho{}_\nu = \Lam_\nu{}^\rho \;.
\label{inv}
\end{equation}
Another useful version of \eq{prod} is
\begin{equation}
g^{\mu\nu} \Lam^\rho{}_\mu \Lam^\sigma{}_\nu =  g^{\rho\sigma} \;.
\label{prod2}
\end{equation}
To get this, start with \eq{prod}, but with the inverse transformations
$(\Lam^{-1})^\mu{}_\rho$ and $(\Lam^{-1})^\nu{}_\sigma$.
Then use \eq{inv}, raise all down indices, and lower all up indices.
The result is \eq{prod2}.

For an infinitesimal Lorentz transformation, we can write
\begin{equation}
\Lam^\mu{}_\nu = \delta^\mu{}_\nu + \delta\w^\mu{}_\nu \;.
\label{inf}
\end{equation}
\Eq{prod} can be used to show that $\delta\omega$ with both indices
down (or up) is antisymmetric:
\begin{equation}
\delta\w_{\rho\sigma} = -\delta\w_{\sigma\rho} \;.
\label{dw}
\end{equation}
Thus there are six independent infinitesimal Lorentz transformations 
(in four spacetime dimensions).  These can be divided into three rotations
($\delta\w_{ij} = -\e_{ijk}\hat n_k\delta\theta$ for a rotation by
angle $\delta\theta$ about the unit vector $\bf\hat n$)
and three boosts ($\delta\w_{i0}=\hat n_i\delta\eta$ for a boost
in the direction $\bf \hat n$ by {\it rapidity\/} $\delta\eta$).

Not all Lorentz transformations can be reached by compounding infinitesimal
ones.  If we take the determinant of \eq{inv}, we get
$(\det\Lam)^{-1}=\det\Lam$, which implies $\det\Lam=\pm 1$.
Transformations with $\det\Lam=+1$ are {\it proper}, and 
transformations with $\det\Lam=-1$ are {\it improper}.
Note that the product of any two proper Lorentz transformations is
also proper.  Also, infinitesimal transformations of the form
$\Lam=1+\delta\w$ are proper.
Therefore, any transformation that can be reached by compounding
infinitesimal ones is proper.
The proper transformations form a {\it subgroup\/} of the Lorentz group.

Another subgroup is that of the {\it orthochronous} Lorentz transformations:
those for which $\Lam^0{}_0\ge+1$.  
Note that \eq{prod} implies 
$(\Lam^0{}_0)^2 - \Lam^i{}_0\Lam^i{}_0=1$; thus,
either $\Lam^0{}_0>+1$ or $\Lam^0{}_0<-1$.  
An infinitesimal transformation is clearly orthochronous,
and it is straightforward to show that
the product of two orthochronous transformations is also
orthochronous.

Thus, the Lorentz transformations that can be reached by compounding
infinitesimal ones are both proper and orthochronous, and they form
a subgroup.  
We can introduce two discrete transformations that take us out of this
subgroup: {\it parity\/} and {\it time reversal}.  The parity transformation is
\begin{equation}
\P^\mu{}_\nu = (\P^{-1}){}^\mu{}_\nu
             =\pmatrix{ +1 & & & \cr
                        & -1 & & \cr
                        & & -1 & \cr
                        & & & -1 \cr}.
\label{p}
\end{equation}
It is orthochronous, but improper.  The time-reversal transformation is
\begin{equation}
\T^\mu{}_\nu = (\T^{-1}){}^\mu{}_\nu
             =\pmatrix{ -1 & & & \cr
                        & +1 & & \cr
                        & & +1 & \cr
                        & & & +1 \cr}.
\label{tmu}
\end{equation}
It is nonorthochronous and improper.

Generally, when a theory is said to be {\it Lorentz invariant}, this
means under the proper orthochronous subgroup only.  Parity and time reversal
are treated separately.  It is possible for a quantum field theory to
be invariant under the proper orthochronous subgroup, 
but not under parity and/or time-reversal.

From here on, in this section, we will treat the proper orthochronous
subgroup only.  Parity and time reversal will be treated in section 23.

In quantum theory, symmetries are represented by unitary (or antiunitary)
operators.  This means that we associate a unitary operator $U(\Lam)$ to 
each proper, orthochronous Lorentz transformation $\Lam$.
These operators must obey the composition rule
\begin{equation}
U(\Lam'\Lam)=U(\Lam')U(\Lam)\;.
\label{uuu}
\end{equation}
For an infinitesimal transformation, we can write
\begin{equation}
U(1{+}\delta\w)=I + {\ts{i\over2}}\delta\w_{\mu\nu}M^{\mu\nu}\;,
\label{udw}
\end{equation}
where $M^{\mu\nu}=-M^{\nu\mu}$ is a set of hermitian operators
called the {\it generators of the Lorentz group}.  If we start with
$U(\Lam)^{-1}U(\Lam')U(\Lam)=U(\Lam^{-1}\Lam'\Lam)$ and let
$\Lam'=1+\delta\w'$, we can show that
\begin{equation}
U(\Lam)^{-1} M^{\mu\nu} U(\Lam) = \Lam^\mu{}_\rho\Lam^\nu{}_\sigma
M^{\rho\sigma}\;.
\label{umu}
\end{equation}
Thus, 
each vector index on $M^{\mu\nu}$ undergoes its own Lorentz transformation.
This is a general result: any operator carrying one or more vector indices
should behave similarly.  For example, consider the energy-momentum four-vector
$P^\mu$, where $P^0$ is the hamiltonian $H$ and $P^i$ are the components 
of the total three-momentum operator.  We expect
\begin{equation}
U(\Lam)^{-1} P^\mu U(\Lam) = \Lam^\mu{}_\nu P^\nu \;.
\label{upu}
\end{equation}

If we now let $\Lam=1+\delta\w$ in \eq{umu}, we get the commutation relations
\begin{equation}
[M^{\mu\nu},M^{\rho\sigma}]=i\Bigl(g^{\mu\rho}M^{\nu\sigma}
                            -(\mu{\leftrightarrow}\nu)\Bigr)
                            -(\rho{\leftrightarrow}\sigma)\;.
\label{comm0}
\end{equation}
We can identify the components of the angular momentum operator $\bf J$ 
with $J_i\equiv\half\e_{ijk}M^{jk}$,
and the components of the boost operator $\bf K$
with $K_i\equiv M^{i0}$.  We then find from \eq{comm0} that
\begin{eqnarray}
{[}J_i,J_j{]} &=& +i\e_{ijk}J_k \;,
\nonumber \\
{[}J_i,K_j{]} &=& +i\e_{ijk}K_k \;,
\nonumber \\
{[}K_i,K_j{]} &=& -i\e_{ijk}J_k \;.
\label{comm2}
\end{eqnarray}
The first of these is the usual set of commutators for angular momentum,
and the second says that $\bf K$ transforms as a three-vector under rotations.
The third implies that a series of boosts can be equivalent to a rotation.

Similarly, we can let $\Lam=1+\delta\w$ in \eq{upu} to get
\begin{equation}
[P^\mu,M^{\rho\sigma}]=i\Bigl(g^{\mu\sigma}P^\rho
                            -(\rho{\leftrightarrow}\sigma)\Bigr)\;,
\label{comm3}
\end{equation}
which becomes
\begin{eqnarray}
{[}J_i,H{]} &=& 0 \;,
\nonumber \\
{[}J_i,P_j{]} &=& +i\e_{ijk}P_k \;,
\nonumber \\
{[}K_i,H{]} &=& +iP_i \;,
\nonumber \\
{[}K_i,P_j{]} &=& +i\delta_{ij}H \;,
\label{comm4}
\end{eqnarray}
Also, the components of $P^\mu$ should commute with each other:
\begin{eqnarray}
{[}P_i,P_j{]} &=& 0 \;,
\nonumber \\
{[}P_i,H{]} &=& 0 \;.
\label{comm5}
\end{eqnarray}
Together, \eqs{comm2}, 
(\ref{comm4}), and (\ref{comm5}) form the {\it Poincar\'e algebra}.

Let us now consider what should happen to a quantum scalar field $\ph(x)$ under
a Lorentz transformation.  We begin by recalling how
time evolution works in the Heisenberg picture:
\begin{equation}
e^{+iHt}\ph(\x,0)e^{-iHt}= \ph(\x,t)\;.
\label{heis}
\end{equation}
Obviously, this should have a relativistic generalization,
\begin{equation}
e^{-iPx}\ph(0)e^{+iPx}= \ph(x)\;,
\label{heis2}
\end{equation}
where $Px=P^\mu x_\mu={\bf P}\cd\x-Ht$.  
We can make this a little fancier by defining
the unitary {\it spacetime translation operator} 
\begin{equation}
T(a) \equiv \exp(-iP^\mu a_\mu)\;.
\label{ta}
\end{equation}
Then we have
\begin{equation}
T(a)^{-1}\ph(x)T(a) = \ph(x-a)\;.
\label{taphta}
\end{equation}
For an infinitesimal translation,
\begin{equation}
T(\delta a) = I - i\delta a_\mu P^\mu \;.
\label{tda}
\end{equation}
Comparing \eqs{udw} and (\ref{tda}), we see that \eq{taphta}
leads us to expect
\begin{equation}
U(\Lam)^{-1} \ph(x) U(\Lam) = \ph(\Lam^{-1}x)\;.
\label{uphu}
\end{equation}
Derivatives of $\ph$ then 
carry vector indices that transform in the appropriate way, e.g.,
\begin{equation}
U(\Lam)^{-1} \d^\mu\ph(x) U(\Lam) 
= \Lam^\mu{}_\rho\bar\d^\rho\ph(\Lam^{-1}x)\;,
\label{udphu}
\end{equation}
where the bar on a derivative means 
that it is with respect to the argument $\bar x=\Lam^{-1}x$.
\Eq{udphu} also implies
\begin{equation}
U(\Lam)^{-1} \d^2\ph(x) U(\Lam) = \bar\d^2\ph(\Lam^{-1}x)\;,
\label{ud2phu}
\end{equation}
so that the Klein-Gordon equation, $(-\d^2+m^2)\ph=0$, is Lorentz invariant.

\vskip0.5in

\begin{center}
Problems
\end{center}

\vskip0.25in

2.1) Verify that \eq{dw} follows from \eq{prod}.

2.2) Verify that \eq{umu} follows from
$U(\Lam)^{-1}U(\Lam')U(\Lam)=U(\Lam^{-1}\Lam'\Lam)$.

2.3) Verify that \eq{comm0} follows from \eq{umu}.

2.4) Verify that \eq{comm2} follows from \eq{comm0}. 

2.5) Verify that \eq{comm3} follows from \eq{upu}.

2.6) Verify that \eq{comm4} follows from \eq{comm3}.

2.7) What property should the translation operator $T(a)$ 
have that could be used to prove \eq{comm5}?

2.8) Let us write
\begin{equation}
\Lam^\rho{}_\tau = \delta^\rho{}_\tau
+{\ts{i\over2}}\delta\w_{\mu\nu}(S^{\mu\nu}_{\sss\rm V})^\rho{}_\tau \;,
\label{lamrhotau2}
\end{equation}
where 
\begin{equation}
(S^{\mu\nu}_{\sss\rm V})^\rho{}_\tau 
\equiv {\ts{1\over i}}( g^{\mu\rho}\delta^\nu{}_\tau
        -g^{\nu\rho}\delta^\mu{}_\tau)
\label{svmunu}
\end{equation}
are matrices which constitute the {\it vector representation\/}
of the Lorentz generators.

a) Argue that the matrices
$S^{\mu\nu}_{\sss\rm V}$ have the same commutation relations
as the the operators $M^{\mu\nu}$.

b) For a rotation by an angle $\theta$ about the $z$ axis, we have
\begin{equation}
\Lam^\mu{}_\nu =\pmatrix{  1 & 0 & \phantom{+}0 & 0 \cr
                          0 & \cos\theta & -\sin\theta & 0 \cr
                          0 & \sin\theta & \phantom{+}\cos\theta & 0 \cr
                          0 & 0 & \phantom{+}0 & 1 \cr} .
\label{lamrot}
\end{equation}
Show that
\begin{equation}
\Lam = \exp(-i\theta S^{12}_{\rm\sss V}) \;.
\label{lamrot2}
\end{equation}

c) For a boost by {\it rapidity\/} $\eta$ in the $z$ direction, we have
\begin{equation}
\Lam^\mu{}_\nu =\pmatrix{ \cosh\eta & 0 & 0 & \sinh\eta \cr
                          0 & 1 & 0 & 0 \cr
                          0 & 0 & 1 & 0 \cr
                         \sinh\eta & 0 & 0 & \cosh\eta \cr } .
\label{lamboost}
\end{equation}
Show that
\begin{equation}
\Lam = \exp(+i\eta S^{30}_{\rm\sss V}) \;.
\label{lamboost2}
\end{equation}

\vfill\eject

%% file: ch003.tex
\noindent Quantum Field Theory \hfill Mark Srednicki

\vskip0.5in

\begin{center}
\large{3: Relativistic Quantum Fields and Canonical Quantization}
\end{center}
\begin{center}
Prerequisite: 2
\end{center}

\vskip0.5in

Let us go back and drastically simplify the hamiltonian we constructed
in section 1, reducing it to that of free particles:
\begin{eqnarray}
H &=& \int \dtx\;\ad(\x)\!
      \left(-{\textstyle{1\over2m}\nabla^2}\right)\!a(\x) 
\nonumber \\
&=& \int \dtp\; {\textstyle{1\over2m}}\p^2\;\adt(\p)\at(\p)  \;,
\label{hamp} 
\end{eqnarray}
where 
\begin{equation}
\at(\p) = \int{\dtx\over(2\pi)^{3/2}}\;e^{-i\p\cdot\x}\;a(\x)\;. 
\label{ft}
\end{equation}
Note that we have simplified our notation by setting $\hbar=1$;
the appropriate factors of $\hbar$ can always be restored in any of our
formulas via dimensional analysis.  The commutation (or anticommutation)
relations of the $\at(\p)$ and $\adt(\p)$ operators are 
\begin{eqnarray}
{[\at({\bf p}),\at({\bf p'})]_\mp} &=& 0 \; ,
\nonumber \\
{[\adt({\bf p}),\adt({\bf p'})]_\mp} &=& 0 \;,
\nonumber \\
{[\at({\bf p}),\adt({\bf p'})]_\mp} &=& \delta^3({\bf p-p'})\;,
\label{atc}
\end{eqnarray}
where $[A,B]_\mp$ is either the commutator (if we want a theory of bosons)
or the anticommutator (if we want a theory of fermions).
Thus $\adt(\p)$ can be interpreted as creating a state of 
definite momentum $\p$, and \eq{hamp} describes a theory of free particles.  
The ground state is the {\it vacuum\/} $|0\ra$; it is annihilated by $\at(\p)$,
\begin{equation}
\at(\p)|0\ra = 0\;,
\label{avac}
\end{equation}
and so its energy eigenvalue is zero.
The other eigenstates of $H$ are all of the form
$\adt(\p_1)\ldots\adt(\p_n)|0\ra$, and the corresponding energy eigenvalue
is $E(\p_1)+\ldots+E(\p_n)$, where $E(\p)={1\over2m}\p^2$.  

It is easy to see how to generalize this theory to a relativistic one;
all we need to do is use the relativistic energy formula
$E(\p)=+(\p^2c^2 + m^2c^4)^{1/2}$:
\begin{equation}
H = \int \dtp\;(\p^2c^2 + m^2c^4)^{1/2}\;\adt(\p)\at(\p)  \;.
\label{hampr} 
\end{equation}
Now we have a theory of free {\it relativistic\/} spin-zero particles,
and they can be either bosons or fermions.

Is this theory really Lorentz invariant?  
We will answer this question (in the affirmative) in a very roundabout
way: by constructing it again, from a rather different point of view,
a point of view that emphasizes Lorentz invariance from the beginning.

We will start with the {\it classical\/} physics of a {\it real scalar
field\/} $\ph(x)$.  {\it Real\/} means that $\ph(x)$
assigns a real number to every point in spacetime.  {\it Scalar\/}
means that Alice [who uses coordinates $x^\mu$ and calls the field $\ph(x)$]
and Bob [who uses coordinates $\bar x^\mu$, related to Alice's coordinates
by $\bar x^\mu=\Lam^\mu{}_\nu x^\nu+a^\nu$, 
and calls the field $\bar\ph(\bar x)$], agree on the numerical
value of the field: $\ph(x)=\bar\ph(\bar x)$. 
This then implies that the equation of motion for $\ph(x)$ must be
the same as that for $\bar\ph(\bar x)$.  We have already
met an equation of this type: the Klein-Gordon equation,
$(-\d^2+m^2)\ph(x)=0$.  (Here, to simplify the notation, we have set
$c=1$ in addition to $\hbar=1$.
As with $\hbar$, factors of $c$ can restored,
if desired, by dimensional analysis.)
Let us adopt this as the equation of motion we would like $\ph(x)$ to obey.

It should be emphasized at this point that we are doing {\it classical
physics\/} of a {\it real scalar field}.  We are {\it not\/} to think
of $\ph(x)$ as a quantum wave function.  Thus, there should not be
any factors of $\hbar$ in this version of the Klein-Gordon equation.
This means that the parameter $m$ must have dimensions of inverse length;
$m$ is {\it not\/} (yet) to be thought of as a mass.  

The equation of motion can be derived from variation of an action
$S=\int dt\,L$, where $L$ is the lagrangian.  Since the Klein-Gordon
equation is local, we expect that the lagrangian can be written as
the space integral of a {\it lagrangian density\/} $\L$: $L=\int \dtx\,\L$.
Thus, $S=\int \dfx\,\L$.  The integration measure $\dfx$ is Lorentz
invariant: if we change to coordinates $\bar x^\mu=\Lam^\mu{}_\nu x^\nu$,
we have $d^{4\!}\bar x=\left|\det\Lam\right|\dfx=\dfx$.  
Thus, for the action to be Lorentz invariant, the
lagrangian density must be a Lorentz scalar: $\L(x)=\bar\L(\bar x)$.
Then we have $\bar S =\int d^{4\!}\bar x\,\bar\L(\bar x)=\int \dfx\,\L(x)=S$.
Any simple function of $\ph$ is a Lorentz scalar, and so are
products of derivatives with all indices contracted, such as
$\d^\mu\ph\d_\mu\ph$.  We will take for $\cal L$
\begin{equation}
\L = -\half\d^\mu\ph\d_\mu\ph -\half m^2\ph^2 + \Lambda_0\;,
\label{calell}
\end{equation}
where $\Lambda_0$ is an arbitrary constant.
We find the equation motion (also known as the {\it Euler-Lagrange equation\/})
by making an infinitesimal variation $\delta\ph(x)$ in $\ph(x)$, 
and requiring the corresponding variation of the action to vanish:
\begin{eqnarray}
0 &=& \delta S
\nonumber \\
\noalign{\medskip}
  &=& \int \dfx\,\left[ -\half\d^\mu\delta\ph\d_\mu\ph
                     -\half\d^\mu\ph\d_\mu\delta\ph
                     -m^2\ph\,\delta\ph\right]
\nonumber \\
&=& \int \dfx\,\left[ +\d^\mu\d_\mu\ph - m^2\ph\right]\delta\ph \;.
\label{el}
\end{eqnarray}
In the last line, we have integrated by parts in each of the first
two terms, putting both derivatives on $\ph$.  We assume $\delta\ph(x)$
vanishes at infinity in any direction (spatial or temporal), so that
there is no surface term.  Since $\delta\ph$ has an arbitrary $x$
dependence, \eq{el} can be true if and only if $(-\d^2+m^2)\ph=0$.

One solution of the Klein-Gordon equation is a plane wave of the form
$\exp(i\k\!\cdot\!\x\pm i\omega t)$, where $\k$ is an arbitrary real
wave-vector, and 
\begin{equation}
\omega=+(\k^2+m^2)^{1/2} \;.  
\label{omega}
\end{equation}
The general solution
(assuming boundary conditions that do not allow $\ph$ to become
infinite at spatial infinity) is then
\begin{equation}
\ph(\x,t) = \int {\dtk\over f(k)}\left[
                 a(\k)e^{i\k\cdot\x-i\omega t}
                +b(\k)e^{i\k\cdot\x+i\omega t}\right]\;,
\label{phi}
\end{equation}
where $a(\k)$ and $b(\k)$ are arbitrary functions of the 
wave vector $\k$, and $f(k)$ is a redundant function of the magnitude of $\k$
which we have inserted for later convenience. 
Note that, {\it if\/} we were attempting to interpret $\ph(x)$ as a quantum
wave function (which we most definitely are {\it not\/}), then the
second term would constitute the ``negative energy'' contributions
to the wave function.  This is because a plane-wave solution of
the nonrelativistic Schr\"odinger equation for a single particle looks like
$\exp(i\p\!\cdot\!\x - iE(\p)t)$, with $E(\p)={1\over 2m}\p^2$;
there is a minus sign in front of the positive energy.  We {\it are\/}
trying to interpret \eq{phi} as a {\it real\/} classical field,
but this formula does not generically result in $\ph$ being real.
We must impose $\ph^*(x)=\ph(x)$, where
\begin{eqnarray}
\ph^*(\x,t) &=& \int {\dtk\over f(k)}\left[
                 a^*(\k)e^{-i\k\cdot\x+i\omega t}
                +b^*(\k)e^{-i\k\cdot\x-i\omega t}\right]
\nonumber \\
\noalign{\smallskip}
            &=& \int {\dtk\over f(k)}\left[
                 a^*(\k)e^{-i\k\cdot\x+i\omega t}
                +b^*(-\k)e^{+i\k\cdot\x-i\omega t}\right]\;.
\label{phistar}
\end{eqnarray}
In the second line, we have changed the dummy integration variable
$\k$ (in the second term only) to $-\k$.  Comparing \eqs{phi}
and (\ref{phistar}), we see that $\ph^*(x)=\ph(x)$ requires $b^*(-\k)=a(\k)$.
Imposing this condition, we can rewrite $\ph$ as
\begin{eqnarray}
\ph(\x,t) &=& \int {\dtk\over f(k)}\left[
                 a(\k)e^{i\k\cdot\x-i\omega t}
                +a^*(-\k)e^{i\k\cdot\x+i\omega t}\right]
\nonumber \\
\noalign{\smallskip}
            &=& \int {\dtk\over f(k)}\left[
                 a(\k)e^{i\k\cdot\x-i\omega t}
                +a^*(\k)e^{-i\k\cdot\x+i\omega t}\right]
\nonumber \\
\noalign{\smallskip}
            &=& \int {\dtk\over f(k)}\left[
                 a(\k)e^{ikx} + a^*(\k)e^{-ikx}\right]\;,
\label{phir}
\end{eqnarray}
where $kx=\k\!\cdot\!\x-\omega t$ is the Lorentz-invariant product
of the four-vectors $x^\mu=(t,\x)$ and $k^\mu=(\omega,\k)$: 
$kx=k^\mu x_\mu=g_{\mu\nu}k^\mu x^\nu$.  Note that 
\begin{equation}
k^2 = k^\mu k_\mu = \k^2-\omega^2 = -m^2 \;.
\label{ksq}
\end{equation}

It is now convenient to choose $f(k)$ so that $\dtk/f(k)$ is
Lorentz invariant.  An integration measure that is manifestly invariant
under orthochronous Lorentz transformations is
$\dfk\,\delta(k^2{+}m^2)\,\theta(k^0)$, where 
$\theta(x)$ is the unit step function, and $k^0$ is treated as
an independent integration variable.  We then have
\begin{equation}
\int_{-\infty}^{+\infty}dk^0\,\delta(k^2{+}m^2)\,\theta(k^0)
={1\over 2\omega}\;.
\label{ps}
\end{equation}
Here we have used the rule
\begin{equation}
\int_{-\infty}^{+\infty}dx\;\delta(g(x))=\sum\limits_i {1\over |g'(x_i)|}\;,
\label{intdelta}
\end{equation}
where $g(x)$ is any smooth function of $x$ with simple zeros at $x=x_i$;
in our case, the only zero is at $k^0=\w$.

Thus we see that if we take $f(k)\propto\w$, then $\dtk/\!f(k)$ will be
Lorentz invariant.  We will take $f(k)=(2\pi)^3 2\omega$.
It is then convenient to give the corresponding Lorentz-invariant differential
its own name:
\begin{equation}
\dk \equiv {\dtk\over(2\pi)^3 2\omega}\;.
\label{dk}
\end{equation}
Thus we finally have
\begin{equation}
\ph(x) = \int \dk\,\left[a(\k)e^{ikx}+a^*(\k)e^{-ikx}\right]\;.
\label{phi2}
\end{equation}

We can also invert this formula to get $a(\k)$ in terms of $\ph(x)$.
We have
\begin{eqnarray}
\int \dtx\;e^{-ikx}\ph(x) &=& 
{\ts{1\over2\w}}a(\k) + {\ts{1\over2\w}}e^{2i\w t}a(-\k)\;,
\nonumber \\
\noalign{\smallskip}
\int \dtx\;e^{-ikx}\d_0\ph(x) &=& 
-{\ts{i\over2}}a(\k) + {\ts{i\over2}}e^{2i\w t}a(-\k)\;.
\label{0a}
\end{eqnarray}
We can combine these to get
\begin{eqnarray}
a(\k) &=& \int \dtx\;e^{-ikx}
          \Bigl[i\d_0\ph(x)+\omega\ph(x)\Bigr]
\nonumber \\
\noalign{\smallskip}
&=& i\int \dtx \;e^{-ikx}\!\buildrel\leftrightarrow\over{\d_0}\!\ph(x)\;,
\label{a}
\end{eqnarray}
where $f\!\buildrel\leftrightarrow\over{\d_\mu}\!g=f(\d_\mu g)-(\d_\mu f)g$,
and $\d_0\ph=\d\ph/\d t=\dot\ph$. 
Note that $a(\k)$ is time independent.

Now that we have the lagrangian, we can construct the hamiltonian by
the usual rules.  Recall that, given a lagrangian $L(q_i,\dot q_i)$
as a function of some coordinates $q_i$ and their time derivatives
$\dot q_i$, the conjugate momenta are given by $p_i = \d L/\d\dot q_i$,
and the hamiltonian by $H=\sum_i p_i \dot q_i-L$.  In our case, the
role of $q_i(t)$ is played by $\ph(\x,t)$, with $\x$ playing the
role of a (continuous) index.  The appropriate generalizations are then 
\begin{equation}
\Pi(x) = {\d\L\over\d\dot\ph(x)}
\label{Pi}
\end{equation}
and
\begin{equation}
\H = \Pi\dot\ph-\L\;,
\label{h0}
\end{equation}
where $\H$ is the {\it hamiltonian density}, and the hamiltonian itself is
$H=\int \dtx\;\H$.  In our case, we have
\begin{equation}
\Pi(x) = \dot\ph(x)
\label{Piphidot}
\end{equation}
and
\begin{equation}
\H = \half\Pi^2 + \half(\nabla\ph)^2 +\half m^2\ph^2 - \Lambda_0\;.
\label{h4}
\end{equation}
Using \eq{phi2}, we can write $H$ in terms of the $a(\k)$ and $a^*(\k)$ 
coefficients:
\begin{eqnarray}
H &=& -\Lambda_0 V + \half\int \dk\;\dk'\;\dtx\;\Bigl[
\nonumber \\
      && \quad\; 
\left(-i\omega\,a(\k)e^{ikx}+i\omega\,a^*(\k)e^{-ikx}\right)
\left(-i\omega'\,a(\k')e^{ik'x}+i\omega'\,a^*(\k')e^{-ik'x}\right)
\nonumber \\
      && {} + 
\left(+i\k\,a(\k)e^{ikx}-i\k\,a^*(\k)e^{-ikx}\right)\cd
\left(+i\k'\,a(\k')e^{ik'x}-i\k'\,a^*(\k')e^{-ik'x}\right)
\nonumber \\
      && {} +
      m^2\left(a(\k)e^{ikx}+a^*(\k)e^{-ikx}\right)
      \left(a(\k')e^{ik'x}+a^*(\k')e^{-ik'x}\right)\Bigr]
\nonumber \\
\noalign{\smallskip}
&=& -\Lambda_0 V + \half(2\pi)^3\int \dk\;\dk'\;\Bigl[
\nonumber \\
&& \quad\; 
\delta^3(\k-\k') (+\omega\omega' + \k\!\cdot\!\k' + m^2)
\nonumber \\
      && \qquad 
\times\left(a^*(\k)a(\k')e^{-i(\omega-\omega')t}
+a(\k)a^*(\k')e^{+i(\omega-\omega')t} \right)
\nonumber \\
      && {}+
\delta^3(\k+\k')(-\omega\omega' - \k\!\cdot\!\k' + m^2) 
\nonumber \\
      && \qquad 
\times\left(a(\k)a(\k')e^{-i(\omega+\omega')t)}
+a^*(\k)a^*(\k')e^{+i(\omega+\omega')t)} \right)
\nonumber \\
\noalign{\smallskip}
&=& -\Lambda_0 V + \half\int \dk\;{\ts{1\over 2\omega}}\Bigl[
\nonumber \\
      && \quad\;
(+\omega^2 + \k^2 + m^2)\Bigl(a^*(\k)a(\k) + a(\k)a^*(\k)\Bigr) 
\nonumber \\
&& {}+(-\omega^2 + \k^2 + m^2)\Bigl(a(\k)a(-\k) + a^*(\k)a^*(-\k)\Bigr)\Bigr]
\nonumber \\
\noalign{\smallskip}
&=& -\Lambda_0 V 
+ \half\int \dk\;\omega \Bigl(a^*(\k)a(\k) + a(\k)a^*(\k)\Bigr)\;,
\label{h5}
\end{eqnarray}
where $V$ is the volume of space.
We have made use of $\omega=(\k^2{+}m^2)^{1/2}$ and 
$\dk=\dtk/(2\pi)^32\omega$ at various points.
Also, we have been careful to keep the ordering of $a(\k)$ and $a^*(\k)$
unchanged throughout, in anticipation of passing to the quantum theory
where these classical functions will become operators that may not commute.

Let us take up the quantum theory now.  We can go from classical to quantum
mechanics via {\it canonical quantization}.  This means that we promote
$q_i$ and $p_i$ to operators, with commutation relations 
$[q_i,q_j]=0$, $[p_i,p_j]=0$, and $[q_i,p_j]=i\hbar\delta_{ij}$.  
In the Heisenberg picture, these operators should be taken at equal times.
In our case, where the ``index'' is continuous (and we have set $\hbar=1$),
this becomes
\begin{eqnarray}
{[\ph(\x,t),\ph(\x',t)]} &=& 0 \;, 
\nonumber \\
\noalign{\smallskip}
{[\Pi(\x,t),\Pi(\x',t)]} &=& 0 \;,
\nonumber \\
\noalign{\smallskip}
{[\ph(\x,t),\Pi(\x',t)]} &=& i\delta^3(\x-\x') \;.
\label{etcr}
\end{eqnarray}
From these, and from \eqs{a} and (\ref{Piphidot}), we can deduce
\begin{eqnarray}
{[a({\bf k}),a({\bf k'})]} &=& 0 \; ,
\nonumber \\
\noalign{\smallskip}
{[a^\dagger({\bf k}),a^\dagger({\bf k'})]} &=& 0 \;,
\nonumber \\
\noalign{\smallskip}
{[a({\bf k}),a^\dagger({\bf k'})]} &=& (2\pi)^3 2\omega\,\delta^3({\bf k-k'})\;.
\label{ac}
\end{eqnarray}
We are now denoting $a^*(\k)$ as $\ad(\k)$, since $\ad(\k)$
is now the hermitian conjugate (rather than the complex conjugate)
of the operator $a(\k)$.
We can now rewrite the hamiltonian as
\begin{equation}
H = \int \dk\;\omega\;\ad(\k)a(\k) + ({\cal E}_0-\Lambda_0)V \;,
\label{h6}
\end{equation}
where ${\cal E}_0 = \half\int \dtk\,\omega$ is the total zero-point energy
of all the oscillators per unit volume, 
and, using $(2\pi)^3\delta^3(\k)=\int \dtx\,e^{i\k\cdot\x}$,
we have interpreted $(2\pi)^3\delta^3({\bf 0})$ as the volume of space $V$.
If you try to evaluate ${\cal E}_0$, you will find that it is infinite.
However, 
$\Lambda_0$ was arbitrary, so we are free to choose $\Lambda_0={\cal E}_0$,
whether or not ${\cal E}_0$ is infinite.  And that is what we will do.
With this choice, the ground state has energy eigenvalue zero.

The hamiltonian of \eq{h6} is now the same as that of \eq{hampr}, with
$a(\k)=[(2\pi)^3 2\omega]^{1/2}\,\at(\k)$.  The commutation relations
(\ref{atc}) and (\ref{ac}) are also equivalent, if we choose commutators
(rather than anticommutators) in \eq{atc}.
Thus, {\it we have re-derived the hamiltonian of free relativistic bosons
by quantization of a scalar field whose equation of motion is the Klein-Gordon
equation.}

What if we want fermions?  Then we should use anticommutators in \eqs{etcr}
and (\ref{ac}).  There is a problem, though; \eq{h5} does not then become
\eq{h6}.  Instead, we get $H=-\Lambda_0 V$, a constant!  Clearly there is
something wrong with using anticommutators.  This is another hint of the
spin-statistics theorem, which we will take up in section 4.

Next, we would like to add Lorentz-invariant interactions to our theory.
With the formalism we have developed, this is easy to do.  Any local function
of $\ph(x)$ is a Lorentz scalar, and so if we add a term like 
$\ph^3$ or $\ph^4$ to the lagrangian density $\L$, the resulting action will
still be Lorentz invariant.  Now, however, we will have interactions 
among the particles.  Our next task is to deduce the consequences of these
interactions.

However, we already have enough tools at our disposal to prove the
spin-statistics theorem for spin-zero particles, and that is what
we turn to next.

\vskip0.5in

\begin{center}
Problems
\end{center}

\vskip0.25in

3.1) Derive \eq{ac} from \eqs{a} and (\ref{Piphidot}).

3.2) Use the commutation relations, \eq{ac}, to
show explicitly that a state of the form 
\begin{equation}
|k_1\ldots k_n\ra \equiv \ad(\k_1) \ldots \ad(\k_n)|0\ra
\label{k1kn}
\end{equation}
is an eigenstate of the hamiltonian, \eq{h6}, with eigenvalue
$\w_1+\ldots+\w_n$.  The vacuum $|0\ra$ is annihilated by $a(\k)$,
$a(\k)|0\ra=0$, and we take $\Lam_0={\cal E}_0$ in \eq{h6}.

3.3) Use $U(\Lam)^{-1}\ph(x)U(\Lam) = \ph(\Lam^{-1}x)$ to show that
\begin{eqnarray}
U(\Lam)^{-1}a(\k)U(\Lam) &=& a(\Lam^{-1}\k) \;,
\nonumber \\
\noalign{\smallskip}
U(\Lam)^{-1}\ad(\k)U(\Lam) &=& \ad(\Lam^{-1}\k) \;,
\label{uau00}
\end{eqnarray}
and hence that
\begin{equation}
U(\Lam)|k_1\ldots k_n\ra = |\Lam k_1 \ldots \Lam k_n\ra\;,
\label{ulamkra}
\end{equation}
where $|k_1\ldots k_n\ra = \ad(\k_1) \ldots \ad(\k_n)|0\ra$ is a
state of $n$ particles with momenta $k_1,\ldots,k_n$.

3.4) Consider a complex (that is, nonhermitian) scalar field $\ph$ with
lagrangian density
\begin{equation}
\L = -\d^\mu\ph^\dagger\d_\mu\ph - m^2\ph^\dagger\ph + \Lambda_0\;.
\label{calell3}
\end{equation}

a) Show that $\ph$ obeys the Klein-Gordon equation.

b) Treat $\ph$ and $\ph^\dagger$ as independent fields, and find the
conjugate momentum for each.  Compute the hamiltonian density in
terms of these conjugate momenta and the fields themselves
(but not their time derivatives).

c) Write the mode expansion of $\ph$ as 
\begin{equation}
\ph(x) = \int \dk\left[a(\k)e^{ikx}+b^\dagger(\k)e^{-ikx}\right]\;.
\label{phi2p}
\end{equation}
Express $a(\k)$ and $b(\k)$ in terms of $\ph$ and $\ph^\dagger$
and their time derivatives.

d) Assuming canonical commutation relations for the fields and
their conjugate momenta, 
find the commutation relations obeyed by $a(\k)$ and $b(\k)$ 
and their hermitian conjugates. 

e) Express the hamiltonian in terms of $a(\k)$ and $b(\k)$ 
and their hermitian conjugates. 
What value must $\Lam_0$ have in order for the ground state
to have zero energy?

\vfill\eject

%% file: ch004.tex
\noindent Quantum Field Theory  \hfill   Mark Srednicki

\vskip0.5in

\begin{center}
\large{4: The Spin-Statistics Theorem}
\end{center}
\begin{center}
Prerequisite: 3
\end{center}

\vskip0.5in

Let us consider a theory of free, spin-zero particles specified by the 
hamiltonian
\begin{equation}
H_0 = \int \dk\;\omega\;\ad(\k)a(\k) \;,
\label{h03}
\end{equation}
where $\w=(\k^2+m^2)^{1/2}$,
and either the commutation or anticommutation relations 
\begin{eqnarray}
	{[a({\bf k}),a({\bf k'})]_\mp} &=& 0 \; ,
\nonumber \\
{[a^\dagger({\bf k}),a^\dagger({\bf k'})]_\mp} &=& 0 \;,
\nonumber \\
{[a({\bf k}),a^\dagger({\bf k'})]_\mp} &=& (2\pi)^3 2\omega\,\delta^3({\bf k-k'})\;.
\label{ac3}
\end{eqnarray}
Of course, if we want a theory of bosons, we should use commutators, and if we
want fermions, we should use anticommutators.  

Now let us consider adding terms to the hamiltonian that will result in
{\it local, Lorentz invariant\/} interactions.  In order to do this, it
is convenient to define a non-hermitian field,
\begin{equation}
\ph^+(\x,0) \equiv \int\dk\;e^{i\k\cdot\x}\;a(\k) 
\label{php}
\end{equation}
and its hermitian conjugate
\begin{equation}
\ph^-(\x,0) \equiv \int\dk\;e^{-i\k\cdot\x}\;\ad(\k) \;.
\label{phm}
\end{equation}
These are then time-evolved with $H_0$:
\begin{eqnarray}
\ph^+(\x,t) &=& e^{iH_0t}\ph^+(\x,0)e^{-iH_0t} = \int\dk\;e^{ikx}\;a(\k) \;,
\nonumber \\
\ph^-(\x,t) &=& e^{iH_0t}\ph^-(\x,0)e^{-iH_0t} = \int\dk\;e^{-ikx}\;\ad(\k) \;.
\label{phpm2}
\end{eqnarray}
Note that the usual hermitian free field $\ph(x)$ is just the sum of these:
$\ph(x)=\ph^+(x)+\ph^-(x)$.  

For a proper orthochronous Lorentz transformation $\Lam$, we have
\begin{equation}
U(\Lam)^{-1}\ph(x)U(\Lam) = \ph(\Lam^{-1}x) \;.
\label{uphu4}
\end{equation}
This implies that the particle creation and annihilation operators
transform as
\begin{eqnarray}
U(\Lam)^{-1}a(\k)U(\Lam) &=& a(\Lam^{-1}\k) \;,
\nonumber \\
\noalign{\smallskip}
U(\Lam)^{-1}\ad(\k)U(\Lam) &=& \ad(\Lam^{-1}\k) \;.
\label{uau}
\end{eqnarray}
This, in turn, implies that $\ph^+(x)$ and $\ph^-(x)$ are Lorentz scalars:
\begin{equation}
U(\Lam)^{-1}\ph^\pm(x)U(\Lam) = \ph^\pm(\Lam^{-1}x) \;.
\label{uphpmu}
\end{equation}
We will then have local, Lorentz invariant interactions if we take
the interaction lagrangian density ${\cal L}_1$ to be a hermitian
function of $\ph^+(x)$ and $\ph^-(x)$.

To proceed we need to recall some facts about time-dependent
perturbation theory in quantum mechanics.  
The transition amplitude $\Tfi$
to start with an initial state $|i\ra$ at time $t=-\infty$
and end with a final state $|f\ra$ at time $t=+\infty$ is
\begin{equation}
\Tfi = \la f|\,{\rm T}\exp\!\left[-i\int_{-\infty}^{+\infty}dt\;
                                  H_I(t)\right]|i\ra \;,
\label{tfi}
\end{equation}
where $H_I(t)$ is the perturbing hamiltonian in the {\it interaction picture},
\begin{equation}
H_I(t) = \exp(+iH_0t)H_1\exp(-iH_0t) \;,
\label{hi}
\end{equation}
$H_0$ is the unperturbed hamiltonian,
and $\rm T$ is the {\it time ordering symbol\/}: a product of operators
to its right is to be ordered, not as written, but with operators at
later times to the left of those at earlier times.
Using \eq{phpm2}, we can write
\begin{equation}
H_I(t) = \int \dtx\;\H_I(x) \;,
\label{h23}
\end{equation}
where $\H_I(x)$ is an ordinary function of $\ph^+(x)$ and $\ph^-(x)$.

Here is the key point:
for the transition amplitude $\Tfi$
to be Lorentz invariant, the time ordering must be {\it frame independent}.
The time ordering of two spacetime points $x$ and $x'$ is frame independent if
their separation is {\it timelike}; this means that $(x-x')^2 < 0$.  
Two spacetime
points whose separation is {\it spacelike}, $(x-x')^2>0$, can have different
temporal ordering in different frames.
In order to avoid $\Tfi$ being different in different frames,
we must then require
\begin{equation}
\Bigl[\H_I(x),\H_I(x')\Bigr] = 0 \quad {\rm whenever} \quad (x-x')^2>0\;.
\label{hh}
\end{equation}
Obviously, 
$[\ph^+(x),\ph^+(x')]_\mp=[\ph^-(x),\ph^-(x')]_\mp=0$.  However,
\begin{eqnarray}
[\ph^+(x),\ph^-(x')]_\mp &=& \int\dk\;\dk'\;e^{i(kx-k'x')}[a(\k),\ad(\k')]_\mp
\nonumber \\
&=& \int\dk\;e^{ik(x-x')}
\nonumber \\
&=& {m\over 4\pi^2 r}\,K_1(mr) 
\nonumber \\
\noalign{\medskip}
&\equiv& C(r)\;.
\label{badcomm}
\end{eqnarray}
In the next-to-last line, 
we have taken $(x-x')^2 = r^2 > 0$, and $K_1(z)$ is the modified
Bessel function.  (This Lorentz-invariant integral is most easily
evaluated in the frame where $t'=t$.)
The function $C(r)$ is {\it not\/} zero for any $r>0$.
(Not even when $m=0$; in this case, $C(r)=1/4\pi^2r^2$.)  
On the other hand, $\H_I(x)$
must involve both $\ph^+(x)$ and $\ph^-(x)$, by hermiticity.  
Thus, generically, we will not be able to satisfy \eq{hh}.  

To resolve this problem, let us try using only particular linear combinations of 
$\ph^+(x)$ and $\ph^-(x)$.  Define
\begin{eqnarray}
\ph_\lam(x) &\equiv& \ph^+(x) + \lam\ph^-(x) \;,
\nonumber \\
\ph^\dagger_\lam(x) &\equiv& \ph^-(x) + \lam^*\ph^+(x) \;,
\label{philam}
\end{eqnarray}
where $\lam$ is an arbitrary complex number.  We then have
\begin{eqnarray}
[\ph_\lam(x),\ph^\dagger_\lam(x')]_\mp 
&=& [\ph^+(x),\ph^-(x')]_\mp + |\lam|^2 [\ph^-(x),\ph^+(x')]_\mp
\nonumber \\
\noalign{\medskip}
&=& (1 \mp |\lam|^2)\,C(r) 
\label{c1lam}
\end{eqnarray}
and
\begin{eqnarray}
[\ph_\lam(x),\ph_\lam(x')]_\mp 
&=& \lam[\ph^+(x),\ph^-(x')]_\mp + \lam[\ph^-(x),\ph^+(x')]_\mp
\nonumber \\
\noalign{\medskip}
&=& \lam(1 \mp 1)\,C(r) \;. 
\label{c2lam}
\end{eqnarray}
Thus, if we want $\ph_\lam(x)$ to either commute or anticommute
with both $\ph_\lam(x')$ and $\ph^\dagger_\lam(x')$ at spacelike separations,
we must choose $|\lam|=1$, {\it and\/} we must choose commutators.
Then (and only then), we can build a suitable $\H_I(x)$ by making it a 
hermitian function of $\ph_\lam(x)$.  

But this has simply returned us to the theory of a real scalar field, because,
for $\lam=e^{i\alpha}$, $e^{-i\alpha/2}\ph_\lam(x)$ is hermitian.  In fact,
if we make the replacement $a(\k)\to e^{i\alpha/2}a(\k)$ (which does not
change the commutation relations of these operators), then
$e^{-i\alpha/2}\ph_\lam(x)=\ph(x)=\ph^+(x)+\ph^-(x)$.
Thus, our attempt to start with the creation and annihilation operators
$a^\dagger(\k)$ and $a(\k)$ as the fundamental objects has simply led us back
to the real, commuting, scalar field $\ph(x)$ as the fundamental object.

Let us return to thinking of $\ph(x)$ as fundamental, with a lagrangian density
specified by some function of the Lorentz scalars $\ph(x)$ and 
$\d^\mu\ph(x)\d_\mu\ph(x)$.  Then, quantization will result in 
$[\ph(x),\ph(x')]_\mp=0$ for $t=t'$.  If we choose anticommutators,
then $[\ph(x)]^2=0$ and $[\d_\mu\ph(x)]^2=0$, resulting in $\L=0$.
This clearly does not make sense.

This situation turns out to generalize to fields of higher spin, in any
number of spacetime dimensions.  One choice of quantization (commutators
or anticommutators) always leads to vanishing $\L$ (or to an $\L$
that is a total derivative), and this choice is disallowed.  Furthermore, the 
allowed choice is always commutators for fields of integer spin,
and anticommutators for fields of half-integer spin.  
If we try treating the particle creation
and annihilation operators as fundamental, rather than the fields, we find
a situation similar to that of the spin-zero case, and are led to the
reconstruction of a field that must obey the appropriate quantization scheme.

\vfill\eject

%% file: ch005.tex
\noindent Quantum Field Theory  \hfill   Mark Srednicki

\vskip0.5in

\begin{center}
\large{5: The LSZ Reduction Formula}
\end{center}
\begin{center}
Prerequisite: 3
\end{center}

\vskip0.5in

Let us now consider how to construct appropriate initial and final states for
scattering experiments.  
In the free theory, we can create a state of one particle
by acting on the vacuum state with the creation operator:
\begin{equation}
|k\ra = \ad(\k)|0\ra\;,
\label{1k}
\end{equation}
where 
\begin{equation}
\ad(\k) = -i\int\dtx\;e^{ikx}\!\buildrel\leftrightarrow\over{\d_0}\!\ph(x)\;.
\label{adag}
\end{equation}
Recall that $\ad(\k)$ is time independent in the free theory.
The state $|k\ra$ has the Lorentz-invariant normalization
\begin{equation}
\la k|k'\ra = (2\pi)^3\,2\w\,\delta^3(\k-\k')\;,
\label{norm}
\end{equation}
where $\w=(\k^2+m^2)^{1/2}$.

Let us consider an operator that (in the free theory) creates a particle 
localized in
momentum space near $\k_1$, and localized in position space near the origin:
\begin{equation}
\ad_1 \equiv \int \dtk \; f_1(\k) \ad(\k)\;,
\label{adag1}
\end{equation}
where
\begin{equation}
f_1(\k) \propto \exp[-(\k-\k_1)^2/4\sigma^2]
\label{f1}
\end{equation}
is an appropriate wave packet, and $\sigma$ is its width in momentum space.
If we time evolve (in the Schr\"odinger picture)
the state created by this time-independent operator,
then the wave packet will propagate (and spread out).  
The particle will thus be localized
far from the origin as $t\to\pm\infty$.  
If we consider instead an initial state of the form
$|i\ra = \ad_1\ad_2|0\ra$, where
$\k_1\ne\k_2$, then the two particles are widely separated in the 
far past.  

Let us guess that this still works in the interacting theory.
One complication is that $\ad(\k)$ will no longer be time independent,
and so $\ad_1$, \eq{adag1}, becomes time dependent as well.
Our guess for a suitable initial state of a scattering experiment is then
\begin{equation}
|i\ra = \lim_{t\to-\infty}\ad_1(t)\ad_2(t)|0\ra \;.
\label{init}
\end{equation}
By appropriately normalizing the wave packets, we can make $\la i|i\ra=1$,
and we will assume that this is the case.
Similarly, we can consider a final state
\begin{equation}
|f\ra = \lim_{t\to+\infty}\ad_{1'}(t)\ad_{2'}(t)|0\ra \;,
\label{final}
\end{equation}
where $\k'_1\ne\k'_2$, and $\la f|f\ra=1$.
This describes two widely separated particles in the far future.
(We could also consider acting with more creation operators, if 
we are interested in the production of some
extra particles in the collision of two.)
Now the scattering amplitude is simply given by $\la f|i\ra$.

We need to find a more useful expression for $\la f|i\ra$.  
To this end, let us note that
\begin{eqnarray}
\ad_1(+\infty) - \ad_1(-\infty) &=& \int_{-\infty}^{+\infty} dt\;\d_0\ad_1(t) 
\nonumber \\
&=& -i\int \dtk\;f_1(\k) \int \dfx\;\d_0\Bigl(
e^{ikx}\!\buildrel\leftrightarrow\over{\d_0}\!\ph(x)\Bigr)
\nonumber \\
&=& -i\int \dtk\;f_1(\k) 
\int \dfx\; e^{ikx}(\d_0^2+\w^2)\ph(x)
\nonumber \\
&=& -i\int \dtk\;f_1(\k) 
\int \dfx\; e^{ikx}(\d_0^2+\k^2+m^2)\ph(x)
\nonumber \\
&=& -i\int \dtk\;f_1(\k) \int \dfx\; e^{ikx}
(\d_0^2 - {\buildrel\leftarrow\over\nabla}{}^2 +m^2)\ph(x)
\nonumber \\
&=& -i\int \dtk\;f_1(\k) \int \dfx\; e^{ikx}
(\d_0^2 - {\buildrel\rightarrow\over\nabla}{}^2 +m^2)\ph(x)
\nonumber \\
&=& -i\int \dtk\;f_1(\k) \int \dfx\; e^{ikx}(-\d^2 +m^2)\ph(x) \;.\qquad
\label{addiff}
\end{eqnarray}
The first equality is just the fundamental theorem of calculus.
To get the second, we substituted the definition of $\ad_1(t)$,
and combined the $\dtx$ from this definition with the $dt$ to get $\dfx$.
The third comes from straightforward evaluation of the time derivatives.  
The fourth uses $\w^2=\k^2+m^2$.
The fifth writes $\k^2$ as $-\nabla^2$ acting on $e^{i\k\cdot\x}$.
The sixth uses integration by parts to move the $\nabla^2$ onto
the field $\ph(x)$; here the wave packet is needed to avoid
a surface term.  The seventh simply identifies $\d_0^2-\nabla^2$
as $-\d^2$.  

In free-field theory, the right-hand side of \eq{addiff} is zero,
since $\ph(x)$ obeys the Klein-Gordon equation.  In an interacting theory,
with (say) $\L_1 = {1\over6}g\ph^3$, we have instead
$(-\d^2+m^2)\ph={1\over2}g\ph^2$.  Thus the right-hand side of \eq{addiff}
is not zero in an interacting theory.

Rearranging \eq{addiff}, we have
\begin{equation}
\ad_1(-\infty) = \ad_1(+\infty) + i\int \dtk\;f_1(\k) 
\int \dfx\; e^{ikx}(-\d^2 +m^2)\ph(x) \;.
\label{addiff2}
\end{equation}
We will also need the hermitian conjugate of this formula, which (after
a little more rearranging) reads
\begin{equation}
a_1(+\infty) = a_1(-\infty) + i\int \dtk\;f_1(\k) 
\int \dfx\; e^{-ikx}(-\d^2 +m^2)\ph(x) \;.
\label{adiff}
\end{equation}

Let us return to the scattering amplitude,
\begin{equation}
\la f|i\ra = \la0|a_{1'}(+\infty)a_{2'}(+\infty)
                  \ad_1(-\infty)\ad_2(-\infty)|0\ra \;.
\label{fi}
\end{equation}
Note that the operators are in time order.  Thus, if we feel like it, we can 
put in a {\it time-ordering symbol\/} without changing anything:
\begin{equation}
\la f|i\ra = \la0|{\rm T}a_{1'}(+\infty)a_{2'}(+\infty)
                  \ad_1(-\infty)\ad_2(-\infty)|0\ra \;.
\label{fit}
\end{equation}
The symbol T means the product of operators
to its right is to be ordered, not as written, but with operators at
later times to the left of those at earlier times.  

Now let us use \eqs{addiff2} and (\ref{adiff}) in \eq{fit}.  The time-ordering
symbol automatically moves all $a_{i'}(-\infty)$'s to the right, where they
annihilate $|0\ra$.  Similarly, all $\ad_i(+\infty)$'s move to the left,
where they annihilate $\la 0|$.  

The wave packets no longer play a key role, and we can take the $\sigma\to0$
limit in \eq{f1}, so that $f_1(\k)=\delta^3(\k-\k_1)$.  
The initial and final states now have a delta-function normalization,
the multiparticle generalization of \eq{norm}.  We are left with 
\begin{eqnarray}
\la f|i\ra &=& 
i^{n+n'}\int d^{4\!}x_1\,e^{ik_1x_1}(-\d^2_1+m^2)\ldots
\nonumber \\
&&\qquad\quad\,
d^{4\!}x'_1\,e^{-ik'_1x'_1}(-\d^2_{1'}+m^2)\ldots
\nonumber \\
\noalign{\smallskip}
&&\qquad\quad\,\times
\la0|{\rm T}\ph(x_1)\ldots\ph(x'_1)\ldots|0\ra \;.
\label{lsz0}
\end{eqnarray}
This formula has been written to apply to the 
more general case of $n$ incoming particles and $n'$ outgoing particles;
the ellipses stand for similar factors for each of the other incoming and
outgoing particles.  

\Eq{lsz0} is the {\it Lehmann-Symanzik-Zimmerman reduction formula}, or LSZ 
formula for short.  It is one of the key equations of quantum field theory.

However, we cheated a little in our derivation of the LSZ formula, because
we assumed that the creation operators of {\it free\/} field theory would
work comparably in the {\it interacting\/} theory.  This is a rather suspect
assumption, and so we must review it. 

\begin{figure}
\begin{center}
\epsfig{file=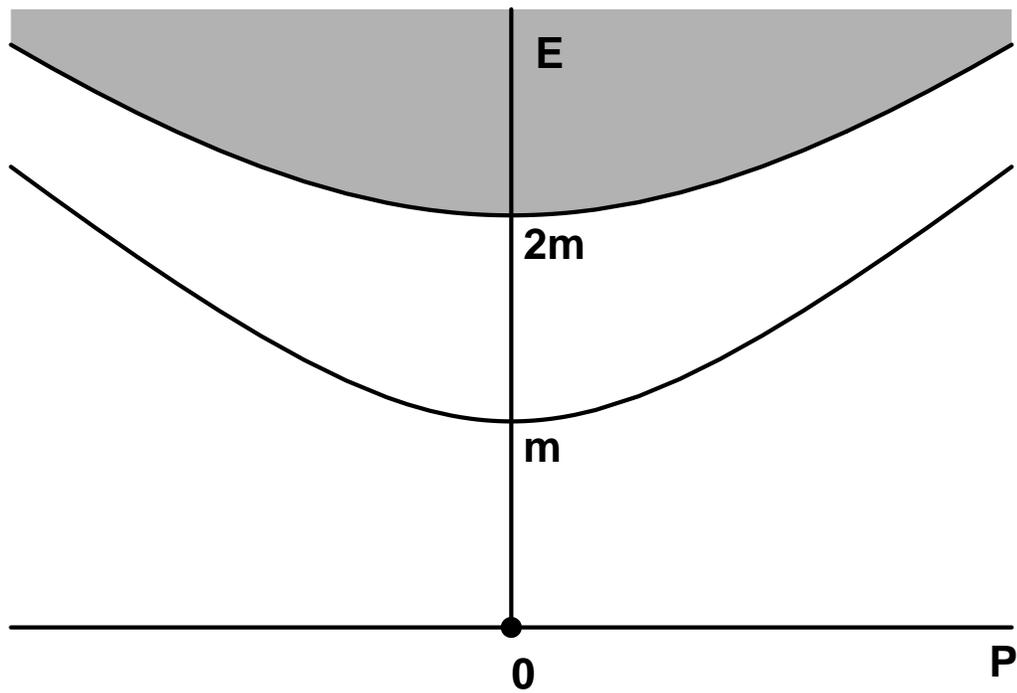}
\end{center}
\caption{The exact energy eigenstates in the $({\bf P},E)$ plane.
The ground state is isolated at $({\bf 0},0)$, the one-particle
states form an isolated hyperbola that passes through 
$({\bf 0},m)$, and the multi-particle continuum lies at and 
above the hyperbola that passes through $({\bf 0},2m)$.}
\label{lszfig}
\end{figure}

Let us consider what we can deduce about the energy and momentum eigenstates
of the interacting theory on physical grounds.  
First, we assume that there is a unique ground state $|0\ra$, with zero
energy and momentum.
The first excited state is a state of a single particle with mass $m$.
This state can have an arbitrary three-momentum $\k$, and then has energy
$E=\w=(\k^2+m^2)^{1/2}$.   
The next excited state is that of two particles.  These two particles could
form a bound state with energy {\it less\/} than $2m$ (like the hydrogen
atom in quantum electrodynamics), 
but, to keep things simple, let us assume that there
are no such bound states.  Then the lowest possible energy of a two-particle
state is $2m$.  However, a two-particle state with zero total
three-momentum can have {\it any\/}  energy above $2m$, because
the two particles could have some {\it relative\/} momentum that contributes
to their total energy.
Thus we are led to a picture of the states of theory as shown in \fig{lszfig}.

Now let us consider what happens when we act on the ground state with
the field operator $\ph(x)$.  To this end, it is helpful to write
\begin{equation}
\ph(x)=\exp(-iP^\mu x_\mu)\ph(0)\exp(+iP^\mu x_\mu)\;,
\label{ev}
\end{equation}
where $P^\mu$ is the energy-momentum four-vector.  (This equation,
introduced in section 2, is just the relativistic generalization 
of the Heisenberg equation.)  Now let us sandwich $\ph(x)$ between
the ground state (on the right), and other possible states (on the left).
For example, let us put the ground state on the left as well.  Then we have
\begin{eqnarray}
\la0|\ph(x)|0\ra &=&\la0|e^{-iPx}\ph(0)e^{+iPx}|0\ra
\nonumber \\
\noalign{\medskip}
&=&\la0|\ph(0)|0\ra \;.
\label{vev}
\end{eqnarray}
To get the second line, we used $P^\mu|0\ra=0$.  The final expression is
just a Lorentz-invariant number.  Since $|0\ra$ is the exact ground
state of the interacting theory, we have (in general) no idea what
this number is.

We would like $\la0|\ph(0)|0\ra$ to be zero.  This is because we would like
$\ad_1(\pm\infty)$, when acting on $|0\ra$, to create a single particle
state.  We do {\it not\/} want $\ad_1(\pm\infty)$ to create a linear
combination of a single particle state and the ground state.  But this
is precisely what will happen if $\la0|\ph(0)|0\ra$ is not zero.

So, if $v \equiv \la0|\ph(0)|0\ra$ is not zero, we will shift the field
$\ph(x)$ by the constant $v$.  This means that we go back to the lagrangian, 
and replace $\ph(x)$ everywhere by $\ph(x)+v$.  This is just a change of 
the name of the operator of interest, and does not affect the physics.  
However, the shifted $\ph(x)$ obeys $\la 0|\ph(x)|0\ra = 0$.  

Let us now consider $\la p|\ph(x)|0\ra$, where $|p\ra$ is a one-particle
state with four-momentum $p$, normalized according to \eq{norm}.
Again using \eq{ev}, we have
\begin{eqnarray}
\la p|\ph(x)|0\ra &=&\la p|e^{-iPx}\ph(0)e^{+iPx}|0\ra
\nonumber \\
\noalign{\medskip}
&=&e^{-ipx}\la p|\ph(0)|0\ra \;,
\label{1ev}
\end{eqnarray}
where $\la p|\ph(0)|0\ra$ is a Lorentz-invariant number.  It is
a function of $p$, but the only Lorentz-invariant functions of $p$
are functions of $p^2$, and $p^2$ is just the constant $-m^2$.  So
$\la p|\ph(0)|0\ra$ is just some number that depends on $m$ and
(presumably) the other parameters in the lagrangian.

We would like $\la p|\ph(0)|0\ra$ to be one.  
That is what it is in free-field theory, 
and we know that, in free-field theory, $\ad_1(\pm\infty)$
creates a correctly normalized one-particle state.  Thus, for
$\ad_1(\pm\infty)$ to create a correctly normalized one-particle state
in the interacting theory, we must have
$\la p|\ph(0)|0\ra=1$.

So, if $v \equiv \la0|\ph(0)|0\ra$ is not zero, we will shift the field
$\ph(x)$ by the constant $v$.  This means that we go back to the lagrangian, 
and replace $\ph(x)$ everywhere by $\ph(x)+v$.  This is just a change of 
the name of the operator of interest, and does not affect the physics.  
However, the shifted $\ph(x)$ obeys $\la 0|\ph(x)|0\ra = 0$.

So, if $\la p|\ph(0)|0\ra$ is not equal to one, we will rescale 
(or, one might say, {\it renormalize\/}) $\ph(x)$ by a multiplicative
constant.  This is just a change of the name of the operator of interest,
and does not affect the physics.  However, the rescaled
$\ph(x)$ obeys $\la p|\ph(0)|0\ra=1$.

Finally, consider $\la p,n|\ph(x)|0\ra$, where $|p,n\ra$ is a 
multiparticle state with total four-momentum $p$, and $n$ is short
for all other labels (such as relative momenta) needed to specify
this state.  We have
\begin{eqnarray}
\la p,n|\ph(x)|0\ra &=&\la p,n|e^{-iPx}\ph(0)e^{+iPx}|0\ra
\nonumber \\
\noalign{\smallskip}
&=&e^{-ipx}\la p,n|\ph(0)|0\ra 
\nonumber \\
\noalign{\smallskip}
&=&e^{-ipx}A_n(\p) \;,
\label{mvev}
\end{eqnarray}
where $A_n(\p)$ is a function of Lorentz invariant products of
the various (relative and total) four-momenta needed to specify the state.
Note that, from \fig{lszfig}, $p^0=(\p^2+M^2)^{1/2}$ 
with $M\ge 2m$.  The invariant mass $M$ is one of the parameters
included in the set $n$.

We would like 
$\la p,n|\ph(x)|0\ra$ to be zero, because we would like
$\ad_1(\pm\infty)$, when acting on $|0\ra$, to create a single particle
state.  We do {\it not\/} want $\ad_1(\pm\infty)$ to create any multiparticle
states.  But this is precisely what may happen if $\la p,n|\ph(x)|0\ra$
is not zero.

Actually, we are being a little too strict.  We really need
$\la p,n|\ad_1(\pm\infty)|0\ra$ to be zero, and perhaps it will be zero
even if $\la p,n|\ph(x)|0\ra$ is not.
Also, we really should test 
$\ad_1(\pm\infty)|0\ra$ only against {\it normalizable\/} states.
Mathematically, non-normalizable states cause all sorts of trouble;
mathematicians don't consider them to be states at all.
In physics, this usually doesn't bother us, but here we must be
especially careful.  So let us write 
\begin{equation}
|\psi\ra = \sum_n\int \dtp\;\psi_n(\p)|p,n\ra \;,
\label{mpsi}
\end{equation}
where the $\psi_n(\p)$'s are wave packets for the total three-momentum $\p$.
Note that \eq{mpsi} is highly schematic; the sum over $n$ is shorthand
for integrals over various continuous parameters (relative momenta).

Now we want to examine 
\begin{equation}
\la\psi|\ad_1(t)|0\ra
= -i\sum_n\int\dtp\;\psi^*_n(\p)\int\dtk\;f_1(\k)
\int\dtx\;e^{ikx} \!\buildrel\leftrightarrow\over{\d_0}\!
\la p,n|\ph(x)|0\ra \;.
\label{mpsi0}
\end{equation}
We will take the limit $t\to\pm\infty$ in a moment.
Using \eq{mvev}, \eq{mpsi0} becomes
\begin{eqnarray}
\la\psi|\ad_1(t)|0\ra
&=& 
-i\sum_n\int\dtp\;\psi^*_n(\p)\int\dtk\;f_1(\k)
\int\dtx\Bigl(e^{ikx}\!\buildrel\leftrightarrow\over{\d_0}\!e^{-ipx}\Bigr)
A_n(\p)
\nonumber \\
&=& 
\sum_n\int\dtp\;\psi^*_n(\p)
\int\dtk\;f_1(\k)
\int\dtx\,(p^0{+}k^0)e^{i(k-p)x}A_n(\p) \;.
\nonumber \\
&& \label{mpsi1}
\end{eqnarray}
Next we use $\int\dtx\,e^{i(\k-\p)\cdot\x}=(2\pi)^3\delta^3(\k-\p)$ to get
\begin{equation}
\la\psi|\ad_1(t)|0\ra =
\sum_n\int\dtp\;(2\pi)^3(p^0{+}k^0)\psi^*_n(\p)f_1(\p)A_n(\p)
e^{i(p^0-k^0)t} \;,
\label{mpsi2}
\end{equation}
where $p^0=(\p^2+M^2)^{1/2}$ and $k^0=(\p^2+m^2)^{1/2}$.

Now comes the key point.  Note that $p^0$ is strictly
greater than $k^0$, because $M\ge 2m$.  Thus the integrand of
\eq{mpsi2} contains a phase factor that oscillates more and more
rapidly as $|\p|\to\infty$.  Therefore, by the
{\it Riemann-Lebesgue lemma}, the right-hand side of \eq{mpsi2} vanishes
as $t\to\pm\infty$.  

Physically, this means that a one-particle wave packet spreads out
differently than a multiparticle wave packet, and the overlap between
them goes to zero as the elapsed time goes to infinity.  Thus,
even though our operator $\ad_1(t)$ creates some multiparticle states
that we don't want, we can ``follow'' the one-particle state that we
do want by using an appropriate wave packet.  By waiting long enough,
we can make the multiparticle contribution to the scattering amplitude
as small as we like.

Let us recap.  The basic formula for a scattering amplitude in terms
of the fields of an interacting quantum field theory is the LSZ formula,
which is worth writing down again:
\begin{eqnarray}
\la f|i\ra &=& 
i^{n+n'}\int \dfx_1\,e^{ik_1x_1}(-\d^2_1+m^2)\ldots
\nonumber \\
&&\qquad\quad\,
\dfx_{1'}\,e^{-ik'_1x'_1}(-\d^2_{1'}+m^2)\ldots
\nonumber \\
\noalign{\smallskip}
&&\qquad\quad\,\times
\la0|{\rm T}\ph(x_1)\ldots\ph(x'_1)\ldots|0\ra \;.
\label{lsz}
\end{eqnarray}
The LSZ formula is valid {\it provided\/} that the field obeys 
\begin{equation}
\la 0|\ph(x)|0\ra = 0 \qquad \hbox{and} \qquad
\la k|\ph(x)|0\ra = e^{-ikx} \;.
\label{cond}
\end{equation}
These normalization conditions may conflict with our original choice
of field and parameter normalization in the lagrangian.  
Consider, for example, a lagrangian originally specified as
\begin{equation}
\L=-\half\d^\mu\ph\d_\mu\ph-\half m^2\ph^2 + {\ts{1\over6}}g\ph^3 \;.
\label{lorg}
\end{equation}
After shifting and rescaling (and renaming some parameters), we will have
instead
\begin{equation}
\L=-\half Z_\ph\d^\mu\ph\d_\mu\ph-\half Z_m m^2\ph^2 + 
{\ts{1\over6}}Z_g  g\ph^3 + Y\ph\;.
\label{lnew}
\end{equation}
Here the three $Z$'s and $Y$ are as yet unknown constants.
They must be chosen to ensure the validity of \eq{cond};
this gives us two conditions in four unknowns.
We also require that the parameter $m$ in $\L$ be the
actual, physical mass of the particle. 
Finally, the parameter $g$ is fixed in terms of some
particular scattering cross section by some definite forumla.
(For example, in quantum electrodynamics, the parameter
analogous to $g$ is the electron charge $e$.  The low-energy
Coulomb scattering cross section is proportional to $e^2$,
with a definite constant of proportionality and
no higher-order corrections; this relationship defines $e$.)
Thus we have four conditions
in four unknowns, and it is possible to calculate $Y$ and the
three $Z$'s order by order in perturbation theory.

Next, we must develop the tools needed to compute the correlation functions
$\la0|{\rm T}\ph(x_1)\ldots|0\ra$ in an interacting quantum field theory.

\vskip0.5in

\begin{center}
Problems
\end{center}

\vskip0.25in

5.1) Work out the LSZ reduction formula for the complex scalar field that
was introduced in problem 3.3.  Note that we must specify the type ($a$ or $b$) of
each incoming and outgoing particle. 

\vfill\eject

%% file: ch006.tex
\noindent Quantum Field Theory  \hfill   Mark Srednicki

\vskip0.5in

\begin{center}
\large{6: Path Integrals in Quantum Mechanics}
\end{center}
\begin{center}
Prerequisite: none
\end{center}

\vskip0.5in

Consider the nonrelativistic quantum mechanics of one particle in one dimension;
the hamiltonian is
\begin{equation}
H(P,Q) = {\ts{1\over2m}}P^2 + V(Q)\;,
\label{h}
\end{equation}
where $P$ and $Q$ are operators obeying $[Q,P]=i$.  (We set $\hbar=1$ for
notational convenience.)  
We wish to evaluate the probability amplitude for the particle to start at
position $q'$ at time $t'$, and end at position $q''$ at time $t''$.  
This amplitude is
$\la q''|e^{-iH(t''-t')}|q'\ra$,
where $|q'\ra$ and $|q''\ra$ are eigenstates of the position operator $Q$.

We can also formulate this question in the Heisenberg picture, where operators
are time dependent and the state of the system is time independent, as opposed
to the more familiar Schr\"odinger picture.  In the Heisenberg picture, 
we write 
$Q(t) = e^{iHt}Qe^{-iHt}$.
We can then define an instantaneous eigenstate of $Q(t)$ via
$Q(t)|q,t\ra = q|q,t\ra$.
These instantaneous eigenstates can be expressed explicitly as
$|q,t\ra = e^{+iHt}|q\ra$,
where $Q|q\ra=q|q\ra$.  Then our transition amplitude can be written as
$\la q'',t''|q',t'\ra$ 
in the Heisenberg picture.

To evaluate $\la q'',t''|q',t'\ra$,
we begin by dividing the time interval $T\equiv t''-t'$ into
$N+1$ equal pieces of duration $\delta t=T/(N+1)$.
Then introduce $N$ complete sets of position eigenstates to get
\begin{equation}
\la q'',t''|q',t'\ra
= \int\prod_{j=1}^N dq_j\;
           \la q''|e^{-iH\delta t}|q_N\ra
           \la q_N|e^{-iH\delta t}|q_{N-1}\ra 
           \ldots
           \la q_1|e^{-iH\delta t}|q'\ra \;.
\label{A2}
\end{equation}
The integrals over the $q$'s all run from $-\infty$ to $+\infty$.

Now consider $\la q_2|e^{-iH\delta t}|q_1\ra$.  We can use the 
Campbell-Baker-Hausdorf formula
\begin{equation}
\exp(A+B) = \exp(A)\exp(B)\exp(-\half[A,B] + \ldots)
\label{cbh}
\end{equation}
to write
\begin{equation}
\exp(-iH\delta t) = \exp[-i(\delta t/2m)P^2]
                    \exp[-i\delta tV(Q)]
                    \exp[O(\delta t^2)] \;.
\label{cbh2}
\end{equation}
Then, in the limit of small $\delta t$, we should be able to ignore the 
final exponential.
Inserting a complete set of momentum states then gives 
\begin{eqnarray}
\la q_2|e^{-iH\delta t}|q_1\ra &=& \int dp_1 \; 
          \la q_2|e^{-i(\delta t/2m)P^2}|p_1\ra
          \la p_1|e^{-i\delta tV(Q)}|q_1\ra
\nonumber \\
&=& \int dp_1 \; e^{-i(\delta t/2m)p_1^2}\,e^{-i\delta tV(q_1)}\,
             \la q_2|p_1\ra\la p_1|q_1\ra
\nonumber \\
&=& \int {dp_1\over2\pi} \; e^{-i(\delta t/2m)p_1^2}\,e^{-i\delta tV(q_1)}
         \,e^{ip_1(q_2-q_1)} \;.
\nonumber \\
&=& \int {dp_1\over2\pi} \; e^{-iH(p_1,q_1)\delta t}\,e^{ip_1(q_2-q_1)} \;.
\label{A3}
\end{eqnarray}
To get the third line, we used $\la q|p\ra = (2\pi)^{\sss -1/2}\exp(ipq)$.  

If we happen to be interested in more general hamiltonians than \eq{h}, 
then we must worry about the ordering of the $P$ and $Q$ operators in any
term that contains both.  If we adopt {\it Weyl ordering}, where
the quantum hamiltonian $H(P,Q)$ is given in terms of the classical 
hamiltonian $H(p,q)$ by
\begin{equation}
H(P,Q) \equiv \int {dx\over2\pi}\,{dk\over2\pi}\,
e^{ixP + ikQ} \int dp\,dq\,e^{-ixp-ikq}\,H(p,q)\;,
\label{hweyl}
\end{equation}
then \eq{A3} is not quite correct; in the last line,
$H(p_1,q_1)$ should be replaced with $H(p_1,\bar q)$,
where $\bar q=\half(q_1+q_2)$.  For the hamiltonian of \eq{h},
which is Weyl ordered,
this replacement makes no difference in the limit $\delta t\to 0$.

Adopting Weyl ordering for the general case, we now have
\begin{equation}
\la q'',t''|q',t'\ra
= \int dp_0\prod_{j=1}^N {dp_j dq_j\over2\pi}\;
                e^{ip_j(q_{j+1}-q_j)}\,e^{-iH(p_j,\bar q_j)\delta t} \;,
\label{A4}
\end{equation}
where $\bar q_j=\half(q_{j+1}+q_j)$.  
If we now define $\dot q_j \equiv (q_{j+1}-q_j)/\delta t$, and take the
formal limit of $\delta t\to0$, then 
\begin{equation}
\la q'',t''|q',t'\ra
= \int \D p\,\D q\;\exp\!\left[i\int_{t'}^{t''} dt
                \Bigl(p(t)\dot q(t) - H(p(t),q(t))\Bigr)\right].
\label{A5}
\end{equation}
The integration is to be understood as over all paths in phase space that 
start at $q(t')=q'$ (with an arbitrary value of the initial momentum) and end at
$q(t'')=q''$ (with an arbitrary value of the final momentum).  

If $H(p,q)$ is no more than quadratic in the momenta 
[as is the case for \eq{h}],
then the integral over $p$ is gaussian, and can be done in closed form.
If the term that is quadratic in $p$ is independent of $q$ 
[as is the case for \eq{h}], then
the prefactors generated by the gaussian integrals are all constants, and
can be absorbed into the definition of $\D q$.  
The result of integrating out $p$ is then
\begin{equation}
\la q'',t''|q',t'\ra
= \int \D q\;\exp\!\left[i\int_{t'}^{t''} dt\;L(\dot q(t),q(t))\right],
\label{A6}
\end{equation}
where $L(\dot q,q)$ is computed by first finding
the stationary point of the $p$ integral by solving
\begin{equation}
0 = {\d\over\d p}\Bigl(p\dot q - H(p,q)\Bigr) = \dot q - {\d H(p,q)\over\d p}
\label{p2}
\end{equation}
for $p$ in terms of $\dot q$ and $q$, and then plugging this solution back into
$p\dot q-H$ to get $L$.  We recognize this procedure from classical mechanics:
we are passing from the hamiltonian formulation to the lagrangian formulation. 

Now that we have \eqs{A5} and (\ref{A6}), what are we going to do with them?
Let us begin by considering some generalizations; let us examine, for example,
$\la q'',t''|Q(t_1)|q',t'\ra$, where $t'<t_1<t''$.  This is given by
\begin{equation}
\la q'',t''|Q(t_1)|q',t'\ra=\la q''|e^{-iH(t''-t_1)}Qe^{-iH(t_1-t')}|q'\ra\;. 
\label{qqq}
\end{equation}
In the path integral formula, the extra operator $Q$ inserted at time $t_1$
will simply result in an extra factor of $q(t_1)$.  Thus 
\begin{equation}
\la q'',t''|Q(t_1)|q',t'\ra = \int \D p\,\D q\;q(t_1)\,e^{iS}\;,
\label{qqq2}
\end{equation}
where $S=\int_{t'}^{t''}dt\,(p\dot q - H)$.
Now let us go in the other direction; consider
$\int \D p\,\D q\,q(t_1)q(t_2)e^{iS}$.  This clearly requires the operators
$Q(t_1)$ and $Q(t_2)$, but their order depends on whether $t_1<t_2$
or $t_2<t_1$.  Thus we have
\begin{equation}
\int\D p\,\D q\;q(t_1)q(t_2)\,e^{iS} 
= \la q'',t''|{\rm T}Q(t_1)Q(t_2)|q',t'\ra \;.
\label{qqq3}
\end{equation}
where T is the {\it time ordering symbol\/}: a product of operators
to its right is to be ordered, not as written, but with operators at
later times to the left of those at earlier times.
This is significant, because time-ordered products enter into
the LSZ formula for scattering amplitudes.

To further develop these methods, 
we need another trick: {\it functional derivatives}.
We define the functional derivative $\delta/\delta f(t)$ via
\begin{equation}
{\delta\over\delta f(t_1)}\,f(t_2) = \delta(t_1-t_2) \;,
\label{fd}
\end{equation}
where $\delta(t)$ is the Dirac delta function.  Also, functional derivatives
are defined to satisfy all the usual rules of derivatives (product rule,
chain rule, etc).  \Eq{fd} can be thought of as the continuous generalization
of $(\d/\d x_i)x_j=\delta_{ij}$.

Now, consider modifying the lagrangian of our theory by including
external forces acting on the particle: 
\begin{equation}
H(p,q)\to H(p,q) - f(t)q(t) - h(t)p(t)\;,
\label{ellfh}
\end{equation}
where $f(t)$ and $h(t)$ are specified functions. 
In this case we will write
\begin{equation}
\la q'',t''|q',t'\ra_{f,h} = \int\D p\, \D q\;
\exp\!\left[i\int_{t'}^{t''} dt\,\Bigl(p\dot q - H + fq + hp\Bigr)\right].
\label{qqfh}
\end{equation}
where $H$ is the original hamiltonian.
Then we have
\begin{eqnarray}
{1\over i}\,{\delta\over\delta f(t_1)}\,
\la q'',t''|q',t'\ra_{f,h} 
&=& \int\D p\,\D q\;q(t_1)\,
e^{i\int dt\,[p\dot q-H+fq+hp]} \;,
\nonumber \\
\noalign{\medskip}
{1\over i}\,{\delta\over\delta f(t_1)}\,
{1\over i}\,{\delta\over\delta f(t_2)}\,
\la q'',t''|q',t'\ra_{f,h} 
&=& \int\D p\, \D q\;q(t_1)q(t_2)\,
e^{i\int dt\,[p\dot q-H+fq+hp]}\;, 
\nonumber \\
\noalign{\medskip}
{1\over i}\,{\delta\over\delta h(t_1)}\,
\la q'',t''|q',t'\ra_{f,h} 
&=& \int\D p\,\D q\;p(t_1)\,
e^{i\int dt\,[p\dot q-H+fq+hp]}\;,
\label{fd2}
\end{eqnarray}
and so on.  After we are done bringing down as many factors of $q(t_i)$
or $p(t_i)$ as we like, we can set $f(t)=h(t)=0$, 
and return to the original hamiltonian.  Thus,
\begin{eqnarray}
&& \la q'',t''|{\rm T}Q(t_1)\ldots P(t_n)\ldots|q',t'\ra 
\nonumber \\
&& \qquad {} =
{1\over i}\,{\delta\over\delta f(t_1)}\ldots 
{1\over i}\,{\delta\over\delta h(t_n)}\ldots 
\la q'',t''|q',t'\ra_{f,h}\biggr|_{f=h=0}\;.
\label{qq}
\end{eqnarray}

Suppose we are also interested in initial and final states other than position
eigenstates.  Then we must multiply by the wave functions for these states, 
and integrate.  We will be interested, in particular, in the ground state as both
the initial and final state.  Also, we will take the limits
$t'\to-\infty$ and $t''\to+\infty$.  The object of our attention is then
\begin{equation}
\la 0|0\ra_{f,h}
= \lim_{{t'\to-\infty \atop t''\to+\infty}}
\int dq''\,dq'\;\psi_0^*(q'')\;\la q'',t''|q',t'\ra_{f,h}\;\psi_0(q')\;,
\label{00f}
\end{equation}
where $\psi_0(q)=\la q|0\ra$ is the ground-state wave function.
\Eq{00f} is a rather cumbersome formula, however.  We will, therefore,
employ a trick to simplify it.

Let $|n\ra$ denote an eigenstate of $H$ with eigenvalue $E_n$.  We will
suppose that $E_0=0$; if this is not the case, we will shift $H$ by an
appropriate constant.  Next we write
\begin{eqnarray}
|q',t'\ra &=& e^{iHt'}|q'\ra
\nonumber \\
\noalign{\smallskip}
&=& \sum_{n=0}^\infty e^{iHt'}|n\ra\la n|q'\ra
\nonumber \\
&=& \sum_{n=0}^\infty \psi^*_n(q')e^{iE_n t'}|n\ra \;,
\label{qtp}
\end{eqnarray}
where $\psi_n(q)=\la q|n\ra$ is the wave function of the $n$th eigenstate.
Now, replace $H$ with $(1{-}i\eps)H$ in \eq{qtp}, where $\eps$ is a small
positive infinitesimal.
Then, take the limit $t'\to-\infty$ of \eq{qtp} with $\eps$ held fixed.
Every state except the ground state is then multiplied by a vanishing
exponential factor, and so the limit is simply $\psi^*_0(q')|0\ra$.
Next, multiply by an arbitrary function $\chi(q')$, and integrate over $q'$.
The only requirement is that $\la0|\chi\ra\ne0$.  We then have a constant
times $|0\ra$, and this constant can be absorbed into the normalization of
the path integral.
A similar analysis of $\la q'',t''|=\la q''|e^{-iHt''}$ shows that
the replacement $H\to(1{-}i\eps)H$ also picks out the ground state as the
final state in the $t''\to+\infty$ limit.

What all this means is that if we use
$(1{-}i\eps)H$ instead of $H$, we can be cavalier about the boundary conditions
on the endpoints of the path.  Any reasonable boundary conditions will
result in the ground state as both the initial and final state.  Thus we have
\begin{equation}
\la 0|0\ra_{f,h}
= \int \D p\,\D q\;
\exp\!\left[i\int_{-\infty}^{+\infty} dt\,
\Bigl(p\dot q - (1{-}i\eps)H + fq + hp\Bigr)\right].
\label{00f2}
\end{equation}

Now let us suppose that $H=H_0+H_1$, where we can solve for the eigenstates 
and eigenvalues of $H_0$, and $H_1$ can be treated as a perturbation.  
Suppressing the $i\eps$, \eq{00f2} can be written as
\begin{eqnarray}
\la 0|0\ra_{f,h}
&=& \int \D p\,\D q\;
\exp\!\left[i\int_{-\infty}^{+\infty} dt\,
\Bigl(p\dot q - H_0(p,q)-H_1(p,q) + fq + hp\Bigr)\right]
\nonumber \\
\noalign{\medskip}
&=& \exp\biggl[-i\int_{-\infty}^{+\infty} dt\,
H_1\!\biggl({1\over i}{\delta\over\delta h(t)},
            {1\over i}{\delta\over\delta f(t)}\biggr)\biggr]
\nonumber \\
&&\times 
\int \D p\,\D q\;
\exp\biggl[i\int_{-\infty}^{+\infty} dt\,
\Bigl(p\dot q - H_0(p,q)+ fq + hp\Bigr)\biggr] \;.
\label{00f3}
\end{eqnarray}
To understand the second line of this equation, take the exponential prefactor
inside the path integral.  Then
the functional derivatives (that appear as the arguments of $H_1$)
just pull out appropriate factors of $p(t)$ and $q(t)$, generating
the right-hand side of the first line.
We presumably can compute the functional integral in the second line,
since it involves only the solvable hamiltonian $H_0$.  The exponential prefactor
can then be expanded in powers of $H_1$ to generate a perturbation series.

If $H_1$ depends only on $q$ (and not on $p$), and if we are only
interested in time-ordered products of $Q$'s (and not $P$'s), 
and if $H$ is no more than quadratic in $P$, and if the term
quadratic in $P$ does not involve $Q$, {\it then\/} \eq{00f3}
can be simplified to
\begin{eqnarray}
\la 0|0\ra_f
&=& \exp\biggl[i\int_{-\infty}^{+\infty} dt\,
  L_1\!\biggl({1\over i}{\delta\over\delta f(t)}\biggr)\biggr]
\nonumber \\
&&\times  \int \D q\;
\exp\biggl[i\int_{-\infty}^{+\infty} dt\,
\Bigl(L_0(\dot q,q)+ fq\Bigr)\biggr] \;.
\label{00f4}
\end{eqnarray}
where $L_1(q)=-H_1(q)$.

\vskip0.5in

\begin{center}
Problems
\end{center}

\vskip0.25in

6.1a)  Find an explicit formula for $\D q$ in \eq{A6}.  Your formula
should be of the form $\D q = C\prod_{j=1}^N dq_j$, where $C$ is
a constant that you should compute.

b) For the case of a free particle, $V(Q)=0$, evaluate the
path integral of \eq{A6} explicitly.  Hint: integrate over $q_1$,
then $q_2$, etc, and look for a pattern.  Express you final
answer in terms of $q'$, $t'$, $q''$, $t''$, and $m$.  Restore
$\hbar$ by dimensional analysis.

c) Compute $\la q'',t''|q',t'\ra=\la q''|e^{-iH(t''-t')}|q'\ra$ by inserting
a complete set of momentum eigenstates, and performing the
integral over the momentum.  Compare with your result in part (b).  

\vfill\eject

%% file: ch007.tex
\noindent Quantum Field Theory  \hfill   Mark Srednicki

\vskip0.5in

\begin{center}
\large{7: The Path Integral for the Harmonic Oscillator}
\end{center}
\begin{center}
Prerequisite: 6
\end{center}

\vskip0.5in

Consider a harmonic oscillator with hamiltonian
\begin{equation}
H(P,Q) = {\ts{1\over2m}}P^2 + \half m\omega^2 Q^2\;.
\label{sho}
\end{equation}
We begin with the formula from section 6 
for the ground state to ground state transition
amplitude in the presence of an external force,
specialized to the case of a a harmonic oscillator:
\begin{equation}
\la 0|0\ra_f 
= \int \D p\,Dq\;\exp i\int_{-\infty}^{+\infty} dt
                \Bigl[p\dot q - (1{-}i\eps)H + fq\Bigr] \;.
\label{00f4b}
\end{equation}
Looking at \eq{sho}, we see that multiplying $H$ by $1{-}i\eps$ is equivalent
to the replacements $m^{-1}\to(1{-}i\eps)m^{-1}$ [or, equivalently,
$m\to(1{+}i\eps)m$] 
and $m\w^2\to(1{-}i\eps)m\w^2$.  
Passing to the lagrangian formulation then gives
\begin{equation}
\la 0|0\ra_f 
= \int Dq\;\exp i\int_{-\infty}^{+\infty} dt
           \Bigl[\half(1{+}i\eps)m{\dot q}^2 - \half(1{-}i\eps)m\w^2 q^2 
           +fq\Bigr] \;.
\label{00f5}
\end{equation}
From now on, we will simplify the notation by setting $m=1$.

Next, let us use Fourier-transformed variables,
\begin{equation}
{\widetilde q}(E) = \int_{-\infty}^{+\infty} dt\;e^{iEt}\;q(t)\;,
\qquad
q(t) = \int_{-\infty}^{+\infty}{dE\over2\pi}\;e^{-iEt}\;{\widetilde q}(E)\;.
\label{ft27}
\end{equation}
The expression in square brackets in \eq{00f5} becomes
\begin{eqnarray}
&&\hskip-0.3in\Bigl[\cdots\Bigr] =
{1\over2}\int_{-\infty}^{+\infty}{dE\over2\pi}\,{dE'\over2\pi}\,
e^{-i(E+E')t}
\Bigl[\Bigl(
-(1{+}i\eps)EE' - (1{-}i\eps)\w^2\Bigr){\widetilde q}(E){\widetilde q}(E')
\nonumber \\
&&\hskip2in
{}+{\widetilde f}(E){\widetilde q}(E')
  +{\widetilde f}(E'){\widetilde q}(E)\Bigr].
\label{box}
\end{eqnarray}
Note that the only $t$ dependence is now in the prefactor.
Integrating over $t$ then generates a factor of $2\pi\delta(E+E')$.
Then we can easily integrate over $E'$ to get
\begin{eqnarray}
S &=& \int_{-\infty}^{+\infty}dt\;\Bigl[\cdots\Bigr]
\nonumber \\
\noalign{\medskip}
&=& {1\over2} \int_{-\infty}^{+\infty}{dE\over2\pi}\,
\Bigl[\Bigl(
(1{+}i\eps)E^2 - (1{-}i\eps)\w^2\Bigr){\widetilde q}(E){\widetilde q}(-E)
\nonumber \\
&&\hskip1in
{}+{\widetilde f}(E){\widetilde q}(-E)
  +{\widetilde f}(-E){\widetilde q}(E)\Bigr].
\label{box2}
\end{eqnarray}
The factor in large parentheses is equal to
$E^2-\w^2+i(E^2+\w^2)\eps$, and we can absorb the positive coefficient
into $\eps$ to get  $E^2-\w^2+i\eps$.

Now it is convenient to change integration variables to
\begin{equation}
{\widetilde x}(E) =  {\widetilde q}(E) 
                  + {{\widetilde f}(E)\over E^2-\w^2+i\eps}\;.
\label{xtilde}
\end{equation}
Then we get
\begin{equation}
S = {1\over2} \int_{-\infty}^{+\infty}{dE\over2\pi}\,
\left[{\widetilde x}(E)(E^2-\w^2+i\eps){\widetilde x}(-E) 
-{{\widetilde f}(E){\widetilde f}(-E)\over E^2-\w^2+i\eps}\right].
\label{s2}
\end{equation}
Furthermore, because \eq{xtilde} is just a shift by a constant, $\D q=\D x$.
Now we have
\begin{eqnarray}
\la 0|0\ra_f 
&=& \exp\!\left[{i\over2}\int_{-\infty}^{+\infty}{dE\over2\pi}\,
{{\widetilde f}(E){\widetilde f}(-E)\over{}-E^2+\w^2-i\eps}\right]
\nonumber \\
&& \times
\int \D x\;\exp\!\left[{i\over2}\int_{-\infty}^{+\infty}{dE\over2\pi}\,
{\widetilde x}(E)(E^2-\w^2+i\eps){\widetilde x}(-E)\right]. 
\label{00f6}
\end{eqnarray}

Now comes the key point.  The path integral on the second line of \eq{00f6} 
is what we get for $\la 0|0\ra_f$ in the case $f=0$.  On the other hand,
if there is no external force, a system in its ground state will remain
in its ground state, and so $\la 0|0\ra_{f=0}=1$.  Thus we find
\begin{equation}
\la 0|0\ra_f 
= \exp\!\left[{i\over2}\int_{-\infty}^{+\infty}{dE\over2\pi}\,
{{\widetilde f}(E){\widetilde f}(-E)\over{}-E^2+\w^2-i\eps}\right].
\label{00f7}
\end{equation}
We can also rewrite this in terms of time-domain variables as
\begin{equation}
\la 0|0\ra_f 
= \exp\!\left[{i\over2}\int_{-\infty}^{+\infty}dt\,dt'\,
f(t)G(t-t')f(t')\right],
\label{00f8}
\end{equation}
where 
\begin{equation}
G(t-t') = \int_{-\infty}^{+\infty}{dE\over2\pi}\,
{e^{-iE(t-t')}\over{}-E^2+\w^2-i\eps} \;.
\label{g}
\end{equation}
Note that $G(t-t')$ is a Green's function for the oscillator 
equation of motion:
\begin{equation}
\left({\d^2\over\d t^2}+\w^2\right)G(t-t') = \delta(t-t')\;.
\label{green}
\end{equation}
This can be seen directly by plugging \eq{g} into \eq{green} and then taking
the $\eps\to0$ limit.  We can also evaluate $G(t-t')$ explicitly by contour
integration; the result is
\begin{equation}
G(t-t') = {i\over 2\w}\exp\Bigl(-i\w|t-t'|\Bigr)\;.
\label{g22}
\end{equation}

Consider now the formula from section 6 
for the time-ordered product of operators.
In the case of initial and final ground states, it becomes
\begin{equation}
\la 0|{\rm T}Q(t_1)\ldots|0\ra =
{1\over i}\,{\delta\over\delta f(t_1)}\ldots 
\la 0|0\ra_f\Bigr|_{f=0}\;.
\label{qqb}
\end{equation}
Using our explicit formula, \eq{00f8}, we have
\begin{eqnarray}
\la 0|{\rm T}Q(t_1)Q(t_2)|0\ra &=&
{1\over i}\,{\delta\over\delta f(t_1)}\,
{1\over i}\,{\delta\over\delta f(t_2)}\,
\la 0|0\ra_f\Bigr|_{f=0}
\nonumber \\
\noalign{\medskip}
&=& {1\over i}\,{\delta\over\delta f(t_1)}
\left[\int_{-\infty}^{+\infty}dt'\,G(t_2-t')f(t')\right]\!
\la 0|0\ra_f\Bigr|_{f=0}
\nonumber \\
\noalign{\medskip}
&=& \left[{\ts{1\over i}}G(t_2-t_1) + (\hbox{term with $f$'s})\right]\!
\la 0|0\ra_f\Bigr|_{f=0}
\nonumber \\
\noalign{\medskip}
&=& {\ts{1\over i}}G(t_2-t_1) \;.
\label{qqc}
\end{eqnarray}
We can continue in this way to compute the ground-state expectation value
of the time-ordered product of more $Q(t)$'s.  If the number of $Q(t)$'s
is odd, then there is always a left-over $f(t)$ in the prefactor, and
so the result is zero.  If the number of $Q(t)$'s is even, then we must
pair up the functional derivatives in an appropriate way to get a nonzero
result.  Thus, for example,
\begin{eqnarray}
\la 0|{\rm T}Q(t_1)Q(t_2)Q(t_3)Q(t_4)|0\ra &=&
{1\over i^2}\Bigl[G(t_1{-}t_2)G(t_3{-}t_4)
\nonumber \\
&& {} + G(t_1{-}t_3)G(t_2{-}t_4)
\nonumber \\
&& {} + G(t_1{-}t_4)G(t_2{-}t_3)\Bigr].
\label{qqd}
\end{eqnarray}
More generally,
\begin{equation}
\la 0|{\rm T}Q(t_1)\ldots Q(t_{2n})|0\ra =
{1\over i^n}\sum_{\rm pairings}
G(t_{i_1}{-}t_{i_2}) \ldots G(t_{i_{2n-1}}{-}t_{i_{2n}})\;.
\label{qqe}
\end{equation}

\vskip0.5in

\begin{center}
Problems
\end{center}

\vskip0.25in

7.1) Starting with \eq{g}, verify \eq{g22}.

7.2) Starting with \eq{g22}, verify \eq{green}.

7.3a) Use the Heisenberg equation of motion, $\dot A = i[H,A]$,
to find explicit expressions for $\dot Q$ and $\dot P$.
Solve these to get the Heisenberg-picture operators $Q(t)$ and $P(t)$
in terms of the Schr\"odinger picture operators $Q$ and $P$.

b) Write the Schr\"odinger picture operators $Q$ and $P$
in terms of the creation and annihilation operators $a$ and $\ad$,
where $H=\hbar\w(\ad a + \half)$.   Then, using your result from part (a),
write the Heisenberg-picture operators $Q(t)$ and $P(t)$
in terms of $a$ and $\ad$.

c) Using your result from part (b), and $a|0\ra =\la 0|\ad=0$, 
verify \eq{qqc}.

\vfill\eject

%% file: ch008.tex
\noindent Quantum Field Theory  \hfill   Mark Srednicki

\vskip0.5in

\begin{center}
\large{8: The Path Integral for Free Field Theory}
\end{center}
\begin{center}
Prerequisite: 3, 7
\end{center}

\vskip0.5in

Our results for the harmonic oscillator can be straightforwardly
generalized to a free field theory with hamiltonian density
\begin{equation}
\H_0 = \half\Pi^2 + \half(\nabla\ph)^2 +\half m^2\ph^2 \;.
\label{h7}
\end{equation}
The dictionary we need is 
\begin{eqnarray}
q(t) &\longrightarrow& \ph(\x,t)\quad\hbox{(classical field)}
\nonumber \\
Q(t) &\longrightarrow& \ph(\x,t)\quad\hbox{(operator field)}
\nonumber \\
f(t) &\longrightarrow& J(\x,t)\quad\hbox{(classical {\it source})}
\label{dic}
\end{eqnarray}
The distinction between the classical field $\ph(x)$ and the 
corresponding operator field should be clear from context.

To employ the $\eps$ trick, we multiply $\H_0$ by $1-i\eps$.  The results
are equivalent to replacing $m^2$ in $\H_0$ with $m^2-i\eps$.  From now
on, for notational simplicity, we will write $m^2$ when we really mean
$m^2-i\eps$.

Let us write down the path integral (also called the {\it functional
integral\/}) for our free field theory:
\begin{equation}
Z_0(J) \equiv \la 0|0\ra_J = \int \D\ph\;e^{i\int \dfx[\L_0+J\ph]}\;,
\label{z0j}
\end{equation}
where
\begin{equation}
\L_0 = -\half\d^\mu\ph\d_\mu\ph - \half m^2\ph^2\;.
\label{l0}
\end{equation}
Note that when we say {\it path integral\/}, we now mean  
a path in the space of field configurations.

We can evaluate $Z_0(J)$ by mimicking what we did for the harmonic
oscillator in section 7.  We introduce four-dimensional Fourier transforms,
\begin{equation}
{\widetilde\ph}(k) = \int \dfx\,e^{-ikx}\;\ph(x)\;,
\qquad
\ph(x) = \int{\dfk\over(2\pi)^4}\;e^{ikx}\;{\widetilde\ph}(k)\;,
\label{ft2}
\end{equation}
where $kx=-k^0t+\k\!\cdot\!\x$, and $k^0$ is an integration variable.
Then, starting with $S_0=\int\dfx\,[\L_0+J\ph]$, we get
\begin{equation}
S_0 = {1\over2}\int{\dfk\over(2\pi)^4}\Bigl[
-{\widetilde\ph}(k)(k^2+m^2){\widetilde\ph}(-k)
+{\widetilde J}(k){\widetilde\ph}(-k)
+{\widetilde J}(-k){\widetilde\ph}(k)\Bigr],
\label{s8}
\end{equation}
where $k^2=\k^2-(k^0)^2$.  
We now change path integration variables to
\begin{equation}
{\widetilde\chi}(k) = {\widetilde\ph}(k)-{{\widetilde J}(k)\over k^2+m^2}\;.
\label{chit}
\end{equation}
Since this is merely a shift by a constant, we have $\D\ph=\D\chi$.
The action becomes
\begin{equation}
S_0 = {1\over2}\int{\dfk\over(2\pi)^4}\left[
{{\widetilde J}(k){\widetilde J}(-k)\over k^2+m^2}
-{\widetilde\chi}(k)(k^2+m^2){\widetilde\chi}(-k)\right].
\label{s9}
\end{equation}
Just as for the harmonic oscillator, the integral over $\chi$ simply
yields a factor of $Z_0(0)=\la 0|0\ra_{J=0}=1$.  Therefore
\begin{eqnarray}
Z_0(J) &=& \exp\!\left[{i\over2}\int{\dfk\over(2\pi)^4}\,
{{\widetilde J}(k){\widetilde J}(-k)\over k^2+m^2-i\eps}\right]
\nonumber \\
\noalign{\medskip}
&=& \exp\!\left[{i\over2}\int \dfx\,\dfx'\,J(x)\Delta(x-x')J(x')\right].
\label{z02}
\end{eqnarray}
Here we have defined the {\it Feynman propagator},
\begin{equation}
\Delta(x-x') = \int{\dfk\over(2\pi)^4}\,
{e^{ik(x-x')}\over k^2+m^2-i\eps}\;.
\label{feyn}
\end{equation}
The Feynman propagator is a Green's function for the Klein-Gordon equation,
\begin{equation}
(-\d^2_x+m^2)\Delta(x-x') = \delta^4(x-x') \;.
\label{green2}
\end{equation}
This can be seen directly by plugging \eq{feyn} into \eq{green2} and then 
taking the $\eps\to0$ limit.  
We can also evaluate $\Delta(x-x')$ explicitly by treating the $k^0$ integral
on the right-hand side of \eq{feyn} as a contour integration in the
complex $k^0$ plane, and then evaluating the contour integral via
the residue theorem.  The result is
\begin{eqnarray}
\Delta(x-x') &=& \int\dk\;e^{i\k\cdot(\x-\x')-i\w|t-t'|}
\nonumber \\
&=& i\theta(t{-}t')\int\dk\;e^{ik(x-x')} 
+ i\theta(t'{-}t)\int\dk\,e^{-ik(x-x')} \;, 
\label{feyn2}
\end{eqnarray}
where $\theta(t)$ is the unit step function.  The integral over $\dk$ can
also be performed in terms of Bessel functions; see section 4.

Now, by analogy with the formula for the ground-state expectation
value of a time-ordered product of operators 
for the harmonic oscillator, we have 
\begin{equation}
\la 0|{\rm T}\ph(x_1)\ldots|0\ra =
{1\over i}\,{\delta\over\delta J(x_1)}\ldots 
Z_0(J)\Bigr|_{J=0}\;.
\label{qqb2}
\end{equation}
Using our explicit formula, \eq{z02}, we have
\begin{eqnarray}
\la 0|{\rm T}\ph(x_1)\ph(x_2)|0\ra &=&
{1\over i}\,{\delta\over\delta J(x_1)}\,
{1\over i}\,{\delta\over\delta J(x_2)}\,
Z_0(J)\Bigr|_{J=0}
\nonumber \\
&=& {1\over i}\,{\delta\over\delta J(x_1)}
\left[\int \dfx'\,\Delta(x_2-x')J(x')\right]
Z_0(J)\Bigr|_{J=0}
\nonumber \\
&=& \left[{\ts{1\over i}}\Delta(x_2-x_1) + (\hbox{term with $J$'s})\right]
Z_0(J)\Bigr|_{J=0}
\nonumber \\
\noalign{\medskip}
&=& {\ts{1\over i}}\Delta(x_2-x_1) \;.
\label{qqc2}
\end{eqnarray}
We can continue in this way to compute the ground-state expectation value
of the time-ordered product of more $\ph$'s.  If the number of $\ph$'s
is odd, then there is always a left-over $J$ in the prefactor, and
so the result is zero.  If the number of $\ph$'s is even, then we must
pair up the functional derivatives in an appropriate way to get a nonzero
result.  Thus, for example,
\begin{eqnarray}
\la 0|{\rm T}\ph(x_1)\ph(x_2)\ph(x_3)\ph(x_4)|0\ra &=&
{1\over i^2}\Bigl[\Delta(x_1{-}x_2)\Delta(x_3{-}x_4)
\nonumber \\
&& {} + \Delta(x_1{-}x_3)\Delta(x_2{-}x_4)
\nonumber \\
&& {} + \Delta(x_1{-}x_4)\Delta(x_2{-}x_3)\Bigr].
\label{qqd2}
\end{eqnarray}
More generally,
\begin{equation}
\la 0|{\rm T}\ph(x_1)\ldots \ph(x_{2n})|0\ra =
{1\over i^n}\sum_{\rm pairings}
\Delta(x_{i_1}{-}x_{i_2}) \ldots \Delta(x_{i_{2n-1}}{-}x_{i_{2n}})\;.
\label{qqe2}
\end{equation}
This result is known as {\it Wick's theorem}.

\vskip0.5in

\begin{center}
Problems
\end{center}

\vskip0.25in

8.1) Starting with \eq{feyn}, verify \eq{green2}.

8.2) Starting with \eq{feyn}, verify \eq{feyn2}.

8.3) Use \eq{phi2}, the commutation relations \eq{ac}, and
$a(\k)|0\ra = 0$, $\la 0|\ad(\k) = 0$ to verify the last line of \eq{qqc2}.

8.4) The retarded and advanced Green's functions for the Klein-Gordon wave
operator satisfy $\Delta_{\rm ret}(x-y)=0$ for $x^0\ge y^0$ and 
$\Delta_{\rm adv}(x-y)=0$ for $x^0\le y^0$.  Find the pole prescriptions
on the right-hand side of \eq{feyn} that yield these Green's functions.

8.5) Let $Z_0(J)=\exp i W_0(J)$, and evaluate the real and imaginary parts
of $W_0(J)$.

8.6) Repeat the analysis of this section for the complex scalar
field that was introduced in problem 3.3, and further studied in problem 5.1.  
Write your source term in the form $J^\dagger\ph + J\ph^\dagger$, and 
find an explicit formula, analogous to \eq{z02}, for
$Z_0(J^\dagger,J)$.  Write down the appropriate generalization of 
\eq{qqb2}, and use it to compute 
$\la 0|{\rm T}\ph(x_1)\ph(x_2)|0\ra$,
$\la 0|{\rm T}\ph^\dagger(x_1)\ph(x_2)|0\ra$, and
$\la 0|{\rm T}\ph^\dagger(x_1)\ph^\dagger(x_2)|0\ra$.
Then verify your results by using the method of problem 8.3.
Finally, give the appropriate generalization of \eq{qqe2}.

\vfill\eject

%% file: ch009.tex
\noindent Quantum Field Theory  \hfill   Mark Srednicki

\vskip0.5in

\begin{center}
\large{9: The Path Integral for Interacting Field Theory}
\end{center}
\begin{center}
Prerequisite: 8
\end{center}

\vskip0.5in

Let us consider an interacting quantum field theory specified by a
lagrangian of the form 
\begin{equation}
\L=-\half Z_\ph\d^\mu\ph\d_\mu\ph-\half Z_m m^2\ph^2 +
{\ts{1\over6}}Z_g g\ph^3 + Y\ph\;.
\label{lnew2}
\end{equation}
As we discussed at the end of section 5,
we fix the parameter $m$ by requiring it to be equal to the actual mass of the 
particle (equivalently, the energy of the first excited state relative 
to the ground state), and we fix the parameter $g$ by requiring some particular 
scattering cross section to depend on $g$ in some particular way.
(We will have more to say about this after we have learned to
calculate cross sections.)
We also assume that the field is normalized by
\begin{equation}
\la 0|\ph(x)|0\ra = 0 \qquad \hbox{and} \qquad
\la k|\ph(x)|0\ra = e^{-ikx} \;.
\label{cond2}
\end{equation}
Here $|0\ra$ is the ground state, normalized via $\la 0|0\ra=1$,
and $|k\ra$ is a state of one
particle with four-momentum $k^\mu$, where $k^2=k^\mu k_\mu=-m^2$,
normalized via 
\begin{equation}
\la k'|k\ra = (2\pi)^3 2k^0\delta^3(\k'-\k)\;.
\label{norm9}
\end{equation}
Thus we have four conditions (the specified values of 
$m$, $g$, $\la 0|\ph|0\ra$, and $\la k|\ph|0\ra$), and we will
use these four conditions to determine the values of 
the four remaining parameters ($Y$ and the three $Z$'s) that appear in $\L$.

Before going further, we should note that this theory (known as $\ph^3$ theory,
pronounced ``phi-cubed'') actually has a fatal flaw.  
The hamiltonian density is
\begin{equation}
\H=\half Z_\ph^{-1}\Pi^2 - Y\ph +\half Z_m m^2\ph^2 
-{\ts{1\over6}}Z_g g\ph^3 \;. 
\label{ham9}
\end{equation}
Classically, we can make this arbitrarily negative by choosing an
arbitrarily large value for $\ph$.  Quantum mechanically, this means
that this hamiltonian has no ground state.
If we start off near $\ph=0$, we can tunnel through the potential
barrier to large $\ph$, and then ``roll down the hill''.
However, this process is invisible in perturbation theory in $g$.
The situation is exactly analogous to the problem of a harmonic oscillator
perturbed by an $x^3$ term.  This system has no ground state, but perturbation
theory (both time dependent and time independent) does not ``know'' this.
We will be interested in \eq{lnew2} only as an example of how to do
perturbation expansions in a simple context, and so we will overlook
this problem.

We would like to evaluate the path integral for this theory,
\begin{equation}
Z(J) \equiv \la 0|0\ra_J = \int \D\ph\;e^{i\int \dfx[\L_0+\L_1+J\ph]}\;.
\label{z}
\end{equation}
We can evaluate $Z(J)$ by mimicking what we did for quantum mechanics.
Specifically, we can rewrite \eq{z} as
\begin{eqnarray}
Z(J) &=& 
e^{i\int\dfx\;\L_1\left({1\over i}{\delta\over\delta J(x)}\right)}
\int \D\ph\;e^{i\int \dfx[\L_0+J\ph]}\;.
\nonumber \\
&\propto&
e^{i\int\dfx\;\L_1\left({1\over i}{\delta\over\delta J(x)}\right)}\;Z_0(J)\;,
\label{z1}
\end{eqnarray}
where $Z_0(J)$ is the result in free-field theory, 
\begin{equation}
Z_0(J) =
\exp\!\left[{i\over2}\int \dfx\,\dfx'\,J(x)\Delta(x-x')J(x')\right].
\label{z03}
\end{equation}
We have written $Z(J)$ as proportional to (rather than equal to)
the right-hand side of \eq{z1} 
because the $\eps$ trick does not give us the correct overall normalization;
instead, we must require $Z(0)=1$, and enforce this by hand.

Note that, in \eq{z03}, we have implicitly assumed that
\begin{equation}
\L_0 = -\half\d^\mu\ph\d_\mu\ph-\half m^2\ph^2 \;,
\label{ell00}
\end{equation}
since this is the $\L_0$ that gives us \eq{z03}.
Therefore, the rest of $\L$ must be included in $\L_1$.
We write
\begin{eqnarray}
\L_1 &=& {\ts{1\over6}}Z_g g\ph^3 + \L_{\rm ct} \;, 
\nonumber \\
\noalign{\medskip}
\L_{\rm ct} &=&
       -\half(Z_\ph{-}1)\d^\mu\ph\d_\mu\ph
       -\half(Z_m{-}1)m^2\ph^2
       + Y\ph\;,
\label{ell1}
\end{eqnarray}
where $\L_{\rm ct}$ is called the {\it counterterm\/} lagrangian.
We expect that, as $g\to0$, $Y\to0$ and $Z_i\to1$.
In fact, as we will see, $Y=O(g)$ and $Z_i=1+O(g^2)$.

In order to make use of \eq{z03}, we will have to compute lots and lots
of functional derivatives of $Z_0(J)$.  Let us begin by ignoring the
counterterms, and computing
\begin{eqnarray}
Z_1(J) &\propto& 
\exp\left[{i\over6}Z_g g\int\dfx
\left({1\over i}{\delta\over\delta J(x)}\right){\vphantom{\bigg|}}^{\!\!{3}}
\right]Z_0(J)
\nonumber \\
&\propto&
\sum_{V=0}^\infty{1\over V!}\left[{iZ_g g\over6}\int\dfx
\left({1\over i}{\delta\over\delta J(x)}\right){\vphantom{\bigg|}}^{\!\!{3}}
\right]^V
\nonumber \\
&&\times
\sum_{P=0}^\infty{1\over P!}\left[
{i\over2}\int\dfy\,\dfz\,J(y)\Delta(y{-}z)J(z)\right]^P .
\label{z2}
\end{eqnarray}
If we focus on a term in \eq{z2} with particular values of $V$ and $P$,
then the number of surviving sources (after we take all the functional
derivatives) is $E=2P-3V$.  
(Here $E$ stands for {\it external\/}, a terminology that should become 
clearer by the end of the next section;
$V$ stands for {\it vertex\/} and $P$ for {\it propagator\/}.)
The overall phase factor of such a term is then
$i^V(1/i)^{3V}i^P=i^{V+E-P}$, and 
the $3V$ functional derivatives can 
act on the $2P$ sources in $(2P)!/(3V)!$ different combinations.
However, many of resulting expressions are algebraically identical.

\begin{figure}
\begin{center}
\epsfig{file=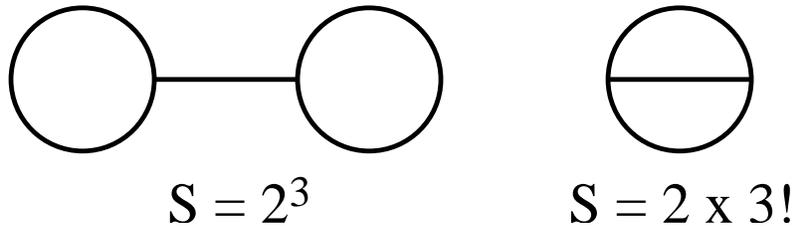}
\end{center}
\caption{All connected diagrams with $E=0$ and $V=2$.}
\label{e0v2}
\end{figure}

\begin{figure}
\begin{center}
\epsfig{file=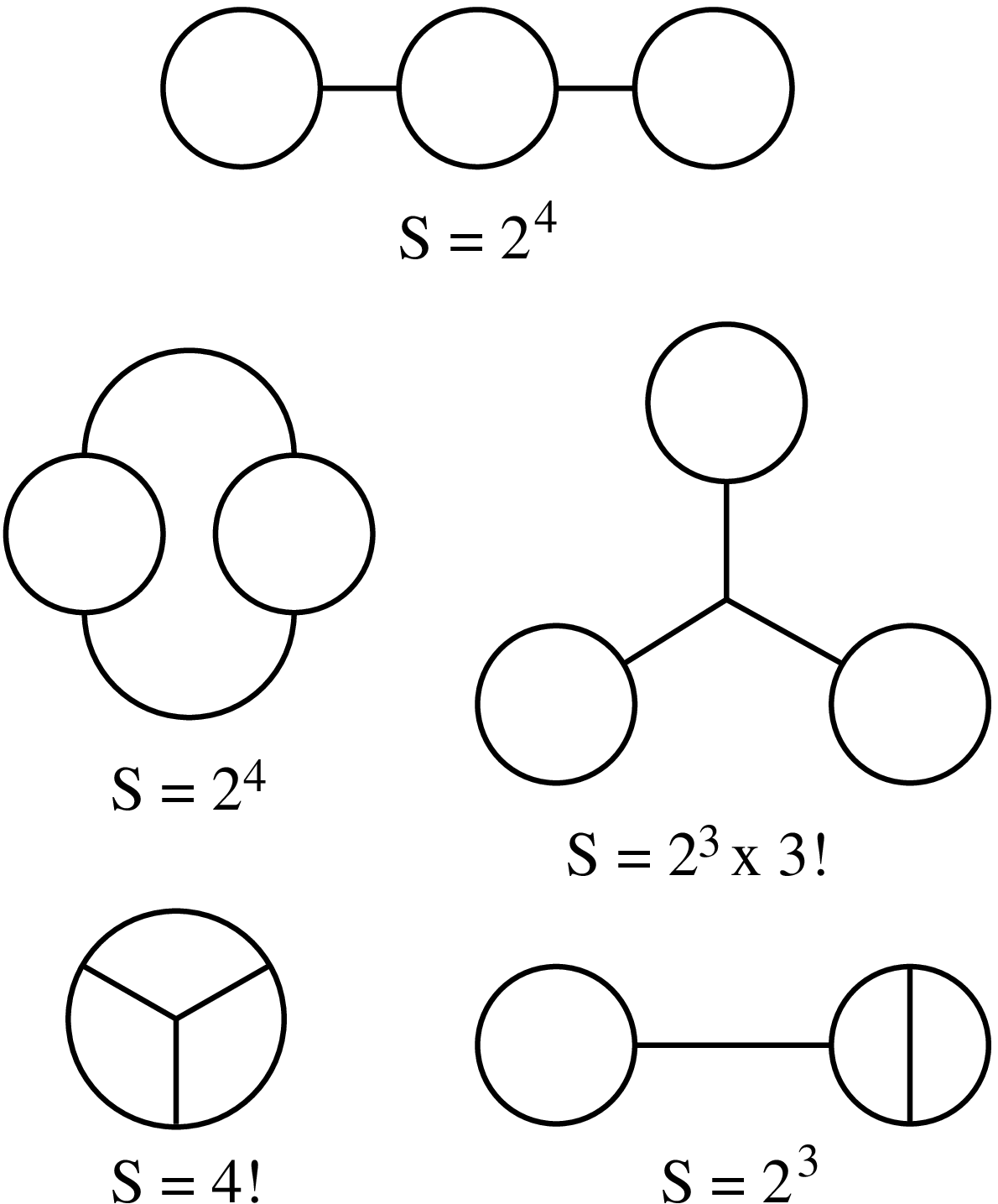}
\end{center}
\caption{All connected diagrams with $E=0$ and $V=4$.}
\label{e0v4}
\end{figure}

\begin{figure}
\begin{center}
\epsfig{file=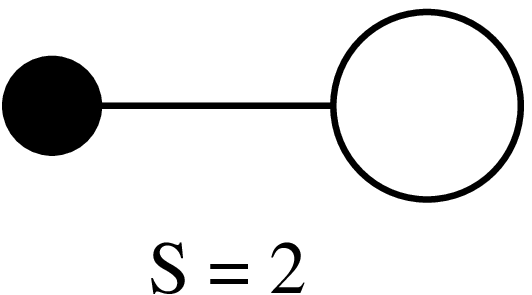}
\end{center}
\caption{All connected diagrams with $E=1$ and $V=1$.}
\label{e1v1}
\end{figure}

\begin{figure}
\begin{center}
\epsfig{file=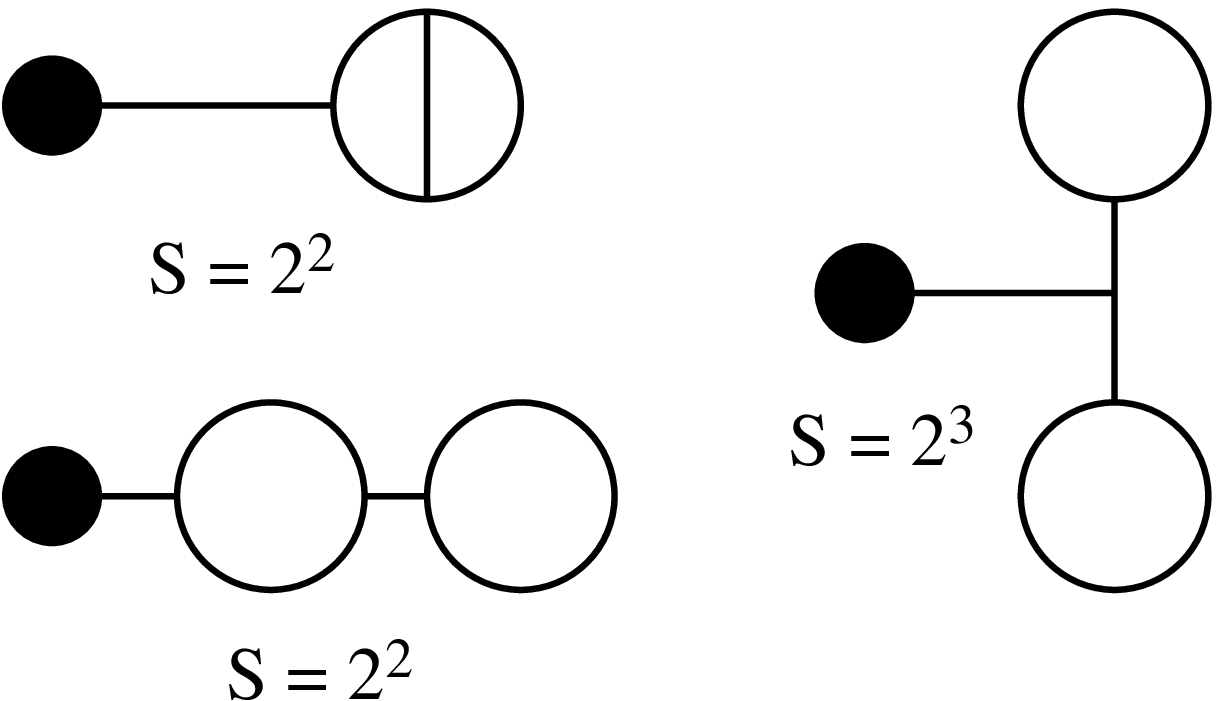}
\end{center}
\caption{All connected diagrams with $E=1$ and $V=3$.}
\label{e1v3}
\end{figure}

\begin{figure}
\begin{center}
\epsfig{file=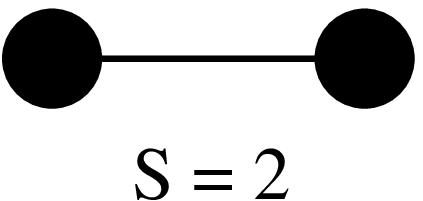}
\end{center}
\caption{All connected diagrams with $E=2$ and $V=0$.}
\label{e2v0}
\end{figure}

\begin{figure}
\begin{center}
\epsfig{file=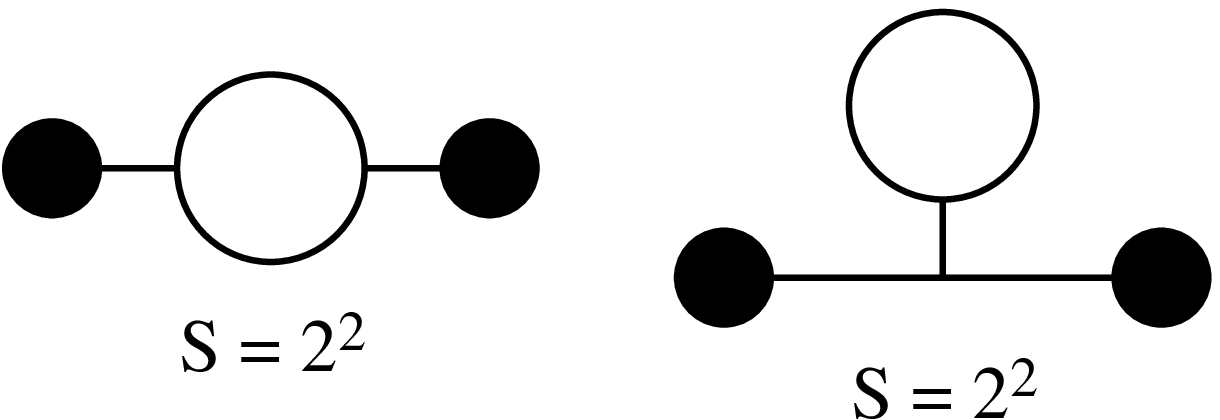}
\end{center}
\caption{All connected diagrams with $E=2$ and $V=2$.}
\label{e2v2}
\end{figure}

\begin{figure}
\begin{center}
\epsfig{file=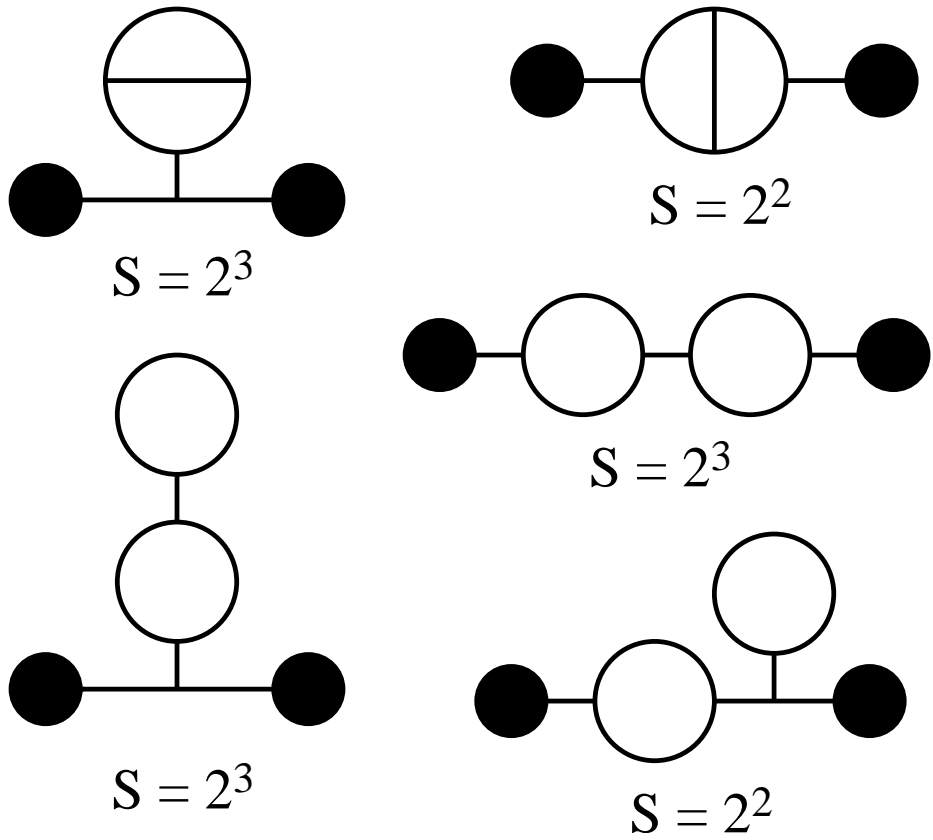}
\end{center}
\caption{Diagrams with $E=2$ and $V=4$ (continued in
the next figure).}
\label{e2v4a}
\end{figure}

\begin{figure}
\begin{center}
\epsfig{file=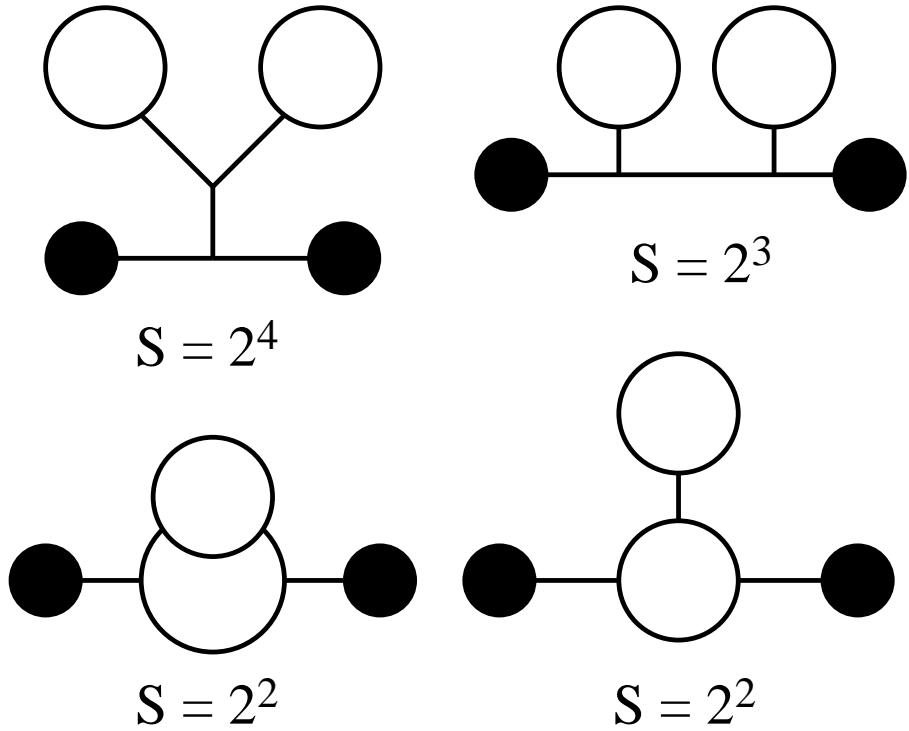}
\end{center}
\caption{Diagrams with $E=2$ and $V=4$ (continued from
the previous figure).}
\label{e2v4b}
\end{figure}

\begin{figure}
\begin{center}
\epsfig{file=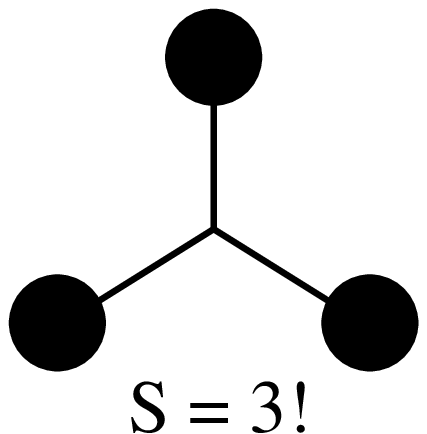}
\end{center}
\caption{All connected diagrams with $E=3$ and $V=1$.}
\label{e3v1}
\end{figure}

\begin{figure}
\begin{center}
\epsfig{file=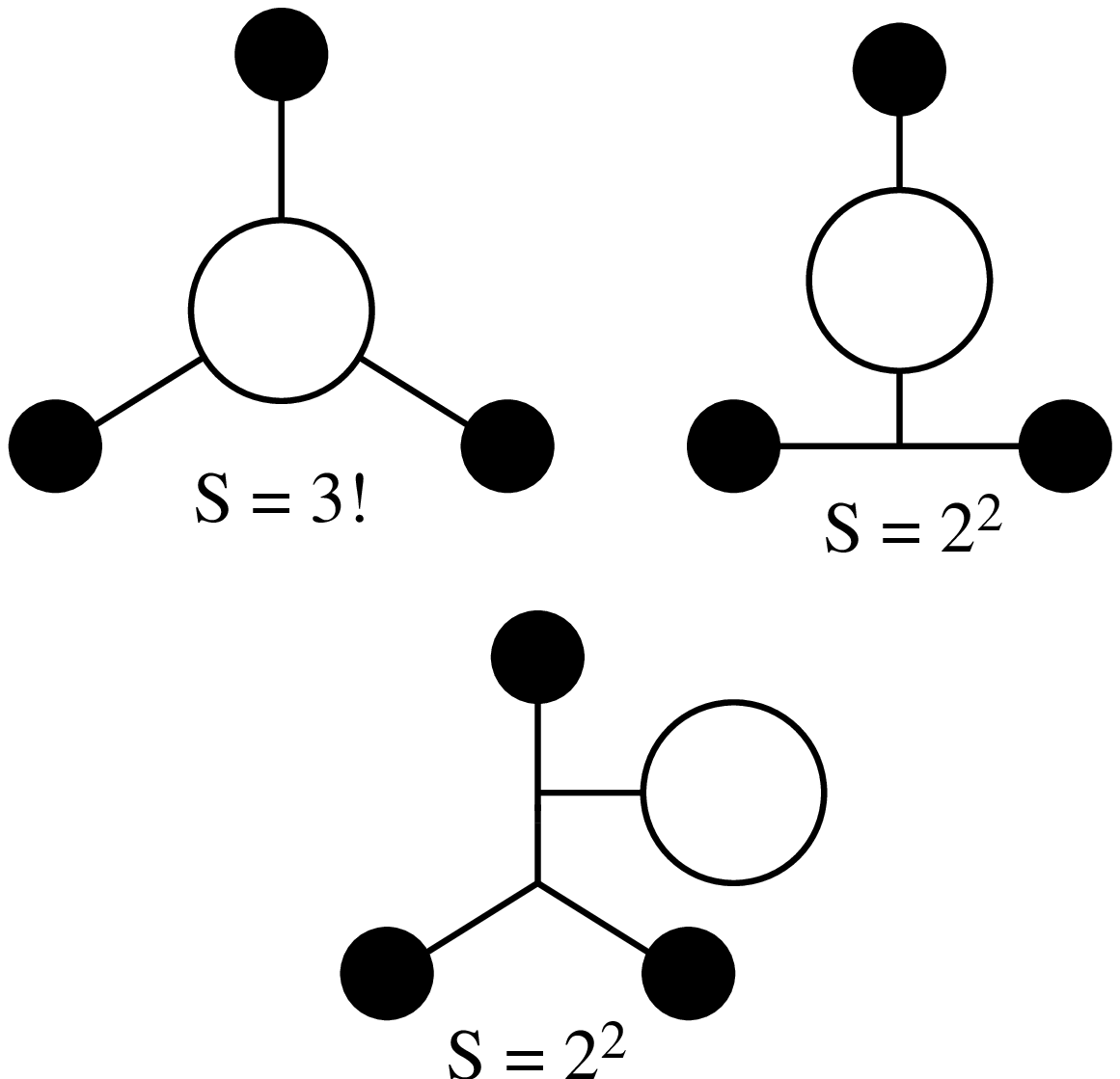}
\end{center}
\caption{All connected diagrams with $E=3$ and $V=3$.}
\label{e3v3}
\end{figure}

\begin{figure}
\begin{center}
\epsfig{file=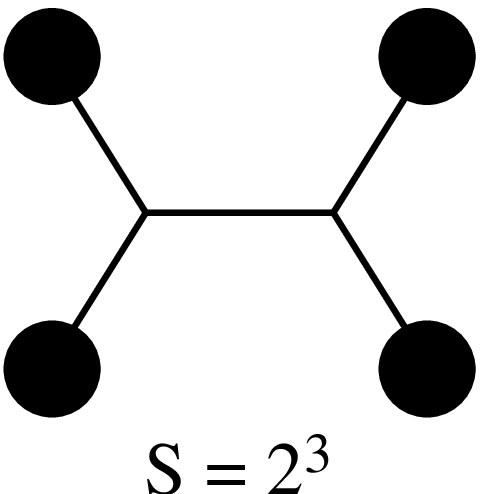}
\end{center}
\caption{All connected diagrams with $E=4$ and $V=2$.}
\label{e4v2}
\end{figure}

\begin{figure}
\begin{center}
\epsfig{file=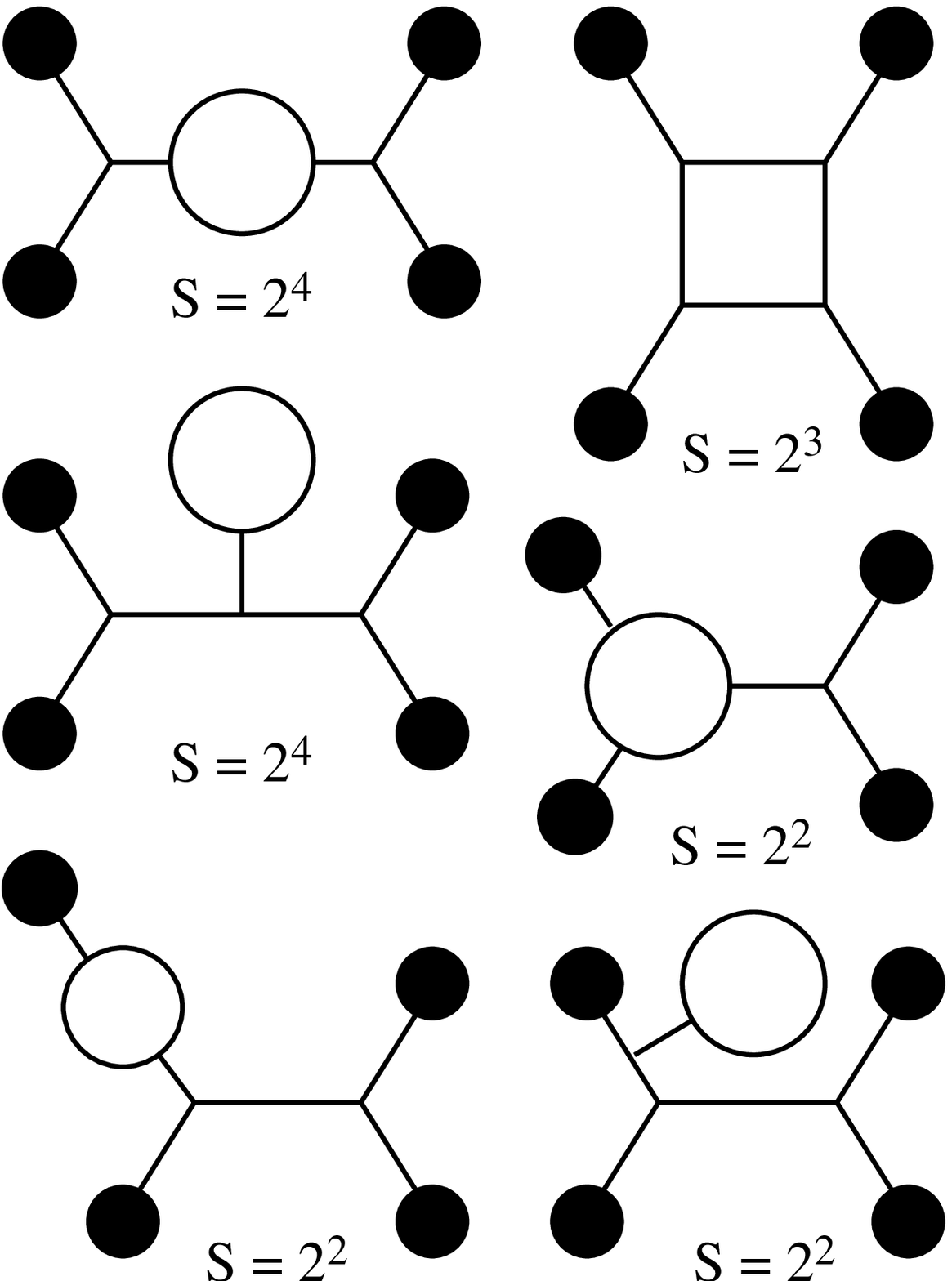}
\end{center}
\caption{All connected diagrams with $E=4$ and $V=4$.}
\label{e4v4}
\end{figure}

To organize them, we introduce {\it Feynman diagrams}.
In these diagrams, a line segment (straight or curved) stands for a propagator
${1\over i}\Delta(x{-}y)$, 
a filled circle at one end of a line segment for a source $i\int\dfx\,J(x)$,
and a vertex joining three line segments for $iZ_g g\int\dfx$.  
The complete set of diagrams for different values of $E$ and $V$ are
shown in figs.$\,$(\ref{e0v2}--\ref{e4v4}).  

To count the number of terms on the right-hand side of \eq{z2} that result
in a particular diagram, we first note that, in each diagram,
the number of lines is $P$, the number of lines connected to a source is $E$, 
and the number of vertices is $V$.  In a given diagram, there are $2^P P!$ 
ways of rearranging the sources (before we take the functional derivatives) 
without changing the resulting diagram.  Similarly, there are $(3!)^V V!$ 
ways of rearranging the functional derivatives (before they act on the 
sources) without changing the resulting diagram.  
These counting factors neatly cancel the
numbers from the dual Taylor expansions in \eq{z2}.

However, this procedure generally results in an overcounting of the number
of terms that give equal results.  This happens when some
rearrangement of derivatives gives the {\it same\/} match-up to sources as
some rearrangement of sources.  This possibility is always connected to
some symmetry property of the diagram, and so the factor by which we
have overcounted is called the {\it symmetry factor}.
The figures show the symmetry factor $S$ of each diagram.
 
Consider, for example, the second diagram of \fig{e0v2}.  
The propagators can be rearranged in
$3!$ ways, and all these rearrangements
can be duplicated by exchanging derivatives.  
Furthermore the endpoints of each propagator can be swapped, 
and the effect duplicated by swapping the two vertices.  
Thus, $S=2\times 3!=12$.

Let us consider two more examples.
In the first diagram of \fig{e2v2}, 
the exchange of the two external propagators (along with their
attached sources) can be duplicated by exchanging all the derivatives
at one vertex for those at the other.  Also, the effect of swapping the 
top and bottom semicircular propagators can be duplicated by swapping
the corresponding derivatives within each vertex. 
Thus, the symmetry factor is $S=2\times2=4$.

In the diagram of \fig{e4v2}, we can exchange derivatives to match swaps of 
the top and bottom external propagators on the left,
or the top and bottom external propagators on the right,
or the set of external propagators on the left with the set of 
external propagators on the right.  
Thus, the symmetry factor is $S=2\times2\times2=8$.

The diagrams in figs.$\,$(\ref{e0v2}--\ref{e4v4}) are all simply connected
(or just {\it connected\/} for short), but these are not
the only contributions to $Z(J)$.  The most general diagram consists of
a product of several connected diagrams.  
Let $C_I$ stand for a particular connected diagram, including
its symmetry factor.  A general diagram $D$ can then be expressed as
\begin{equation}
D = {1\over S_D}\prod_I \left( C_I \right)^{n_I} \;,
\label{diag}
\end{equation}
where $n_I$ is an integer that counts the number of $C_I$'s in $D$,
and $S_D$ is the {\it additional\/} symmetry factor for $D$ (that is,
the part of the symmetry factor that is not already accounted for by
the symmetry factors already included in each of the connected diagrams).
We now need to determine $S_D$.  

Since we have already accounted for propagator and vertex rearrangements
{\it within\/} each $C_I$, we need to consider only exchanges of propagators
and vertices among {\it different\/} connected diagrams.  These can leave the 
total diagram $D$ unchanged only if (1) the exchanges are made among
different but {\it identical\/} connected diagrams, and only if (2)
the exchanges involve {\it all\/} of the propagators and vertices in a
given connected diagram. 
If there are $n_I$ factors of $C_I$ in $D$, there are $n_I!$ ways to
make these rearrangements.  Overall, then, we have
\begin{equation}
S_D = \prod_I {n_I}! \;.
\label{sd}
\end{equation}
Now $Z_1(J)$ is given (up to an overall normalization) by summing all 
diagrams $D$, and each $D$ is labeled by the integers $n_I$.  Therefore
\begin{eqnarray}
Z_1(J) &\propto& \sum_{\{n_I\}}D
\nonumber \\
&\propto& \sum_{\{n_I\}}\prod_I {1\over{n_I}!} \left(C_I\right)^{n_I} 
\nonumber \\
&\propto& \prod_I \sum_{n_I=0}^\infty {1\over{n_I}!}\left(C_I\right)^{n_I} 
\nonumber \\
&\propto& \prod_I \exp\left(C_I\right) 
\nonumber \\
&\propto& \exp\left({\ts\sum_I C_I}\right) \;.
\label{z1j}
\end{eqnarray}
Thus we have a remarkable result: $Z_1(J)$ is given by the exponential
of the sum of {\it connected\/} diagrams.
This makes it easy to impose the normalization $Z_1(0)=1$:
we simply omit the {\it vacuum diagrams\/} (those with no sources),
like those of \figs{e0v2} and (\ref{e0v4}).
We then have
\begin{equation}
Z_1(J)=\exp[iW_1(J)] \;,
\label{z1jw1j}
\end{equation}
where we have defined
\begin{equation}
iW_1(J) \equiv \sum_{I\ne\{0\}} C_I \;,
\label{w1j}
\end{equation}
and the notation $I\ne\{0\}$ means that the vacuum diagrams are
omitted from the sum, so that $W_1(0)=0$.
[We have included a factor of $i$ on the left-hand side of \eq{w1j}
because then $W_1(J)$ is real in free-field theory; see problem 8.5.]

\begin{figure}
\begin{center}
\epsfig{file=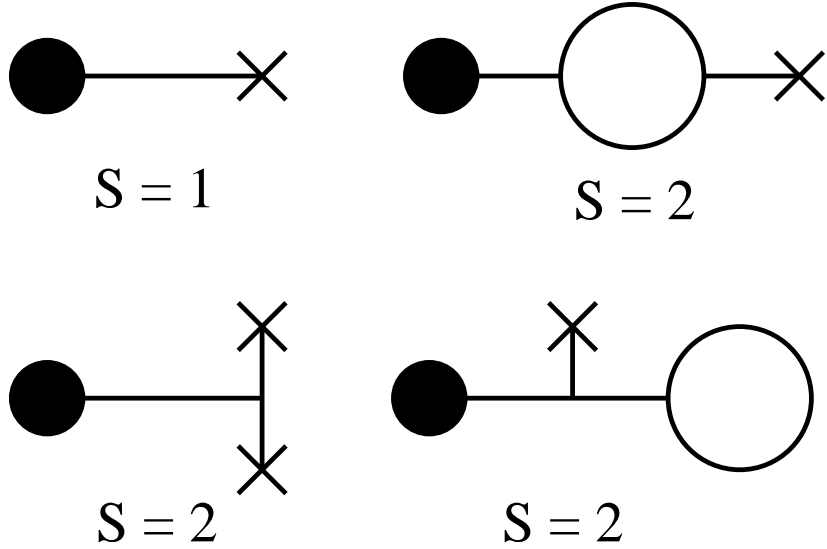}
\end{center}
\caption{All connected diagrams with $E=1$, $X\ge1$
(where $X$ is the number of one-point vertices from the linear counterterm), 
and $V+X\le3$.} 
\label{e1x}
\end{figure}

Were it not for the counterterms in $\L_1$, we would have $Z(J)=Z_1(J)$.
Let us see what we would get if this was, in fact, the case.
In particular, let us compute the vacuum expectation value of the 
field $\ph(x)$, which is given by
\begin{eqnarray}
\la 0|\ph(x)|0\ra &=& 
{1\over i}{\delta\over\delta J(x)}\,Z_1(J)\bigg|_{J=0}
\nonumber \\
\noalign{\medskip}
&=& {\delta\over\delta J(x)}\,W_1(J)\bigg|_{J=0}\;.
\label{vev0}
\end{eqnarray}
This expression is then the sum of all diagrams [such as those in
\figs{e1v1} and (\ref{e1v3})] 
that have a single source, with the source removed:
\begin{equation}
\la 0|\ph(x)|0\ra = {\ts{1\over2}}ig 
\int\dfy\,{\ts{1\over i}}\Delta(x{-}y){\ts{1\over i}}\Delta(y{-}y) +O(g^3)\;.
\label{vev2}
\end{equation}
Here we have set $Z_g=1$ in the first term, since $Z_g=1+O(g^2)$.
We see the vacuum-expectation value of $\ph(x)$ is not zero, as is required
for the validity of the LSZ formula.  To fix this, we must introduce
the counterterm $Y\ph$.  Including this term in the interaction lagrangian
$\L_1$ introduces a new kind of vertex, one where a single line segment ends;
the corresponding vertex factor is $iY\int\dfy$.  
The simplest diagrams including this new vertex
are shown in \fig{e1x}, with an X standing for the vertex.  

Assuming $Y=O(g)$, only the first diagram in 
\fig{e1x} contributes at $O(g)$, and we have
\begin{equation}
\la 0|\ph(x)|0\ra = 
\left(iY + {\ts{1\over2}}(ig){\ts{1\over i}}\Delta(0)\right)
\int\dfy\,{\ts{1\over i}}\Delta(x{-}y) + O(g^3)\;.
\label{vev3}
\end{equation}
Thus, in order to have $\la 0|\ph(x)|0\ra =0$, we should choose
\begin{equation}
Y = \half ig\Delta(0) + O(g^3)\;.
\label{y}
\end{equation}
The factor of $i$ is disturbing, because $Y$ must be a real number:
it is the coefficient of a hermitian operator in the hamiltonian,
as seen in \eq{ham9}.  Therefore, $\Delta(0)$ must be purely imaginary,
or we are in trouble.  We have
\begin{equation}
\Delta(0) = \int{\dfk\over(2\pi)^4}\,{1\over k^2+m^2-i\eps} \;.
\label{d0}
\end{equation}
\Eq{d0} does not clearly show whether or not $\Delta(0)$ is purely
imaginary, but it does reveal another problem: 
the integral diverges at large $k$, and $\Delta(0)$ is infinite.

To make some progress, we will modify the propagator in an ad hoc way:
\begin{equation}
\Delta(x-y) \to \int{\dfk\over(2\pi)^4}\,{e^{ik(x-y)}\over k^2+m^2-i\eps}
              \left( {\Lam^2\over k^2+\Lam^2-i\eps}\right)^{\!\! 2} .
\label{feynreg}
\end{equation}
Here $\Lambda$ is a new parameter called the {\it ultraviolet cutoff}.
It has dimensions of energy, and we assume that it is much larger than
any energy of physical interest.
Note that the modified propagator has the same Lorentz-transformation
properties as the original, so the Lorentz invariance of the theory
should not be affected.
In the limit $\Lam\to\infty$, the modified $\Delta(x{-}y)$ goes back
to the original one.

We can now evaluate the modified $\Delta(0)$ with the
methods of section 14; the result is
\begin{equation}
\Delta(0) = {i\over 16\pi^2}\,\Lam^2  \;.
\label{d0reg}
\end{equation}
Thus $Y$ is real, as required.  We can now formally take the 
limit $\Lam\to\infty$.
The parameter $Y$ becomes infinite, but $\la0|\ph(x)|0\ra$ remains zero,
at least to this order in $g$.

It may be disturbing to have a parameter in
the lagrangian which is formally infinite.  However, such parameters are
not directly measurable, and so need not obey our preconceptions about
their magnitudes.  Also, it is important to remember
that $Y$ includes a factor of $g$; this means that we can expand
in powers of $Y$ as part of our general expansion in powers of $g$.  
When we compute something measurable (like a scattering cross section),
all the formally infinite numbers will cancel in a well-defined way, 
leaving behind finite coefficients for the various powers of $g$.  
We will see how this works in detail in sections 14--20.

As we go to higher orders in $g$,
things become more complicated, but in principle the procedure is the same.
Thus, at $O(g^3)$, we sum up the diagrams of \figs{e1v3} and (\ref{e1x}), 
and then add to $Y$ whatever $O(g^3)$ term is needed
to maintain $\la0|\ph(x)|0\ra=0$.  In this way we can determine the value
of $Y$ order by order in powers of $g$.

\begin{figure}
\begin{center}
\epsfig{file=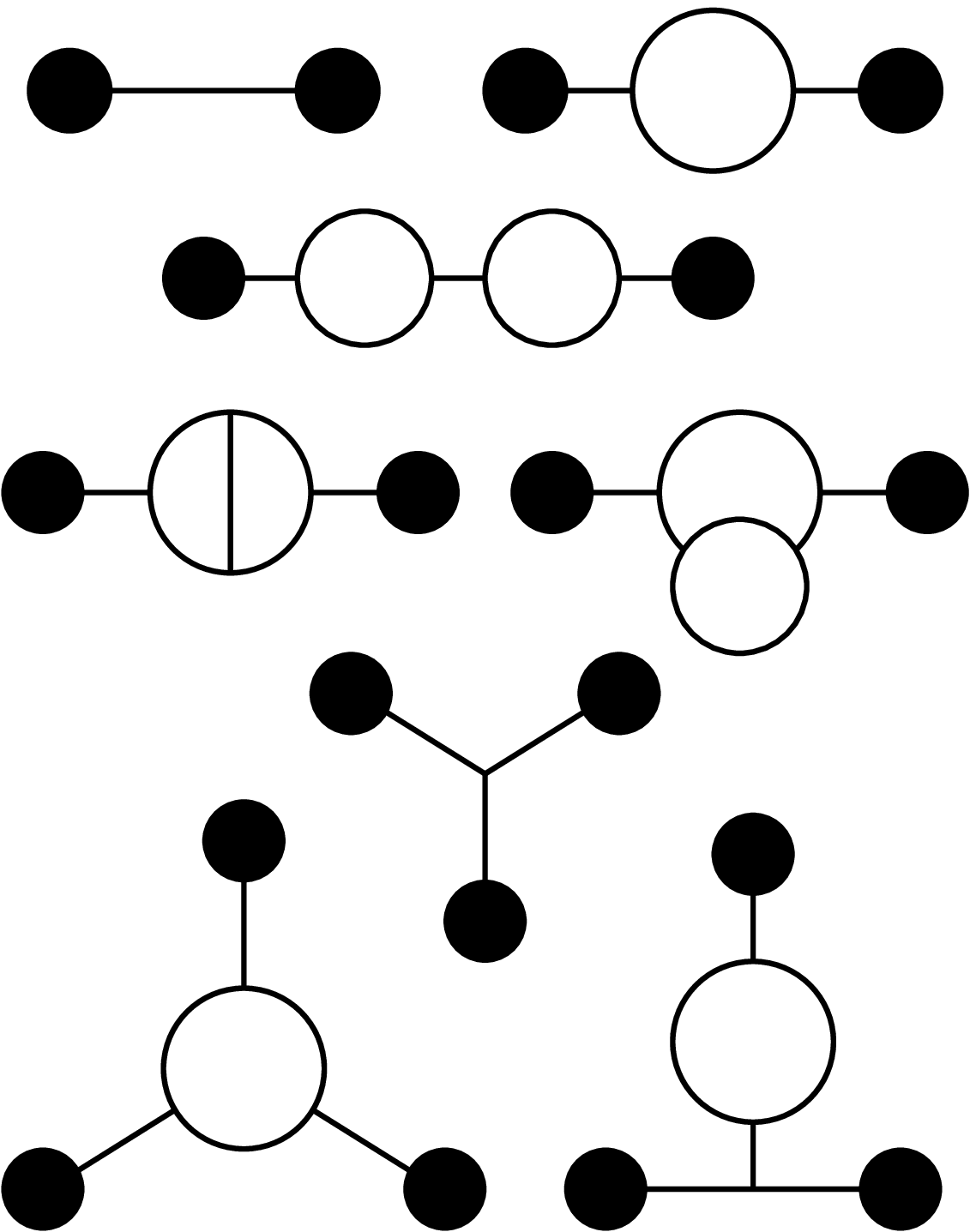}
\end{center}
\caption{All connected diagrams without tadpoles with $E\le 3$ and $V\le4$.}
\label{notad-e3}
\end{figure}

\begin{figure}
\begin{center}
\epsfig{file=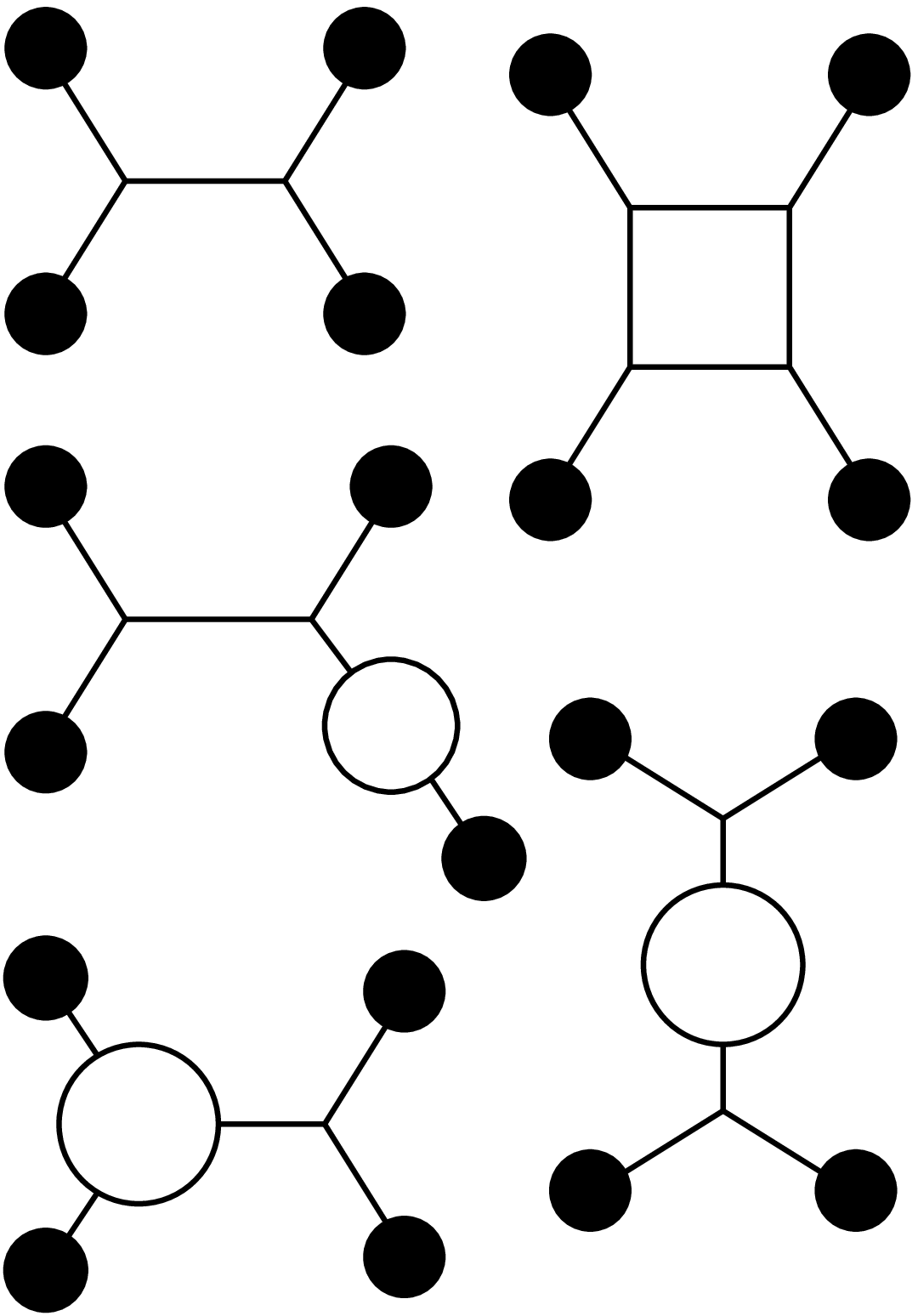}
\end{center}
\caption{All connected diagrams without tadpoles with $E=4$ and $V\le4$.}
\label{notad-e4}
\end{figure}

Once this is done, there is a remarkable simplification.  
Our adjustment of $Y$ to keep $\la0|\ph(x)|0\ra=0$
means that the sum of all connected diagrams with a single source is zero. 
Consider now that same infinite set of diagrams, but replace the source
in each of them with some other subdiagram.  Here is the point: 
{\it no matter what this replacement subdiagram is, 
the sum of all these diagrams is still zero.}  Therefore,
we need not bother to compute any of them!  The rule is this: 
ignore any diagram that, when a single line is cut,
falls into two parts, {\it one of which has no sources}. 
All of these diagrams (known as {\it tadpoles\/}) 
are canceled by the $Y$ counterterm, no matter what subdiagram they
are attached to.
The diagrams that remain (and need to be computed!) are shown in 
\figs{notad-e3} and (\ref{notad-e4}).

We turn next to the remaining two counterterms.  For notational simplicity
we define
\begin{equation}
A = Z_\ph-1 \;, \qquad
B = Z_m-1 \;, 
\label{abc} 
\end{equation}
and recall that we expect each of these to be $O(g^2)$.
We now have
\begin{equation}
Z(J) = \exp\left[-{i\over2}\int\dfx
\left({1\over i}{\delta\over\delta J(x)}\right)
\Bigl(-A\d_x^2+Bm^2\Bigr)
\left({1\over i}{\delta\over\delta J(x)}\right)\right]Z_1(J)\;.
\label{zj}
\end{equation}
We have integrated by parts to put both $\d_x$'s onto one $\delta/\delta J(x)$.
Also, we have cheated a little:
the time derivatives in this interaction really need to be treated by including
an extra source term for the conjugate momentum $\Pi=\dot\ph$. 
However, the terms with space derivatives are correctly treated, and
the time derivatives must work out comparably by Lorentz invariance.

\Eq{zj} results in a new vertex at which two lines meet.  
The corresponding vertex
factor is $(-i)\int\dfx\,(A\d_x^2+Bm^2)$; the $\d^2_x$ acts on the $x$ in one
or the other (but not both) propagators.  (Which one does not matter, and
can be changed via integration by parts.)  Diagramatically, all we need do is
sprinkle these new vertices onto the propagators in our existing diagrams.
How many of these vertices we need to add depends on the order in $g$ we
are working to achieve.

This completes our calculation of $Z(J)$ in $\ph^3$ theory.
We express it as
\begin{equation}
Z(J)=\exp[iW(J)] \;,
\label{zjwj10}
\end{equation}
where $W(J)$ is given by the sum of all connected diagrams
with no tadpoles and at least two sources, and including the
counterterm vertices just discussed.

Now that we have $Z(J)$,
we must find out what we can do with it.

\vskip0.5in

\begin{center}
Problems
\end{center}

\vskip0.25in

9.1) Compute the symmetry factor for each diagram in \figs{notad-e3}
and (\ref{notad-e4}).  (You can then check your answers by consulting
the earlier figures.)

9.2) Consider a real scalar field with $\L=\L_0+\L_1$, where
\begin{eqnarray}
\L_0 &=& -\half\d^\mu\ph\d_\mu\ph-\half m^2\ph^2 \;,
\nonumber \\
\L_1 &=& -{\ts{1\over24}}Z_\lam\lam\ph^4 + \L_{\rm ct} \;, 
\nonumber \\
\L_{\rm ct} &=&
       -\half(Z_\ph{-}1)\d^\mu\phd\d_\mu\ph
       -\half(Z_m{-}1)m^2\phd\ph \;.
\nonumber
\end{eqnarray}

a) What kind of vertex appears in the diagrams for
this theory (that is, how many line segments does it join?), 
and what is the associated vertex factor?

b) Ignoring the counterterms, draw all the connected 
diagrams with $1\le E \le 4$ and 
$0 \le V \le 2$, and find their symmetry factors.

c) Explain why we did not have to include a counterterm
linear in $\ph$ to cancel tadpoles.

9.3) Consider a complex scalar field (see problems
3.3, 5.1, and 8.5) with $\L=\L_0+\L_1$, where
\begin{eqnarray}
\L_0 &=& -\d^\mu\phd\d_\mu\ph - m^2\phd\ph \;,
\nonumber \\
\L_1 &=& -{\ts{1\over4}}Z_\lam\lam(\phd\ph)^2 + \L_{\rm ct} \;, 
\nonumber \\
\L_{\rm ct} &=&
       -(Z_\ph{-}1)\d^\mu\phd\d_\mu\ph
       -(Z_m{-}1)\phd\ph \;.
\nonumber
\end{eqnarray}
This theory has two kinds of sources,
$J$ and $J^\dagger$, and so we need a way to tell
which is which when we draw the diagrams.  
Rather than 
labeling the source blobs with a $J$ or $J^\dagger$, 
we will indicate which is which
by putting an arrow on the attached propagator that
points {\it towards\/} the source if it is a $J^\dagger$,
and {\it away\/} from the source if it is a $J$.

a) What kind of vertex appears in the diagrams for
this theory, and what is the associated vertex factor?
Hint: your answer should involve those arrows!

b) Ignoring the counterterms, draw all the connected 
diagrams with $1\le E \le 4$ and 
$0 \le V \le 2$, and find their symmetry factors.
Hint: the arrows are important!Ä

9.4) Consider the integral
\begin{equation}
\exp W(g,J) \equiv {1\over\sqrt{2\pi}}\int_{-\infty}^{+\infty}
dx\,\exp\Bigl[-\half x^2 + {\ts{1\over6}}gx^3 + Jx\Bigr] \;.
\label{IJg}
\end{equation}
This integral does not converge, but it can be used to
generate a joint power series in $g$ and $J$,
\begin{equation}
W(g,J) = \sum_{V=0}^\infty\sum_{E=0}^\infty C_{V,E}\,g^V\! J^E \;.
\label{IJg2}
\end{equation}

a) Show that 
\begin{equation}
C_{V,E}=\sum_I {1\over S_I} \;,
\label{CVE}
\end{equation}
where the sum is over all connected Feynman diagrams with $E$ sources and
$V$ three-point vertices, and $S_I$ is the symmetry factor for each diagram.

b) Use \eqs{IJg} and (\ref{IJg2}) to compute $C_{V,E}$ for $V\le 4$ and
$E\le 4$.  (This is most easily done with a symbolic manipulation program
like Maple or Mathematica.)  Verify that the symmetry factors given in 
figs.$\,$(\ref{e0v2}--\ref{e4v4}) satisfy the sum rule of \eq{CVE}.

c) Now consider $W(g,J{+}Y)$, with $Y$ fixed by the ``no tadpole'' condition
\begin{equation}
{\d\over\d J}\,W(g,J{+}Y)\biggr|_{J=0} = 0 \;.
\label{DWJ}
\end{equation}
Then write
\begin{equation}
W(g,J{+}Y) 
= \sum_{V=0}^\infty\sum_{E=0}^\infty {\widetilde C}_{V,E}\,g^V\! J^E \;.
\label{IJYg2}
\end{equation}
Show that
\begin{equation}
{\widetilde C}_{V,E}=\sum_I {1\over S_I} \;,
\label{CVEY}
\end{equation}
where the sum is over all connected Feynman diagrams with $E$ sources and
$V$ three-point vertices and {\it no tadpoles}, 
and $S_I$ is the symmetry factor for each diagram.

d) Let $Y=a_1 g + a_3 g^3 + \ldots\,$, 
and use \eq{DWJ} to determine $a_1$ and $a_3$.  
Compute ${\widetilde C}_{V,E}$ for $V\le 4$ and $E\le 4$.  
Verify that the symmetry factors for the diagrams in 
figs.$\,$(\ref{notad-e3}--\ref{notad-e4}) satisfy the sum rule of \eq{CVEY}.

\vfill\eject

%% file: ch010.tex
\noindent Quantum Field Theory  \hfill   Mark Srednicki

\vskip0.5in

\begin{center}
\large{10: Scattering Amplitudes and the Feynman Rules}
\end{center}
\begin{center}
Prerequisite: 5, 9
\end{center}

\vskip0.5in

Now that we have an expression for $Z(J)=\exp iW(J)$, 
we can take functional derivatives
to compute vacuum expectation values of time-ordered products of fields.  
Consider the case of two fields; we define the exact propagator via
\begin{equation}
{\ts{1\over i}}{\bf\Delta}(x_1-x_2) \equiv
\la0|{\rm T}\ph(x_1)\ph(x_2)|0\ra \;.
\label{bfd0}
\end{equation}
For notational simplicity let us define
\begin{equation}
\delta_j \equiv {1\over i}\,{\delta\over \delta J(x_j)} \;.
\label{dj}
\end{equation}
Then we have
\begin{eqnarray}
\la0|{\rm T}\ph(x_1)\ph(x_2)|0\ra 
&=& \delta_1\delta_2 Z(J)\Big|_{J=0}
\nonumber \\
\noalign{\smallskip}
&=& \delta_1\delta_2 iW(J)\Big|_{J=0}
-\delta_1 iW(J)\Big|_{J=0}\,\delta_2 iW(J)\Big|_{J=0}
\nonumber \\
\noalign{\smallskip}
&=& \delta_1\delta_2 iW(J)\Big|_{J=0} \;.
\label{prop}
\end{eqnarray}
To get the last line we used $\delta_x W(J)|_{J=0}=\la0|\ph(x)|0\ra=0$.
Diagramatically, $\delta_1$ removes a source, 
and labels the propagator endpoint $x_1$.  
Thus ${1\over i}{\bf\Delta}(x_1{-}x_2)$ is given by the sum
of diagrams with two sources, with those sources removed and the endpoints
labeled $x_1$ and $x_2$.  (The labels must be applied in both ways.  If the
diagram was originally symmetric on exchange of the two sources, the
associated symmetry factor of 2 is then canceled by the double labeling.)
At lowest order, the only contribution is the ``barbell'' diagram of
\fig{e2v0} with the sources removed.
Thus we recover the obvious fact that
${1\over i}{\bf\Delta}(x_1{-}x_2)={1\over i}\Delta(x_1{-}x_2) + O(g^2)$.
We will take up the subject of the $O(g^2)$ corrections in section 14.

For now, let us go on to compute
\begin{eqnarray}
\la0|{\rm T}\ph(x_1)\ph(x_2)\ph(x_3)\ph(x_4)|0\ra 
&=& \delta_1\delta_2\delta_3\delta_4 Z(J)
\nonumber \\
&=& \Bigl[\;\delta_1\delta_2\delta_3\delta_4 iW 
\nonumber \\
&&   {}\; +(\delta_1\delta_2 iW)(\delta_3\delta_4 iW)
\nonumber \\
&&   {}\; +(\delta_1\delta_3 iW)(\delta_2\delta_4 iW)
\nonumber \\
&&   {}\; +(\delta_1\delta_4 iW)(\delta_2\delta_3 iW)\; \Bigr]_{J=0} \;.
\label{4ph}
\end{eqnarray}
We have dropped terms that contain a factor of $\la 0|\ph(x)|0\ra=0$.
According to \eq{prop},
the last three terms in \eq{4ph} simply give products of the
exact propagators.  Let us see what happens when these terms are inserted
into the LSZ formula for two incoming and two outgoing particles,
\begin{eqnarray}
\la f|i\ra &=&
i^4\int \dfx_1\,\dfx_2\,\dfx'_1\,\dfx'_2\,
e^{i(k_1x_1+k_2x_2-k'_1x'_1-k'_2x'_2)}
\nonumber \\
&&\quad\times
(-\d^2_1+m^2)(-\d^2_2+m^2)(-\d^2_{1'}+m^2)(-\d^2_{2'}+m^2)
\nonumber \\
&&\quad\times
\la0|{\rm T}\ph(x_1)\ph(x_2)\ph(x'_1)\ph(x'_2)|0\ra \;.
\label{lsz4}
\end{eqnarray}
If we consider, for example, 
${1\over i}\bfd(x_1{-}x'_1){1\over i}\bfd(x_2{-}x'_2)$
as one term in the correlation function in \eq{lsz4}, we get from this term
\begin{eqnarray}
&&\int \dfx_1\,\dfx_2\,\dfx'_1\,\dfx'_2\,
e^{i(k_1x_1+k_2x_2-k'_1x'_1-k'_2x'_2)}
\;F(x_{11'})F(x_{22'})
\nonumber \\
&&\qquad = (2\pi)^4\delta^4(k_1{-}k'_1)\,(2\pi)^4\delta^4(k_2{-}k'_2)\,
           {\widetilde F}(\bar k_{11'})\,
           {\widetilde F}(\bar k_{22'})\;,
\label{4ph33}
\end{eqnarray}
where $F(x_{ij})\equiv(-\d_i^2+m^2)(-\d^2_j+m^2)\bfd(x_{ij})$,
${\widetilde F}(k)$ is its Fourier transform, 
$x_{ij'}\equiv x_i{-}x'_j$, 
and $\bar k_{ij'}\equiv\half(k_i{+}k'_j)$.
The important point is the two delta functions: these
tell us that the four-momenta of the two outgoing particles
($1'$ and $2'$) are equal to the four-momenta of the two incoming
particles (1 and 2).  In other words, no scattering has occurred.
This is not the event whose probability we wish to compute!
The other two similar terms in \eq{4ph}
either contribute to ``no scattering'' events, or vanish due to factors like
$\delta^4(k_1{+}k_2)$ (which is zero because $k_1^0{+}k_2^0\ge 2m>0$).
In general, the diagrams that contribute to the scattering process of interest
are only those that are {\it fully connected\/}: every endpoint can be
reached from every other endpoint by tracing through the diagram.  
These are the diagrams that arise from all the $\delta$'s acting on
a single factor of $W$.  
Therefore, from here on, we restrict our attention to those diagrams alone.
We define the {\it connected\/} correlation functions via
\begin{equation}
\la0|{\rm T}\ph(x_1)\ldots\ph(x_E)|0\ra_{\rm C}
\equiv \delta_1 \ldots \delta_E iW(J)\Big|_{J=0}\;,
\label{cc}
\end{equation}
and use these instead of 
$\la0|{\rm T}\ph(x_1)\ldots\ph(x_E)|0\ra$ in the LSZ formula.

\begin{figure}
\begin{center}
\epsfig{file=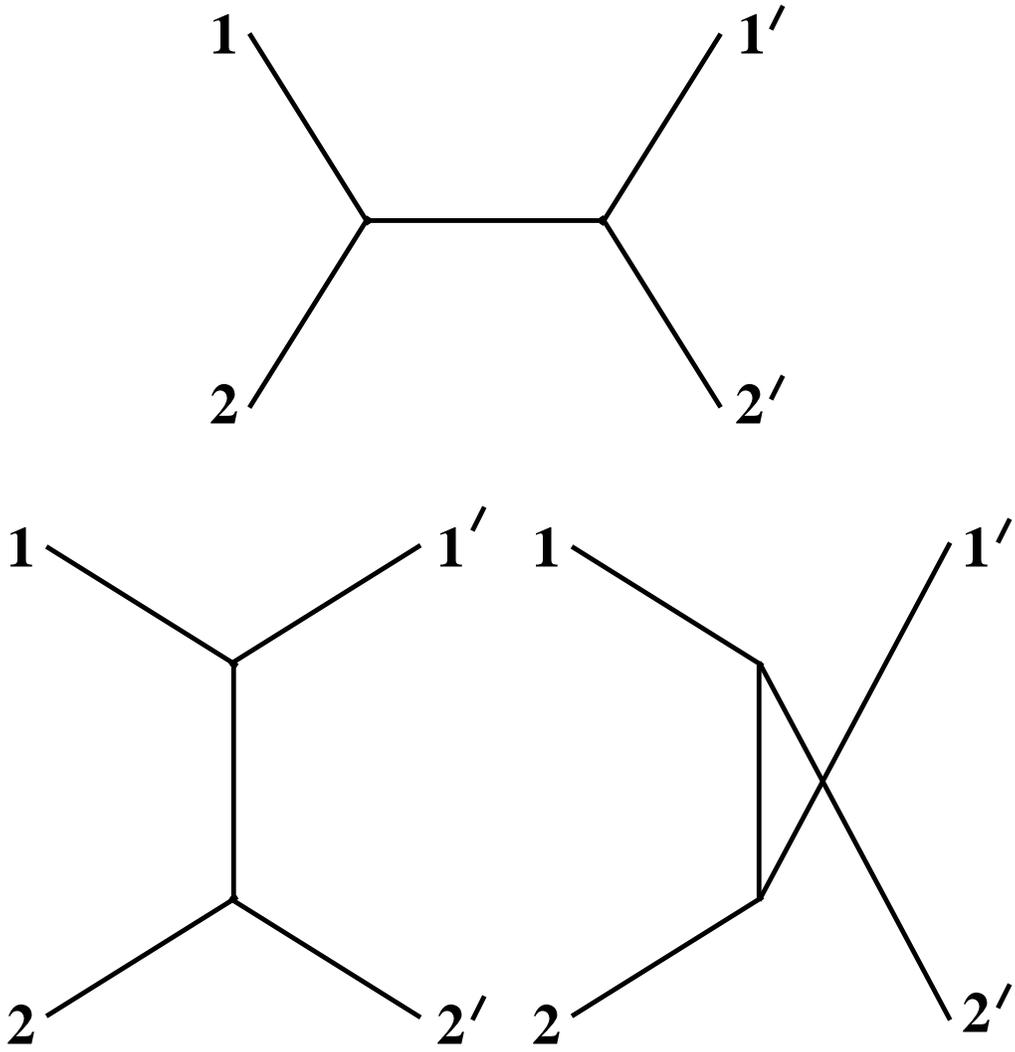}
\end{center}
\caption{The three tree-level Feynman diagrams that contribute 
to the connected correlation function
$\la 0|{\rm T}\ph(x_1)\ph(x_2)\ph(x'_1)\ph(x'_2)|0\ra_{\rm C}$.}
\label{scat4}
\end{figure}

Returning to \eq{4ph}, we have
\begin{equation}
\la0|{\rm T}\ph(x_1)\ph(x_2)\ph(x'_1)\ph(x'_2)|0\ra_{\rm C} =
    \delta_1\delta_2\delta_{1'}\delta_{2'} iW \Big|_{J=0} \;.
\label{4ph2}
\end{equation}
The lowest-order (in $g$) nonzero contribution to this comes from
the diagram of \fig{e4v2}, which has four sources and two vertices.
The four $\delta$'s remove the four sources; there are $4!$ ways
of matching up the $\delta$'s to the sources.  These 24 diagrams can
then be collected into 3 groups of 8 diagrams each; the 8 diagrams
in each group are identical.  The 3 distinct diagrams are shown
in \fig{scat4}.  Note that the factor of 8 neatly cancels the symmetry
factor $S=8$ of this diagram.  

This is a general result for {\it tree diagrams\/} (those with no
closed loops): once the sources have been stripped off and the endpoints
labeled, each diagram with a distinct endpoint labeling has an overall
symmetry factor of one.  The tree diagrams for a given process represent
the lowest-order (in $g$) nonzero contribution to that process. 

We now have
\begin{eqnarray}
&& \la0|{\rm T}\ph(x_1)\ph(x_2)\ph(x'_1)\ph(x'_2)|0\ra_{\rm C} 
\nonumber \\
&&\qquad \qquad \qquad 
= (ig)^2\left({\ts{1\over i}}\right)^5 \int\dfy\,\dfz\,\Delta(y{-}z)
\nonumber \\
&&\qquad \qquad \qquad \quad 
\times\Bigl[\;
\Delta(x_1{-}y) \Delta(x_2{-}y) \Delta(x'_1{-}z) \Delta(x'_2{-}z)
\nonumber \\
&&\qquad \qquad \qquad \quad \;
{} + \Delta(x_1{-}y) \Delta(x'_1{-}y) \Delta(x_2{-}z) \Delta(x'_2{-}z)
\nonumber \\
&&\qquad \qquad \qquad \quad \;
{} + \Delta(x_1{-}y) \Delta(x'_2{-}y) \Delta(x_2{-}z) \Delta(x'_1{-}z)\; \Bigr]
\nonumber \\
&&\qquad \qquad \qquad \qquad \;
{} + O(g^4)\;. 
\label{4ph3}
\end{eqnarray}
Next, we use \eq{4ph3} in the LSZ formula, \eq{lsz4}.  Each Klein-Gordon wave
operator acts on a propagator to give
\begin{equation}
(-\d^2_i+m^2)\Delta(x_i - y) = \delta^4(x_i - y) \;.
\label{kgprop}
\end{equation}
The integrals over the external spacetime labels $x_{1,2,1',2'}$ are then
trivial, and we get
\begin{eqnarray}
\la f|i\ra &=&
(ig)^2\,\left({\ts{1\over i}}\right)\int\dfy\,\dfz\,\Delta(y{-}z)
\;\Bigl[\;e^{i(k_1y+k_2y-k'_1z-k'_2z)}
\nonumber \\
&& \qquad\qquad\qquad\qquad\qquad\quad\;
{}+e^{i(k_1y+k_2z-k'_1y-k'_2z)}
\nonumber \\
&& \qquad\qquad\qquad\qquad\qquad\quad\;
{}+e^{i(k_1y+k_2z-k'_1z-k'_2y)}\;\Bigr]+O(g^4)\;.
\label{lsz5}
\end{eqnarray}
This can be simplified by substituting
\begin{equation}
\Delta(y-z) = \int{\dfk\over(2\pi)^4}\,{e^{ik(y-z)}\over k^2+m^2-i\eps}
\label{prop2}
\end{equation}
into \eq{4ph3}.  Then the spacetime arguments appear only in phase factors,
and we can integrate them to get delta functions: 
\begin{eqnarray}
\la f|i\ra &=& 
ig^2 \int{\dfk\over(2\pi)^4}\,{1\over k^2+m^2-i\eps}
\nonumber \\
&&\qquad\quad\times
\Bigl[\,(2\pi)^4\delta^4(k_1{+}k_2{+}k)\,(2\pi)^4\delta^4(k'_1{+}k'_2{+}k)
\nonumber \\
&&\qquad\quad\,
   {} +(2\pi)^4\delta^4(k_1{-}k'_1{+}k)\,(2\pi)^4\delta^4(k'_2{-}k_2{+}k)
\nonumber \\
&&\qquad\quad\,
   {} +(2\pi)^4\delta^4(k_1{-}k'_2{+}k)\,(2\pi)^4\delta^4(k'_1{-}k_2{+}k)
\;\Bigr] + O(g^4)
\label{lsz5a} \\
\noalign{\medskip}
&=&ig^2\,(2\pi)^4\delta^4(k_1{+}k_2{-}k'_1{-}k'_2)
\nonumber \\
&&\;\times\left[
{1\over(k_1{+}k_2)^2+m^2} +{1\over(k_1{-}k'_1)^2+m^2}
+{1\over(k_1{-}k'_2)^2+m^2}
\right]\qquad{}
\nonumber \\
&&{}+O(g^4)\vphantom{\Bigg|}\;.
\label{lsz6}
\end{eqnarray}
In \eq{lsz6}, we have left out the $i\eps$'s for notational convenience only;
$m^2$ is really $m^2-i\eps$.
The overall delta function in \eq{lsz6} tells that that four-momentum is
conserved in the scattering process, which we should, of course, expect.
For a general scattering process, 
it is then convenient to define a scattering matrix element $\T$ via
\begin{equation}
\la f|i\ra = i(2\pi)^4\delta^4(k_{\rm in}{-}k_{\rm out})\,\T \;,
\label{tscat}
\end{equation}
where $k_{\rm in}$ and $k_{\rm out}$ are the total four-momenta of
the incoming and outgoing particles, respectively.

Examining the calculation which led to \eq{lsz6}, we can take away
some universal features that lead to a simple set of {\it Feynman rules\/}
for computing contributions
to $\T$ for a given scattering process.  The Feynman rules are:

\vskip0.1in

1) Draw lines (called {\it external lines\/}) 
for each incoming and each outgoing particle.

2) Leave one end of each external line free,
and attach the other to a vertex at which exactly three lines meet.
Include extra {\it internal lines\/} in order to do this.  
In this way, draw all possible diagrams that are 
{\it topologically inequivalent}.

3) On each incoming line, draw an arrow pointing towards the vertex.
On each outgoing line, draw an arrow pointing away from the vertex.
On each internal line, draw an arrow with an arbitrary direction.

4) Assign each line its own four-momentum.  
The four-momentum of an external line should be the four-momentum
of the corresponding particle.

5) Think of the four-momenta as flowing along the arrows,
and conserve four-momentum at each vertex. 
For a tree diagram, this fixes the momenta on all the internal lines.

6) The value of a diagram consists of the following factors:
for each external line, 1; for each vertex, $iZ_g g$; for each internal line,
$-i/(k^2+m^2-i\eps)$, where $k$ is the four-momentum of that line.

7) A diagram with $L$ closed loops will have $L$ internal momenta
that are not fixed by Rule~4.  Integrate over each of these momenta $\ell_i$
with measure $d^4\ell_i/(2\pi)^4$

8) A loop diagram may have some leftover symmetry factors
if there are exchanges of {\it internal\/} propagators and vertices that
leave the diagram unchanged; in this case, divide the value of the diagram
by the symmetry factor associated with
exchanges of internal propagators and vertices.

9) Include diagrams with the {\it counterterm vertex\/}
that connects two propagators, each with the same four-momentum $k$.
The value of this vertex is $-i(Ak^2+Bm^2)$, where $A=Z_\ph-1$ and $B=Z_m-1$,
and each is $O(g^2)$.

10) The value of $i\T$ is given by a sum over the values of all these diagrams.

\vskip0.1in

\begin{figure}
\begin{center}
\epsfig{file=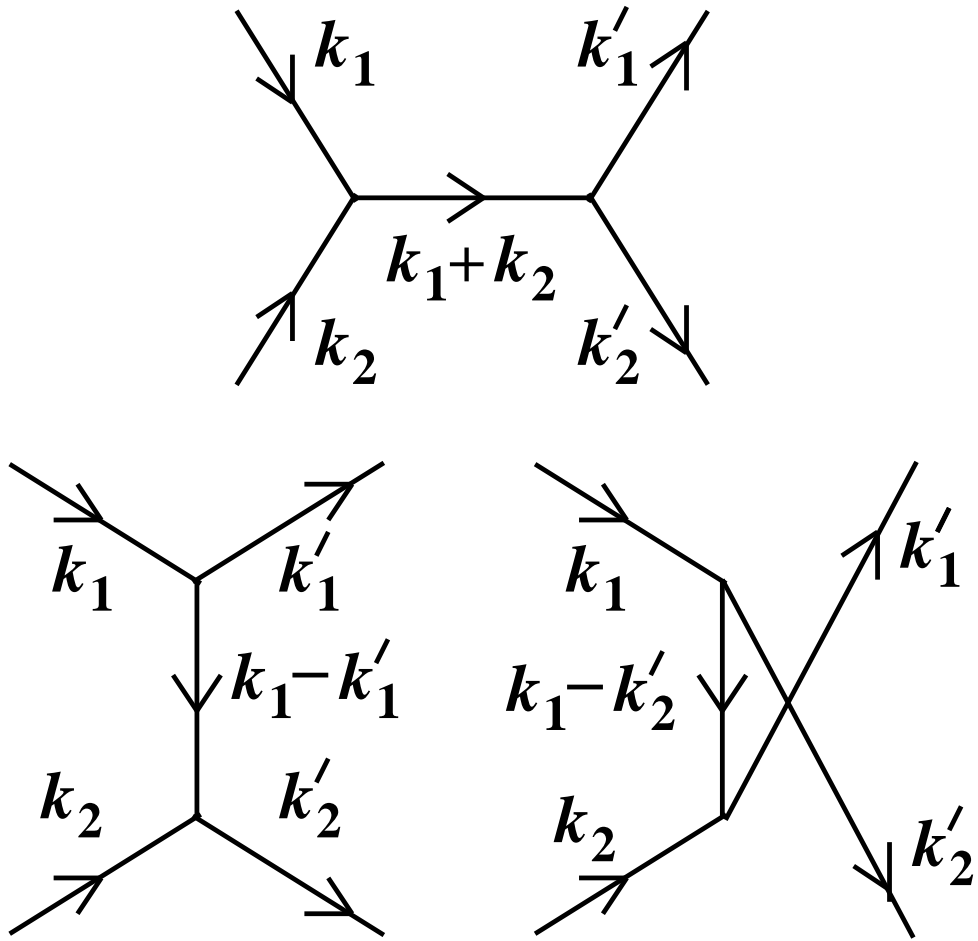}
\end{center}
\caption{The tree-level $s$-, $t$-, and $u$-channel diagrams contributing to
$i\T$ for two particle scattering.}
\label{scat4mom}
\end{figure}

For the two-particle scattering process, the tree diagrams
resulting from these rules are shown in \fig{scat4mom}.

Now that we have our procedure for computing the scattering amplitude
$\T$, we must see how to relate it to a measurable cross section.

\vskip0.5in

\begin{center}
Problems
\end{center}

\vskip0.25in

10.1) Write down the Feynman rules for the complex scalar field
of problem 9.3.  Remember that there are two kinds of particles
now (which we can think of as positively and negatively charged),
and that your rules must have a way of distinguishing them.
Hint: the most direct approach requires 
two kinds of arrows: momentum arrows (as discussed in this
section) and what we might call ``charge'' arrows (as discussed
in problem 9.3).  Try to find a more elegant approach that requires
only one kind of arrow.

10.2) Consider a complex scalar field $\ph$ that interacts with a real
scalar field $\chi$ via $\L_1 =  g \chi\d^\mu\phd\d_\mu\ph$.
Use a solid line for the $\ph$ propagator and a dashed line for
the $\chi$ propagator.  Draw the vertex (remember the arrows!), 
and find the associated vertex factor.  

10.3) Consider a real scalar field with $\L_1 = \half g \ph\d^\mu\ph\d_\mu\ph$.
Find the associated vertex factor.  

10.4) Consider a theory of three real scalar fields ($A$, $B$, and $C$)
with  
\begin{eqnarray}
\L &=& {} - \half\d^\mu\! A\d_\mu A - \half m^2_{\sss A}A^2
\nonumber \\
   && {}  - \half\d^\mu\! B\d_\mu B - \half m^2_{\sss B}B^2
\nonumber \\
   && {}  - \half\d^\mu   C\d_\mu C - \half m^2_{\sss C}C^2
\nonumber \\
   && {}  + gABC \;.
\label{10.2}
\end{eqnarray}
Write down the tree-level scattering amplitude (given by the sum of
the contributing tree diagrams) for each of the
following processes:
\begin{eqnarray}
AA &\to& AA \;,
\nonumber \\
AA &\to& AB \;,
\nonumber \\
AA &\to& BB \;,
\nonumber \\
AA &\to& BC \;,
\nonumber \\
AB &\to& AB \;,
\nonumber \\
AB &\to& AC \;.
\label{10.2a}
\end{eqnarray}
Your answers should take the form
\begin{equation}
\T = g^2\left[{c_s\over m^2_s-s}+{c_t\over m^2_t-t}+{c_u\over m^2_u-u}\right],
\label{10.2b}
\end{equation}
where, in each case, each $c_i$ is a positive integer, 
and each $m^2_i$ is $m^2_{\sss A}$ or $m^2_{\sss B}$ or $m^2_{\sss C}$.
Hint: $\T$ may be zero for some processes.

\vfill\eject

%% file: ch011.tex
\noindent Quantum Field Theory  \hfill   Mark Srednicki

\vskip0.5in

\begin{center}
\large{11: Cross Sections and Decay Rates}
\end{center}
\begin{center}
Prerequisite: 10
\end{center}

\vskip0.5in

Now that we have a method for computing the scattering amplitude $\T$,
we must convert it into something that could be measured in an experiment.  

In practice, we are almost always concerned with
one of two generic cases: one incoming particle,
for which we compute a {\it decay rate},
or two incoming particles,
for which we compute a {\it cross section}.
We begin with the latter.

Let us also specialize, for now, to the case of two outgoing particles as well
as two incoming particles.  In $\ph^3$ theory, we found in section 10 that
in this case we have 
\begin{equation}
\T = g^2\!\left[{1\over(k_1{+}k_2)^2+m^2} +{1\over(k_1{-}k'_1)^2+m^2}
+{1\over(k_1{-}k'_2)^2+m^2} \right]  +O(g^4)  \;,
\label{t}
\end{equation}
where $k_1$ and $k_2$ are the four-momenta of the two incoming particles,
$k'_1$ and $k'_2$ are the four-momenta of the two outgoing particles,
and $k_1+k_2=k'_1+k'_2$.  Also, these particles are all {\it on shell\/}:
$k^2_i=-m^2_i$.  (Here, for later use, we allow for the possibility 
that the particles all have different masses.)  

Let us think about the kinematics of this process.  In the 
{\it center-of-mass frame}, or {\it CM frame\/} for short, we take
$\k_1+\k_2={\bf 0}$, and choose $\k_1$ to be in the $+z$ direction.
Now the only variable left to specify about the initial state 
is the magnitude of $\k_1$.  
Equivalently, we could specify the total energy in the CM frame, 
$E_1+E_2$.  However, it is even more convenient to define a Lorentz 
scalar $s\equiv -(k_1+k_2)^2$.  In the CM frame, $s$ reduces
to $(E_1+E_2)^2$; $s$ is therefore called the 
{\it center-of-mass energy squared}.  Then, since 
$E_1 = (\k_1^2+m_1^2)^{1/2}$ and $E_2 = (\k_1^2+m_2^2)^{1/2}$,
we can solve for $|\k_1\!|$ in terms of $s$, with the result
\begin{equation}
|\k_1\!|={1\over2\sqrt{s}}\sqrt{s^2-2(m_1^2+m_2^2)s+(m_1^2-m_2^2)^2}
\quad\hbox{(CM frame)}\;.
\label{k1}
\end{equation}
Now consider the two outgoing particles.  
Since momentum is conserved,
we must have $\k'_1+\k'_2={\bf 0}$, and
since energy is conserved,
we must also have $(E'_1+E'_2)^2=s$.  
Then we find
\begin{equation}
|\k'_1\!|= {1\over2\sqrt{s}}\sqrt{s^2-2(m_{1'}^2+m_{2'}^2)s
                                  +(m_{1'}^2-m_{2'}^2)^2}
\quad\hbox{(CM frame)}\;.
\label{k3}
\end{equation}
Now the only variable left to specify about the final state 
is the angle $\theta$ between $\k_1$ and $\k'_1$.  
However, it is often more convenient to work with the Lorentz scalar
$t\equiv-(k_1-k'_1)^2$, which is related to $\theta$ by
\begin{equation}
t=m_1^2+m_{1'}^2-2E_1 E'_1 + 2|\k_1\!||\k'_1\!|\cos\theta \;.
\label{tth}
\end{equation}
This formula is valid in any frame.

The Lorentz scalars $s$ and $t$ are two of the three {\it Mandelstam
variables}, defined as
\begin{eqnarray}
s &\equiv& -(k_1{+}k_2)^2 = -(k'_1{+}k'_2)^2 \;,
\nonumber \\
\noalign{\smallskip}
t &\equiv& -(k_1{-}k'_1)^2 = -(k_2{-}k'_2)^2 \;,
\nonumber \\
\noalign{\smallskip}
u &\equiv& -(k_1{-}k'_2)^2 = -(k_2{-}k'_1)^2 \;.
\label{mand}
\end{eqnarray}
The three Mandelstam variables are not independent; 
they satisfy the linear relation
\begin{equation}
s+t+u=m_1^2+m_2^2+m_{1'}^2+m_{2'}^2\;.
\label{stu}
\end{equation}
In terms of $s$, $t$, and $u$, we can rewrite \eq{t} as
\begin{equation}
\T = g^2\left[{1\over m^2-s}+{1\over m^2-t}+{1\over m^2-u}\right]+O(g^4)\;,
\label{tscat11}
\end{equation}
which demonstrates the notational utility of the Mandelstam variables.

Now let us consider a different frame, the {\it fixed target\/} or 
{\it FT frame\/} (also sometimes called the {\it lab frame\/}), 
in which particle \#2 is initially at rest: $\k_2={\bf 0}$.
In this case we have
\begin{equation}
|\k_1\!|= {1\over2m_2}\sqrt{s^2-2(m_1^2+m_2^2)s+(m_1^2-m_2^2)^2}
\quad\hbox{(FT frame)} \;.
\label{k1ft}
\end{equation}
Note that, from \eqs{k1ft} and (\ref{k1}),
\begin{equation}
m_2|\k_1\!|_{\rm\sss FT} = \sqrt{s}\,|\k_1\!|_{\rm\sss CM} \;.
\label{k1k1}
\end{equation}
This will be useful later.

We would now like to derive a formula for the differential scattering
cross section.  In order to do so, we assume that the whole experiment
is taking place in a big box of volume $V$, and lasts for a large time $T$.
We should really think about wave packets coming together, but we will use
some simple shortcuts instead.
Also, to get a more general answer, we will let the number of outgoing
particles be arbitrary.

Recall from section 10
that the overlap between the initial and final states is given by
\begin{equation}
\la f|i\ra = i(2\pi)^4\delta^4(k_{\rm in}{-}k_{\rm out})\,\T \;.
\label{fit2}
\end{equation}
To get a probability, we must square $\la f|i\ra$, 
and divide by the norms of the initial and final states: 
\begin{equation}
P = { |\la f|i\ra|^2 \over \la f|f\ra \la i|i\ra } \;.
\label{prob}
\end{equation}
The numerator of this expression is
\begin{equation}
|\la f|i\ra|^2 =
[(2\pi)^4 \delta^4(k_{\rm in}{-}k_{\rm out})]^2\,|\T|^2 \;.
\label{num2}
\end{equation}
We write the square of the delta function as 
\begin{equation}
[(2\pi)^4 \delta^4(k_{\rm in}{-}k_{\rm out})]^2
=(2\pi)^4 \delta^4(k_{\rm in}{-}k_{\rm out})\times(2\pi)^4 \delta^4(0)\;,
\label{sqd}
\end{equation}
and note that
\begin{equation}
(2\pi)^4 \delta^4(0)=\int \dfx\,e^{i0\cdot x}= VT\;.
\label{sqd2}
\end{equation}
Also, the norm of a single particle state is given by
\begin{equation}
\la k|k\ra = (2\pi)^3 2k^0\delta^3({\bf 0})=2k^0 V\;.
\label{norm2}
\end{equation}
Thus we have
\begin{eqnarray}
\la i|i\ra &=& 4E_1 E_2 V^2 \;,
\label{ii} \\
\la f|f\ra &=& \prod_{j=1}^{n'}2k'_j{}^0 V\;,
\label{ff}
\end{eqnarray}
where $n'$ is the number of outgoing particles.

If we now divide \eq{prob} by the elapsed time $T$, we get a probability per
unit time
\begin{equation}
\dot P ={
(2\pi)^4 \delta^4(k_{\rm in}{-}k_{\rm out})\,V\,|\T|^2
\over
4E_1 E_2 V^2 \prod_{j=1}^{n'}2k^{'0}_j V} \;.
\label{dpdt}
\end{equation}
This is the probability per unit time to scatter into a set of outgoing
particles with precise momenta.  To get something measurable, we should
sum each outgoing three-momentum $\k_j'$ over some small range.
Due to the box, all three-momenta are quantized: $\k'_j=(2\pi/L){\bf n}'_j$,
where $V=L^3$, and ${\bf n}'_j$ is a three-vector with integer entries.  
(Here we have assumed periodic boundary conditions, 
but this choice does not affect the final result.)
In the limit of large $L$, we have
\begin{equation}
\sum_{{\bf n}_j'}\to {V\over(2\pi)^3}\int\dtk{}'_j \;.
\label{sumtoint}
\end{equation}
Thus we should multiply $\dot P$ by a factor of $V\dtk{}'_j/(2\pi)^3$
for each outgoing particle.  Then we get
\begin{equation}
{\dot P}={
(2\pi)^4 \delta^4(k_{\rm in}{-}k_{\rm out})
\over 4E_1 E_2 V} \, |\T|^2 \,
\prod_{j=1}^{n'}\dk{}'_j \;,
\label{dpdt2}
\end{equation}
where we have identified the Lorentz-invariant phase-space differential
\begin{equation}
\dk \equiv {\dtk\over(2\pi)^3 2k^0}
\label{dk2}
\end{equation}
that we first introduced in section 3.

To convert $\dot P$ to a differential cross section $d\sigma$, 
we must divide by the incident flux.
Let us see how this works in the FT frame, where particle \#2 is at rest.
The incident flux is the number of particles per unit
volume that are striking the target particle (\#2), times their speed.  
We have one incident particle (\#1) in a volume $V$
with speed $v=|\k_1\!|/E_1$, and so the incident flux is $|\k_1\!|/E_1 V$.  
Dividing \eq{dpdt2} by this flux cancels the last factor of $V$, 
and replaces $E_1$ in the denominator with $|\k_1\!|$.
We also set $E_2=m_2$ and 
note that \eq{k1ft} gives $|\k_1\!|m_2$ as a function of $s$; 
$d\sigma$ will be Lorentz invariant if, in other frames, 
we simply use this function as the value of $|\k_1\!|m_2$.  
Adopting this convention, and using \eq{k1k1}, we have
\begin{equation}
d\sigma = {1 \over 4 |\k_1\!|_{\sss\rm CM} \sqrt{s}} 
          \;|\T|^2\;d\hbox{LIPS}_{n'} \;,
\label{ds0}
\end{equation}
where $|\k_1\!|_{\sss\rm CM}$ is given as a function of $s$ by \eq{k1},
and we have defined the $n'$-body Lorentz-invariant phase-space measure
\begin{equation}
d\hbox{LIPS}_{n'} \equiv 
(2\pi)^4 \delta^4(k_1{+}k_2{-}{\ts{\sum_{j=1}^{n'}k'_i}})
                                        \prod_{j=1}^{n'}\dk{}'_j\;.
\label{dlips}
\end{equation}
\Eq{ds0} is our final result for the differential cross section
for the scattering of two incoming particles into $n'$ outgoing particles.

Let us now specialize to the case of two outgoing particles.  
We need to evaluate
\begin{equation}
d\hbox{LIPS}_2=(2\pi)^4 \delta^4(k_1{+}k_2{-}k'_1{-}k'_2)\,\dk{}'_1\dk{}'_2\;.
\label{dlips2}
\end{equation}
Since $d\hbox{LIPS}_2$ is Lorentz invariant, 
we can compute it in any convenient frame.  
Let us work in the CM frame, where $\k_1+\k_2={\bf 0}$ and
$E_1+E_2=\sqrt{s}$; then we have
\begin{equation}
d\hbox{LIPS}_2={1\over 4(2\pi)^2 E'_1 E'_2}\,
\delta(E'_1{+}E'_2{-}\sqrt{s}\,)\,
\delta^3(\k'_1{+}\k'_2)\,\dtk{}'_1\dtk{}'_2\;.
\label{dlips3}
\end{equation}
We can use the spatial part of the delta function to integrate over 
$\dtk{}'_2$, with the result
\begin{equation}
d\hbox{LIPS}_2={1\over 4(2\pi)^2 E'_1 E'_2}\,
\delta(E'_1{+}E'_2{-}\sqrt{s}\,)\,\dtk{}'_1 \;,
\label{dlips4}
\end{equation}
where now
\begin{equation}
E'_1 = \sqrt{\k'_1{}^2+m_{1'}^2} 
\quad \hbox{and} \quad
E'_2 = \sqrt{\k'_1{}^2+m_{2'}^2}\;.
\label{ep}
\end{equation}
Next, let us write
\begin{equation}
\dtk{}'_1=|\k'_1\!|^2 \,d|\k'_1\!|\, d\Omega_{\sss\rm CM} \;,
\label{dtk}
\end{equation}
where $d\Omega_{\sss\rm CM} = \sin\theta\,d\theta\,d\phi$ 
is the differential solid angle,
and $\theta$ is the angle between $\k_1$ and $\k'_1$ in the CM frame.  
We can carry out the integral over the magnitude of $\k'_1$ in
\eq{dlips4} using $\int dx\,\delta(f(x))=\sum_i |f'(x_i)|^{-1}$,
where $x_i$ satisfies $f(x_i)=0$.  In our case, the argument of the
delta function vanishes at just one value of
$|\k'_1\!|$, the value given by \eq{k3}.  Also, the derivative of that
argument with respect to $|\k'_1\!|$ is 
\begin{eqnarray}
{\d\over\d |\k'_1\!|}\Bigl(E'_1+E'_2-\sqrt{s}\Bigr)
&=&{ |\k'_1\!| \over E'_1 } +{ |\k'_1\!| \over E'_2 }
\nonumber \\
\noalign{\smallskip}
&=& |\k'_1\!| \! \left( {E'_1+E'_2 \over E'_1 \, E'_2} \right)
\nonumber \\
\noalign{\smallskip}
&=& { |\k'_1\!|\sqrt{s} \over E'_1 E'_2} \;.
\label{derivdelta}
\end{eqnarray}
Putting all of this together, we get
\begin{equation}
d\hbox{LIPS}_2={ |\k'_1\!| \over 16\pi^2 \sqrt{s}}\;d\Omega_{\sss\rm CM}\;.
\label{dlips5}
\end{equation}
Combining this with \eq{ds0}, we have
\begin{equation}
{d\sigma\over d\Omega}_{\sss\rm \!CM}
 = {1 \over 64\pi^2 s}\,{|\k'_1\!|\over |\k_1\!|}\;|\T|^2\;,
\label{ds}
\end{equation}
where $|\k_1\!|$ and $|\k'_1\!|$ are the functions of $s$ given by
\eqs{k1} and (\ref{k3}), and $d\Omega_{\sss\rm CM}$ is the differential
solid angle in the CM frame.

The differential cross section can also be expressed in a frame-independent
manner by noting that, in the CM frame, we can take the differential of 
\eq{tth} at fixed $s$ to get
\begin{eqnarray}
dt &=& 2\,|\k_1\!|\, |\k'_1\!|\,d\cos\theta
\label{dt} \\
\noalign{\medskip}
&=&  2\,|\k_1\!|\, |\k'_1\!|\,{d\Omega_{\sss\rm CM}\over2\pi}\;.
\label{dt2}
\end{eqnarray}
Now we can rewrite \eq{ds} as
\begin{equation}
{d\sigma\over dt} = {1 \over 64\pi s |\k_1\!|^2} \;|\T|^2 \;,
\label{dsdt}
\end{equation}
where $|\k_1\!|$ is given as a function of $s$ by \eq{k1}.

We can now transform $d\sigma/dt$ into $d\sigma/d\Omega$ in any frame
we might like (such as the FT frame) by taking the differential of \eq{tth}
in that frame.
In general, though, $|\k'_1\!|$ depends on $\theta$ as well as $s$,
so the result is more complicated than it is in \eq{dt} for the CM frame.

Returning to the general case of $n'$ outgoing particles,
we can define a Lorentz invariant {\it total cross section\/} 
by integrating completely over all the outgoing momenta, and
dividing by an appropriate {\it symmetry factor} $S$.
If there are $n'_i$ identical outgoing particles of type $i$, then
\begin{equation}
S=\prod_i n'_i ! \;,
\label{s}
\end{equation}
and
\begin{equation}
\sigma = {1\over S}\int d\sigma \;,
\label{stot}
\end{equation}
where $d\sigma$ is given by \eq{ds0}.
We need the symmetry factor because merely integrating over all the outgoing
momenta in $d$LIPS${}_{n'}$ treats the final state as being labeled by an 
{\it ordered\/} list of these momenta.  But if some outgoing particles
are identical, this is not correct; 
the momenta of the identical particles should
be specified by an {\it unordered\/} list
[because, for example, the state $\ad_1\ad_2|0\ra$ is identical to
the state $\ad_2\ad_1|0\ra$].  The symmetry factor provides the
appropriate correction.

In the case of two outgoing particles, \eq{stot} becomes
\begin{eqnarray}
\sigma &=& {1\over S}\int d\Omega_{\sss\rm CM}\,
                        {d\sigma\over d\Omega}_{\sss\rm \!CM} 
\label{stotdom} \\
\noalign{\medskip}
&=& {2\pi\over S}\int_{-1}^{+1} d\cos\theta\,
                        {d\sigma\over d\Omega}_{\sss\rm \!CM} \;,  
\label{stotdcos}
\end{eqnarray}
where $S=2$ if the two outgoing particles are identical, and $S=1$
if they are distinguishable.  Equivalently, we can compute $\sigma$
from \eq{dsdt} via
\begin{equation}
\sigma = {1\over S}\int_{t_{\rm min}}^{t_{\rm max}}dt\,{d\sigma\over dt} \;,
\label{stotdt}
\end{equation}
where $t_{\rm min}$ and $t_{\rm max}$ are given by \eq{tth} in the
CM frame with $\cos\theta=-1$ and $+1$, respectively.
To compute $\sigma$ with \eq{stotdcos}, we should first express
$t$ and $u$ in terms of $s$ and $\theta$ via \eqs{tth} and (\ref{stu}),
and then integrate over $\theta$ at fixed $s$.
To compute $\sigma$ with \eq{stotdt}, we should first express
$u$ in terms of $s$ and $t$ via \eq{stu},
and then integrate over $t$ at fixed $s$.

Let us see how all this works for the scattering amplitude of $\ph^3$ theory,
\eq{tscat11}.  In this case, all the masses are equal, and so, in the CM frame,
$E=\half\sqrt{s}$ for all four particles, 
and $|\k'_1\!|=|\k_1\!|=\half(s-4m^2)^{1/2}$.
Then \eq{tth} becomes
\begin{equation}
t=-\half(s-4m^2)(1-\cos\theta) \;.
\label{t3}
\end{equation}
From \eq{stu}, we also have
\begin{equation}
u=-\half(s-4m^2)(1+\cos\theta) \;.
\label{u3}
\end{equation}
Thus $|\T|^2$ is quite a complicated function of  $s$ and $\theta$.
In the nonrelativistic limit, $|\k_1\!|\ll m$ or equivalently
$s-4m^2\ll m^2$, we have
\begin{eqnarray}
\T &=& {5g^2\over 3m^2} 
\left[\,1-{8\over 15}\left({s-4m^2\over m^2}\right)
         +{5\over 18}\left(1+{27\over25}\cos^2\theta\right)
          \left({s-4m^2\over m^2}\right)
          {\vphantom{m}^{\!\!\!2}\atop\vphantom{m^2}} 
         + \ldots\,\right]
\nonumber \\
\noalign{\medskip}
&& {} +O(g^4) \;.
\label{nonrel}
\end{eqnarray}
Thus the differential cross section is nearly isotropic.  In the extreme
relativistic limit, $|\k_1\!|\gg m$ or equivalently
$s\gg m^2$, we have
\begin{eqnarray}
\T &=& {g^2\over s\sin^2\theta} 
\left[\,{3+\cos^2\theta}
       -\left({(3+\cos^2\theta)^2\over\sin^2\theta}-16\right){m^2\over s}
       + \ldots\,\right]
\nonumber \\
\noalign{\medskip}
&& {} +O(g^4) \;.
\label{exrel}
\end{eqnarray}
Now the differential cross section is sharply peaked in the forward
($\theta=0$) and backward ($\theta=\pi$) directions.

We can compute the total cross section $\sigma$ from \eq{stotdt}.
We have in this case $t_{\rm min}=-(s-4m^2)$ and $t_{\rm max}=0$.
Since the two outgoing particles are identical, the symmetry factor is $S=2$.
Then setting $u=4m^2-s-t$, and performing the integral in \eq{stotdt}
over $t$ at fixed $s$, we get 
\begin{eqnarray}
\sigma &=& {g^4\over 32\pi s(s-4m^2)}
\Biggl[\,{2\over m^2}
      + {s-4m^2\over(s-m^2)^2}
      - {2\over s -3m^2 }
\nonumber \\
&& \qquad \quad {} + {4 m^2 \over (s - m^2)(s - 2m^2)}
        \ln\!\left({s - 3m^2\over m^2}\right)\Biggr] +O(g^6) \;.
\label{stottot}
\end{eqnarray}
In the nonrelativistic limit, this becomes
\begin{equation}
\sigma = {25g^4\over 1152\pi m^6}
           \left[\, 1 - {79\over60} \left({s-4m^2\over m^2}\right)
           + \ldots\,\right] + O(g^6) \;.
\label{stotnonrel}
\end{equation}
In the extreme relativistic limit, we get
\begin{equation}
\sigma = {g^4\over 16\pi m^2 s^2}
           \left[\, 1 + {7\over2}\,{m^2\over s}
           + \ldots\,\right] + O(g^6) \;.
\label{stotexrel}
\end{equation}
These results illustrate how even a very simple quantum field theory
can yield specific predictions for cross sections that could be
tested experimentally.

Let us now turn to the other basic problem mentioned at the beginning 
of this section: the case of a single incoming particle that {\it decays\/}
to $n'$ other particles.

We have an immediate conceptual problem.  According to our development of
the LSZ formula in section 5, each incoming and outgoing particle should
correspond to a single-particle state that is an exact eigenstate of the
exact hamiltonian.  This is clearly not the case for a particle 
that can decay.
Referring to \fig{lsz}, the hyperbola of such a particle must lie above the
continuum threshold.  Strictly speaking, then,
the LSZ formula is not applicable.

A proper understanding of this issue requires a study of loop corrections
that we will undertake in section 24.  For now, we will simply assume that
the LSZ formula continues to hold for a single incoming particle.
Then we can retrace the steps from \eq{prob} to \eq{dpdt2};
the only change is that the norm of the initial state is now
\begin{equation}
\la i|i\ra = 2E_1V
\label{ii2}
\end{equation}
instead of \eq{ii}.  Identifying the differential decay rate $d\Gamma$
with $\dot P$ then gives
\begin{equation}
d\Gamma = {1\over2E_1}\,|\T|^2\,d\hbox{LIPS}_{n'} \;,
\label{dg}
\end{equation}
where now $k_2=0$ and $s=m_1^2$.
In the CM frame (which is now the rest frame of the initial particle),
we have $E_1=m_1$; 
in other frames, the relative factor of $E_1/m_1$ in $d\Gamma$ 
accounts for relativistic time dilation of the decay rate.

We can also define a total decay rate by integrating over all the outgoing
momenta, and dividing by the symmetry factor of \eq{s}:
\begin{equation}
\Gamma = {1\over S}\int d\Gamma \;.
\label{gam}
\end{equation}

\vskip0.5in

\begin{center}
Problems
\end{center}

\vskip0.25in

11.1a) Consider a theory of a two real scalar fields $A$ and $B$ with
an interaction $\L_1 = gAB^2$.  Assuming that $m_{\sss A}>2m_{\sss B}$, 
compute the total decay rate of the $A$ particle at tree level.

b) Consider a theory of a real scalar field $\ph$ and a complex scalar
field $\chi$ with $\L_1 = g\ph\chi^\dagger\chi$.
Assuming that $m_\ph>2m_\chi$, compute the total decay rate of the $\ph$
particle at tree level.


%
%

\vfill\eject

%% file: ch012.tex
\noindent Quantum Field Theory  \hfill   Mark Srednicki

\vskip0.5in

\begin{center}
\large{12: The \LK Form of the Exact Propagator}
\end{center}
\begin{center}
Prerequisite: 9
\end{center}

\vskip0.5in

Before turning to the subject of loop corrections to scattering amplitudes,
it will be helpful to consider what we can learn about the exact propagator
$\bfd(x-y)$ from general principles.  We define the exact propagator via
\begin{equation}
\bfd(x-y)\equiv i\la 0|{\rm T}\ph(x)\ph(y)|0\ra \;.
\label{bfd}
\end{equation}
We take the field $\ph(x)$ to be normalized so that
\begin{equation}
\la 0|\ph(x)|0\ra = 0 \qquad \hbox{and} \qquad
\la k|\ph(x)|0\ra = e^{-ikx} \;,
\label{cond0}
\end{equation}
where the one-particle state $|k\ra$ has the normalization
\begin{equation}
\la k|k'\ra = (2\pi)^3\,2\w\,\delta^3(\k-\k')\;,
\label{norm3}
\end{equation}
with $\omega=(\k^2+m^2)^{1/2}$.
The corresponding completeness statement is
\begin{equation}
\int\dk\;|k\ra\la k| = I_1\;,
\label{i1}
\end{equation}
where $I_1$ is the identity operator in the one-particle subspace, and
\begin{equation}
\dk \equiv {\dtk\over(2\pi)^3 2\w}
\label{dtk2}
\end{equation}
is the Lorentz invariant phase-space differential.
We also define the exact momentum-space propagator $\tbfd(k^2)$ via
\begin{equation}
\bfd(x-y)\equiv \int {\dfk\over(2\pi)^4}\,e^{ik(x-y)}\,\tbfd(k^2) \;.
\label{tbfd}
\end{equation}

In free-field theory, the momentum-space propagator is
\begin{equation}
\td(k^2) = {1\over k^2+m^2-i\eps}\;.
\label{td}
\end{equation}
It has an isolated pole at $k^2=-m^2$ with residue one;  $m$ is the actual,
physical mass of the particle, the mass that enters into the energy-momentum relation.  

Now let us return to the exact propagator, \eq{bfd}, take $x^0>y^0$,
and insert a complete set of energy eigenstates between the two fields.
Recall from section 5 that there are three general classes of energy
eigenstates:  (1) The ground state or vacuum $|0\ra$, which is a single state
with zero energy and momentum.  (2) The one particle states $|k\ra$, specified
by a three-momentum $\k$ and with energy $\omega=(\k^2+m^2)^{1/2}$.
(3) States in the multiparticle continuum $|k,n\ra$, 
specified by a three-momentum $\k$
and other parameters (such as relative momenta among the
different particles) that we will collectively denote as $n$.
The energy  of one of these states is $\omega=(\k^2+M^2)^{1/2}$, 
where $M\ge 2m$; $M$ is one of the parameters in the set $n$.
Thus we get
\begin{eqnarray}
\la0|\ph(x)\ph(y)|0\ra 
&=& \la0|\ph(x)|0\ra\la0|\ph(y)|0\ra 
\nonumber \\
\noalign{\medskip}
&& {} + \int\dk\,\la0|\ph(x)|k\ra\la k|\ph(y)|0\ra
\nonumber \\
&& {} + \sum_n\int\dk\,\la0|\ph(x)|k,n\ra\la k,n|\ph(y)|0\ra \;.
\label{comp}
\end{eqnarray}
The first two terms can be simplified via \eq{cond0}.  
Also, writing the field as
$\ph(x)=\exp(-iP^\mu x_\mu)\ph(0)\exp(+iP^\mu x_\mu)$,
where $P^\mu$ is the energy-momentum operator, gives us
\begin{equation}
\la k,n|\ph(x)|0\ra = e^{-ikx} \la k,n|\ph(0)|0\ra \;,
\label{cond20}
\end{equation}
where $k^0=(\k^2+M^2)^{1/2}$.  We now have
\begin{equation}
\la0|\ph(x)\ph(y)|0\ra 
=  \int\dk\,e^{ik(x-y)}
 + \sum_n\int\dk\,e^{ik(x-y)}|\la k,n|\ph(0)|0\ra|^2 \;.
\label{comp2}
\end{equation}
Next, we define the {\it spectral density}
\begin{equation}
\rho(s) \equiv \sum_n |\la k,n|\ph(0)|0\ra|^2\,\delta(s-M^2) \;.
\label{rho}
\end{equation}
Obviously, $\rho(s)\ge 0$ for $s\ge 4m^2$, and $\rho(s)=0$ for $s<4m^2$.
Now we have
\begin{equation}
\la0|\ph(x)\ph(y)|0\ra 
=  \int\dk\,e^{ik(x-y)}
 + \int_{4m^2}^\infty ds\,\rho(s)\int\dk\,e^{ik(x-y)} \;.
\label{comp3}
\end{equation}
In the first term, $k^0=(\k^2+m^2)^{1/2}$, and in the second term,
$k^0=(\k^2+s)^{1/2}$.  Clearly we can also swap $x$ and $y$ to get
\begin{equation}
\la0|\ph(y)\ph(x)|0\ra 
=  \int\dk\,e^{-ik(x-y)}
 + \int_{4m^2}^\infty ds\,\rho(s)\int\dk\,e^{-ik(x-y)} \;.
\label{comp4}
\end{equation}
as well.  We can then combine \eqs{comp3} and (\ref{comp4}) into a formula for the
time-ordered product 
\begin{equation}
\la0|{\rm T}\ph(x)\ph(y)|0\ra =
\theta(x^0-y^0)\la0|\ph(x)\ph(y)|0\ra + 
\theta(y^0-x^0)\la0|\ph(y)\ph(x)|0\ra,
\label{to}
\end{equation}
where $\theta(t)$ is the unit step function, by means of the identity
\begin{equation}
\int {\dfk\over(2\pi)^4}\,{e^{ik(x-y)}\over k^2+m^2-i\eps}
=i\theta(x^0{-}y^0)\int\dk\,e^{ik(x-y)}
+i\theta(y^0{-}x^0)\int\dk\,e^{-ik(x-y)}\;;
\label{cont}
\end{equation}
the derivation of \eq{cont} was sketched in section 8.
Combining \eqs{comp3}, (\ref{comp4}), (\ref{to}), and (\ref{cont}), we get
\begin{eqnarray}
i\la0|{\rm T}\ph(x)\ph(y)|0\ra 
&=&  \int {\dfk\over(2\pi)^4}\,e^{ik(x-y)}\Biggl[\,{1\over k^2+m^2-i\eps}
\nonumber \\
&&\quad\, {} +  \int_{4m^2}^\infty ds\,\rho(s)\,{1\over k^2+{\;s\;}-i\eps}\,\Biggr] .
\label{comp5}
\end{eqnarray}
Comparing \eqs{bfd}, (\ref{tbfd}), and (\ref{comp5}), we see that
\begin{equation}
\tbfd(k^2)=
{1\over k^2+m^2-i\eps} 
+  \int_{4m^2}^\infty ds\,\rho(s)\,{1\over k^2+s-i\eps} \;.
\label{tbfd212}
\end{equation}
This is the 
{\it \LK form\/} of the exact momentum-space propagator $\tbfd(k^2)$.  
We note in particular that 
$\tbfd(k^2)$ {\it has an isolated pole at $k^2=-m^2$ with residue one},
just like the propagator in free-field theory.  

\vfill\eject

%% file: ch013.tex

\noindent Quantum Field Theory  \hfill   Mark Srednicki

\vskip0.5in

\begin{center}
\large{13: Dimensional Analysis with $\hbar=c=1$}
\end{center}
\begin{center}
Prerequisite: 3
\end{center}

\vskip0.5in

We have set $\hbar=c=1$.  This allows us to convert a time $T$ to a length $L$
via $T=L/c$, and a length $L$ to a mass $M$ via $M=\hbar c^{-1}/L$.  Thus
any quantity $A$ can be thought of as having units of mass to some some power
(positive, negative, or zero) that we will call $[A]$.  For example,
\begin{eqnarray}
[m] &=& +1 \;,
\label{mdm} \\
{[}x^\mu{]} &=& -1 \;,
\label{mdx} \\
{[}\d^\mu{]} &=& +1 \;,
\label{mdd} \\
{[}\ddx{]} &=& -d \;.
\label{mddd}
\end{eqnarray}
In the last line, we have generalized our considerations to theories in $d$
spacetime dimensions.

Let us now consider a scalar field in $d$ spacetime dimensions with 
lagrangian density 
\begin{equation}
\L=-\half\d^\mu\ph\d_\mu\ph-\half m^2\ph^2 
      - \sum_{n=3}^N {\ts{1\over n!}}g_n\ph^n\;.
\label{calell2}
\end{equation}
The action is
\begin{equation}
S = \int \ddx\,\L \;,
\label{calell0}
\end{equation}
and the path integral is
\begin{equation}
Z(J) = \int\D\ph\,\exp\!\left[\,i\!\int \ddx\,(\L+J\ph)\right] \;.
\label{zj13}
\end{equation}
From \eq{zj13}, 
we see that the action $S$ must be dimensionless, because it appears
as the argument of the exponential function.  Therefore
\begin{equation}
[S]=0\;.
\label{mds}
\end{equation}
From \eqs{mds} and (\ref{mddd}), we see that
\begin{equation}
[\L]=+d\;.
\label{mdl}
\end{equation}
Then, from \eqs{mdl} and (\ref{mdd}), and the fact that $\d^\mu\ph\d_\mu\ph$ is
a term in $\L$, we see that we must have
\begin{equation}
[\ph]={1\over 2}(d-2)\;.
\label{mdph}
\end{equation}
Then, since $g_n\ph^n$ is also a term in $\L$, we must have
\begin{equation}
[g_n]=d-{n\over2}(d-2)\;.
\label{mdg}
\end{equation}
In particular, for the $\ph^3$ theory we have been working with, we have
\begin{equation}
[g_3]={1\over 2}(6-d)\;.
\label{mdg3}
\end{equation}
Thus we see that the coupling constant of $\ph^3$ theory is dimensionless
in $d=6$ spacetime dimensions.

Theories with dimensionless couplings tend to be more interesting than theories
with dimensionful couplings.  This is because any nontrivial dependence of
a scattering amplitude on a coupling must be expressed as a function of
a dimensionless parameter.  If the coupling is itself dimensionful, this
parameter must be the ratio of the coupling to the appropriate power of either
the particle mass $m$ (if it isn't zero) or, in the high-energy regime 
$s\gg m^2$, the Mandelstam variable $s$.  
Thus the relevant parameter is  $g\,s^{-[g]/2}$.
If $[g]$ is negative [and it usually is: see \eq{mdg}], 
then $g\,s^{-[g]/2}$ blows up at high energies, and the perturbative
expansion breaks down.  This behavior is connected to the
{\it nonrenormalizability\/} of theories with couplings 
with negative mass dimension,
a subject we will take up in section 18.  
It turns out that, at best, such theories
require an infinite number of input parameters to make sense.  In the 
opposite case, $[g]$ positive, the theory becomes trivial at high energy,
because $g\,s^{-[g]/2}$ goes rapidly to zero.  

Thus the case of $[g]=0$ is just right: scattering amplitudes 
can have a nontrivial dependence on $g$ at all energies.

Therefore, from here on, we will be primarily interested in $\ph^3$ theory 
in $d=6$ spacetime dimensions, where $[g_3]=0$.

\vfill\eject

%% file: ch014.tex

\noindent Quantum Field Theory  \hfill   Mark Srednicki

\vskip0.5in

\begin{center}
\large{14: Loop Corrections to the Propagator}
\end{center}
\begin{center}
Prerequisite: 10, 12, 13
\end{center}

\vskip0.5in

\begin{figure}
\begin{center}
\epsfig{file=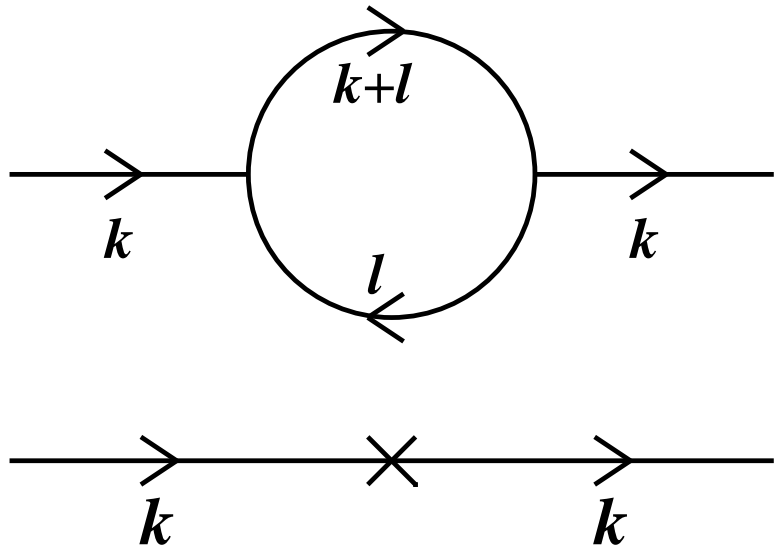}
\end{center}
\caption{The $O(g^2)$ corrections to the propagator.}
\label{propg2}
\end{figure}

In section 10, we wrote the exact propagator as
\begin{equation}
{\ts{1\over i}}\bfd(x_1{-}x_2) \equiv \la0|{\rm T}\ph(x_1)\ph(x_2)|0\ra 
= \delta_1\delta_2 iW(J)\Big|_{J=0} \;,
\label{propagain}
\end{equation}
where $iW(J)$ is the sum of connected diagrams, and $\delta_i$ acts
to remove a source from a diagram and label the corresponding propagator
endpoint $x_i$.  In $\ph^3$ theory, the $O(g^2)$ corrections to 
${\ts{1\over i}}\bfd(x_1{-}x_2)$ come from the diagrams of \fig{propg2}.
To compute them, it is simplest to work directly in momentum space, 
following the Feynman rules of section 10.  
An appropriate assignment of momenta to the lines is shown
in \fig{propg2}; we then have
\begin{equation}
{\ts{1\over i}}\tbfd(k^2) = {\ts{1\over i}}\td(k^2) 
+ {\ts{1\over i}}\td(k^2)\Bigl[i\Pi(k^2)\Bigr]{\ts{1\over i}}\td(k^2)
+O(g^4) \;,
\label{ppip}
\end{equation}
where
\begin{equation}
\td(k^2) = {1\over k^2+m^2-i\eps}
\label{tdtd}
\end{equation}
is the free-field propagator, and
\begin{eqnarray}
i\Pi(k^2) &=& \half(ig)^2\left({\ts{1\over i}}\right)^{\!2}
\int{\ddl\over(2\pi)^d}\,\td((\ell{+}k)^2)\td(\ell^2) 
\nonumber \\
&& {} - i(Ak^2 + Bm^2)  +O(g^4) \;.
\label{pidi}
\end{eqnarray}
Here we have written the integral appropriate for $d$ spacetime dimensions;
for now we will leave $d$ arbitrary, but later we will want to focus on $d=6$,
where the coupling $g$ is dimensionless.  The factor of one-half 
is due to the symmetry factor associated with
exchanging the top and bottom semicircular propagators.
Also, we have written the vertex factor as $ig$ rather than $iZ_g g$
because we expect $Z_g=1+O(g^2)$, and so the $Z_g-1$ contribution
can be lumped into the $O(g^4)$ term.
In the second term, $A=Z_\ph-1$ and $B=Z_m-1$ 
are both expected to be $O(g^2)$.

\begin{figure}
\begin{center}
\epsfig{file=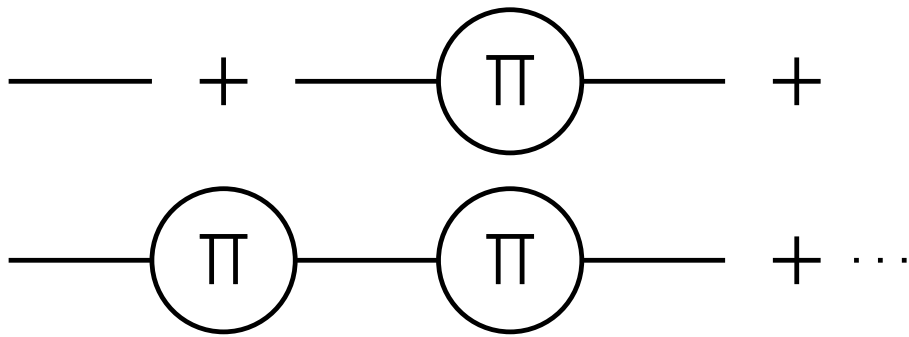}
\end{center}
\caption{The infinite series of insertions of $\Pi(k^2)$.}
\label{piseries}
\end{figure}

\begin{figure}
\begin{center}
\epsfig{file=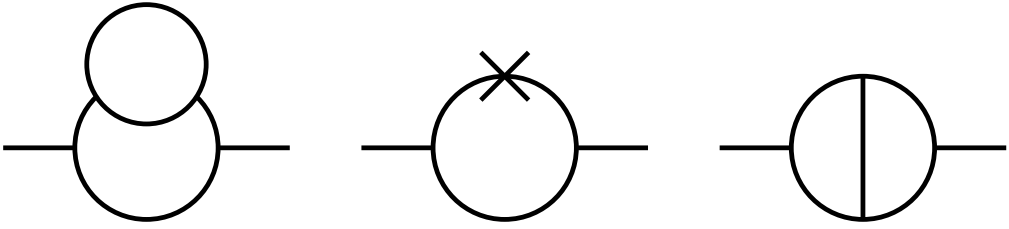}
\end{center}
\caption{The $O(g^4)$ contributions to $\Pi(k^2)$.}
\label{propg4}
\end{figure}

Before evaluating $\Pi(k^2)$,
let us consider the infinite series of diagrams that result from
further insertions of $\Pi(k^2)$, as shown in \fig{piseries}.
We have
\begin{eqnarray}
{\ts{1\over i}}\tbfd(k^2) = {\ts{1\over i}}\td(k^2) 
&+& {\ts{1\over i}}\td(k^2)\Bigl[i\Pi(k^2)\Bigr]{\ts{1\over i}}\td(k^2)
\nonumber \\
&+& {\ts{1\over i}}\td(k^2)
\Bigl[i\Pi(k^2)\Bigr]{\ts{1\over i}}\td(k^2)
\Bigl[i\Pi(k^2)\Bigr]{\ts{1\over i}}\td(k^2)
\nonumber \\
&+& \ldots \;.
\label{dpdpd}
\end{eqnarray}
This sum will include {\it all\/} the diagrams that contribute to $\tbfd(k^2)$
if we define $i\Pi(k^2)$ to be given by the sum of all diagrams that are 
{\it one-particle irreducible}, or 1PI for short.  
A diagram is 1PI if it is still simply connected after any one line is cut.
The $O(g^4)$ contributions to $i\Pi(k^2)$ are shown in \fig{propg4}.
In writing down the value of one of these diagrams, we omit the
two external propagators.

The nice thing about \eq{dpdpd} is that it represents a geometric series
that can be summed up to give
\begin{equation}
\tbfd(k^2) = {1\over k^2+m^2-i\eps-\Pi(k^2)}\;.
\label{tbfd2}
\end{equation}
$\Pi(k^2)$ is called the {\it self-energy\/} of the particle.

In section 12, we learned that the exact propagator has a pole at
$k^2=-m^2$ with residue one.  
This is consistent with \eq{tbfd2} if and only if
\begin{eqnarray}
\Pi(-m^2) &=& 0 \;,
\label{pi0} \\
\Pi'(-m^2) &=& 0 \;,
\label{pip0}
\end{eqnarray}
where the prime denotes a derivative with respect to $k^2$.
{\it We will use \eqs{pi0} and (\ref{pip0}) to fix the values
of $A$ and $B$.}

Next we turn to the evaluation of the $O(g^2)$ contribution to
$i\Pi(k^2)$ in \eq{pidi}.  We have the immediate problem that
the integral on the right-hand side clearly diverges at large $\ell$ 
for $d\ge 4$.
We faced a similar situation in section 9 when we evaluated the lowest-order
tadpole diagram.  There we modified $\td(\ell^2)$ by changing its
behavior at large $\ell^2$.  Here, for now, we will simply restrict
our attention to $d<4$, where the integral in \eq{pidi} is finite.
Later we will see what we can say about larger values of $d$.

We will evaluate the integral in \eq{pidi} with a series of tricks.  
We first use {\it Feynman's formula\/} to combine denominators,
\begin{equation}
{1\over a_1 \ldots a_n} = \int dF_n\,(x_1 a_1 + \ldots + x_n a_n)^{-n} \;,
\label{fff}
\end{equation}
where the integration measure over the {\it Feynman parameters\/} $x_i$ is
\begin{equation}
\int dF_n  = (n{-}1)!\int_0^1 dx_1\ldots dx_n\;
                   \delta(x_1+\ldots+x_n-1)\;.
\label{dfn}
\end{equation}
This measure is normalized so that
\begin{equation}
\int dF_n\,1 = 1\;.
\label{dfn1}
\end{equation}
\Eq{fff} can be proven by direct evaluation for $n=2$, and by induction
for $n>2$.  

In the case at hand, we have
\begin{eqnarray}
\td((k{+}\ell)^2)\td(\ell^2) &=& 
{1\over(\ell^2+m^2)((\ell+k)^2+m^2)}
\nonumber \\
&=& \int_0^1 dx\,\Bigl[\,x((\ell+k)^2+m^2)+(1{-}x)(\ell^2+m^2)\,\Bigr]^{-2}
\nonumber \\
&=& \int_0^1 dx\,\Bigl[\,\ell^2 + 2x\ell\cd k + xk^2 + m^2\,\Bigr]^{-2}
\nonumber \\
&=& \int_0^1 dx\,\Bigl[\,(\ell+xk)^2 + x(1{-}x)k^2 + m^2\,\Bigr]^{-2}
\nonumber \\
&=& \int_0^1 dx\,\Bigl[\,q^2 + D\,\Bigr]^{-2} \;,
\label{ellq}
\end{eqnarray}
where we have suppressed the $i\eps$'s for notational convenience;
they can be restored via the replacement $m^2\to m^2{-}i\eps$.
In the last line we have defined $q\equiv \ell + xk$ and 
\begin{equation}
D \equiv x(1{-}x)k^2 + m^2 \;.
\label{D14}
\end{equation}
We then change the integration variable in \eq{pidi} from $\ell$ to $q$;
the jacobian is trivial, and we have $\ddl=\ddq$.

\begin{figure}
\begin{center}
\epsfig{file=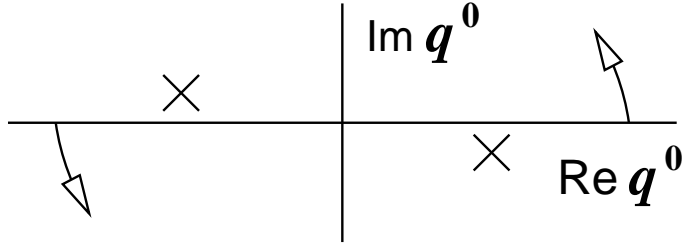}
\end{center}
\caption{The $q^0$ integration contour along the real axis can
be rotated to the imaginary axis without passing through the poles
at $q^0=-\omega+i\eps$ and $q^0=+\omega-i\eps$.}
\label{contour}
\end{figure}

Next, think of the integral over $q^0$ from $-\infty$ to $+\infty$
as a contour integral in the complex $q^0$ plane.  
If the integrand vanishes fast enough as $|q^0|\to\infty$,
we can rotate this contour clockwise by $90^\circ$, as shown in 
\fig{contour}, so that it runs from $-i\infty$ to $+i\infty$. 
In making this {\it Wick rotation}, the contour does not pass over any poles.
(The $i\eps$'s are needed to make this statement unambiguous.)
Thus the value of the integral is unchanged.  It is now convenient to define
a euclidean $d$-dimensional vector $\qbar$ via
$q^0 =  i\qbar_d$ and $q_j = \qbar_j$; then $q^2=\qbar^2$, where 
\begin{equation}
\qbar^2 = \qbar_1^2 + \ldots + \qbar_d^2 \;.
\label{q2}
\end{equation}
Also, $\ddq = i\,\ddqb$.
Therefore, in general,
\begin{equation}
\int \ddq\;f(q^2{-}i\eps)  = i \int \ddqb\; f(\qbar^2) 
\label{qqb14}
\end{equation}
as long as $f(\qbar^2)\to 0$ faster than $1/\qbar^d$ as $\qbar\to\infty$.

Now we can write
\begin{equation}
\Pi(k^2) = \half g^2 I(k^2) - Ak^2 - Bm^2 +O(g^4) \;,
\label{pi2}
\end{equation}
where
\begin{equation}
I(k^2) \equiv \int_0^1 dx \int{\ddqb\over(2\pi)^d}\,{1\over(\qbar^2+D)^2} \;.
\label{ik2}
\end{equation}
It is now straightforward to evaluate the $d$-dimensional integral
over $\qbar$ in spherical coordinates.

Before we perform this calculation, however, let us introduce another trick, 
one that can simplify the task of fixing $A$ and $B$ through the imposition of
\eqs{pi0} and (\ref{pip0}).
Here is the trick:  differentiate $\Pi(k^2)$ twice with respect
to $k^2$ to get
\begin{equation}
\Pi''(k^2) = \half g^2 I''(k^2) +O(g^4) \;,
\label{pidd}
\end{equation}
where, from \eqs{ik2} and (\ref{D14}),
\begin{equation}
I''(k^2) = \int_0^1 dx \;6x^2(1{-}x)^2
\int{\ddqb\over(2\pi)^d}\,{1\over(\qbar^2+D)^4} \;.
\label{idd}
\end{equation}
Then, after we evaluate these integrals, we can get $\Pi(k^2)$ by integrating
with respect to $k^2$, subject to the boundary conditions of 
\eqs{pi0} and (\ref{pip0}).  In this way we can construct $\Pi(k^2)$ without ever
explicitly computing $A$ and $B$.

Notice that this trick does something else for us as well.
The integral over $\qbar$ in \eq{idd} is finite for any $d<8$, whereas the original
integral in \eq{ik2} is finite only for $d<4$.  
This expanded range of $d$ now includes the value
of greatest interest, $d=6$.

How did this happen?  We can gain some insight by making a Taylor expansion of
$\Pi(k^2)$ about $k^2=-m^2$:
\begin{eqnarray}
\Pi(k^2) &=& \left[\,\half g^2 I(-m^2) + (A - B)m^2\,\right] 
\nonumber \\
         && {} +\left[\,\half g^2 I'(-m^2) + A \,\right](k^2+m^2) 
\nonumber \\
         && {} +{\ts{1\over 2!}}\!
                \left[\,\half g^2 I''(-m^2)\,\right](k^2+m^2)^2 + \ldots 
\nonumber \\
         && {} + O(g^4) \;.
\label{pit}
\end{eqnarray}
From \eqs{ik2} and (\ref{D14}), it is straightforward to see that
$I(-m^2)$ is divergent for $d\ge4$, 
$I'(-m^2)$ is divergent for $d\ge6$, and, in general, 
$I^{(n)}(-m^2)$ is divergent for $d\ge4+2n$.
We can use the $O(g^2)$ terms in $A$ and $B$ to cancel 
off the $\half g^2 I(-m^2)$ and $\half g^2 I'(-m^2)$ terms in $\Pi(k^2)$,
whether or not they are divergent.  But if we are to end up with a finite
$\Pi(k^2)$, all of the remaining terms
must be finite, since we have no more free parameters
left to adjust.  This is the case for $d<8$.

Of course, for $4\le d<8$, the values of $A$ and $B$ (and hence
the lagrangian coefficients $Z=1+A$ and $Z_m=1+B$) are formally
infinite, and this may be disturbing.  However, these coefficients are
not directly measurable, and so need not obey our preconceptions about
their magnitudes.  Also, it is important to remember
that $A$ and $B$ each includes a factor of $g^2$; this means that we 
can expand in powers of $A$ and $B$ as part of our general expansion 
in powers of $g$.  When we compute $\Pi(k^2)$ (which enters
into observable cross sections),
all the formally infinite numbers cancel in a well-defined way,
provided $d<8$.

For $d\ge 8$, this procedure breaks down, and we do not obtain
a finite expression for $\Pi(k^2)$.  In this case,
we say that the theory is {\it nonrenormalizable}.  We will discuss
the criteria for renormalizability of a theory in detail in section 18.
It turns out that $\ph^3$ theory is renormalizable for $d\le 6$.
(The problem with $6<d<8$ arises from higher-order corrections,
as we will see in section 18.)

Now let us return to the calculation of $\Pi(k^2)$.  
Rather than using the trick of first computing $\Pi''(k^2)$, we will instead
evaluate $\Pi(k^2)$ directly from \eq{ik2} as a function of $d$ for $d<4$.
Then we will analytically continue the result to arbitrary $d$.
This procedure is known as {\it dimensional regularization}.
Then we will fix $A$ and $B$ by imposing \eqs{pi0} and (\ref{pip0}), and
finally take the limit $d\to 6$.  

We could just as well use the method of section 9.  
Making the replacement
\begin{equation}
\td(p^2) \to {1\over p^2+m^2-i\eps}\;
             {\Lam^2\over p^2+\Lam^2-i\eps} \;, 
\label{pv}
\end{equation}
where $\Lam$ is a parameter with dimensions of mass called the
{\it ultraviolet cutoff\/}, 
renders the $O(g^2)$ term in $\Pi(k^2)$ finite for $d<8$;
This procedure is known as {\it Pauli--Villars regularization}.  
We then evaluate $\Pi(k^2)$ as a function of $\Lam$,
fix $A$ and $B$ by imposing \eqs{pi0} and (\ref{pip0}), and 
take the $\Lam\to\infty$ limit.   
Calculations with Pauli-Villars regularization are 
generally much more cumbersome than they are with dimensional regularization.
However, the final result for $\Pi(k^2)$ is the same.
\Eq{pit} demonstrates that {\it any\/} regularization scheme will
give the same result for $d<8$, 
at least as long as it preserves the Lorentz invariance of the integrals.

We therefore turn to the evaluation of $I(k^2)$, \eq{ik2}.
The angular part of the integral 
over $\qbar$ yields the area $\Omega_d$ of
the unit sphere in $d$ dimensions, which is  
$\Omega_d = 2\pi^{d/2}/\Gamma(\half d)$. (This is most easily verified
by computing the gaussian integral $\int \ddqb\,e^{-\qbar^2}$ in
both cartesian and spherical coordinates.)  Here $\Gamma(x)$ is the
Euler gamma function; for a nonnegative
integer $n$ and small $x$,
\begin{eqnarray}
\Gamma(n{+}1) &=& n! \;,
\label{gn} \\
\noalign{\medskip}
\Gamma(n{+}\half) &=& {(2n)!\over n!2^n}\sqrt{\pi} \;,
\label{gnh} \\
\noalign{\medskip}
\Gamma(-n{+}x)  &=&  {(-1)^n\over n!}\left[\,{1\over x} - \gamma + 
{\sum\nolimits_{k=1}^n k^{-1}}+O(x)\,\right] \;,
\label{gnx}
\end{eqnarray}
where 
$\gamma=0.5772\ldots$ is the Euler-Mascheroni constant.

The radial part of the $\qbar$ integral can also be evaluated in terms of 
gamma functions.  The overall result (generalized slightly for later use) is
\begin{equation}
\int{\ddqb\over(2\pi)^d}\,{(\qbar^2)^a\over(\qbar^2+D)^b}
= {\Gamma(b{-}a{-}\half d)\Gamma(a{+}\half d) \over
   (4\pi)^{d/2}
   \Gamma(b)\Gamma(\half d)}\,D^{-(b-a-d/2)} \;.
\label{int} 
\end{equation}
In the case of interest, \eq{ik2}, we have $a=0$ and $b=2$.  

There is one more complication to deal with.  
Recall that we want to focus on $d=6$ because in that case $g$
is dimensionless.  However, for general $d$, $g$ has mass dimension
$\e/2$, where
\begin{equation}
\e \equiv 6-d \;.
\label{6-e}
\end{equation}
To account for this, we introduce a new parameter $\mut$ with dimensions of
mass, and make the replacement
\begin{equation}
g \to g\mut^{\e/2}\;.
\label{mut}
\end{equation}
In this way $g$ remains dimensionless for all $\e$.  Of course,
$\mut$ is not an actual parameter of the $d=6$ theory.  Therefore,
nothing measurable (like a cross section) can depend on it.

This seemingly innocuous statement is actually quite powerful,
and will eventually serve as the foundation of the renormalization group.

We now return to \eq{ik2}, use \eq{gnx}, and set $d=6-\e$; we get
\begin{equation}
I(k^2) = {\Gamma(-1{+}{\ts{\e\over2}})\over(4\pi)^3}
              \int_0^1 dx\,D\left({4\pi\over D}\right)^{\!\e/2} \;.
\label{ik22}
\end{equation}
Hence, with the substitution of \eq{mut}, and defining
\begin{equation}
\alpha \equiv  {g^2\over(4\pi)^3} 
\label{alpha}
\end{equation}
for notational convenience, we have
\begin{eqnarray}
\Pi(k^2) &=& \half\alpha\,\Gamma(-1{+}{\ts{\e\over2}})
                     \int_0^1 dx\,D\left({4\pi\mut^2\over D}\right)^{\!\e/2}
 \nonumber \\
&& {} - Ak^2 - Bm^2 +O(\alpha^2)\;.
\label{pi3}
\end{eqnarray}
Now we can take the $\e\to 0$ limit, using \eq{gnx} and
\begin{equation}
A^{\e/2} = 1 + {\ts{\e\over2}}\ln A + O(\e^2)\;.
\label{aeps}
\end{equation}
The result is
\begin{eqnarray}
\Pi(k^2) &=& -\half\alpha\!\left[
\left({\ts{2\over\e}}+1\right)\!\left({\ts{1\over6}}k^2 + m^2\right) +
         \int_0^1 dx\,D\ln\left({4\pi\mut^2\over e^{\gamma} D}\right)\right]
\nonumber \\
&& {} - Ak^2 - Bm^2 +O(\alpha^2)\;.
\label{pi4}
\end{eqnarray}
Here we have used $\int_0^1dx\,D={1\over 6}k^2 + m^2$.
It is now convenient to define
\begin{equation}
\mu \equiv \sqrt{4\pi}\,e^{-\gamma/2}\,\mut \;,
\label{mu}
\end{equation}
and rearrange things to get
\begin{eqnarray}
\Pi(k^2) &=&  \half\alpha \int_0^1 dx\,D \ln( D/m^2 )
\nonumber \\
&& {}-\left\{ {\ts{1\over 6}}\alpha\!\left[{\ts{1\over\e}} 
                             + \ln(\mu/m)+\half\right] + A\right\}k^2
\nonumber \\
&&{} -\left\{\phantom{\ts{1\over6}}\alpha\!\left[{\ts{1\over\e}} 
             + \ln(\mu/m)+\half\right] + B\right\}m^2 + O(\alpha^2) \;.
\label{pi5}
\end{eqnarray}
If we take $A$ and $B$ to have the form
\begin{eqnarray}
A &=& -{\ts{1\over 6}}\alpha\!\left[{\ts{1\over\e}} 
                      + \ln(\mu/m) + \half + \kappa_A\right] + O(\alpha^2) \;,
\label{aa} \\
B &=& -\phantom{\ts{1\over6}}\alpha\!\left[{\ts{1\over\e}} 
                      + \ln(\mu/m) + \half + \kappa_B\right] + O(\alpha^2) \;,
\label{b}
\end{eqnarray}
where $\kappa_A$ and $\kappa_B$ are purely numerical constants,
then we get 
\begin{equation}
\Pi(k^2) = \half\alpha \int_0^1 dx\,D \ln( D/m^2 )
 + \alpha\Bigl({\ts{1\over6}}\kappa_A k^2 + \kappa_B m^2\Bigr) 
+ O(\alpha^2) \;.
\label{pi6}
\end{equation}
Thus this choice of $A$ and $B$ renders $\Pi(k^2)$ finite and independent of
$\mu$, as required.

To fix $\kappa_A$ and $\kappa_B$, we must still impose the conditions 
$\Pi(-m^2)=0$ and $\Pi'(-m^2)=0$.
The easiest way to do this is to first note that, schematically,
\begin{equation}
\Pi(k^2) = \half\alpha \int_0^1 dx\,D\ln D 
+ \hbox{linear in $k^2$ and $m^2$} + O(\alpha^2)\;.
\label{pi7}
\end{equation}
We can then impose $\Pi(-m^2)=0$ via
\begin{equation}
\Pi(k^2) = \half\alpha \int_0^1 dx\,D \ln( D/D_0 ) 
+ \hbox{linear in $(k^2+m^2)$}
+ O(\alpha^2)\;.
\label{pi8}
\end{equation}
where 
\begin{equation}
D_0\equiv D\Big|_{k^2=-m^2}=[1{-}x(1{-}x)]m^2 \;.
\label{D0}
\end{equation}
Now it is straightforward to differentiate \eq{pi8} with respect to $k^2$, 
and find that
$\Pi'(-m^2)$ vanishes for
\begin{equation}
\Pi(k^2) = \half\alpha \int_0^1 dx\,D \ln( D/D_0 ) 
-{\ts{1\over 12}}\alpha(k^2+m^2)+ O(\alpha^2)\;.
\label{pi9}
\end{equation}
The integral over $x$ can be done in closed form; the result is
\begin{equation}
\Pi(k^2) = {\ts{1\over12}}\alpha\Bigl[ c_1 k^2 + c_2 m^2 + 2k^2 f(r)\Bigr] 
+ O(\alpha^2)\;,
\label{pi10} 
\end{equation}
where $c_1 = 3{-}\pi\sqrt3$, $c_2 = 3{-}2\pi\sqrt3$, and
\begin{eqnarray}
f(r) &=& r^{-3/2}\tanh^{-1}(r^{1/2}) \;,
\label{fofr} \\
\noalign{\medskip}
r &=& k^2/(k^2 + 4m^2)\;.
\label{r} 
\end{eqnarray}
There is a square-root branch point at $k^2=-4m^2$, and $\Pi(k^2)$
acquires an imaginary part for $k^2<-4m^2$; 
we will discuss this further in the next section.  

\begin{figure}
\begin{center}
\epsfig{file=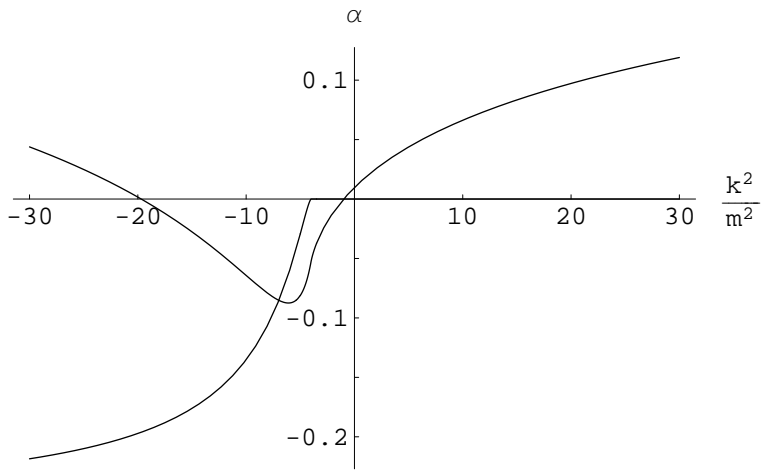}
\end{center}
\caption{The real and imaginary parts of $\Pi(k^2)$ in units of $\alpha$.}
\label{piksq}
\end{figure}

We can write the exact propagator as
\begin{equation}
\tbfd(k^2) 
= \left({1\over 1 - \Pi(k^2)/(k^2+m^2)}\right){1\over k^2 + m^2 -i\eps}\;.
\label{tbfd0}
\end{equation}
In \fig{piksq}, we plot the real and imaginary parts of $\Pi(k^2)/(k^2 + m^2)$
in units of $\alpha$.  We see that its values are quite modest for the 
plotted range.  For much larger values of $|k^2|$, we have
\begin{equation}
{\Pi(k^2)\over k^2+m^2} \simeq 
{\ts{1\over12}}\alpha\Bigl[\ln(k^2/m^2)+c_1\Bigr]
+ O(\alpha^2)\;.
\label{pikk}
\end{equation}
If we had kept track of the $i\eps$'s, $k^2$ 
would be $k^2-i\eps$; when $k^2$ is negative, we have
$\ln(k^2-i\eps) = \ln|k^2| -i\pi$.
The imaginary part of $\Pi(k^2)/(k^2+m^2)$ therefore 
approaches the asymptotic value of 
$-{\ts{1\over12}}\pi\alpha + O(\alpha^2)$ when $k^2$ is large and negative.  
The real part of $\Pi(k^2)/(k^2+m^2)$, 
however, continues to increase logarithmically
with $|k^2|$ when $|k^2|$ is large.  
We will begin to address the meaning of this in section 25.

\vskip0.5in

\begin{center}
Problems
\end{center}

\vskip0.25in

14.1) Compute the numerical values of $\kappa_A$ and $\kappa_B$.

14.2) Compute the $O(\lambda)$ correction to the propagator in
$\ph^4$ theory (see problem 9.2).
What are the $O(\lambda)$ terms in $A$ and $B$?

14.3) Repeat problem 14.2 for the theory of problem 9.3.

\vfill\eject

%% file: ch015.tex
\noindent Quantum Field Theory  \hfill   Mark Srednicki

\vskip0.5in

\begin{center}
\large{15: The One-Loop Correction in \LK Form}
\end{center}
\begin{center}
Prerequisite: 14
\end{center}

\vskip0.5in

In section 12, we found that the exact propagator could be written 
in {\LK form\/} as
\begin{equation}
\tbfd(k^2)=
{1\over k^2+m^2-i\eps}
+\int_{4m^2}^\infty ds\,\rho(s)\,{1\over k^2+s-i\eps} \;,
\label{lk}
\end{equation}
where the {\it spectral density\/} $\rho(s)$ is real and nonnegative.
In section 14, on the other hand, we found that the exact propagator could
be written as
\begin{equation}
\tbfd(k^2)=
{1\over k^2+m^2-i\eps-\Pi(k^2)}\;,
\label{dpi}
\end{equation}
and that, to $O(g^2)$ in $\ph^3$ theory in six dimensions,
\begin{equation}
\Pi(k^2) = \half\alpha \int_0^1 dx\,D \ln( D/D_0 )
-{\ts{1\over 12}}\alpha(k^2+m^2)+ O(\alpha^2)\;,
\label{pi}
\end{equation}
where 
\begin{equation}
\alpha\equiv g^2/(4\pi)^3 \;,
\label{aaa}
\end{equation}
\begin{equation}
D=x(1{-}x) k^2 + m^2 -i\eps \;,
\label{D15}
\end{equation}
\begin{equation}
D_0=[1{-}x(1{-}x)]m^2 \;.
\label{D00}
\end{equation}
In this section, we will attempt to reconcile \eqs{dpi} and (\ref{pi})
with \eq{lk}.

Let us begin by considering the imaginary part of the propagator. 
We will always take $k^2$ and $m^2$ to be real, and explicitly
include the appropriate factors of $i\eps$ whenever they are needed.

We can use \eq{lk} and the identity
\begin{eqnarray}
{1\over x-i\eps} &=& {x\over x^2 + \eps^2} + {i\eps\over x^2 + \eps^2}
\nonumber \\
\noalign{\medskip}
&=& P\,{1\over x} + i\pi\delta(x) \;,
\label{id}
\end{eqnarray}
where $P$ means the principal part, to write
\begin{eqnarray}
\Im\tbfd(k^2) &=&
\pi\delta(k^2+m^2) 
+\int_{4m^2}^\infty ds\,\rho(s)\,\pi\delta(k^2+s) \;.
\nonumber \\
\noalign{\medskip}
&=&
\pi\delta(k^2+m^2) +\pi\rho(-k^2) \;,
\label{imlk}
\end{eqnarray}
where $\rho(s)\equiv 0$ for $s<4m^2$.
Thus we have
\begin{equation}
\pi\rho(s) = \Im\tbfd(-s)
\quad\hbox{for}\quad 
s \ge 4m^2 \;.
\label{rhos}
\end{equation}

Let us now suppose that $\Im\Pi(k^2)=0$ for some range of $k^2$.
(In section 14, we saw that 
the $O(\alpha)$ contribution to $\Pi(k^2)$ is purely real for $k^2>-4m^2$.)
Then, from \eqs{dpi} and (\ref{id}), we get
\begin{equation}
\Im\tbfd(k^2)=
\pi\delta(k^2+m^2-\Pi(k^2))
\quad\hbox{for}\quad 
\Im\Pi(k^2)=0\;.
\label{imdpi}
\end{equation}
From $\Pi(-m^2)=0$, we know that the argument of the delta function
vanishes at $k^2=-m^2$, and from $\Pi'(-m^2)=0$, we know that the
derivative of this argument with respect to $k^2$ equals one at $k^2=-m^2$.
Therefore
\begin{equation}
\Im\tbfd(k^2)= \pi\delta(k^2+m^2)
\quad\hbox{for}\quad 
\Im\Pi(k^2)=0\;.
\label{imdpi2}
\end{equation}
Comparing this with \eq{imlk}, we see that $\rho(-k^2)=0$ if $\Im\Pi(k^2)=0$

Now suppose $\Im\Pi(k^2)$ is {\it not\/} zero for some range of $k^2$.
(In section 14, we saw that 
the $O(\alpha)$ contribution to $\Pi(k^2)$ has a nonzero imaginary part for
$k^2<-4m^2$.)
Then we can ignore the $i\eps$ in \eq{dpi}, and
\begin{equation}
\Im\tbfd(k^2)= {\Im\Pi(k^2)\over (k^2+m^2+\Re\Pi(k^2))^2 + (\Im\Pi(k^2))^2}
\quad\hbox{for}\quad 
\Im\Pi(k^2)\ne 0\;.
\label{imdpi3}
\end{equation}
Comparing this with \eq{imlk} we see that
\begin{equation}
\pi\rho(s) = {\Im\Pi(-s)\over (-s+m^2+\Re\Pi(-s))^2 + (\Im\Pi(-s))^2} \;.
\label{imdpi4}
\end{equation}
Since we know $\rho(s)=0$ for $s<4m^2$, this tells us that we must
also have $\Im\Pi(-s)=0$ for $s<4m^2$, or equivalently
$\Im\Pi(k^2)=0$ for $k^2>-4m^2$.  This is just what we found
for the $O(\alpha)$ contribution to $\Pi(k^2)$ in section 14.

We can also see this directly from \eq{pi}, without doing the
integral over $x$.
The integrand in this formula is real as long as the argument of the
logarithm is real and positive.  
From \eq{D15}, we see that $D$ is real and positive if and only if
$x(1{-}x)k^2 > -m^2$.  The maximum value of $x(1{-}x)$ is $1/4$, and so
the argument of the logarithm is real and positive for the whole
integration range $0\le x\le1$ if and only if
$k^2 > -4m^2$.  In this regime, $\Im\Pi(k^2)=0$.  On the other hand, for
$k^2 < -4m^2$, the argument of the logarithm becomes negative for some
of the integration range, and so $\Im\Pi(k^2) \ne 0$ for
$k^2 < -4m^2$.  This is exactly what we need to reconcile
\eqs{dpi} and (\ref{pi}) with \eq{lk}.

\vfill\eject

%% file: ch016.tex

\noindent Quantum Field Theory  \hfill   Mark Srednicki

\vskip0.5in

\begin{center}
\large{16: Loop Corrections to the Vertex}
\end{center}
\begin{center}
Prerequisite: 14
\end{center}

\vskip0.5in

\begin{figure}
\begin{center}
\epsfig{file=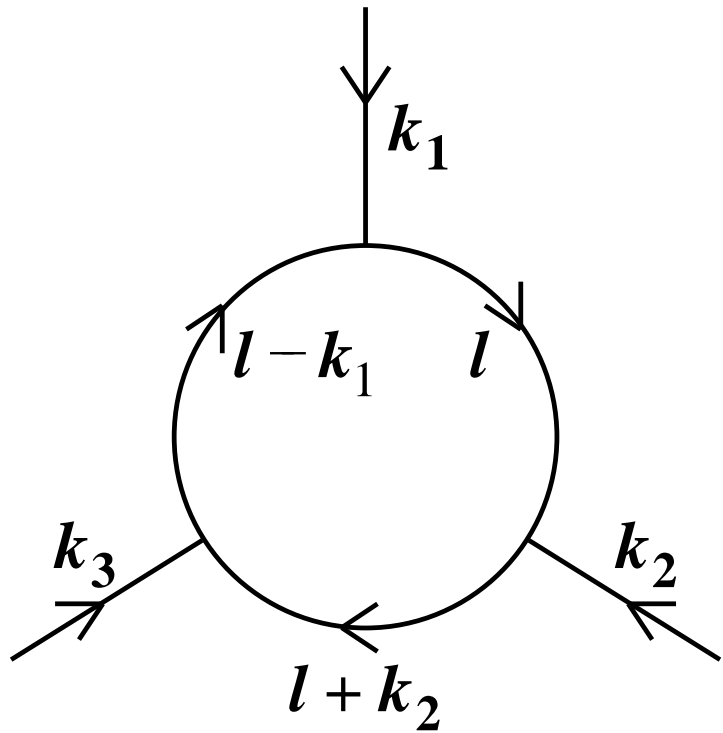}
\end{center}
\caption{The $O(g^3)$ correction to the vertex $i\V_3(k_1,k_2,k_3)$.}
\label{3vertg2}
\end{figure}

Consider the $O(g^3)$ diagram of \fig{3vertg2}, which corrects the
$\ph^3$ vertex.  In this section we will evaluate this diagram.

We can define an exact three-point vertex function 
$i\V_3(k_1,k_2,k_3)$ 
as the sum of one-particle irreducible diagrams with three external lines
carrying momenta $k_1$, $k_2$, and $k_3$, all incoming, with 
$k_1+k_2+k_3=0$ by momentum conservation.  
(In adopting this convention, we allow $k_i^0$ to have either sign;
if $k_i$ is the momentum of an external particle, then the sign of $k^0_i$
is positive if the particle is incoming, and negative if it is outgoing.)  
The original vertex $iZ_g g$ is the first term
in this sum, and the diagram of \fig{3vertg2} is the second.  Thus we have
\begin{eqnarray}
i\V_3(k_1,k_2,k_3) &=& iZ_g g 
+ (ig)^3\left({\ts{1\over i}}\right)^3 \int{\ddl\over(2\pi)^d}\,
\td((\ell{-}k_1)^2) \td((\ell{+}k_2)^2) \td(\ell^2) 
\nonumber \\
&& {} + O(g^5)\;.
\label{v3-1}
\end{eqnarray}
In the second term, we have set $Z_g=1+O(g^2)$.
We proceed immediately to the evaluation of this integral, using
the series of tricks from section 14.

First we use Feynman's formula to write
\begin{eqnarray}
&& \td((\ell{-}k_1)^2) \td((\ell{+}k_2)^2) \td(\ell^2) 
\nonumber \\
&& \quad {} = 
\int dF_3\,\Bigl[\,x_1(\ell{-}k_1)^2
                       +x_2(\ell{+}k_2)^2
                       +x_3\ell^2+m^2\,\Bigr]^{-3} \;,
\label{ellq2}
\end{eqnarray}
where
\begin{equation}
\int dF_3  = 2\int_0^1 dx_1\,dx_2\,dx_3\;
                   \delta(x_1{+}x_2{+}x_3{-}1)\;.
\label{df3}
\end{equation}
We manipulate the right-hand side of \eq{ellq2}
to get
\begin{eqnarray}
&& \td((\ell{-}k_1)^2) \td((\ell{+}k_2)^2) \td(\ell^2) 
\nonumber \\
&& \quad {} = \int dF_3\,
\Bigl[\,\ell^2 - 2\ell\cd(x_1k_1 - x_2k_2) 
      + x_1k_1^2 + x_2k_2^2 + m^2\,\Bigr]^{-3}
\nonumber \\
&& \quad {} = \int dF_3\,
\Bigl[\,(\ell-x_1k_1+x_2k_2)^2 
      + x_1(1{-}x_1)k_1^2 + x_2(1{-}x_2)k_2^2 
\nonumber \\
&& \qquad\qquad\qquad\qquad\qquad\qquad \quad \;\;
    {} + 2x_1x_2k_1\cd k_2 + m^2\,\Bigr]^{-3}
\nonumber \\
&& \quad {} = \int dF_3\, \Bigl[\,q^2 + D\,\Bigr]^{-3} \;.
\label{ellq3}
\end{eqnarray}
In the last line, we have defined $q\equiv \ell-x_1k_1+x_2k_2$, and
\begin{eqnarray}
D &\equiv& x_1(1{-}x_1)k_1^2 + x_2(1{-}x_2)k_2^2 + 2x_1x_2k_1\cd k_2 + m^2
\nonumber \\
\noalign{\medskip}
&=& x_2 x_3 k_1^2
  + x_3 x_1 k_2^2
  + x_1 x_2 k_3^2 + m^2 \;.
\label{D214}
\end{eqnarray}
To get the more symmetric form of $D$, we used $k_3^2=(k_1+k_2)^2$,
$x_1+x_2+x_3=1$, and swapped $x_1\leftrightarrow x_2$.

After making a Wick rotation of the $q^0$ contour, we have
\begin{equation}
\V_3(k_1,k_2,k_3)/g = Z_g 
+ g^2\int dF_3 \int{\ddqb\over(2\pi)^d}\,{1\over(\qbar^2+D)^3} + O(g^4)\;,
\label{v32}
\end{equation}
where $\qbar$ is a euclidean vector.  This integral diverges for $d\ge 6$. 
We therefore evaluate it for $d<6$, using the general formula
from section 14; the result is
\begin{equation}
\int{\ddqb\over(2\pi)^d}\,{1\over(\qbar^2+D)^3}
= {\Gamma(3{-}\half d)\over
   2(4\pi)^{d/2} }\,D^{-(3-d/2)} \;.
\label{intv}
\end{equation}
Now we set $d=6-\e$.  To keep $g$ 
dimensionless, we make the replacement $g\to g\mut^{\e/2}$.
Then we have
\begin{equation}
\V_3(k_1,k_2,k_3)/g = Z_g 
+ \half\alpha\,\Gamma({\ts{\e\over2}}) \int dF_3\;
   \Biggl({4\pi\mut^2\over D}\Biggr)^{\e/2} + O(\alpha^2) \;,
\label{v33}
\end{equation}
where $\alpha=g^2/(4\pi)^3$.
Now we can take the $\e\to 0$ limit.  The result is
\begin{equation}
\V_3(k_1,k_2,k_3)/g = Z_g 
+\half\alpha\!\left[\,
{2\over\e}
+ \int dF_3\;\ln\left({4\pi\mut^2\over e^{\gamma} D}\right)\right]
 +O(\alpha^2)\;,
\label{v34}
\end{equation}
where we have used $\int dF_3 =1$.  
We now let $\mu^2=4\pi e^{-\gamma}\mut^2$, set
\begin{equation}
Z_g = 1 + C \;,
\label{zg}
\end{equation}
and rearrange to get
\begin{eqnarray}
\V_3(k_1,k_2,k_3)/g &=& 1 +
\left\{ \alpha\!\left[{\ts{1\over\e}} + \ln(\mu/m)\right] + C\right\}
\nonumber \\
&&  \phantom{1} - \,\half\alpha \int dF_3\; \ln ( D/m^2 )
\nonumber \\
&& \phantom{1} + \,O(\alpha^2) \;.
\label{v35}
\end{eqnarray}
If we take $C$ to have the form
\begin{equation}
C = -\alpha\!\left[{\ts{1\over\e}} 
                      + \ln(\mu/m) + \kappa_C\right] + O(\alpha^2) \;,
\label{ckap}
\end{equation}
where $\kappa_C$ is a purely numerical constant, we get 
\begin{equation}
\V_3(k_1,k_2,k_3)/g = 1-\half\alpha \int dF_3\, \ln ( D/m^2 )
 -\kappa_C\alpha + O(\alpha^2) \;.
\label{v36}
\end{equation}
Thus this choice of $C$ renders $\V_3(k_1,k_2,k_3)$ finite and 
independent of $\mu$, as required.

We now need a condition, analogous to $\Pi(-m^2)=0$ and $\Pi'(-m^2)=0$,
to fix the value of $\kappa_C$.  These conditions on $\Pi(k^2)$ were mandated
by known properties of the exact propagator, but there is nothing directly
comparable for the vertex.  Different choices of $\kappa_C$ correspond to
different definitions of the coupling $g$.  This is because, in order to measure
$g$, we would measure a cross section that depends on $g$; these cross sections 
also depend on $\kappa_C$.  Thus we can use any value for
$\kappa_C$ that we might fancy, as long as we all agree on that value
when we compare our calculations with experimental measurements.  
It is then most convenient to simply set $\kappa_C=0$.
This corresponds to the condition
\begin{equation}
\V_3(0,0,0) = g \;. 
\label{v30}
\end{equation}
This condition can then also
be used to fix the higher-order (in $g$) terms in $Z_g$.

The integrals over the Feynman parameters in \eq{v36}
cannot be done in closed form, but it is easy to see that if
(for example) $|k_1^2|\gg m^2$, then
\begin{equation}
\V_3(k_1,k_2,k_3)/g \simeq 1-\half\alpha
\Bigl[\ln(k_1^2/m^2)+O(1)\Bigr] +O(\alpha^2)\;.
\label{v3k1}
\end{equation}
Thus the magnitude of the one-loop correction to the vertex function
increases logarithmically with $|k_i^2|$ when $|k_i^2|\gg m^2$.  
This is the same behavior that we found for $\Pi(k^2)/(k^2+m^2)$
in section 14.

\vskip0.5in

\begin{center}
Problems
\end{center}

\vskip0.25in

16.1) 
Compute the $O(\lambda^2)$ correction to $\V_4$ in
$\ph^4$ theory (see problem 9.2).
Take $\V_4=\lam$ when all four external momenta are on shell,
and $s=4m^2$.  What is the $O(\lambda)$ contribution to $C$?

16.2) Repeat problem 16.1 for the theory of problem 9.3.

\vfill\eject

%% file: ch017.tex
\noindent Quantum Field Theory  \hfill   Mark Srednicki

\vskip0.5in

\begin{center}
\large{17: Other 1PI  Vertices}
\end{center}
\begin{center}
Prerequisite: 16
\end{center}

\vskip0.5in

In section 16, we defined the three-point vertex function
$i\V_3(k_1,k_2,k_3)$ as the sum of all one-particle irreducible
diagrams with three external lines, with the external propagators removed.
We can extend this definition to the $n$-point vertex $i\V_n(k_1,\ldots,k_n)$.

There are two key differences between $\V_{n>3}$ and $\V_3$ 
in $\ph^3$ theory.  The first is that there is no tree-level 
contribution to $\V_{n>3}$.  The second is that the one-loop
contribution to $\V_{n>3}$ is finite for $d<2n$.  In particular,
the one-loop contribution to $\V_{n>3}$ is finite for $d=6$.

\begin{figure}
\begin{center}
\epsfig{file=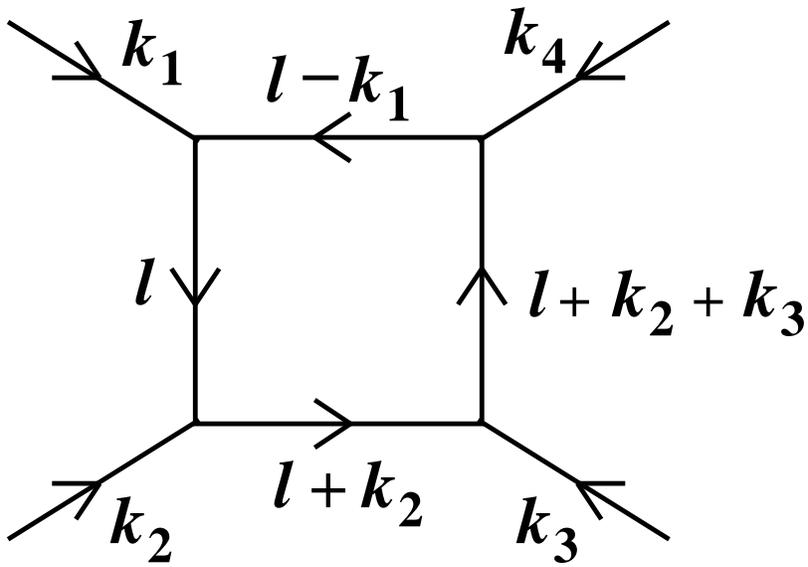}
\end{center}
\caption{One of the three one-loop Feynman diagrams contributing to 
the four-point vertex
$i\V_4(k_1,k_2,k_3,k_4)$; the other two are obtained by swapping
$k_3\leftrightarrow k_2$ and $k_3\leftrightarrow k_4$.}
\label{4vert}
\end{figure}

Let us see how this works for the case $n=4$.
We treat all the external momenta as incoming, so that
$k_1+k_2+k_3+k_4=0$.  
One of the three contributing one-loop diagrams is shown in \fig{4vert};
in this diagram, the $k_3$ vertex is opposite to the $k_1$ vertex.
Two other inequivalent diagrams are then obtained 
by swapping $k_3\leftrightarrow k_2$
and $k_3\leftrightarrow k_4$.  We then have
\begin{eqnarray}
i\V_4 &=& g^4\int{\dsxl\over(2\pi)^6}\,
\td((\ell{-}k_1)^2)\td((\ell{+}k_2)^2)\td((\ell{+}k_2{+}k_3)^2)\td(\ell^2)
\nonumber \\
&& {} + ( k_3\leftrightarrow k_2 ) + ( k_3\leftrightarrow k_4 ) 
\nonumber \\
&& {} + O(g^6)\;.
\label{v4}
\end{eqnarray}
Feynman's formula gives
\begin{eqnarray}
&& \td((\ell{-}k_1)^2)\td((\ell{+}k_2)^2)\td((\ell{+}k_2{+}k_3)^2)\td(\ell^2)
\nonumber \\
&&  \quad {} = \int dF_4\,
\Bigl[\,x_1(\ell{-}k_1)^2
        +x_2(\ell{+}k_2)^2
        +x_3(\ell{+}k_2{+}k_3)^2
        +x_4\ell^2+m^2\,\Bigr]^{-4} 
\nonumber \\
&&  \quad {} = \int dF_4\, \Bigl[\,q^2 + D_{1234}\,\Bigr]^{-4} \;,
\label{ellq4}
\end{eqnarray}
where $q=\ell - x_1k_1 + x_2k_2 + x_3(k_2{+}k_3)$ and,
after making repeated use of $x_1{+}x_2{+}x_3{+}x_4=1$
and $k_1{+}k_2{+}k_3{+}k_4=0$,
\begin{eqnarray}
D_{1234}  &=&    x_1 x_4 k_1^2
       + x_2 x_4 k_2^2
       + x_2 x_3 k_3^2
       + x_1 x_3 k_4^2
\nonumber \\
&& {}       +x_1 x_2(k_1{+}k_2)^2 + x_3 x_4 (k_2{+}k_3)^2  + m^2  \;.
\label{D4}
\end{eqnarray}
We see that the integral over $q$ is finite for $d<8$,
and in particular for $d=6$. After a Wick
rotation of the $q^0$ contour and applying the
general formula of section 14, we find
\begin{equation}
\int{\dsq\over(2\pi)^6}\,{1\over(q^2+D)^4}
= {i\over 6(4\pi)^3 D} \;.
\label{intv4}
\end{equation}
Thus we get
\begin{equation}
\V_4 = {g^4\over 6(4\pi)^3}\int dF_4\left(
{1\over D_{1234}} + {1\over D_{1324}} + {1\over D_{1243}} \right)
 + O(g^6)\;.
\label{v42}
\end{equation}
This expression is finite and well-defined;
the same is true for the one-loop contribution to $\V_n$
for all $n>3$.

\vfill\eject

%% file: ch018.tex
\noindent Quantum Field Theory  \hfill   Mark Srednicki

\vskip0.5in

\begin{center}
\large{18: Higher-Order Corrections and Renormalizability}
\end{center}
\begin{center}
Prerequisite: 17
\end{center}

\vskip0.5in

In sections 14--17, we computed the one-loop diagrams with
two, three, and four external lines for $\ph^3$ theory in six dimensions.  
We found that the first two involved divergent momentum integrals, 
but that these divergences could be absorbed into the coefficients 
of terms in the lagrangian.  If this is true for all higher-order (in $g$)
contributions to the propagator and to the one-particle irreducible
vertex functions (with $n\ge3$ external lines), 
then we say that the theory is {\it renormalizable}.
If this is not the case, and further divergences arise, it may be
possible to absorb them by adding some new terms to the
lagrangian.  If a finite number of such new terms is required,
the theory is still said to be renormalizable.  However, if an infinite
number of new terms is required, then the theory is said to be
{\it nonrenormalizable}.

In this section we wish to consider the circumstances under which
a theory is renormalizable.  As an example,
we will analyze a scalar field theory in
$d$ spacetime dimensions of the form
\begin{equation}
\L=-\half Z_\ph \d^\mu\ph\d_\mu\ph - \half Z_m m^2\ph^2 
- \sum_{n=3}^\infty {\ts{1\over n!}}Z_n g_n \ph^n \;.
\label{ellgn}
\end{equation}

Consider a Feynman diagram with $E$ external lines, $I$ internal lines,
$L$ closed loops, and $V_n$ vertices that connect $n$ lines.
(Here $V_n$ is just a number, not to be confused
with the vertex function $\V_n$.)  Do the momentum
integrals associated with this diagram diverge?

We begin by noting that 
each closed loop gives a factor of $\ddl_i$, and each internal propagator
gives a factor of $1/(p^2+m^2)$, where $p$ is some linear combination
of external momenta $k_i$ and loop momenta $\ell_i$.  The diagram 
would then appear to have an ultraviolet divergence at large $\ell_i$ 
if there are more $\ell$'s in the numerator than there are in the
denominator.  The number of $\ell$'s
in the numerator minus the number of $\ell$'s in the denominator
is the diagram's {\it superficial degree of divergence\/} 
\begin{equation}
D \equiv dL-2I \;,
\label{ddiv}
\end{equation}
and the diagram appears to be divergent if 
\begin{equation}
D\ge 0 \;.
\label{dge0}
\end{equation}

Next we derive a more useful formula for $D$.  The diagram has $E$
external lines, so another contributing diagram is the tree diagram where
all the lines are joined by a single vertex, with vertex factor $-iZ_E g_E$;
this is, in fact, the value of this entire diagram, which then has
mass dimension $[g_E]$.  (The $Z$'s are all dimensionless, by
definition.)  Therefore, the original diagram also has mass dimension
$[g_E]$, since both are contributions to the same scattering amplitude:
\begin{equation}
[\hbox{diagram}] = [g_E]\;.
\label{diad}
\end{equation}
 On the other hand, the mass
dimension of any diagram is given by the sum of the mass
dimensions of its components, namely
\begin{equation}
[\hbox{diagram}] = dL-2I+\sum_{n=3}^\infty V_n [g_n] \;.
\label{diad2}
\end{equation}
From \eqs{ddiv}, (\ref{diad}), and (\ref{diad2}), we get
\begin{equation}
D = [g_E] - \sum_{n=3}^\infty V_n [g_n] \;.
\label{ddiv1}
\end{equation}
This is the formula we need.

From \eq{ddiv1}, it is immediately clear that if any $[g_n]<0$, we expect
uncontrollable divergences, 
since $D$ increases with every added vertex of this type.
Therefore, {\it a theory with any $[g_n]<0$ is nonrenormalizable}.

According to our results in section 13, 
the coupling constants have mass dimension
\begin{equation}
[g_n]=d-{n\over2}(d-2)\;,
\label{mdg2}
\end{equation}
and so we have 
\begin{equation}
[g_n]<0 \quad \hbox{if} \quad n > {2d\over d-2} \;.
\label{gn0}
\end{equation}
Thus we are limited to powers no higher than $\ph^4$ in four dimensions,
and no higher than $\ph^3$ in six dimensions.

The same criterion applies to more complicated theories as well:
{\it a theory is nonrenormalizable if any coefficient of any term in the lagrangian
has negative mass dimension}.

What about theories with couplings with only positive or zero mass dimension?
We see from \eq{ddiv1} that the only dangerous diagrams (those with $D\ge0$)
are those for which $[g_E]\ge 0$. 
But in this case, we can absorb the divergence simply by adjusting
the value of $Z_E$.  This discussion also applies to the propagator;
we can think of $\Pi(k^2)$ as representing the loop-corrected counterterm
vertex $Ak^2 + Bm^2$, with $A$ and $Bm^2$ playing the roles of two
couplings.  We have $[A]=0$ and $[Bm^2]=2$, so the contributing
diagrams are expected to be divergent (as we have already seen
in detail), and the divergences must be absorbed into $A$ and $Bm^2$.

$D$ is called the {\it superficial\/} degree of divergence because a diagram
might diverge even if $D<0$, or might be finite even if $D\ge 0$.  The latter
can happen if there are cancellations among $\ell$'s in the numerator;
quantum electrodynamics provides an example of this phenomenon
that we will encounter in Part III.  For now we turn our attention to the case
of diagrams with $D<0$ that nevertheless diverge.

\begin{figure}
\begin{center}
\epsfig{file=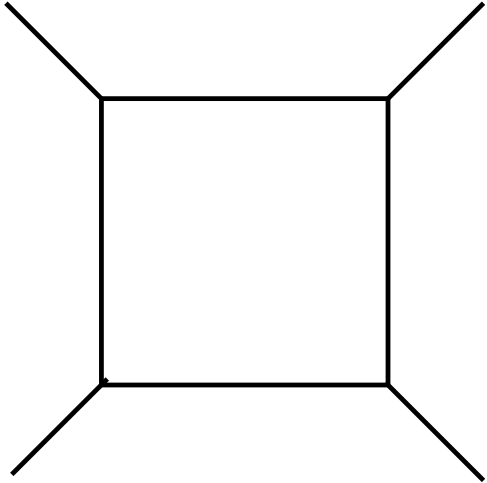}
\end{center}
\caption{The one-loop contribution to $\V_4$.}
\label{4vert1}
\end{figure}

\begin{figure}
\begin{center}
\epsfig{file=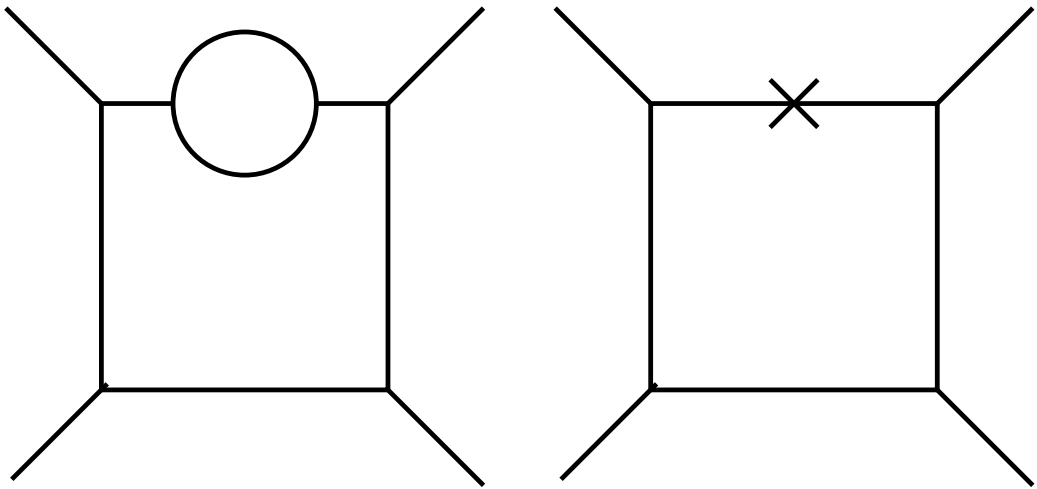}
\end{center}
\caption{A two-loop contribution to $\V_4$,
and the corresponding counterterm insertion.}
\label{4vert2}
\end{figure}

Consider, for example, the diagrams of \fig{4vert1} and (\ref{4vert2}).
The one-loop diagram of \fig{4vert1} with $E=4$ is finite, but
the two-loop correction from the first diagram of \fig{4vert2}
is not: the bubble on the upper
propagator diverges.  This is an example of a 
{\it divergent subdiagram}.  However, this is not
a problem in this case, because this divergence is canceled by
the second diagram of \fig{4vert2}, which has 
a counterterm vertex in place of the bubble.  

This is the generic situation: divergent subdiagrams are 
diagrams that, considered in isolation, have $D\ge0$.
These are precisely the diagrams whose divergences
can be canceled by adjusting the $Z$ factor of the
corresponding tree diagram (in theories where
$[g_n]\ge0$ for all nonzero $g_n$).  

Thus, we expect that theories with couplings whose mass
dimensions are all positive or zero will be renormalizable.
A detailed study of the properties of the momentum
integrals in Feynman diagrams is necessary to give
a complete proof of this.  It turns out to be true without
further restrictions for theories that have spin-zero and
spin-one-half fields only.

Theories  with spin-one fields are renormalizable for $d=4$
if and only if these spin-one fields are associated with a
{\it gauge symmetry}.  We will study this in Part III.

Theories of fields with spin greater than one are never
renormalizable for $d\ge4$.

\vfill\eject

%% file: ch019.tex
\noindent Quantum Field Theory  \hfill   Mark Srednicki

\vskip0.5in

\begin{center}
\large{19: Perturbation Theory to All Orders: the Skeleton Expansion}
\end{center}
\begin{center}
Prerequisite: 18
\end{center}

\vskip0.5in

In section 18, we found that, generally, a theory is renormalizable
if all of its lagrangian coefficients have positive or zero mass dimension.
In this section, using $\ph^3$ theory in six dimensions as our example,
we will see how to construct a finite expression for a scattering 
amplitude to arbitrarily high order in the $\ph^3$ coupling $g$.

We begin by summing all one-particle irreducible diagrams 
with two external lines; 
this gives us the propagator correction $\Pi(k^2)$.
Order by order in $g$, we must adjust the value of the counterterm
coefficients $A=Z_\ph-1$ and $B=Z_m-1$ to maintain the conditions
$\Pi(-m^2)=0$ and $\Pi'(-m^2)=0$.  

We next sum all 1PI diagrams with three external lines; 
this gives us the vertex function $\V_3(k_1,k_2,k_3)$.
Order by order in $g$, we must adjust the value of 
$C=Z_g-1$ to maintain the condition $\V_3(0,0,0)=g$.

Next we consider the other 1PI vertex functions $\V_n(k_1,\ldots,k_n)$
for $4\le n\le E$, where $E$ is the number of external lines in the process 
of interest.  We compute these using a {\it skeleton expansion}.
This means that we draw all the contributing diagrams, but omit diagrams that
include either propagator or vertex corrections.
That is, we consider only diagrams that are not only 1PI, but also 2PI and 3PI:
they remain simply connected when any one, two, or three lines are cut.
(Cutting three lines may isolate a single tree-level vertex, but nothing
more complicated.)   

We take the propagators and vertices
in these diagrams to be given by the {\it exact\/} propagator
$\tbfd(k^2)=(k^2 + m^2 - \Pi(k^2))^{-1}$ and vertex
$\V_3(k_1,k_2,k_3)$, rather than by the tree-level propagator
$\td(k^2)=(k^2+m^2)^{-1}$ and vertex $g$.
(More precisely, by the exact propagator
and vertex computed to however high an order in $g$ we wish to go,
or could manage to do.)  Then we sum these {\it skeleton diagrams}
to get $\V_n$ for $4\le n\le E$.

Next we draw all {\it tree-level\/} diagrams contributing to the process
of interest (which has $E$ external lines), 
including not only three-point vertices, but also $n$-point
vertices for $n=3,4,\ldots,E$.
Then we evaluate these diagrams using the exact propagator
$\tbfd(k^2)$ for internal lines, and the exact 1PI
vertices $\V_n$.  

External lines are assigned a factor of one.
This is because, in the LSZ formula, each Klein-Gordon wave
operator becomes (in momentum space) a factor of
$k_i^2+m^2$ that multiplies each external propagator,
leaving behind only the residue of the pole in that propagator
at $k^2_i=-m^2$.  We have constructed the exact propagator 
so that this residue is precisely one.

A careful examination of this complete procedure will reveal that we have
now included all of the original contributing Feynman diagrams,
with the correct counting factors.

Thus we now know how to compute an arbitrary scattering
amplitude to arbitrarily high order.  The procedure is the
same in any quantum field theory; only the form of the propagators
and vertices change, depending on the spins of the fields.  

The tree-level diagrams of the final step
can be thought of as the Feynman diagrams
of a {\it quantum action\/} (or {\it effective action}, or {\it quantum
effective action\/}) $\Gamma(\ph)$.  There is a simple and interesting
relationship between the effective action $\Gamma(\ph)$ and the
sum of connected diagrams with sources $iW(J)$.  We derive it
in section 21.

\vfill\eject

%% file: ch020.tex

\noindent Quantum Field Theory  \hfill   Mark Srednicki

\vskip0.5in

\begin{center}
\large{20: Two-Particle Elastic Scattering at One Loop}
\end{center}
\begin{center}
Prerequisite: 10, 19
\end{center}

\vskip0.5in

We now illustrate the general rules of section 19 by computing the
two-particle elastic scattering amplitude, including all one-loop corrections,
in $\ph^3$ theory in six dimensions.
{\it Elastic\/} means that the number of outgoing particles (of each
species, in more general contexts) is the same
as the number of incoming particles (of each species).

We computed the amplitude for this process at tree level in
section 10, with the result
\begin{equation}
i\T_{\rm tree} = {\ts{1\over i}}(ig)^2\left[\td(-s)+\td(-t)+\td(-u)\right] \;,
\label{ttree}
\end{equation}
where $\td(-s)=1/(-s+m^2-i\eps)$ is the free-field propagator,
and $s$, $t$, and $u$ are the Mandelstam variables.
Later we will need to remember that $s$ is positive,
that $t$ and $u$ are negative, and that $s+t+u=4m^2$.

\begin{figure}
\begin{center}
\epsfig{file=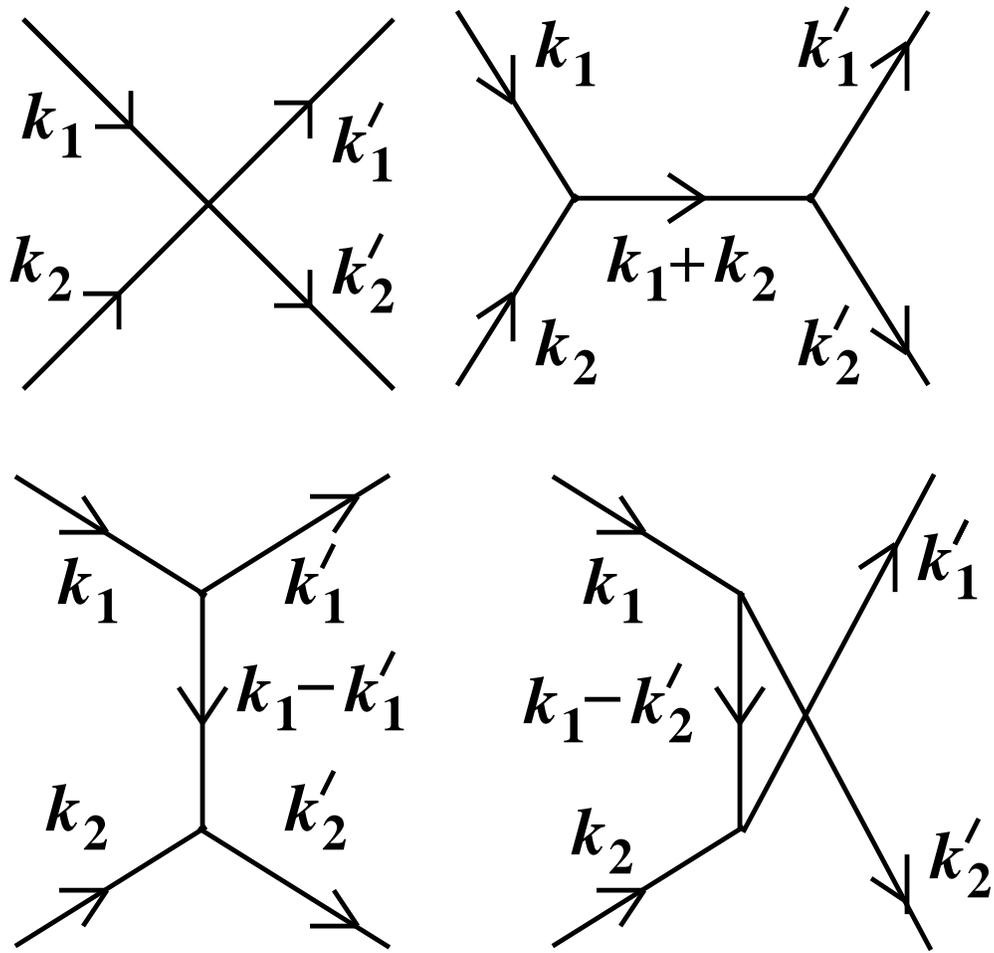}
\end{center}
\caption{The Feynman diagrams contributing to the two-particle
elastic scattering amplitude; in these diagrams, the lines and
points represent the exact propagators and vertices.}
\label{4exact}
\end{figure}

The exact scattering amplitude is given by the
diagrams of \fig{4exact}, with all propagators and vertices 
interpreted as {\it exact\/} propagators and vertices.  
(Recall, however, that each external propagator contributes only 
the residue of the pole at $k^2=-m^2$, and that this residue is one;
thus the factor associated with each external line is simply one.)
We get the one-loop approximation to the
exact amplitude by using the one-loop expressions for the internal propagators 
and vertices.  We thus have
\begin{eqnarray}
i\T_{\rm 1-loop} &=& {\ts{1\over i}}\left\{[
i\V_3(s)]^2\tbfd(-s)+[i\V_3(t)]^2\tbfd(-t)+[i\V_3(u)]^2\tbfd(-u)\right\}
\nonumber \\
&& {} + i\V_4(s,t,u) \;,
\label{t1l}
\end{eqnarray}
where, suppressing the $i\eps$'s,
\begin{eqnarray}
\tbfd(-s) &=& {1\over -s+m^2-\Pi(-s)} \;,
\label{tbfd7} \\
\noalign{\medskip}
\Pi(-s) &=& \half\alpha \int_0^1 dx\,D_2(s) \ln\Bigl( D_2(s)/D_0 \Bigr) 
-{\ts{1\over 12}}\alpha(-s+m^2)\;, \qquad
\label{pi100} \\
\noalign{\medskip}
\V_3(s)/g &=& 1-\half\alpha \int dF_3\, \ln \Bigl( D_3(s)/m^2 \Bigr) \;,
\label{v37} \\
\noalign{\medskip}
\V_4(s,t,u) &=& {\ts{1\over 6}}g^2\alpha\int dF_4
\left[ {1\over D_4({s,t})}+{1\over D_4({t,u})}+{1\over D_4({u,s})} \right].
\label{v43}
\end{eqnarray}
Here $\alpha = g^2/(4\pi)^3$, the Feynman integration measure is
\begin{eqnarray}
\int dF_n\,f(x)
&=& (n{-}1)!\int_0^1 dx_1\ldots dx_n\,\delta(x_1{+}\ldots{+}x_n{-}1)f(x)
\nonumber \\
&=& (n{-}1)!\int_0^1 dx_1 \int_0^{1-x_1} dx_2 \ldots
\int_0^{1-x_1-\ldots-x_{n-2}} dx_{n-1}\, 
\nonumber \\
&&\qquad\qquad\qquad \times 
    f(x)\Big|_{x_n=1-x_1-\ldots-x_{n-1}}\;,
\label{dfn2}
\end{eqnarray}
and we have defined
\begin{eqnarray}
D_2(s) &=& -x(1{-}x)s+m^2 \;,
\label{D2} \\
D_0 &=& +[1{-}x(1{-}x)]m^2 \;,
\label{D01} \\
D_3(s) &=& - x_1x_2 s + [1{-}(x_1{+}x_2)x_3]m^2 \;,
\label{D3} \\
D_4(s,t) &=& - x_1 x_2 s - x_3 x_4 t + [1{-}(x_1{+}x_2)(x_3{+}x_4)]m^2 \;.
\label{dsu}
\end{eqnarray}
We obtain $\V_3(s)$ from the general three-point function $\V_3(k_1,k_2,k_3)$
by setting two of the three $k_i^2$ to $-m^2$, and the third to $-s$.
We obtain $\V_4(s,t,u)$ from
the general four-point function $\V_4(k_1,\ldots,k_4)$
by setting all four
$k_i^2$ to $-m^2$, $(k_1+k_2)^2$ to $-s$,
$(k_1+k_3)^2$ to $-t$, and $(k_1+k_4)^2$ to $-u$.
(Recall that the vertex functions are defined with all momenta
treated as incoming; here we have identified $-k_3$ and
$-k_4$ as the outgoing momenta.)

Eqs.$\,$(\ref{t1l}--\ref{dsu}) are formidable expressions.
To gain some intuition about them, let us consider the
limit of high-energy, fixed angle scattering, where we take
$s$, $|t|$, and $|u|$ all much larger than $m^2$.  Equivalently,
we are considering the amplitude in the limit of zero particle mass.

We can then set $m^2=0$ in $D_2(s)$, $D_3(s)$, and $D_4(s,t)$.
For the self-energy, we get
\begin{eqnarray}
\Pi(-s) &=& -\half\alpha\,s \int_0^1 dx\,x(1{-}x)
\left[\ln\left({-s\over m^2}\right) 
    + \ln\left({x(1{-}x)\over 1{-}x(1{-}x)}\right)\right]
+{\ts{1\over 12}}\alpha\,s
\nonumber \\
\noalign{\medskip}
&=& -{\ts{1\over12}}\alpha\,s \Bigl[\,\ln(-s/m^2) + 3-\pi\sqrt3 \;\Bigr]  \;.
\label{pi11}
\end{eqnarray}
Thus,
\begin{eqnarray}
\tbfd(-s) &=& {1\over -s-\Pi(-s)} 
\nonumber \\
\noalign{\medskip}
&=& -{1\over s}\left(
1+{\ts{1\over12}}\alpha\Bigl[\,\ln(-s/m^2) + 3-\pi\sqrt3 \;\Bigr]
\right) + O(\alpha^2) \;. \qquad
\label{tbfd8}
\end{eqnarray}
The appropriate branch of the logarithm is found by replacing
$s$ by $s+i\eps$.  For $s$ real and positive, $-s$ lies just
below the negative real axis, and so 
\begin{equation}
\ln(-s)=\ln s - i\pi\;.
\label{lns}
\end{equation}
For $t$ (or $u$), which is negative, we have instead
\begin{eqnarray}
\ln(-t) &=& \ln |t| \;, 
\nonumber \\
\noalign{\medskip}
\ln t &=& \ln |t| + i\pi\;. 
\label{lnt}
\end{eqnarray}

For the three-point vertex, we get
\begin{eqnarray}
\V_3(s)/g &=& 1-\half\alpha \int dF_3\,\Bigl[\,\ln(-s/m^2)
 + \ln(x_1x_2)\,\Bigr] \;,
\nonumber \\
\noalign{\medskip}
&=& 1 - \half\alpha\Bigl[\,\ln(-s/m^2) - 3\,\Bigr] \;,
\label{v38}
\end{eqnarray}
where the same comments about the appropriate branch apply.

For the four-point vertex, after some intrigue with
the integral over the Feynman parameters, we get
\begin{eqnarray}
\int{dF_4\over D_4(s,t)}
&=& {}-{3\over s+t}\left(\pi^2+\Bigl[\ln(s/t)\Big]^2\right) 
\nonumber \\
\noalign{\medskip}
&=& {}+{3\over u}\left(\pi^2+\Bigl[\ln(s/t)\Big]^2\right) \;,
\label{v44}
\end{eqnarray}
where the second line follows from $s+t+u=0$.

Putting all of this together, we have
\begin{equation}
\T_{\rm 1-loop} = g^2\Bigl[\,F(s,t,u)+F(t,u,s)+F(u,s,t)\,\Bigr]\;,
\label{t1l2}
\end{equation}
where
\begin{equation}
F(s,t,u) \equiv
{}-{1\over s}\left(1-{\ts{11\over12}}\alpha\Bigl[\,\ln(-s/m^2) + c\,\Bigr]
-\half\alpha\Bigl[\ln(t/u)\Big]^2\right) \;,
\label{F}
\end{equation}
and $c=(6\pi^2+\pi\sqrt3-39)/11 = 2.33$.
This is a typical result of a loop calculation: the original tree-level
amplitude is corrected by powers of logarithms of kinematic variables.

\vfill\eject

%% file: ch021.tex
\noindent Quantum Field Theory  \hfill   Mark Srednicki

\vskip0.5in

\begin{center}
\large{21: The Quantum Action}
\end{center}
\begin{center}
Prerequisite: 19
\end{center}

\vskip0.5in

In section 19, we saw how to compute 
(in $\ph^3$ theory in six dimensions)
the 1PI vertex functions
$\V_n(k_1,\ldots,k_n)$ for $n\ge4$ via the {\it skeleton expansion\/}:
draw all Feynman diagrams with $n$ external lines that 
are one-, two-, and three-particle
irreducible, and compute them using the exact propagator $\tbfd(k^2)$
and three-point vertex function $\V_3(k_1,k_2,k_3)$.

We now define the {\it quantum action} 
(or {\it effective action}, or {\it quantum effective action\/}) 
\begin{eqnarray}
\Gamma(\ph) &\equiv&
{1\over2}\int{\dsk\over(2\pi)^6}\,\tph(-k)\Bigl(k^2+m^2-\Pi(k^2)\Bigr)\tph(k)
\nonumber \\
&& {} + \sum_{n=3}^\infty {1\over n!} \int
{\dsk_1\over(2\pi)^6}\ldots{\dsk_n\over(2\pi)^6}\,
(2\pi)^6\delta^6(k_1{+}\ldots{+}k_n)
\nonumber \\
&& \qquad\qquad\quad 
\times \V_n(k_1,\ldots,k_n)\,\tph(k_1)\ldots\tph(k_n)\;,
\label{qa}
\end{eqnarray}
where $\tph(k)=\int \dsx\,e^{-ikx}\ph(x)$.
The quantum action has the property that the {\it tree-level\/}
Feynman diagrams it generates give the {\it complete\/} scattering
amplitude of the original theory.

In this section, we will determine the relationship between
$\Gamma(\ph)$ and the sum of connected diagrams with sources,
$iW(J)$, introduced in section 9.
Recall that $W(J)$ is related to the path integral
\begin{equation}
Z(J)=\int \D\ph\;\exp\biggl[\,iS(\ph)+i\!\int\!\dsx\,J\ph\,\biggr]\;,
\label{zzjj}
\end{equation}
where $S=\int\dsx\,\L$ is the action, via
\begin{equation}
Z(J)=\exp[iW(J)]\;.
\label{zw}
\end{equation}

Consider now the path integral
\begin{eqnarray}
Z_\Gamma(J) &\equiv& \int \D\ph\;\exp\biggl[\,i\Gamma(\ph)
+i\!\int\!\dsx\,J\ph\,\biggr]
\label{zg2} \\
\noalign{\bigskip}
&=& \exp[iW_\Gamma(J)] \;.
\label{zgw}
\end{eqnarray}
$W_\Gamma(J)$ is given by the sum of connected diagrams (with sources)
in which each line represents the exact propagator, and each $n$-point vertex
represents the exact 1PI vertex $\V_n$.
$W_\Gamma(J)$ would be equal to $W(J)$ if we included only tree diagrams in 
$W_\Gamma(J)$.

We can isolate the tree-level contribution to a path integral by
means of the following trick.  Introduce a dimensionless parameter 
that we will call $\hbar$, and the path integral
\begin{eqnarray}
Z_{\Gamma,\hbar}(J) &\equiv& \int \D\ph\;\exp
\left[\,{i\over\hbar}\left(\Gamma(\ph)+\int\!\dsx\,J\ph\right)\right]
\label{zgh} \\
\noalign{\bigskip}
&=& \exp[iW_{\Gamma,\hbar}(J)] \;.
\label{zgwh}
\end{eqnarray}
In a given connected diagram with sources, every propagator
(including those that connect to sources) is multiplied by $\hbar$, 
every source by $1/\hbar$, and every vertex by $1/\hbar$.  
The overall factor of $\hbar$ is then $\hbar^{P-E-V}$, where 
$V$ is the number of vertices,
$E$ is the number of sources 
(equivalently, the number of external lines after we remove the sources),
and $P$ is the number of propagators (external and internal).
We next note that $P{-}E{-}V$ is equal to $L{-}1$, where $L$ is the number
of closed loops.  This can be seen by counting the number of internal momenta
and the constraints among them.  Specifically, assign an unfixed momentum to 
each internal line; there are $P{-}E$ of these momenta.  Then the $V$ vertices
provide $V$ constraints.  One linear combination of these constraints gives
overall momentum conservation, and so does not constrain the internal momenta.
Therefore, the number of internal momenta left unfixed by the vertex
constraints is $(P{-}E){-}(V{-}1)$, and the number of unfixed momenta
is the same as the number of loops $L$.

So, $W_{\Gamma,\hbar}(J)$ can be expressed as a power series in $\hbar$ of
the form
\begin{equation}
W_{\Gamma,\hbar}(J)
= \sum_{L=0}^\infty \hbar^{L-1}\,W_{\Gamma,L}(J) \;.
\label{wgL}
\end{equation}
If we take the formal limit of $\hbar\to0$, the dominant term is the
one with $L=0$, which is given by the sum of tree diagrams only. 
This is just what we want.  We conclude that
\begin{equation}
W(J) = W_{\Gamma,L=0}(J) \;.
\label{wg0}
\end{equation}

Next we perform the path integral in \eq{zgh} by the method of
stationary phase.  We find the point (actually, the field configuration)
at which the exponent is stationary; this is given by the solution of
the {\it quantum equation of motion}
\begin{equation}
{\delta\over\delta\ph(x)}\Gamma(\ph) = -J(x)\;.
\label{dgdp}
\end{equation}
Let $\ph_J(x)$ denote the solution of \eq{dgdp} with a specified  
source function $J(x)$.   Then the stationary-phase approximation
to $Z_{\Gamma,\hbar}(J)$ is
\begin{equation}
Z_{\Gamma,\hbar}(J) = \exp
\left[\,{i\over\hbar}\left(\Gamma(\ph_J)+\int\!\dsx\,J\ph_J\right)
+O(\hbar^0)\right]\;.
\label{zghsp} 
\end{equation}
Combining the results of \eqs{zgwh}, (\ref{wgL}), (\ref{wg0}), and
(\ref{zghsp}), we find
\begin{equation}
W(J) = \Gamma(\ph_J)+\int\! \dsx\,J\ph_J \;.
\label{wg} 
\end{equation}
This is the main result of this section.

Let us explore it further.  Recall from section 9
that the vacuum expectation value of the field operator
$\ph(x)$ is given by
\begin{equation}
\la 0|\ph(x)|0\ra = {\delta\over\delta J(x)}W(J) \bigg|_{J=0}\;.
\label{vev222} 
\end{equation}
Now consider what we get if we do not set $J=0$ after taking the derivative:
\begin{equation}
\la 0|\ph(x)|0\ra_J \equiv {\delta\over\delta J(x)}W(J)\;.
\label{vevj} 
\end{equation}
This is the vacuum expectation value of $\ph(x)$ in the presence of a nonzero
source function $J(x)$.  We can get some more information about it 
by using \eq{wg} for $W(J)$.  Making use of the product rule for derivatives, 
we have
\begin{equation}
\la 0|\ph(x)|0\ra_J = {\delta\over\delta J(x)}\Gamma(\ph_J) + \ph_J(x)
+\int\!\dsy\,J(y){\delta\ph_J(y)\over\delta J(x)}\;.
\label{vevj2} 
\end{equation}
We can evaluate the first term on the right-hand side by using the chain rule,
\begin{equation}
{\delta\over\delta J(x)}\Gamma(\ph_J) =
\int\!\dsy\,{\delta\Gamma(\ph_J)\over\delta\ph_J(y)}
\,{\delta\ph_J(y)\over\delta J(x)} \;.
\label{dgdj} 
\end{equation}
Then we can combine 
the first and third terms on the right-hand side of \eq{vevj2} to get
\begin{equation}
\la 0|\ph(x)|0\ra_J = \int\!\dsy\,\left[
{\delta\Gamma(\ph_J)\over\delta\ph_J(y)} + J(y)\right]
{\delta\ph_J(y)\over\delta J(x)} + \ph_J(x)\;.
\label{vevj3} 
\end{equation}
Now we note from \eq{dgdp} 
that the factor in large brackets on the right-hand side
of \eq{vevj3} vanishes, and so
\begin{equation}
\la 0|\ph(x)|0\ra_J =  \ph_J(x)\;.
\label{vevj4} 
\end{equation}
That is, the vacuum expectation value of the field operator 
$\ph(x)$ in the presence of a nonzero source function is also the solution 
to the quantum equation of motion, \eq{dgdp}.

We can also write the quantum action in terms of a {\it derivative expansion},
\begin{equation}
\Gamma(\ph)=\int\!\dsx\,\left[{}-{\cal U}(\ph)
-\half{\cal Z}(\ph)\d^\mu\ph\d_\mu\ph + \ldots\,\right]\;,
\label{qa2} 
\end{equation}
where the ellipses stand for an infinite number of terms with
more and more derivatives, and ${\cal U}(\ph)$ and ${\cal Z}(\ph)$
are ordinary functions (not functionals) of $\ph(x)$.
${\cal U}(\ph)$ is called the {\it quantum potential} 
(or {\it effective potential}, or {\it quantum effective potential\/}),
and it plays an important conceptual role in theories with
spontaneous symmetry breaking; see section 28.
However, it is rarely necessary to compute it explicitly,
except in those cases where we are unable to do so.

\vskip0.5in

\begin{center}
Problems
\end{center}

\vskip0.25in

21.1) Show that
\begin{equation}
\Gamma(\ph) = W(J_\ph) - \int \!\dsx\,J_\ph\ph \;,
\label{wjph} 
\end{equation}
where $J_\ph(x)$ is the solution of 
\begin{equation}
{\delta\over \delta J(x)} W(J)  = \ph(x)
\label{wjjx} 
\end{equation}
for a specified $\ph(x)$.

21.2)  Consider performing the path integral in the presence of a
{\it background field\/} $\bar\ph(x)$; we define
\begin{equation}
\exp[iW(J;\bar\ph)]\equiv  
\int \D\ph\;\exp\biggl[\,iS(\ph{+}\bar\ph)+i\!\int\!\dsx\,J\ph\,\biggr]\;.
\label{wjph2} 
\end{equation}
Then $W(J;0)$ is the original $W(J)$ of \eq{zw}.  We also define the 
quantum action in the presence of the background field,
\begin{equation}
\Gamma(\ph;\bar\ph) \equiv W(J_\ph;\bar\ph) - \int \!\dsx\,J_\ph\ph \;,
\label{wjph3} 
\end{equation}
where $J_\ph(x)$ is the solution of 
\begin{equation}
{\delta\over \delta J(x)} W(J;\bar\ph)  = \ph(x)
\label{wjjx2} 
\end{equation}
for a specified $\ph(x)$.  Show that 
\begin{equation}
\Gamma(\ph;\bar\ph)=\Gamma(\ph{+}\bar\ph;0) \;,
\label{GG} 
\end{equation}
where $\Gamma(\ph,0)$ is the original quantum action of \eq{qa}.

\vfill\eject

%% file: ch022.tex

\noindent Quantum Field Theory  \hfill   Mark Srednicki

\vskip0.5in

\begin{center}
\large{22: Continuous Symmetries and Conserved Currents}
\end{center}
\begin{center}
Prerequisite: 8
\end{center}

\vskip0.5in

Suppose we have a set of scalar fields $\ph_a(x)$, and a lagrangian
density $\L(x)=\L(\ph_a(x),\d_\mu\ph_a(x))$.  
Consider what happens to $\L(x)$ if we make an infinitesimal
change $\ph_a(x)\to\ph_a(x)+\delta\ph_a(x)$ in each field.
We have $\L(x)\to\L(x)+\delta\L(x)$, where
$\delta\L(x)$ is given by the chain rule,
\begin{equation}
\delta\L(x)=
{\d\L\over\d\ph_a(x)} \, \delta\ph_a(x) +
{\d\L\over\d(\d_\mu\ph_a(x))} \, \d_\mu\delta\ph_a(x) \;.
\label{dell}
\end{equation}

Next consider the classical equations of motion
(also known as the Euler-Lagrange equations, or the 
{\it field equations\/}), 
given by the action principle
\begin{equation}
{\delta S\over\delta\ph_a(x)} = 0\;,
\label{ds00}
\end{equation}
where $S=\int\dfy\,\L(y)$ is the action, and $\delta/\delta\ph_a(x)$
is a functional derivative.  We have (with repeated indices implicitly summed)
\begin{eqnarray}
{\delta S\over\delta\ph_a(x)} 
&=& \int\dfy\,{\delta\L(y)\over\delta\ph_a(x)}
\nonumber \\
\noalign{\medskip}
&=& \int\dfy\left[{\d\L(y)\over\d\ph_b(y)}\,
                    {\delta\ph_b(y)\over\delta\ph_a(x)} +
                    {\d\L(y)\over\d(\d_\mu\ph_b(y))}\,
                    {\delta(\d_\mu\ph_b(y))\over\delta\ph_a(x)}\right]
\nonumber \\
\noalign{\medskip}
&=& \int\dfy\left[{\d\L(y)\over\d\ph_b(y)}\,
                    \delta_{ba}\delta^4(y{-}x) +
                    {\d\L(y)\over\d(\d_\mu\ph_b(y))}\,
                    \delta_{ba}\d_\mu\delta^4(y{-}x) \right]
\nonumber \\
\noalign{\medskip}
&=& {\d\L(x)\over\d\ph_a(x)} - \d_\mu\,{\d\L(x)\over\d(\d_\mu\ph_a(x))}\;.
\label{ds22}
\end{eqnarray}
We can use this result to make the replacement
\begin{equation}
{\d\L(x)\over\d\ph_a(x)} \to \d_\mu\,{\d\L(x)\over\d(\d_\mu\ph_a(x))}
+{\delta S\over\delta\ph_a(x)} 
\label{eleqs}
\end{equation}
in \eq{dell}.  Then, combining two of the terms, we get
\begin{equation}
\delta\L(x)
= \d_\mu\!\left({\d\L(x)\over\d(\d_\mu\ph_a(x))} \, \delta\ph_a(x)\right)
+{\delta S\over\delta\ph_a(x)} \delta\ph_a(x) \;.
\label{dell2}
\end{equation}
Next we identify the object in large parentheses in \eq{dell2} as the 
{\it Noether current\/}
\begin{equation}
j^\mu(x) \equiv {\d\L(x)\over\d(\d_\mu\ph_a(x))}\,\delta\ph_a(x)\;.
\label{noether}
\end{equation}
\Eq{dell2} then implies
\begin{equation}
\d_\mu j^\mu(x)=\delta\L(x) - {\delta S\over\delta\ph_a(x)}\delta\ph_a(x) \;.
\label{divj}
\end{equation}
If the classical field equations are satisfied, then the second term
on the right-hand side of \eq{divj} vanishes.

The Noether current plays a special role if we can find a set
of infinitesimal
field transformations that leaves the lagrangian unchanged,
or {\it invariant}.  In this case, we have
$\delta\L=0$, and we say that 
the lagrangian has a {\it continuous symmetry}. 
From \eq{divj}, we then have 
$\d_\mu j^\mu=0$ whenever the field equations are satisfied,
and we say that the Noether current is {\it conserved}.  
In terms of its space and time components, this means that
\begin{equation}
{\d\over\d t}\;j^0(x) + \nabla\cdot{\bf j}(x) = 0\;.
\label{dj0}
\end{equation}
If we interpret $j^0(x)$ as a {\it charge density\/}, and
${\bf j}(x)$ as the corresponding {\it current density\/}, then
\eq{dj0} expresses the local conservation of this charge.

Let us see an example of this.
Consider a theory of a {\it complex\/} scalar field with lagrangian
\begin{equation}
\L=-\d^\mu\ph^\dagger\d_\mu\ph-m^2\ph^\dagger\ph 
- {\ts{1\over4}}\lam(\ph^\dagger\ph)^2\;.
\label{ellu1}
\end{equation}
We can also rewrite $\L$ in terms of two real scalar fields by setting
$\ph=(\ph_1+i\ph_2)/\sqrt2$ to get
\begin{equation}
\L=-\half\d^\mu \ph_1\d_\mu \ph_1 -\half\d^\mu \ph_2\d_\mu \ph_2
-\half m^2(\ph_1^2+\ph_2^2)
- {\ts{1\over16}}\lam(\ph_1^2+\ph_2^2)^2\;.
\label{ellab}
\end{equation}
In the form of \eq{ellu1}, it is obvious that $\L$ is left invariant
by the transformation
\begin{equation}
\ph(x)\to e^{-i\alpha}\ph(x) \;,
\label{u1}
\end{equation}
where $\alpha$ is a real number.  
This is called a {\it U(1) transformation}, a transformation by a
unitary $1\times1$ matrix.
In terms of $\ph_1$ and $\ph_2$, this transformation reads
\begin{equation}
\pmatrix{ 
\ph_1(x) \cr 
\noalign{\medskip}
\ph_2(x) \cr}
\to
\pmatrix{ \phantom{-}\cos\alpha & \sin\alpha \cr
\noalign{\medskip}
           {-}\sin\alpha & \cos\alpha \cr}
\pmatrix{ 
\ph_1(x) \cr 
\noalign{\medskip}
\ph_2(x) \cr}.
\label{ab}
\end{equation}
If we think of $(\ph_1,\ph_2)$ as a two-component vector, 
then \eq{ab} is just a rotation
of this vector in the plane by angle $\alpha$.  
\Eq{ab} is called an {\it SO(2) transformation}, a transformation
by an orthogonal $2\times2$ matrix with a special value of the 
determinant (namely $+1$, as opposed to $-1$, the only other
possibility for an orthogonal matrix).  
We have learned that
a U(1) transformation can be mapped into an SO(2) transformation. 

The infinitesimal form of \eq{u1} is
\begin{eqnarray}
\ph(x) &\to& \phantom{{}^\dagger}\ph(x)-i\alpha\ph(x) \;,
\nonumber \\
\noalign{\medskip}
\ph^\dagger(x) &\to& \ph^\dagger(x)+i\alpha\ph^\dagger(x) \;,
\label{u12}
\end{eqnarray}
where $\alpha$ is now infinitesimal.  In \eq{noether},
we should treat $\ph$ and $\phd$ as independent fields.
The Noether current is then
\begin{eqnarray}
j^\mu &=& {\d\L\over\d(\d_\mu\ph)}\,\delta\ph
        + {\d\L\over\d(\d_\mu\ph^\dagger)}\,\delta\ph^\dagger
\nonumber \\
\noalign{\medskip}
&=& \Bigl(-\d^\mu\ph^\dagger\Bigr)
    \Bigl(-i\alpha\ph\Bigr)
  + \Bigl(-\d^\mu\ph\Bigr)
    \Bigl(+i\alpha\ph^\dagger\Bigr)
\nonumber \\
\noalign{\medskip}
&=&\alpha
\Im\Bigl(\ph^\dagger{\buildrel\leftrightarrow\over{\d^\mu}}\ph\Bigr) \;,
\label{noether2}
\end{eqnarray}
where $A{\buildrel\leftrightarrow\over{\d^\mu}}\!B \equiv
A\d^\mu\!B-(\d^\mu\!A)B$.
It is conventional to drop the infinitesimal parameter
on the right-hand side in the final expression for the
Noether current.

We can also repeat this exercise using the SO(2) form of the transformation.
For infinitesimal $\alpha$, \eq{ab} becomes
$\delta \ph_1=+\alpha \ph_2$ and $\delta\ph_2=-\alpha \ph_1$.
Then the Noether current is
\begin{eqnarray}
j^\mu &=& {\d\L\over\d(\d_\mu \ph_1)}\,\delta \ph_1
        + {\d\L\over\d(\d_\mu \ph_2)}\,\delta \ph_2
\nonumber \\
\noalign{\medskip}
&=& \Bigl(-\d^\mu\ph_1\Bigr)
    \Bigl(+\alpha\ph_2\Bigr)
  + \Bigl(-\d^\mu\ph_2\Bigr)
    \Bigl(-\alpha\ph_1\Bigr)
\nonumber \\
\noalign{\medskip}
&=&\alpha\,\Bigl(\ph_1{\buildrel\leftrightarrow\over{\d^\mu}}\ph_2\Bigr)\;,
\label{noether3}
\end{eqnarray}
which is (hearteningly) equivalent to \eq{noether2}.

Let us define the total charge
\begin{equation}
Q\equiv\int \dtx\,j^0(x) =\int \dtx\,
\Im\Bigl(\ph^\dagger{\buildrel\leftrightarrow\over{\d^0}}\ph\Bigr) \;,
\label{Q}
\end{equation}
and investigate its properties.
If we integrate \eq{dj0} over $\dtx$, use Gauss's law to write the
volume integral of $\nabla\cd{\bf j}$ as a surface integral,
and assume that the boundary
conditions at infinity fix ${\bf j}(x)=0$ on that surface, 
then we find that $Q$ is constant in time.
To get a better idea of the physical implications of this,
let us rewrite $Q$ using the free-field expansions
\begin{eqnarray}
\ph(x) &=& \int \dk\left[a(\k)e^{ikx}+b^*(\k)e^{-ikx}\right]\;,
\nonumber \\
\noalign{\medskip}
\ph^\dagger(x) &=& \int \dk\left[b(\k)e^{ikx}+a^*(\k)e^{-ikx}\right]\;.
\label{abexp}
\end{eqnarray}
We have written $a^*(\k)$ and $b^*(\k)$ rather than $\ad(\k)$ and $\bd(\k)$
because so far our discussion has been about the classical field theory.
In a theory with interactions, these formulae (and their first time
derivatives) are valid at any one particular time (say, $t=-\infty$).
Then, we can plug them into \eq{Q}, and find (after some manipulation
similar to what we did for the hamiltonian in section 3)
\begin{equation}
Q=\int\dk\,\Bigl[a^*(\k)a(\k)-b(\k)b^*(\k)\Bigr]\;.
\label{Qab}
\end{equation}
In the quantum theory,
this becomes an operator that
counts the number of $a$ particles minus the number of $b$ particles.
This number is then time-independent, and so
the scattering amplitude vanishes identically
for any process that changes the value of $Q$.
This can be seen directly from the Feynman rules,
which conserve $Q$ at every vertex.

To better understand the implications of the 
Noether current in the quantum theory,
we begin by considering the infinitesimal transformation
$\ph_a(x)\to\ph_a(x)+\delta\ph_a(x)$ as a change of integration variable
in the path integral,
\begin{equation}
Z(J) = \int\D\ph\,e^{i[S + \int\!\!\dfy\,J_a\ph_a]} \;. 
\label{zj23}
\end{equation}
As with any integral, its value is unchanged by a change of integration
variable.  In our case, this change is just a shift, with unit jacobian,
and so the measure $\D\ph$ is unchanged.  Thus we have
\begin{eqnarray}
0 &=& \delta Z(J) 
\nonumber \\
\noalign{\medskip}
&=& i\int \D\ph\,e^{i[S+\int\!\!\dfy\,J_b\ph_b]}
\int\dfx\left( {\delta S\over\delta\ph_a(x)}+J_a(x)\right)\delta\ph_a(x) \;.
\label{qnoether}
\end{eqnarray}
Since this is true for arbitrary $\delta\ph_a(x)$, we can remove it (and
the integral over $\dfx$) from the right-hand side.  We can also
take $n$ functional derivatives with respect to $J_{a_j}(x_j)$, and then set
$J=0$, to get
\begin{eqnarray}
0 &=& \int \D\ph\,e^{iS}\Biggl[
\,i\,{\delta S\over\delta\ph_a(x)}\,\ph_{a_1}(x_1)\ldots\ph_{a_n}(x_n) 
\nonumber \\
&& \qquad\qquad\; {} + 
\sum_{j=1}^n\ph_{a_1}(x_1) \ldots \delta_{aa_j} \delta^4(x{-}x_j) \ldots     
\ph_{a_n}(x_n)\Biggr] 
\label{schd0} \\
\noalign{\medskip}
&=& i\la 0|{\rm T}
{\delta S\over\delta\ph_a(x)}\,\ph_{a_1}(x_1)\ldots\ph_{a_n}(x_n)|0\ra 
\nonumber \\
&& \quad\;\; {} + 
\sum_{j=1}^n \la0|{\rm T}\ph_{a_1}(x_1)
\ldots\delta_{aa_j}\delta^4(x{-}x_j)\ldots\ph_{a_n}(x_n)|0\ra \;.
\label{schd}
\end{eqnarray}
These are the {\it Schwinger-Dyson equations\/} for the theory. 

To get a feel for them, let us look at free-field theory for a single
real scalar field, for which
$\delta S/\delta\ph(x)=(\d_x^2-m^2)\ph(x)$.  For $n=1$ we get
\begin{equation}
(-\d_x^2+m^2)i\la 0|{\rm T}\ph(x)\ph(x_1)|0\ra = \delta^4(x{-}x_1)\;.
\label{kgg}
\end{equation}
That the Klein-Gordon wave operator should sit outside
the time-ordered product (and hence act on the time-ordering
step functions) is clear from the path integral form of \eq{schd0}.
We see from \eq{kgg} that the free-field propagator,
$\Delta(x{-}x_1)=i\la 0|{\rm T}\ph(x)\ph(x_1)|0\ra$, is a Green's
function for the Klein-Gordon wave operator, a fact we first learned
in section~8.

More generally, we can write
\begin{equation}
\la 0|{\rm T}{\delta S\over\delta\ph_a(x)}\,
\ph_{a_1}(x_1)\ldots\ph_{a_n}(x_n)|0\ra = 0
\quad\hbox{for}\quad 
x\ne x_{1,\ldots,n}\;. 
\label{schd2}
\end{equation}
We see that the classical equation of motion is satisfied by a quantum
field inside a correlation function, as long as its spacetime argument
differs from those of all the other fields.  When this is not the case,
we get extra {\it contact terms}.

Let us now consider a theory that has a continuous symmetry and a corresponding
Noether current.   Take \eq{schd} and multiply it by $\delta\ph_a(x)$,
where $\delta\ph_a(x)$ is the infinitesimal change in $\ph_a(x)$ that results
in $\delta\L(x)=0$.  Now sum over the index $a$, and use \eq{divj}.  The result
is the {\it Ward identity}
\begin{eqnarray}
0 &=& \d_\mu\la 0|{\rm T}j^\mu(x)\ph_{a_1}(x_1)\ldots\ph_{a_n}(x_n)|0\ra 
\nonumber \\
&& {} + 
i\sum_{j=1}^n \la0|{\rm T}\ph_{a_1}(x_1)
\ldots\delta\ph_{a_j}(x)\delta^4(x{-}x_j)\ldots\ph_{a_n}(x_n)|0\ra \;.
\label{divjq}
\end{eqnarray}
Thus, conservation of the Noether current holds in the quantum
theory, with the current inside a correlation function,
up to contact terms with a specific form that depends on the
details of the infinitesimal transformation that leaves $\L$ invariant.

The Noether current is also useful in a slightly more general context.
Suppose we have a transformation of the fields such that $\delta\L(x)$
is not zero, but instead is a total divergence:
$\delta\L(x)=\d_\mu K^\mu(x)$ for some $K^\mu(x)$.
Then there is still a conserved current, now given by
\begin{equation}
j^\mu(x) = {\d\L(x)\over\d(\d_\mu\ph_a(x))}\,\delta\ph_a(x)-K^\mu(x)\;.
\label{noether4}
\end{equation}
An example of this is provided by the symmetry of {\it spacetime translations}.
We transform the fields via $\ph_a(x)\to\ph_a(x+a)$, where $a^\mu$
is a constant four-vector.  The infinitesimal version of this is
$\ph_a(x)\to\ph_a(x)+a^\nu\d_\nu\ph_a(x)$, and so we have
$\delta\ph_a(x)=a^\nu\d_\nu\ph_a(x)$.  
Under this transformation, we obviously
have $\L(x)\to\L(x+a)$, and so
$\delta\L(x)=a^\nu\d_\nu\L(x)=\d_\nu(a^\nu\L(x))$.
Thus in this case $K^\nu(x)=a^\nu\L(x)$, and the conserved current is
\begin{eqnarray}
j^\mu(x) &=& {\d\L(x)\over\d(\d_\mu\ph_a(x))}\,a^\nu\d_\nu\ph_a(x) -a^\mu\L(x) 
\nonumber \\
\noalign{\medskip}
&=& -a_\nu T^{\mu\nu}(x)\;,
\label{noether5}
\end{eqnarray}
where we have defined the {\it stress-energy\/} or
{\it energy-momentum tensor\/}
\begin{equation}
T^{\mu\nu}(x) \equiv {}-{\d\L(x)\over\d(\d_\mu\ph_a(x))}\,\d^\nu\!\ph_a(x)
            +g^{\mu\nu}\L(x) \;.
\label{tmunu}
\end{equation}

For a renormalizable theory of a set of real scalar fields $\ph_a(x)$, 
the lagrangian takes the form
\begin{equation}
\L = -\half\d^\mu\ph_a\d_\mu\ph_a-V(\ph) \;,
\label{ellagain}
\end{equation}
where $V(\ph)$ is a polynomial in the $\ph_a$'s.
In this case
\begin{equation}
T^{\mu\nu} = \d^\mu\ph_a\d^\nu\!\ph_a+g^{\mu\nu}\L \;.
\label{tmunu2}
\end{equation}
In particular,
\begin{equation}
T^{00} = \half\Pi^2_a + \half(\nabla\ph_a)^2 + V(\ph) \;,
\label{t00}
\end{equation}
where $\Pi_a=\d_0\ph_a$ is the canonical
momentum conjugate to the field $\ph_a$.
We recognize $T^{00}$ as the {\it hamiltonian density $\cal H$\/} that
corresponds to the lagrangian density of \eq{ellagain}.
Then, by Lorentz symmetry, $T^{0j}$ must be the corresponding momentum density.  We have
\begin{equation}
T^{0j} = \d^0\ph_a\d^j\!\ph_a = -\Pi_a\nabla^j\ph_a \;.
\label{t0j}
\end{equation}
If we use the free-field expansion for a set of real scalar fields
[the same as \eq{abexp} but with $b(\k)=a(\k)$ for each field], we find
that the momentum operator is given by
\begin{equation}
P^j = \int\dtx\,T^{0j}(x) = \int\dk\; k^j\,\ad_a(\k)a^{\phantom{\dagger}}_a(\k) \;.
\label{pj}
\end{equation}
We therefore identify the {\it energy-momentum four-vector} as
\begin{equation}
P^\mu = \int\dtx\,T^{0\mu}(x) \;.
\label{pmu}
\end{equation}

Recall that in section 2 we defined the 
{\it spacetime translation operator\/} as
\begin{equation}
T(a) \equiv \exp(-iP^\mu a_\mu)\;,
\label{ta2}
\end{equation}
and announced that it had the property that
\begin{equation}
T(a)^{-1}\ph(x)T(a) = \ph(x-a)\;.
\label{tpht2}
\end{equation}
Now that we have an explicit formula for $P^\mu$, we can check this.
This is easiest to do for infinitesimal $a^\mu$; then \eq{tpht2} becomes
\begin{equation}
[\ph(x),P^\mu] = {\ts{1\over i}}\d^\mu\ph(x)\;.
\label{tpht3}
\end{equation}
This can indeed be verified by using the canonical commutation relations
for $\ph(x)$ and $\Pi(x)$.

One more symmetry we can investigate is Lorentz symmetry.  If we make
an infinitesimal Lorentz transformation, we have
$\ph_a(x)\to\ph_a(x+\delta\w\cd x)$, where $\delta\omega\cd x$ is
shorthand for $\delta\omega^\nu{}_\rho x^\rho$.  This case is very
similar to that of spacetime translations; the only difference is
that the translation parameter $a^\nu$ is now $x$ dependent,
$a^\nu\to\delta\omega^\nu{}_\rho x^\rho$. 
The resulting conserved current is
\begin{equation}
{\cal M}^{\mu\nu\rho}(x) 
= x^\nu T^{\mu\rho}(x) - x^\rho T^{\mu\nu}(x) \;,
\label{mmunurho}
\end{equation}
and it obeys $\d_\mu{\cal M}^{\mu\nu\rho}=0$, with the derivative 
contracted with the first index.
${\cal M}^{\mu\nu\rho}$ is antisymmetric on its second two indices;
this comes about because $\delta\w^{\nu\rho}$ is antisymmetric.
The conserved charges associated with this current are
\begin{equation}
M^{\nu\rho} = \int\dtx\,{\cal M}^{0\nu\rho}(x) \;, 
\label{mnurho}
\end{equation}
and these are the {\it generators of the Lorentz group\/} that
were introduced in section~3.  Again, we can use the canonical
commutation relations for the fields to check that the Lorentz
generators have the right commutation relations, both with
the fields and with each other.

\vfill\eject

%% file: ch023.tex

\noindent Quantum Field Theory  \hfill   Mark Srednicki

\vskip0.5in

\begin{center}
\large{23: Discrete Symmetries: $P$, $T$, $C$, and $Z$}
\end{center}
\begin{center}
Prerequisite: 22
\end{center}

\vskip0.5in

In section 2, 
we studied the {\it proper orthochronous\/} Lorentz transformations,
which are continuously connected to the identity.  In this section, we will
consider the effects of {\it parity},
\begin{equation}
\P^\mu{}_\nu = (\P^{-1}){}^\mu{}_\nu
             =\pmatrix{ +1 & & & \cr
                        & -1 & & \cr
                        & & -1 & \cr
                        & & & -1 \cr}.
\label{p3}
\end{equation}
and {\it time reversal},
\begin{equation}
\T^\mu{}_\nu = (\T^{-1}){}^\mu{}_\nu
             =\pmatrix{ -1 & & & \cr
                        & +1 & & \cr
                        & & +1 & \cr
                        & & & +1 \cr}.
\label{tmu2}
\end{equation}
We will also consider certain other discrete transformations that are not Lorentz
transformations, but are usefully treated together.

Recall from section 2 that for every
proper orthochronous Lorentz transformation $\Lam^\mu{}_\nu$ there
is an associated unitary operator $U(\Lam)$ with the property that
\begin{equation}
U(\Lam)^{-1} \ph(x) U(\Lam) = \ph(\Lam^{-1}x)\;.
\label{uphu23}
\end{equation}
Thus for parity and time-reversal, we expect that there are corresponding 
unitary operators
\begin{eqnarray}
P &\equiv& U(\P) \;,
\label{pop} \\
T &\equiv& U(\T) \;,
\label{top}
\end{eqnarray}
such that
\begin{eqnarray}
P^{-1} \ph(x) P &=& \ph(\P x)\;,
\label{pphp} \\
T^{-1} \ph(x) T &=& \ph(\T x)\;.
\label{tpht}
\end{eqnarray}

There is, however, an extra possible complication.  If we make a second
parity or time-reversal transformation, we get
\begin{eqnarray}
P^{-2} \ph(x) P^2 &=& \ph(x)\;,
\label{p2php2} \\
T^{-2} \ph(x) T^2 &=& \ph(x)\;,
\label{t2pht2}
\end{eqnarray}
and so the field returns to itself.  Since the field is in principle an 
observable---it is a hermitian operator---this is required.  
However, another possibility,
different from \eqs{pphp} and (\ref{tpht}) but nevertheless consistent with
\eqs{p2php2} and (\ref{t2pht2}), is
\begin{eqnarray}
P^{-1} \ph(x) P &=& -\ph(\P x)\;,
\label{pphp-1} \\
T^{-1} \ph(x) T &=& -\ph(\T x)\;.
\label{tpht-1}
\end{eqnarray}
This possible extra minus sign cannot arise for proper orthochronous Lorentz
transformations, because they are continuously connected to the identity,
and for the identity transformation (that is, no transformation at all), we
must obviously have the plus sign.

If the minus sign appears on the right-hand side, we say that the field
is {\it odd under parity\/} (or time reversal).  If a scalar field is odd under parity,
we sometimes say that it is a {\it pseudoscalar}.  [It is still a scalar under
proper orthochronous Lorentz transformations; that is, \eq{uphu23} still holds.
Thus the appellation {\it scalar\/} often means \eq{uphu23}, and
{\it either\/} \eq{pphp} {\it or\/} \eq{pphp-1}, and that is how we will
use the term.]

So, how do we know which is right, \eqs{pphp} and (\ref{tpht}), or
\eqs{pphp-1} and (\ref{tpht-1})?
The general answer is that we get to choose, but there is a key
principle to guide our choice:  if at all possible, we want to define
$P$ and $T$ so that the lagrangian density is even,
\begin{eqnarray}
P^{-1} \L(x) P &=& +\L(\P x)\;,
\label{plp} \\
T^{-1} \L(x) T &=& +\L(\T x)\;.
\label{tlt}
\end{eqnarray}
Then, after we integrate over $\dfx$ to get the action $S$, the action
will be invariant.  
This means that parity and time-reversal are {\it conserved\/}.

For theories with spin-zero fields only, it is clear that the choice of 
\eqs{pphp} and (\ref{tpht}) always leads to \eqs{plp} and (\ref{tlt}),
and so there is no reason to flirt with \eqs{pphp-1} and (\ref{tpht-1}).
For theories that also include spin-one-half fields, certain scalar
bilinears in these fields are necessarily odd under parity and time reversal,
as we will see in section 39.
If a scalar field couples to such a bilinear, then \eqs{plp} and (\ref{tlt})
will hold if and only if 
we choose \eqs{pphp-1} and (\ref{tpht-1}) for that scalar,
and so that is what we must do.

There is one more interesting fact about the time-reversal operator $T$:
it is {\it antiunitary}, rather than unitary.  {\it Antiunitary\/} means that
$T^{-1}iT = -i$.

To see why this must be the case, consider a Lorentz 
transformation of the energy-momentum four-vector,
\begin{equation}
U(\Lam)^{-1} P^\mu U(\Lam) = \Lam^\mu{}_\nu P^\nu \;.
\label{upu2}
\end{equation}
For parity and time-reversal, we therefore expect
\begin{eqnarray}
P^{-1}  P^\mu P &=& \P^\mu{}_\nu P^\nu\;,
\label{ppmup} \\
T^{-1}  P^\mu T &=& \T^\mu{}_\nu P^\nu\;.
\label{tpmut}
\end{eqnarray}
In particular, for $\mu=0$, we expect
$P^{-1}HP=+H$ and $T^{-1}HT=-H$.  The first of these is fine;
it says the hamiltonian is invariant under parity, which is what
we want.  [It may be that no operator exists that satisfies
either \eq{pphp} or \eq{pphp-1}, and also \eq{ppmup}; in this case
we say that parity is {\it explicitly broken}.]   However, \eq{tpmut}
is a disaster: it says that the hamiltonian is invariant under time-reversal
if and only if $H=-H$.  This is clearly untrue for a system whose
energy is bounded below and unbounded above, as we always
have in a realistic quantum field theory.

Can we just toss in an extra minus sign on the right-hand side
of \eq{tpmut}, as we did for \eq{tpht-1}?  The answer is no. 
We constructed $P^\mu$ explicitly in terms of the fields
in section 22, and it is easy to check that
choosing \eq{tpht-1} for the fields 
does not yield an extra minus sign in \eq{tpmut}
for the energy-momentum four-vector. 

Let us reconsider the origin of \eq{upu2}.  We can derive it from
\begin{equation}
U(\Lam)^{-1} T(a) U(\Lam) = T(\Lam^{-1}a) \;,
\label{utu}
\end{equation}
where $T(a)=\exp(-iP\cd a)$ is the spacetime translation operator
(not to be confused with the time-reversal operator!),
which transforms the field via 
\begin{equation}
T(a)^{-1}\ph(x)T(a)=\ph(x-a)\;.
\label{taphita}
\end{equation}
We can get \eq{utu} (up to a possible 
phase that turns out to be irrelevant) from
\begin{eqnarray}
U(\Lam)^{-1} T(a)^{-1} U(\Lam) \ph(x)U(\Lam)^{-1} T(a) U(\Lam)
\hskip-1.5in 
\nonumber \\
&=& U(\Lam)^{-1} T(a)^{-1}\ph(\Lam x) T(a) U(\Lam)
\nonumber \\
&=& U(\Lam)^{-1} \ph(\Lam x - a) U(\Lam)
\nonumber \\
&=& \ph(x - \Lam^{-1}a)
\nonumber \\
&=& T(\Lam^{-1}a)^{-1} \ph(x) T(\Lam^{-1}a) \;.
\label{utu3}
\end{eqnarray}
Now, treat $a^\mu$ as infinitesimal in \eq{utu} to get
\begin{eqnarray}
U(\Lam)^{-1} (I-ia_\mu P^\mu) U(\Lam) 
&=&  I-i(\Lam^{-1})_\nu{}^\mu a_\mu P^\nu
\nonumber \\
&=&  I-i\Lam^\mu{}_\nu a_\mu P^\nu \;.
\label{upmuu3}
\end{eqnarray}
For time-reversal, this becomes
\begin{equation}
T^{-1} (I-ia_\mu P^\mu) T =  I-i\T^\mu{}_\nu a_\mu P^\nu\;.
\label{utu4}
\end{equation}
If we now identify the coefficients of $-ia_\mu$ on each side,
we get \eq{tpmut}.  {\it But}, we will get that extra minus sign
that we need if we impose the antiunitary condition
\begin{equation}
T^{-1} i T =  -i \;.
\label{tit2}
\end{equation}
And so that is what we must do.

We turn now to other unitary operators that change the signs of
scalar fields, but do nothing to their spacetime arguments.
Suppose we have a theory with real scalar fields $\ph_a(x)$,
and a unitary operator $Z$ that obeys
\begin{equation}
Z^{-1} \ph_a(x) Z =  \eta_a\ph_a(x) \;,
\label{zphz}
\end{equation}
where $\eta_a$ is either $+1$ or $-1$ for each field.
We will call $Z$ a {\it Z${}_2$ operator}, because
Z${}_2$ is the additive group of the integers modulo 2,
which is equivalent to the multiplicative group of $+1$ and $-1$.
This also implies that $Z^2=1$, and so $Z^{-1}=Z$.
(For theories with spin-zero fields only, the same is also true of
$P$ and $T$, but things are more subtle for higher spin,
as we will see in Parts II and III.)

Consider the theory of a complex scalar field
$\ph=(\ph_1+i\ph_2)/\sqrt2$ that was introduced in section 22, with lagrangian
\begin{eqnarray}
\L&=&-\d^\mu\ph^\dagger\d_\mu\ph-m^2\ph^\dagger\ph 
- {\ts{1\over4}}\lam(\ph^\dagger\ph)^2
\label{ellu123} \\
\noalign{\medskip}
&=& -\half\d^\mu \ph_1\d_\mu \ph_1 -\half\d^\mu \ph_2\d_\mu \ph_2
-\half m^2(\ph_1^2+\ph_2^2)
- {\ts{1\over16}}\lam(\ph_1^2+\ph_2^2)^2.\qquad\quad
\label{ellab23}
\end{eqnarray}
In the form of \eq{ellu123}, $\L$ is obviously invariant under the U(1)
transformation
\begin{equation}
\ph(x)\to e^{-i\alpha}\ph(x) \;.
\label{u123}
\end{equation}
In the form of \eq{ellab23}, $\L$ is obviously invariant under the equivalent
SO(2) transformation,
\begin{equation}
\pmatrix{ 
\ph_1(x) \cr 
\noalign{\medskip}
\ph_2(x) \cr}
\to
\pmatrix{ \phantom{-}\cos\alpha & \sin\alpha \cr
\noalign{\medskip}
           {-}\sin\alpha & \cos\alpha \cr}
\pmatrix{ 
\ph_1(x) \cr 
\noalign{\medskip}
\ph_2(x) \cr}.
\label{ab23}
\end{equation}
However, it is also obvious that $\L$ has an additional 
discrete symmetry,
\begin{equation}
\ph(x) \leftrightarrow \phd(x) 
\label{phphd}
\end{equation}
in the form of \eq{ellu123}, or equivalently 
\begin{equation}
\pmatrix{ 
\ph_1(x) \cr 
\noalign{\medskip}
\ph_2(x) \cr}
\to
\pmatrix{ +1 & 0 \cr
\noalign{\medskip}
                 0 & -1 \cr}
\pmatrix{ 
\ph_1(x) \cr 
\noalign{\medskip}
\ph_2(x) \cr}.
\label{121-2}
\end{equation}
in the form of \eq{ellab23}.
This discrete symmetry is called {\it charge conjugation}.
It always occurs as a companion to a continuous U(1)
symmetry.  In terms of the two real fields, it enlarges
the group from SO(2) (the group of $2\times2$ orthogonal
matrices with determinant $+1$) to O(2) 
(the group of $2\times2$ orthogonal matrices).

We can implement charge conjugation by means of
a particular Z${}_2$ operator $C$ that obeys
\begin{equation}
C^{-1}\ph(x)C = \phd(x) \;,
\label{cphc}
\end{equation}
or equivalently
\begin{eqnarray}
C^{-1}\ph_1(x)C &=& +\ph_1(x) \;,
\label{cph1c} \\
C^{-1}\ph_2(x)C &=& -\ph_2(x) \;.
\label{cph2c}
\end{eqnarray}
We then have
\begin{equation}
C^{-1}\L(x)C = \L(x) \;,
\label{cellc}
\end{equation}
and so charge conjugation is a symmetry of the theory.
Physically, it implies that the scattering amplitudes are
unchanged if we exchange all the $a$-type particles (which have
charge $+1$) with all the $b$-type particles (which have charge $-1$).
This means, in particular, that the $a$ and $b$ particles
must have exactly the same mass.  We say that $b$ is
$a$'s {\it antiparticle}.

More generally, we can also have Z${}_2$ symmetries
that are not related to antiparticles.  
Consider, for example,
$\ph^4$ theory, where $\ph$ is a real scalar field with lagrangian
\begin{equation}
\L=-\half\d^\mu \ph\d_\mu \ph -\half m^2\ph^2
- {\ts{1\over24}}\lam\ph^4\;.
\label{ellph4}
\end{equation}
If we define the Z${}_2$ operator $Z$ via
\begin{equation}
Z^{-1} \ph(x) Z =  -\ph(x) \;,
\label{zphz2}
\end{equation}
then $\L$ is obviously invariant.  We therefore have
$Z^{-1} H Z =  H$, or equivalently $[Z,H]=0$,
where $H$ is the hamiltonian.
If we assume that (as usual) the ground state is
unique, then, since $Z$ commutes with $H$,
the ground state must also be an eigenstate of $Z$.
We can fix the phase of $Z$ [which is undetermined
by \eq{zphz2}] via 
\begin{equation}
Z|0\ra=Z^{-1}|0\ra=+|0\ra \;.  
\label{z0}
\end{equation} 
Then, using \eqs{zphz2} and (\ref{z0}), we have
\begin{eqnarray}
\la 0|\ph(x)|0\ra
&=& \la 0|ZZ^{-1}\ph(x)ZZ^{-1}|0\ra
\nonumber \\
&=& -\la 0|\ph(x)|0\ra \;.
\label{z023}
\end{eqnarray}
Since $\la 0|\ph(x)|0\ra$ is equal to minus itself,
it must be zero.  Thus, as long as the ground state
is unique, {\it the Z${}_2$ symmetry
of $\ph^4$ theory guarantees that the field
has zero vacuum expectation value.}
We therefore do not need to enforce this condition with a
counterterm $Y\ph$, as we did in $\ph^3$ theory.
(The assumption of a unique ground state does not
necessarily hold, however, as we will see in section 28.)

\vfill\eject

%% file: ch024.tex

\noindent Quantum Field Theory  \hfill   Mark Srednicki

\vskip0.5in

\begin{center}
\large{24: Unstable Particles and Resonances}
\end{center}
\begin{center}
Prerequisite: 10, 14
\end{center}

\vskip0.5in

Consider a theory of two real scalar fields, $\ph$ and $\chi$,
with lagrangian
\begin{equation}
\L=-\half\d^\mu\ph\d_\mu\ph
-\half m_\ph^2\ph^2
-\half\d^\mu\chi\d_\mu\chi
-\half m_\chi^2\chi^2
+{\ts{1\over2}}g\ph\chi^2 
+{\ts{1\over6}}h\ph^3 \;, 
\label{ellpc}
\end{equation}
This theory is renormalizable in six dimensions, where $g$ and $h$
are dimensionless coupling constants.  

\begin{figure}
\begin{center}
\epsfig{file=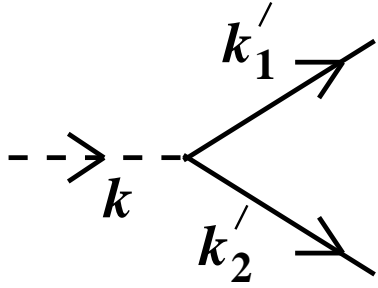}
\end{center}
\caption{The tree-level Feynman diagram for the decay of
a $\ph$ particle (dashed line) into two $\chi$ particles
(solid lines).}
\label{decay}
\end{figure}

Let us assume that $m_\ph > 2m_\chi$.  Then it is kinematically possible 
for the $\ph$ particle to decay into two $\chi$ particles.  The amplitude 
for this process is given at tree level by the Feynman diagram of \fig{decay},
and is simply $\T = g$.  
We can also choose to work in an on-shell renormalization
scheme in which we define $g$ by saying that the exact 
$\ph\chi^2$ vertex function $\V_3(k,k'_1,k'_2)$ equals $g$ 
when all three particles are on shell: $k^2=-m_\ph^2$,
$k'_1{}^{\!2}=k'_2{}^{\!2}=-m_\chi^2$.  This implies that
\begin{equation}
\T=g
\label{t=g}
\end{equation}
exactly.

According to the formulae of section 11, the differential decay rate (in the
rest frame of the initial $\ph$ particle) is
\begin{equation}
d\Gamma = {1\over2m_\ph}\,d\hbox{LIPS}_{2} \, |\T|^2\;,
\label{dg2}
\end{equation}
where  $d\hbox{LIPS}_{2}$ 
is the Lorentz invariant phase space
differential for two outgoing particles, introduced in section 11.
We must make a slight adaptation for six dimensions:
\begin{equation}
d\hbox{LIPS}_2
\equiv (2\pi)^6 \delta^6(k'_1{+}k'_2{-}k)\, \dk{}'_1\,\dk{}'_2\;.
\label{dlips26}
\end{equation}
Here $k=(m_\ph,{\bf 0})$ is the energy-momentum of the decaying particle, and
\begin{equation}
\dk = {d^5\!k\over(2\pi)^5 2\omega} 
\label{dk5}
\end{equation}
is the Lorentz-invariant phase-space differential for one particle.
Recall that we can also write it as
\begin{equation}
\dk = {\dsk\over(2\pi)^6}\,2\pi\delta(k^2+m_\chi^2)\,\theta(k^0)\;,
\label{dk6} 
\end{equation}
where $\theta(x)$ is the unit step function.
Performing the integral over $k^0$ turns \eq{dk6} into \eq{dk5}.

Repeating for six dimensions what we did in section 11 for four dimensions,
we find
\begin{equation}
d\hbox{LIPS}_2
= {|\k'_{1\!}|^3\over4(2\pi)^4 m_\ph}\,d\Omega \;,
\label{dlips27}
\end{equation}
where $|\k'_{1\!}|=\half(m_\ph^2-4m_\chi^2)^{1/2}$ is the magnitude
of the spatial momentum of one of the outgoing particles.
We can now plug this into \eq{dg2}, and integrate 
$\int d\Omega=\Omega_5=2\pi^{5/2}/\Gamma({5\over2})={8\over3}\pi^2$.
We also need a symmetry factor of one half,
due to the presence of two identical particles in the final state.  
The result is
\begin{eqnarray}
\Gamma &=& {1\over2}\cdot{1\over2m_\ph}\int
d\hbox{LIPS}_{2} \, |\T|^2
\label{g30} \\
\noalign{\medskip}
&=& {\ts{1\over12}}\pi\alpha
(1-4m_\chi^2/m_\ph^2)^{3/2}\,m_\ph\;,
\label{g3}
\end{eqnarray}
where $\alpha=g^2/(4\pi)^3$.

However, as we discussed in section 11,
we have a conceptual problem.  According to our development of
the LSZ formula in section 5, each incoming and outgoing particle should
correspond to a single-particle state that is an exact eigenstate of the
exact hamiltonian.  
This is clearly not the case for a particle that can decay.

\begin{figure}
\begin{center}
\epsfig{file=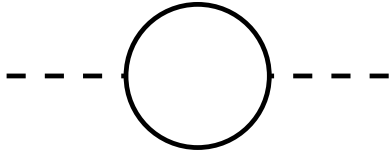}
\end{center}
\caption{A loop of $\chi$ particles correcting
the $\ph$ propagator.}
\label{chiloop}
\end{figure}

Let us, then, compute something else instead: the correction to the
$\ph$ propagator from a loop of $\chi$ particles, as shown in \fig{chiloop}. 
The diagram is the same as the one we already analyzed in section 14,
except that the internal propagators contain $m_\chi$ instead of $m_\ph$.
(There is also a contribution from a loop of $\ph$ particles, 
but we can ignore it if we assume that $h\ll g$.)  We have
\begin{equation}
\Pi(k^2) = \half\alpha \int_0^1 dx\,D\ln D
- A' k^2 - B' m_\ph^2 \;,
\label{pi240}
\end{equation}
where 
\begin{equation}
D = x(1{-}x)k^2+m_\chi^2-i\eps \;,
\label{d10}
\end{equation}
and $A'$ and $B'$ are the finite counterterm 
coefficients that remain after the infinities have been absorbed.
We now try to fix $A'$ and $B'$ by imposing
the usual on-shell conditions
$\Pi(-m_\ph^2)=0$ and $\Pi'(-m_\ph^2)=0$.

But, we have a problem.  For $k^2=-m_\ph^2$ and $m_\ph > 2m_\chi$,
$D$ is negative for part of the range of $x$.  Therefore $\ln D$
has an imaginary part.  This imaginary part cannot be canceled
by $A'$ and $B'$, since $A'$ and $B'$ must be real:
they are coefficients of hermitian operators in the lagrangian.
The best we can do is $\Re\Pi(-m_\ph^2)=0$ and $\Re\Pi'(-m_\ph^2)=0$.
Imposing these gives
\begin{equation}
\Pi(k^2) = \half\alpha\int_0^1 dx\,D\ln(D/|D_0|)
-{\ts{1\over12}}\alpha(k^2+m_\ph^2) \;,
\label{pi25}
\end{equation}
where
\begin{equation}
D_0 = -x(1{-}x)m_\ph^2+m_\chi^2 \;.
\label{d010}
\end{equation}

Now let us compute the imaginary part of $\Pi(k^2)$.
This arises from the integration range
$x_-<x<x_+$, where 
$x_\pm = \half\pm\half(1+m_\chi^2/k^2)^{1/2}$ 
are the roots of $D=0$ when $k^2<-4m^2_\chi$.  
In this range, $\Im\ln D=-i\pi$;
the minus sign arises because, according to \eq{d10},
$D$ has a small negative imaginary part.  Now we have
\begin{eqnarray}
\Im\Pi(k^2) &=& -\half\pi\alpha\int_{x_-}^{x_+} dx\,D
\nonumber \\ 
\noalign{\medskip}
&=& -{\ts{1\over12}}\pi\alpha(1+4m_\chi^2/k^2)^{3/2}\,k^2 
\label{impi} 
\end{eqnarray}
when $k^2<-4m_\chi^2$.  Evaluating \eq{impi} at $k^2=-m_\ph^2$, we get
\begin{equation}
\Im\Pi(-m_\ph^2) = {\ts{1\over12}}\pi\alpha
                   (1-4m_\chi^2/m_\ph^2)^{3/2}\,m_\ph^2 \;.
\label{impi2}
\end{equation}
From this and \eq{g3}, we see that 
\begin{equation}
\Im\Pi(-m_\ph^2) = m_\ph\Gamma \;.
\label{impig}
\end{equation}

This is not an accident.  Instead, it is a general rule.
We will argue this in two ways:  first, from the mathematics
of Feynman diagrams, and second, from the physics of resonant
scattering in quantum mechanics.

We begin with the mathematics of Feynman diagrams.  Return to
the diagrammatic expression for $\Pi(k^2)$, before we evaluated any
of the integrals:
\begin{eqnarray}
\Pi(k^2) &=& -\half i g^2\int{\dsxl_1\over(2\pi)^6}\,{\dsxl_2\over(2\pi)^6}\,
(2\pi)^6\delta^6(\ell_1{+}\ell_2{-}k)
\nonumber \\
&& \qquad \qquad {} \times
{1\over \ell_1^2+m_\chi^2-i\eps}\,
{1\over \ell_2^2+m_\chi^2-i\eps}
\nonumber \\
\noalign{\medskip}
&& {} - (Ak^2 + Bm_\ph^2) \;.
\label{pidi24}
\end{eqnarray}
Here, for later convenience, we have assigned the internal lines
momenta $\ell_1$ and $\ell_2$, and explicitly included the
momentum-conserving delta function that fixes one of them. 
We can take the imaginary part of $\Pi(k^2)$ by using the identity
\begin{equation}
{1\over x-i\eps} = P{1\over x} + i\pi\delta(x) \;,
\label{id2}
\end{equation}
where $P$ means the principal part.    
We then get, in a shorthand notation,
\begin{equation}
\Im\Pi(k^2)= -\half g^2\int\Bigl(P_1P_2-\pi^2\delta_1\delta_2\Bigr) \;.
\label{ppdd}
\end{equation}

Next, we notice that
the integral in \eq{pidi24} is the real part of the Fourier transform
of $[\Delta(x{-}y)]^2$, where
\begin{equation}
\Delta(x{-}y) = \int{\dsk\over(2\pi)^6}\,{e^{ik(x-y)}\over k^2+m_\chi^2-i\eps} 
\label{Dxy6}
\end{equation} 
is the Feynman propagator. 
Recall (from problem 8.4) that we can get the retarded or advanced propagator
(rather than the Feynman propagator) by replacing the $\eps$ in \eq{Dxy6} with,
respectively,  $-s\eps$  or $+s\eps$, where $s\equiv \sign(k^0)$.  
Therefore, in \eq{ppdd},  replacing $\delta_1$ with 
$-s_1\delta_1$ and $\delta_2$ with $+s_2\delta_2$ yields an integral that is
the real part of the Fourier transform of  
$\Delta_{\rm ret}(x{-}y)\Delta_{\rm adv}(x{-}y)$.  But this product is zero,
because the first factor vanishes when $x^0\ge y^0$, and the second when
$x^0\le y^0$.   So we can subtract the modified integrand from the original 
without changing the value of the integral.  Thus we have
\begin{equation}
\Im\Pi(k^2)= \half g^2\pi^2\int(1+s_1 s_2)\delta_1\delta_2 \;.
\label{ssdd}
\end{equation}
The factor of $1+s_1 s_2$ vanishes if $\ell^0_1$ and $\ell^0_2$
have opposite signs, and equals 2 if they have the same sign.
Because the delta function in \eq{pidi24} enforces 
$\ell^0_1+\ell^0_2=k^0$, and $k^0=m_\ph$ is positive, both
$\ell^0_1$ and $\ell^0_2$ must be positive.  So we can replace the
factor of  $1+s_1 s_2$ in \eq{ssdd} with $2\theta_1\theta_2$, where
$\theta_i=\theta(\ell^0_i)$.  Rearranging the numerical factors, we have
\begin{eqnarray}
\Im\Pi(k^2) &=& {\ts{1\over4}}g^2
\int{\dsxl_1\over(2\pi)^6}\,{\dsxl_2\over(2\pi)^6}\,
(2\pi)^6\delta^6(\ell_1{+}\ell_2{-}k)
\nonumber \\
&& \qquad \;\; {} \times
2\pi\delta(\ell_1^2+m_\chi^2)\theta(\ell^0_1)\,
2\pi\delta(\ell_2^2+m_\chi^2)\theta(\ell^0_2) \;.
\label{impi3}
\end{eqnarray}
If we now set $k^2=-m^2_\ph$, use \eqs{dlips26} and (\ref{dk6}), 
and recall that $\T=g$ is the decay amplitude, 
we can rewrite \eq{impi3} as
\begin{equation}
\Im\Pi(-m^2_\ph) = {\ts{1\over 4}}\int d\hbox{LIPS}_2\,|\T|^2 \;.
\label{impilips}
\end{equation}
Comparing \eqs{g30} and (\ref{impilips}), we see that
we indeed have
\begin{equation}
\Im\Pi(-m_\ph^2) = m_\ph\Gamma \;.
\label{impig2}
\end{equation}
This relation persists at higher orders in perturbation theory.
Our analysis can be generalized to give the {\it Cutkosky rules\/}
for computing the imaginary part of any Feynman diagram,
but this is beyond the scope of our current interest.

To get a more physical understanding of this result, 
recall that in nonrelativistic
quantum mechanics, a metastable state with energy $E_0$ and 
angular momentum quantum number $\ell$ shows up as a {\it resonance} 
in the partial-wave scattering amplitude,
\begin{equation}
f_\ell(E) \sim {1\over E-E_0+i\Gamma/2} \;.
\label{fe}
\end{equation}
If we imagine convolving this amplitude with a wave packet 
$\widetilde\psi(E)e^{-iEt}$,
we will find a time dependence
\begin{eqnarray}
\psi(t) &\sim& \int dE\,{1\over E-E_0+i\Gamma/2}\,
\widetilde\psi(E)e^{-iEt}
\nonumber \\
\noalign{\medskip}
&\sim& e^{-iE_0t-\Gamma t/2} \;.
\label{ft3}
\end{eqnarray}
Therefore $|\psi(t)|^2 \sim e^{-\Gamma t}$, and we identify 
$\Gamma$ as the inverse lifetime of the metastable state.

\begin{figure}
\begin{center}
\epsfig{file=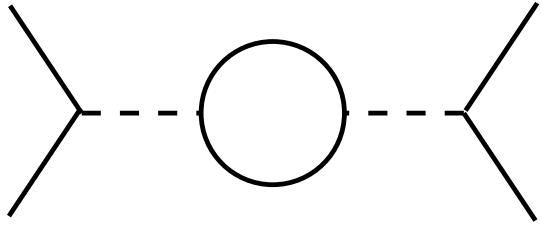}
\end{center}
\caption{$\chi\chi$ scattering with an intermediate $\ph$
propagator in the $s$-channel.}
\label{chiscat}
\end{figure}

In the relativistic case, consider the scattering $\chi\chi\to\chi\chi$
with an intermediate $\ph$ propagator, as shown in \fig{chiscat}.
In this case we have
\begin{equation}
\T = {g^2\over -s + m_\ph^2 - \Pi(-s)} + (s\to t) + (s\to u) \;.
\label{tres}
\end{equation}
If $s$ is close $m_\ph^2$, we can ignore vertex corrections,
because we have chosen a renormalization scheme in which these
vanish when all particles are on shell.

Let us, then, tune the center-of-mass energy squared $s$ to
be close to $m_\ph^2$.  Specifically, let
\begin{equation}
s=(m_\ph+\e)^2\simeq m_\ph^2+2m_\ph\e \;,
\label{sres}
\end{equation}
where $\e\ll m_\ph$ is the amount of energy by which our
incoming particles are {\it off resonance}.  We then have
\begin{equation}
\T \simeq {-g^2/2m_\ph\over \e + \Pi(-m_\ph^2)/2m_\ph} \;.
\label{tres2}
\end{equation}
Recalling that $\Re\Pi(-m_\ph^2)=0$, and comparing with \eq{fe}, 
we see that we should make
the identification of \eq{impig2}.

\vfill\eject

%% file: ch025.tex
\noindent Quantum Field Theory  \hfill   Mark Srednicki

\vskip0.5in

\begin{center}
\large{25: Infrared Divergences}
\end{center}
\begin{center}
Prerequisite: 20
\end{center}

\vskip0.5in

In section 20, we computed the $\ph\ph\to\ph\ph$ scattering amplitude in
$\phi^3$ theory in six dimensions in the 
high-energy limit ($s$, $|t|$, and $|u|$ all much larger than $m^2$).
We found that
\begin{equation}
\T = \T_0
    \left[1 - {\textstyle{11\over12}}\alpha
    \Bigl(\ln(s/m^2)+O(m^0)\Bigr)
+O(\alpha^2)
\right] ,
\label{tt}
\end{equation} 
where $\T_0=-g^2(s^{-1}+t^{-1}+u^{-1})$ is the tree-level result, and
the $O(m^0)$ term includes everything without 
a {\it large logarithm\/} that blows up in the limit $m\to0$.
[In writing $\T$ in this form, we have traded factors of
$\ln t$ and $\ln u$ for $\ln s$ by first using
$\ln t=\ln s + \ln(t/s)$, and then hiding the $\ln(t/s)$
terms in the $O(m^0)$ catchall.]

Suppose we are interested in the limit of massless particles.
The large log is then problematic, since it blows up in this limit.
What does this mean?

It means we have made a mistake.  Actually, two mistakes.
In this section, we will remedy one of them.

Throughout the physical sciences, it is necessary to make various idealizations
of problems in order to make progress (recall the ``massless springs''
and ``frictionless planes'' of freshman mechanics).  Sometimes these
idealizations can lead us into trouble, and that is one of the 
things that has gone wrong here.

We have assumed that we can isolate individual particles.  The reasoning
behind this was carefully explained in section 5.  However, our
reasoning breaks down in the massless limit.  In this case, it is possible
that the scattering process involved the creation of some extra very low
energy (or {\it soft\/}) particles that escaped detection.  
Or, there may have been some extra soft
particles hiding in the initial state that discreetly participated
in the scattering process.  
Or, what was seen as a single high-energy particle may actually have been 
two or more particles that were moving collinearly and sharing the energy.  

Let us, then, correct our idealization of a perfect detector 
and account for these possibilities.
We will work with $\ph^3$ theory, 
initially in $d$ spacetime dimensions.

\begin{figure}
\begin{center}
\epsfig{file=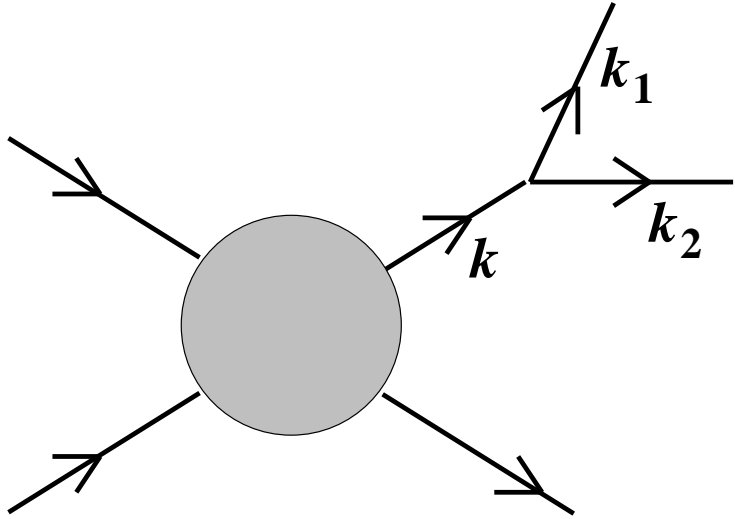}
\end{center}
\caption{An outgoing particle splits into two.  The gray circle
stands for the sum of all diagrams contributing to the original
amplitude $i\T$.}
\label{splitfig}
\end{figure}

Let $\T$ be the amplitude for some scattering process in $\ph^3$ theory.
Now consider the possibility 
that one of the outgoing particles in this process splits into two,
as shown in \fig{splitfig}.  
The amplitude for this new process is given in terms of $\T$ by 
\begin{equation}
\T_{\rm split} = ig\,{-i\over k^2+m^2}\,\T \;,
\label{split}
\end{equation}
where $k=k_1+k_2$, and $k_1$ and $k_2$ are the on-shell four-momenta of the
two particles produced by the split.
(For notational convenience, we drop our usual primes on the
outgoing momenta.)
The key point is this: in the massless limit, it is possible for
$1/(k^2+m^2)$ to diverge.  

To understand the physical consequences of this possibility,
we should compute an appropriate cross-section.  To get the cross section
for the original process (without the split), we multiply $|\T|^2$ by $\dk$
(as well as by similar differentials for
other outgoing particles, and by an overall energy-momentum delta function).
For the process with the split, 
we multiply $|\T_{\rm split}|^2$ by
$\half\dk_1\dk_2$ instead of $\dk$.  
(The factor of one-half is for counting of identical particles.)
If we assume that (due to some imperfection) 
our detector cannot tell whether or
not the one particle actually split into two, 
then we should (according to the usual rules of quantum mechanics) 
add the probabilities for the two events,
which are distinguishable in principle. 
We can therefore define an effectively observable squared-amplitude via
\begin{equation}
|\T|^2_{\rm obs}\,\dk = |\T|^2\,\dk + |\T_{\rm split}|^2\,\half\dk_1\dk_2 
                         +\ldots\;.
\label{t2obs}
\end{equation}
Here the ellipses stand for all other similar processes involving 
emission of one or more extra particles in the final state, 
or absorption of one or more extra particles in the initial state.
We can simplify \eq{t2obs} by including a factor of
\begin{equation}
1=(2\pi)^{d-1}\,2\w\,\delta^{d-1}(\k_1{+}\k_2{-}\k)\,\dk 
\label{oneagain}
\end{equation}
in the second term.
Now all terms in \eq{t2obs} include a factor of $\dk$, so we can drop it.
Then, using \eq{split}, we get
\begin{equation}
|\T|^2_{\rm obs} 
\equiv |\T|^2\left[1 + {g^2\over (k^2+m^2)^2}\,
(2\pi)^{d-1}\,2\w\,\delta^{d-1}(\k_1{+}\k_2{-}\k)
\,\half\dk_1\dk_2 + \ldots\right].
\label{obs}
\end{equation}
Now we come to the point: in the massless limit, the phase space integral
in the second term in \eq{obs} can diverge.  This is because, for $m=0$,
\begin{equation}
k^2=(k_1+k_2)^2 = -4\w_1\w_2\sin^2(\theta/2) \;,
\label{k2}
\end{equation}
where $\theta$ is the angle between the spatial 
momenta $\k_1$ and $\k_2$, and 
$\w_{1,2}=|\k_{1,2}|$.
Also, for $m=0$,
\begin{equation}
\dk_1\dk_2 \sim (\w_1^{d-3}\,d\w_1)\,(\w_2^{d-3}\,d\w_2)\,
                (\sin^{d-3}\theta\,d\theta)\;.
\label{dk1dk2}
\end{equation}
Therefore, for small $\theta$,
\begin{equation}
{\dk_1\dk_2 \over (k^2)^2} \sim 
{d\w_1\over\w_1^{5-d}}\;
{d\w_2\over\w_2^{5-d}}\;
{d\theta\over\theta^{7-d}}\;.
\label{dk1dk2k4}
\end{equation}
Thus the integral over each $\w$ diverges at the low end
for $d\le 4$, and the integral over $\theta$ diverges at the low end
for $d\le 6$.  These divergent integrals would be
cut off (and rendered finite) if we kept the mass $m$ nonzero,
as we will see below.

Our discussion leads us to expect that 
the $m\to0$ divergence in the second term of \eq{obs} should {\it cancel\/} the
$m\to0$ divergence in the loop correction to $|\T|^2$.  We will now
see how this works (or fails to work) in detail for the familiar case
of two-particle scattering in six spacetime dimensions, where $\T$
is given by \eq{tt}.  For $d=6$, there is no problem with 
soft particles (corresponding to the small-$\w$ divergence),
but there is a problem with collinear
particles (corresponding to the small-$\theta$ divergence).

Let us assume that our imperfect detector cannot tell one particle
from two nearly collinear particles if the angle $\theta$ between
their spatial momenta is less than some small angle $\delta$.
Since we ultimately want to take the $m\to0$ limit,
we will evaluate \eq{obs} with $m^2/\k^2 \ll \delta^2 \ll 1$.

We can immediately integrate over $d^5\!k_2$ using the delta function, which
results in setting $\k_2=\k-\k_1$ everywhere.  Let $\beta$ then be the
angle between $\k_1$ (which is still to be integrated over) and $\k$ (which
is fixed).  For two-particle scattering, $|\k|=\half\sqrt s$ in the limit
$m\to 0$.  We then have
\begin{equation}
(2\pi)^5\,2\w\,\delta^5(\k_1{+}\k_2{-}\k)\,\half\dk_1\dk_2 \to
{\Omega_4\over 4(2\pi)^5}\,{\w\over\w_1\w_2}\,
|\k_1\!|^4\,d|\k_1\!|\,\sin^3\!\beta\,d\beta \;,
\label{diff}
\end{equation}
where $\Omega_4=2\pi^2$ is the area of the unit four-sphere.
Now let $\gamma$ be the angle between $\k_2$ and $\k$.     
The geometry of this trio of vectors implies 
$\theta = \beta+\gamma$,
$|\k_1\!| = (\sin\gamma/\!\sin\theta)|\k|$, and
$|\k_2| = (\sin\beta/\!\sin\theta)|\k|$.
All three of the angles are small and positive, and it then is useful to write
$\beta = x\theta$ and $\gamma = (1{-}x)\theta$, with $0\le x\le 1$
and $\theta\le\delta\ll 1$. 

In the low mass limit, we can safely set $m=0$ everywhere in \eq{obs}
except in the propagator, $1/(k^2+m^2)$.  
Then, expanding to
leading order in both $\theta$ and $m$, we find (after some algebra)
\begin{equation}
k^2+m^2 \simeq -x(1{-}x) \k^2 \left[\theta^2 + (m^2/\k^2)f(x)\right],
\label{k2m2}
\end{equation}
where $f(x)=(1{-}x{+}x^2)/(x{-}x^2)^2$.
Everywhere else in \eq{obs}, we can safely set $\w_1=|\k_1\!|=(1{-}x)|\k|$
and $\w_2=|\k_2|=x|\k|$.  Then, changing the integration variables in \eq{diff}
from $|\k_1\!|$ and $\beta$ to $x$ and $\theta$, we get
\begin{equation}
|\T|^2_{\rm obs} = |\T|^2\left[1 +
{g^2\Omega_4\over4(2\pi)^5}\int_0^1x(1{-}x)dx\int_0^\delta
{\theta^3\,d\theta\over[\theta^2+(m^2/\k^2)f(x)]^2}
+ \ldots\right].
\label{obs2}
\end{equation}
Performing the integral over $\theta$ yields
\begin{equation}
\half\ln\Bigl(\delta^2\k^2/m^2\Bigr)-\half\ln f(x)-\half\;.
\label{inttheta}
\end{equation}
Then, performing the integral over $x$ and using
$\Omega_4=2\pi^2$ and $\alpha=g^2/(4\pi)^3$, we get
\begin{equation}
|\T|^2_{\rm obs} = |\T|^2\left[1 +
{\textstyle{1\over12}}\alpha\Bigl(\ln(\delta^2\k^2/m^2)+c\Bigr)
+ \ldots\right],
\label{obs3}
\end{equation}
where $c=(4-3\sqrt3\pi)/36=-0.34$.

The displayed correction term accounts for the possible splitting
of {\it one\/} of the two outgoing particles.  Obviously, there is an identical
correction for the other outgoing particle.  Less obviously (but still true),
there is an identical correction for each of the two incoming particles.
(A glib explanation is that we are computing an effective
amplitude-squared, and this is the same for the reverse
process, with in and outgoing particles switched.  So in and out
particles should be treated symmetrically.)  Then, since we have a total
of four in and out particles (before accounting for any splitting),
\begin{equation}
|\T|^2_{\rm obs} = |\T|^2\left[1 +
{\textstyle{4\over12}}\alpha\Bigl(\ln(\delta^2\k^2/m^2)+c\Bigr)
+ O(\alpha^2) \right].
\label{obs4}
\end{equation}

We have now accounted for the $O(\alpha)$ corrections due to the failure
of our detector to separate 
two particles whose spatial momenta are nearly parallel.
Combining this with \eq{tt}, and recalling that $\k^2={1\over4}s$, we get 
\begin{eqnarray}
|\T|^2_{\rm obs} &=& |\T_0|^2 
\left[1 - {\textstyle{11\over6}}\alpha\Bigl(\ln(s/m^2)+O(m^0)\Bigr)
        + O(\alpha^2) \right]
\nonumber \\
&& \quad\, \times \left[1 + {\textstyle{ 1\over3}}\alpha
                     \Bigl(\ln(\delta^2 s/m^2)+O(m^0)\Bigr) 
        + O(\alpha^2) \right]
\nonumber \\
\noalign{\medskip}
&=& |\T_0|^2 
\Bigl[1 - \alpha \Bigl({\textstyle{3\over2}}\ln(s/m^2) 
                   + {\textstyle{1\over3}}\ln(1/\delta^2)+O(m^0)\Bigr) 
\nonumber \\
&& \qquad\quad     {} + O(\alpha^2) \Bigr].
\label{obs5}
\end{eqnarray}
We now have two kinds of large logs.  One is $\ln(1/\delta^2)$; 
this factor depends on the properties of our detector.  
If we build a very good detector, one 
for which $\alpha\ln(1/\delta^2)$ is not small, then we will have to do more
work, and calculate higher-order corrections to \eq{obs5}.  

The other large log is our original nemesis $\ln(s/m^2)$.  
This factor blows up in the massless limit.  
This means that there is still a mistake hidden somewhere in our analysis.

\vfill\eject

%% file: ch026.tex
\noindent Quantum Field Theory  \hfill   Mark Srednicki

\vskip0.5in

\begin{center}
\large{26: Other Renormalization Schemes}
\end{center}
\begin{center}
Prerequisite: 25
\end{center}

\vskip0.5in

To find the remaining mistake in \eq{obs5}, we must review our renormalization
procedure.  Recall our result from section 14 for 
the one-loop correction to the propagator,
\begin{eqnarray}
\Pi(k^2) &=& -\left[A + {\ts{1\over6}}\alpha\Bigl({\ts{1\over\e}}+\half\Bigr)
                  \right]\!k^2 
            -\left[B + \alpha\Bigl({\ts{1\over\e}}+\half\Bigr)
                  \right]\!m^2 
\nonumber \\
        && {}+\half\alpha\int_0^1 dx\,D\ln(D/\mu^2) + O(\alpha^2)\;,
\label{pi27}
\end{eqnarray}
where $\alpha=g^2/(4\pi)^3$ and $D=x(1{-}x)k^2{+}m^2$.  
The derivative of $\Pi(k^2)$ with respect to $k^2$ is
\begin{eqnarray}
\Pi'(k^2) &=& -\left[A + {\ts{1\over6}}\alpha\Bigl({\ts{1\over\e}}+\half\Bigr)
                  \right]
\nonumber \\
          && {}+\half\alpha\int_0^1 dx\,x(1-x)\Bigl[\ln(D/\mu^2)+1\Bigr]
              + O(\alpha^2)\;.
\label{pip}
\end{eqnarray}
We previously determined $A$ and $B$ via the requirements
$\Pi(-m^2)=0$ and $\Pi'(-m^2)=0$.  The first condition ensures that the exact
propagator $\tbfd(k^2)$ has a pole at $k^2=-m^2$, and the second ensures
that the residue of this pole is one.  Recall that the field must be normalized
in this way for the validity of the LSZ formula.

We now consider the massless limit.  We have $D=x(1{-}x)k^2$, and 
we should apparently try to impose $\Pi(0)=\Pi'(0)=0$.  However,
$\Pi(0)$ is now automatically zero for any values of $A$ and $B$,
while $\Pi'(0)$ is ill defined. 

Physically, the problem is that the one-particle states are no longer
separated from the multiparticle continuum by a finite gap in energy.
Mathematically, the pole in $\tbfd(k^2)$ at $k^2=-m^2$ merges with
the branch point at $k^2=-4m^2$, and is no longer a simple pole.

The only way out of this difficulty is to change the 
{\it renormalization scheme}.
Let us first see what this means in the case $m\ne 0$, where we know what we
are doing.  

Let us try making a different choice of $A$ and $B$.  Specifically, let
\begin{eqnarray}
A &=& -{\ts{1\over6}}\alpha{\ts{1\over\e}} + O(\alpha^2) \;,
\nonumber \\
\noalign{\medskip}
B &=& -\alpha{\ts{1\over\e}} + O(\alpha^2) \;.
\label{abab}
\end{eqnarray}
Here we have chosen $A$ and $B$ to cancel the infinities, and nothing more;
we say that $A$ and $B$ have no {\it finite parts}.
This choice represents a different {\it renormalization scheme}.
Our original choice (which, up until now, we have pretended was inescapable!)
is called the {\it on-shell\/} or OS scheme.  The choice of \eq{abab}
is called the {\it modified minimal-subtraction\/} or $\msb$
(pronounced ``emm-ess-bar'') scheme.
[``Modified'' because we introduced $\mu$ via
$g\to g\mut^{\e/2}$, with
$\mu^2 = 4\pi e^{-\gamma}\mut^2$; 
had we set $\mu=\mut$ instead, the scheme would be just plain
{\it minimal subtraction} or MS.] Now we have
\begin{equation}
\pimsb(k^2) = -{\ts{1\over12}}\alpha(k^2+6m^2)
           +\half\alpha\int_0^1 dx\,D\ln(D/\mu^2) + O(\alpha^2)\;,
\label{pi28}
\end{equation}
as compared to our old result in the on-shell scheme,
\begin{equation}
\Pi_\os(k^2) = -{\ts{1\over12}}\alpha(k^2+m^2)
           +\half\alpha\int_0^1 dx\,D\ln(D/D_0) + O(\alpha^2)\;,
\label{pi29}
\end{equation}
where again $D=x(1{-}x)k^2+m^2$, and $D_0=[{-}x(1{-}x){+}1]m^2$.
Notice that $\pimsb(k^2)$ has a well-defined $m\to0$ limit,
whereas $\pios(k^2)$ does not.  On the other hand,
$\pimsb(k^2)$ depends explicitly on the fake parameter $\mu$,
whereas $\pios(k^2)$ does not. 

What does this all mean?

First, in the $\msb$ scheme, the propagator
$\bfdmsb(k^2)$ will no longer have a pole at $k^2=-m^2$.
The pole will be somewhere else.  However, {\it by definition}, the actual
physical mass $\mph$ of the particle is determined by the location of this pole:
$k^2=-\mph^2$.  Thus, the lagrangian parameter $m$ is no longer the same as
$\mph$.  

Furthermore, the residue of this pole is no longer one.  Let us call
the residue $R$.  The LSZ formula must now be corrected
by multiplying its right-hand side by a factor of $R^{-1/2}$ 
for each external particle (incoming or outgoing).
This is because it is the field $R^{-1/2}\ph(x)$ that now has 
unit amplitude to create a one-particle state.

Note also that, in the LSZ formula, each Klein-Gordon wave operator
should be $-\d^2+\mph^2$, and not $-\d^2+m^2$; also, each external
four-momentum should square to $-\mph^2$, and not $-m^2$.  A review
of the derivation of the LSZ formula clearly shows that each of 
these mass parameters must be the actual particle mass,
and not the parameter in the lagrangian. 

Finally, in the LSZ formula, each external line will contribute
a factor of $R$ when the associated Klein-Gordon wave operator hits
the external propagator and cancels its momentum-space pole, leaving
behind the residue $R$.  Combined with the
correction factor of $R^{-1/2}$ for each field,
we get a net factor of $R^{1/2}$ for each external line when using
the $\overline{\rm MS}$ scheme.  Internal lines each contribute a factor of
$(-i)/(k^2+m^2)$, where $m$ is the lagrangian-parameter mass, and each
vertex contributes a factor of $iZ_g g$, 
where $g$ is the lagrangian-parameter coupling.

Let us now compute the relation between $m$ and $\mph$,
and then compute $R$.  We have 
\begin{equation}
\bfdmsb(k^2)^{-1}=k^2+m^2-\pimsb(k^2) \;,
\label{bfdmsb26}
\end{equation}
and, by definition,
\begin{equation}
\bfdmsb(-\mph^2)^{-1}=0 \;.
\label{bfdmph}
\end{equation}
Setting $k^2=-\mph^2$ in \eq{bfdmsb26}, using \eq{bfdmph}, and rearranging,
we find
\begin{equation}
\mph^2 = m^2 - \pimsb(-\mph^2) \;.
\label{mph}
\end{equation}
Since $\pimsb(k^2)$ is $O(\alpha)$, we see that the difference between
$\mph^2$ and $m^2$ is $O(\alpha)$.  Therefore, on the right-hand side,
we can replace $\mph^2$ with $m^2$, and only make an error of $O(\alpha^2)$.
Thus
\begin{equation}
\mph^2 = m^2 - \pimsb(-m^2) 
+ O(\alpha^2)\;.
\label{mph1}
\end{equation}
Working this out, we get
\begin{equation}
\mph^2 = m^2 - \half\alpha\!\left[{\ts{1\over6}}m^2-m^2
                                +\int_0^1dx\,D_0\ln(D_0/\mu^2)\right]
         + O(\alpha^2)\;,
\label{mph3}
\end{equation}
where $D_0=[1{-}x(1{-}x)]m^2$.  Doing the integrals yields
\begin{equation}
\mph^2 = m^2\left[1 + {\ts{5\over12}}\alpha\Bigl(\ln(\mu^2/m^2) + c'\Bigr)
         + O(\alpha^2)\right].
\label{mph4}
\end{equation}
where $c'=(34-3\pi\sqrt3)/15=1.18$.

Now, physics should be independent of the fake parameter $\mu$.  However,
the right-hand side of \eq{mph4} depends explicitly on $\mu$.  It must, be,
then, that $m$ and $\alpha$ take on different numerical values as $\mu$
is varied, in just the right way 
to leave physical quantities (like $\mph$) unchanged.

We can use this information to 
find differential equations that tell us how $m$ and $\alpha$ change
with $\mu$.  For example, take the logarithm of \eq{mph4}
and divide by two to get
\begin{equation}
\ln\mph = \ln m + {\ts{5\over12}}\alpha\Bigl(\ln(\mu/m) + \half c'\Bigr)
         + O(\alpha^2)\;.
\label{mph5}
\end{equation}
Now differentiate with respect to $\ln\mu$ and require $\mph$ to remain fixed:
\begin{eqnarray}
0 &=& {d\over d\ln\mu}\ln\mph
\nonumber \\
\noalign{\medskip}
  &=& {1\over m}\,{dm\over d\ln\mu} 
+ {\ts{5\over12}}\alpha 
+ O(\alpha^2)\;.
\label{dmdmu}
\end{eqnarray}
To get the second line, we had to assume that
$d\alpha/d\ln\mu=O(\alpha^2)$, which we will verify shortly.  
Then, rearranging \eq{dmdmu} gives
\begin{equation}
{dm\over d\ln\mu} = \left(-{\ts{5\over12}}\alpha + O(\alpha^2)\right)m\;.
\label{mph6}
\end{equation}
The factor in large parentheses on the right is called
the {\it anomalous dimension\/} of the mass parameter,
and it is often given the name $\gamma_m(\alpha)$.

Turning now to the residue $R$, we have
\begin{equation}
R^{-1} = {d\over dk^2}\Bigl[\bfdmsb(k^2)^{-1}\Bigr]
               \Bigg|_{k^2=-\mph^2} \;.
\label{zmsb0}
\end{equation}
Using \eq{bfdmsb26}, we get
\begin{eqnarray}
R^{-1} &=& 1-\pimsb'(-\mph^2)
\nonumber \\
\noalign{\smallskip}
&=& 1-\pimsb'(-m^2) + O(\alpha^2)
\nonumber \\
\noalign{\smallskip}
&=& 1+{\ts{1\over12}}\alpha\Bigl(\ln(\mu^2/m^2)+c''\Bigr)+O(\alpha^2)\;,
\label{zmsb}
\end{eqnarray}
where $c''=(17-3\pi\sqrt3)/3=0.23$.

We can also use $\overline{\rm MS}$ to define the vertex function.  We take
\begin{equation}
C = -\alpha\,{\ts{1\over\e}} + O(\alpha^2) \;,
\label{c26}
\end{equation}
and so
\begin{equation}
\vtmsb(k_1, k_2, k_3) = g\left[1-\half\alpha\int dF_3\,\ln(D/\mu^2)
+ O(\alpha^2)\right]
\label{vtmsb}
\end{equation}
where $D=xyk_1^2+yzk_2^2+zxk_3^2+m^2$.

Let us now compute the $\ph\ph\to\ph\ph$ scattering amplitude in our fancy new
renormalization scheme.  In the low-mass limit, repeating the steps that led
to \eq{tt}, and including the LSZ correction factor $(R^{1/2})^4$, we get
\begin{equation}
\T = R^2\,\T_0
     \left[1 - {\textstyle{11\over12}}\alpha
     \Bigl(\ln(s/\mu^2)+O(m^0)\Bigr)
+ O(\alpha^2)
\right], 
\label{tmsb}
\end{equation}
where $\T_0=-g^2(s^{-1}+t^{-1}+u^{-1})$ is the tree-level result.
Now using $R$ from \eq{zmsb}, we find
\begin{equation}
\T = \T_0
    \left[1 - \alpha\Bigl({\ts{11\over12}}\ln(s/\mu^2)
         +{\ts{ 1\over 6}}\ln(\mu^2/m^2) + O(m^0)\Bigr)
+ O(\alpha^2)
\right]. 
\label{tmsb2}
\end{equation}
To get an observable amplitude-squared with an imperfect detector, 
we must square \eq{tmsb2} and multiply it by the correction factor
we derived in section 25,
\begin{equation}
|\T|^2_{\rm obs} = |\T|^2\left[1 +
{\textstyle{1\over3}}\alpha\Bigl(\ln(\delta^2 s/m^2)+O(m^0)\Bigr)
+ O(\alpha^2)
\right], 
\label{obs4b}
\end{equation}
where $\delta$ is the angular resolution of the detector.
Combining this with \eq{tmsb2}, we get
\begin{equation}
|\T|^2_{\rm obs} = |\T_0|^2
\left[1 - \alpha \Bigl({\textstyle{3\over2}}\ln(s/\mu^2)
                   + {\textstyle{1\over3}}\ln(1/\delta^2)+O(m^0)\Bigr) 
+ O(\alpha^2)
\right].
\label{tmsbobs}
\end{equation}
All factors of $\ln m^2$ have disappeared!  
Finally, we have obtained an expression that has a well-defined $m\to0$ limit.  

Of course, $\mu$ is still a fake parameter, and so $|\T|^2_{\rm obs}$
cannot depend on it.  It must be, then, that the explicit dependence on 
$\mu$ in \eq{tmsbobs} is 
canceled by the implicit $\mu$ dependence of $\alpha$.
We can use this information to figure out how $\alpha$ must vary with $\mu$.  
Noting that $|\T_0|^2 = O(g^4)=O(\alpha^2)$, we have
\begin{equation}
\ln |\T|^2_{\rm obs} = C_1 + 2\ln\alpha + 3\alpha \Bigl(\ln\mu + C_2\Bigr)
                        + O(\alpha^2)\;,
\label{logtsq}
\end{equation}
where $C_1$ and $C_2$ are independent of $\mu$ and $\alpha$
(but depend on the Mandelstam variables).
Differentiating with respect to $\ln\mu$ then gives
\begin{eqnarray}
0 &=& {d\over d\ln\mu}\ln |\T|^2_{\rm obs}
\nonumber \\
\noalign{\medskip}
  &=& {2\over\alpha}\,{d\alpha\over d\ln\mu} + 3\alpha + O(\alpha^2)\;,
\label{dtsqdmu}
\end{eqnarray}
or, after rearranging,
\begin{equation}
{d\alpha\over d\ln\mu} = - {\ts{3\over2}}\alpha^2 +O(\alpha^3)\;.
\label{beta0}
\end{equation}
The right-hand side of this equation is called the {\it beta function}.

Returning to \eq{tmsbobs}, we are free to choose any convenient value
of $\mu$ that we might like.  
To avoid introducing unnecessary large logs, we should choose $\mu^2 \sim s$.  

To compare the results at different values of $s$, we need to solve \eq{beta0}.
Keeping only the leading term in the beta function, the solution is
\begin{equation}
\alpha(\mu_2) = {\alpha(\mu_1)\over 1+{\ts{3\over2}}\alpha(\mu_1)\ln(\mu_2/\mu_1)}\;.
\label{betasol}
\end{equation}
Thus, as $\mu$ increases, $\alpha(\mu)$ decreases.  A theory with this property
is said to be {\it asymptotically free}.  
In this case, the tree-level approximation 
(in the $\msb$ scheme with $\mu^2 \sim s$) becomes better and better 
at higher and higher energies.

Of course, the opposite is true as well: as $\mu$ decreases, $\alpha(\mu)$
increases.  As we go to lower and lower energies, the theory becomes more
and more strongly coupled.  

If the particle mass is nonzero, this process stops at $\mu\sim m$.  This is 
because the minimum value of $s$ is $4m^2$, and so the factor of 
$\ln(s/\mu^2)$ becomes an unwanted large log for $\mu\ll m$.  We should
therefore not use values of $\mu$ below $m$.
Perturbation theory is still good at these low energies if $\alpha(m) \ll 1$.

If the particle mass is zero, $\alpha(\mu)$ continues to increase at lower
and lower energies, and eventually perturbation theory breaks down.
This is a signal that the low-energy physics may be quite different from what
we expect on the basis of a perturbative analysis.

In the case of $\ph^3$ theory, we know what the correct low-energy
physics is: the perturbative ground state is unstable against tunneling
through the potential barrier, and there is no true ground state.
Asymptotic freedom is, in this case, a signal of this impending disaster.  

Much more interesting is asymptotic freedom in a theory that {\it does\/}
have a true ground state, such as quantum chromodynamics.
In this example, the particle excitations
are colorless hadrons, rather than the quarks and gluons
we would expect from examining the lagrangian. 

If the sign of the beta function is positive, then the theory is
{\it infrared free}.  The coupling increases as $\mu$ increases, and,
at sufficiently high energy, perturbation theory breaks down.
On the other hand, the coupling decreases as we go to lower energies.
Once again, though, we should stop this process at $\mu \sim m$ if the
particles have nonzero mass.  Quantum electrodynamics 
with massive electrons (but,
of course, massless photons) is in this category.

Still more complicated behaviors are possible if the beta function has a zero
at a nonzero value of $\alpha$.  We briefly
consider this case in the next section.

\vskip0.5in

\begin{center}
Problems
\end{center}

\vskip0.25in

26.1) In $\ph^4$ theory (see problem 9.2), 
compute the beta function to $O(\lam^2)$, 
the anomalous dimension of $m$ to $O(\lam)$, ,
and the anomalous dimension of $\ph$ to $O(\lam)$.

26.2) Repeat problem 26.1 for the theory of problem 9.3.

\vfill\eject

%% file: ch027.tex
\noindent Quantum Field Theory  \hfill   Mark Srednicki

\vskip0.5in

\begin{center}
\large{27: Formal Development of the Renormalization Group}
\end{center}
\begin{center}
Prerequisite: 26
\end{center}

\vskip0.5in

In section 26 we introduced the $\msb$ renormalization scheme, and
used the fact that physical observables must be independent of the
fake parameter $\mu$ to figure out how the lagrangian parameters
$m$ and $g$ must change with $\mu$.  In this section we re-derive
these results from a much more formal (but calculationally simpler)
point of view, and see how they extend to all orders of perturbation
theory.  Equations that tells us how the lagrangian parameters 
(and other objects that are not directly measurable, like 
correlation functions)
vary with $\mu$ are collectively called the equations of the
{\it renormalization group}.

Let us recall the lagrangian of our theory, 
and write it in two different ways.  In $d=6-\e$ dimensions, we have
\begin{equation}
\L = -\half Z_\ph\d^{\mu\!}\ph\d_\mu\ph -\half Z_m m^2\ph^2
       +{\ts{1\over6}}Z_g g\mut^{\e/2}\ph^3 + Y\ph\
\label{ell}
\end{equation}
and
\begin{equation}
\L = -\half\d^{\mu\!}\ph_0\d_\mu\ph_0 -\half m_0^2\ph_0^2
       +{\ts{1\over6}}g_0\ph_0^3 + Y_0\ph_0 \;.
\label{ell0}
\end{equation}
The fields and parameters in \eq{ell} are the {\it renormalized\/}
fields and parameters.  (And in particular, they are renormalized using
the $\msb$ scheme, with $\mu^2=4\pi e^{-\gamma}\mut^2$.)  
The fields and parameters in \eq{ell0} are the {\it bare\/}
fields and parameters.  Comparing \eqs{ell} and (\ref{ell0}) gives us
the relationships between them:
\begin{eqnarray}
\ph_0(x) &=& Z_\ph^{1/2}\ph(x) \;,
\label{phi0} \\
m_0 &=& Z_\ph^{-1/2}Z_m^{1/2}m \;,
\label{m0} \\
g_0 &=& Z_\ph^{-3/2}Z_g g\mut^{\e/2} \;,
\label{g0} \\
Y_0 &=& Z_\ph^{-1/2}Y \;.
\label{y0}
\end{eqnarray}

Recall that, after using dimensional regularization, the infinities coming
from loop integrals take the form of inverse powers of $\e=6-d$.  
In the $\msb$ renormalization scheme, we choose the $Z$'s to cancel off these
powers of $1/\e$, and nothing more.   Therefore the $Z$'s can be written as
\begin{eqnarray}
Z_\ph &=& 1 + \sum_{n=1}^\infty {a_n(\alpha)\over\e^n} \;,
\label{zzz} \\
Z_m &=& 1 + \sum_{n=1}^\infty {b_n(\alpha)\over\e^n} \;,
\label{zm} \\
Z_g &=& 1 + \sum_{n=1}^\infty {c_n(\alpha)\over\e^n} \;,
\label{zg3}
\end{eqnarray}
where $\alpha=g^2/(4\pi)^3$.
Computing $\Pi(k^2)$ and $\V_3(k_1,k_2,k_3)$ in perturbation theory
in the $\msb$ scheme
gives us Taylor series in $\alpha$ for $a_n(\alpha)$, $b_n(\alpha)$, 
and $c_n(\alpha)$.  So far we have found
\begin{eqnarray}
a_1(\alpha) &=& -{\ts{1\over6}}\alpha  + O(\alpha^2) \;,
\label{a1} \\
\noalign{\smallskip}
b_1(\alpha) &=& -\alpha  + O(\alpha^2) \;,
\label{b1} \\
\noalign{\smallskip}
c_1(\alpha) &=& -\alpha  + O(\alpha^2) \;,
\label{c1}
\end{eqnarray}
and that $a_n(\alpha)$, $b_n(\alpha)$, and $c_n(\alpha)$ are all at least
$O(\alpha^2)$ for $n\ge2$.

Next we turn to the trick that we will employ to compute 
the beta function for $\alpha$, the anomalous dimension of $m$,
and other useful things.  This is the trick: 
{\it bare fields and parameters must be independent of $\mu$.}

Why is this so?  Recall that
we introduced $\mu$ when we found that we had to regularize the theory to
avoid infinities in the loop integrals of Feynman diagrams.  We argued
at the time (and ever since) that physical quantities had to be independent 
of $\mu$.  Thus $\mu$ is not really a parameter of the theory, but just
a crutch that we had to introduce at an intermediate stage of the calculation.
In principle, the theory is completely specified by the values of 
the bare parameters, and, if we were smart enough,
we would be able to compute the exact scattering amplitudes in terms of them,
without ever introducing $\mu$.
Since the exact scattering amplitudes are independent of $\mu$,
the bare parameters must be as well.

Let us start with $g_0$.  It is convenient to define
\begin{equation}
\alpha_0 \equiv {g_0^2/(4\pi)^3} = Z_g^2 Z_\ph^{-3}\mut^\e\alpha \;,
\label{a0}
\end{equation}
and also
\begin{equation}
G(\alpha,\e) \equiv \ln\Bigl(Z_g^2  Z_\ph^{-3}\Bigr) \;.
\label{bigg}
\end{equation}
From the general structure of \eqs{zzz} and (\ref{zg3}), we have
\begin{equation}
G(\alpha,\e) = \sum_{n=1}^\infty {G_n(\alpha)\over \e^n} \;,
\label{biggs}
\end{equation}
where, in particular, 
\begin{eqnarray}
G_1(\alpha) &=& 2c_1(\alpha) - 3 a_1(\alpha)
\nonumber \\
\noalign{\medskip}
&=& -{\ts{3\over2}}\alpha + O(\alpha^2) \;.
\label{bigg1}
\end{eqnarray}
The logarithm of \eq{a0} can now be written as
\begin{equation}
\ln \alpha_0 = G(\alpha,\e) +\ln\alpha + \e\ln\mut \;.
\label{logg0}
\end{equation}

Next, differentiate \eq{logg0} with respect to $\ln\mu$, 
and require $\alpha_0$ to be independent of it:
\begin{eqnarray}
0 &=& {d\over d\ln\mu}\ln\alpha_0
\nonumber \\
\noalign{\medskip}
&=& {\d G(\alpha,\e)\over\d\alpha}\,{d\alpha\over d\ln\mu}
    +{1\over\alpha}\,{d\alpha\over d\ln\mu} + \e \;.
\label{zero}
\end{eqnarray}
Now regroup the terms, multiply by $\alpha$, and use \eq{biggs} to get
\begin{equation}
0 = \left(1+{\alpha G_1'(\alpha)\over\e}+{\alpha G_2'(\alpha)\over\e^2}+\ldots
\right) {d\alpha\over d\ln\mu} + \e\alpha \;.
\label{dadlmu2}
\end{equation}

Next we use some physical reasoning:
$d\alpha/d\ln\mu$ is the rate at which $\alpha$ must change to compensate
for a small change in $\ln\mu$.  If compensation is possible at all,
this rate should be finite in the $\e\to0$ limit.  
Therefore, in a renormalizable theory, we should have
\begin{equation}
{d\alpha\over d\ln\mu} = -\e\alpha + \beta(\alpha) \;.
\label{dadlmu3}
\end{equation}
The first term, $-\e\alpha$, is fixed by matching the $O(\e)$
terms in \eq{dadlmu2}.  The second term, the beta function $\beta(\alpha)$, 
is similarly determined by matching the $O(\e^0)$ terms; the result is
\begin{equation}
\beta(\alpha) = \alpha^2 G'_1(\alpha)\;.
\label{betaa}
\end{equation}
Terms that are higher-order in $1/\e$ must also cancel, and this
determines all the other $G'_n(\alpha)$'s in terms of $G'_1(\alpha)$.
Thus, for example, cancellation of the $O(\e^{-1})$ terms fixes
$G'_2(\alpha)=\alpha G'_1(\alpha)^2$.  These relations among the 
$G'_n(\alpha)$'s can of course be checked order by order in 
perturbation theory.

From \eq{betaa} and \eq{bigg1}, we find that the beta function is
\begin{equation}
\beta(\alpha) = -{\ts{3\over2}}\alpha^2 + O(\alpha^3)\;.
\label{beta2}
\end{equation}
Hearteningly, 
this is the same result we found in section 26 by requiring the observed
scattering cross section $|\T|^2_{\rm obs}$ to be independent of $\mu$.
However, simply as a matter of practical calculation, it is much easier
to compute $G_1(\alpha)$ than it is to compute $|\T|^2_{\rm obs}$.

Next consider the invariance of $m_0$.  We begin by defining
\begin{eqnarray}
M(\alpha,\e) &\equiv& \ln\Bigl(Z_m^{1/2}Z_\ph^{-1/2}\Bigr)
\nonumber \\
\noalign{\medskip}
&=& \sum_{n=1}^\infty {M_n(\alpha)\over \e^n} \;.
\label{bigm}
\end{eqnarray}
From \eqs{a1} and (\ref{c1}) we have
\begin{eqnarray}
M_1(\alpha) &=& \half b_1(\alpha) - \half a_1(\alpha)
\nonumber \\
\noalign{\medskip}
&=& -{\ts{5\over12}}\alpha + O(\alpha^2) \;.
\label{bigm1}
\end{eqnarray}
Then, from \eq{m0}, we have
\begin{equation}
\ln m_0 = M(\alpha,\e) + \ln m \;.
\label{logm0}
\end{equation}
Take the derivative with respect to $\ln\mu$ and require $m_0$ to be unchanged:
\begin{eqnarray}
0 &=& {d\over d\ln\mu}\ln m_0
\nonumber \\
\noalign{\medskip}
&=& {\d M(\alpha,\e)\over\d\alpha}\,{d\alpha\over d\ln\mu}
    +{1\over m}\,{dm\over d\ln\mu} \;.
\nonumber \\
\noalign{\medskip}
&=& {\d M(\alpha,\e)\over\d\alpha}\,\Bigl(-\e\alpha + \beta(\alpha)\Bigr)
    +{1\over m}\,{dm\over d\ln\mu} \;.
\label{zerom}
\end{eqnarray}
Rearranging, we find
\begin{eqnarray}
{1\over m}\,{dm\over d\ln\mu} &=&
\Bigl(\e\alpha - \beta(\alpha)\Bigr)\sum_{n=1}^\infty{M'_n(\alpha)\over\e^n}
\nonumber \\
\noalign{\medskip}
&=& \alpha M'_1(\alpha) + \ldots \;,
\label{dmdlmu}
\end{eqnarray}
where the ellipses stand for terms with powers of $1/\e$.
In a renormalizable theory, $dm/d\ln\mu$ should be finite
in the $\e\to0$ limit, and so these terms must actually all be zero.
Therefore, the anomalous dimension of the mass, defined via
\begin{equation}
\gamma_m(\alpha) \equiv {1\over m}\,{dm\over d\ln\mu}  \;,
\label{gma}
\end{equation}
is given by
\begin{eqnarray}
\gamma_m(\alpha) &=& \alpha M'_1(\alpha)
\nonumber \\
\noalign{\medskip}
&=& -{\ts{5\over12}}\alpha + O(\alpha^2) \;.
\label{dmdlmu2}
\end{eqnarray}
Comfortingly, this is just what we found in section 26.

Let us now consider the propagator in the $\msb$ renormalization scheme,
\begin{equation}
\tbfd(k^2) = i\int d^{6\!}x\,e^{ikx}\langle 0|{\rm T}\ph(x)\ph(0)|0\rangle \;.
\label{propmsb}
\end{equation}
The bare propagator,
\begin{equation}
\tbfd_0(k^2) = 
i\int d^{6\!}x\,e^{ikx}\langle 0|{\rm T}\ph_0(x)\ph_0(0)|0\rangle \;,
\label{prop0}
\end{equation}
should be (by the now-familiar argument) independent of $\mu$.
The bare and renormalized propagators are related by
\begin{equation}
\tbfd_0(k^2) = Z_\ph\tbfd(k^2) \;. 
\label{2ps}
\end{equation}
Taking the logarithm and differentiating with respect to $\ln\mu$, we get
\begin{eqnarray}
0 &=& {d\over d\ln\mu}\ln\tbfd_0(k^2)
\nonumber \\
\noalign{\medskip}
&=& {d\ln Z_\ph\over d\ln\mu} + {d\over d\ln\mu}\ln\tbfd(k^2)
\nonumber \\
\noalign{\medskip}
&=& {d\ln Z_\ph\over d\ln\mu} + {1\over \tbfd(k^2)}\left({\d\over\d\ln\mu} +
               {d\alpha\over d\ln\mu}\,{\d\over\d\alpha} +
               {dm\over d\ln\mu}\,{\d\over\d m}\right)\tbfd(k^2).  \qquad
\label{dpdlmu}
\end{eqnarray}
We can write
\begin{equation}
\ln Z_\ph 
= {a_1(\alpha)\over \e}+{a_2(\alpha)-\half a^2_1(\alpha)\over\e^2}+\ldots \;.
\label{logz}
\end{equation}
Then we have
\begin{eqnarray}
{d\ln Z_\ph\over d\ln\mu} &=&
{\d\ln Z_\ph\over\d\alpha}\,{d\alpha\over d\ln\mu} 
\nonumber \\
\noalign{\medskip}
&=& \left({a'_1(\alpha)\over \e}
+\ldots\right)\Bigl(-\e\alpha + \beta(\alpha)\Bigr)
\nonumber \\
\noalign{\medskip}
&=& -\alpha a'_1(\alpha) + \ldots \;,
\label{dlogz}
\end{eqnarray}
where the ellipses in the last line stand for terms with powers of $1/\e$.  
Since $\tbfd(k^2)$ should vary smoothly with $\mu$ in the $\e\to0$ limit, 
these must all be zero.  We then define
the {\it anomalous dimension of the field}
\begin{equation}
\gamma_\ph(\alpha) \equiv {1\over2}\, {d \ln Z_\ph \over d\ln\mu} \;.
\label{gamphi}
\end{equation}
From \eq{dlogz} we find
\begin{eqnarray}
\gamma_\ph(\alpha) &=& -\half\alpha a'_1(\alpha) 
\nonumber \\
\noalign{\medskip}
&=& +{\ts{1\over12}}\alpha + O(\alpha^2) \;.
\label{gphi}
\end{eqnarray}
\Eq{dpdlmu} can now be written as
\begin{equation}
\left({\d\over\d\ln\mu} + \beta(\alpha){\d\over\d\alpha} + 
\gamma_m(\alpha)m{\d\over\d m} +2\gamma_\ph(\alpha)\right)\tbfd(k^2)=0
\label{cs}
\end{equation} 
in the $\e\to0$ limit.  
This is the {\it Callan-Symanzik equation} for the propagator.

The Callan-Symanzik equation is most interesting in the massless limit,
and for a theory with a zero of the beta function at a 
nonzero value of $\alpha$.
So, let us suppose that $\beta(\alpha_*)=0$ for some $\alpha_*\ne0$.  Then,
for $\alpha=\alpha_*$ and $m=0$, the Callan-Symanzik equation becomes
\begin{equation}
\left({\d\over\d\ln\mu} + 2\gamma_\ph(\alpha_*)\right)\tbfd(k^2)=0\;.
\label{cs2}
\end{equation}
The solution is
\begin{equation}
\tbfd(k^2)={C(\alpha_*)\over k^2}
\left(\mu^2\over k^2\right)^{-\gamma_\ph(\alpha_*)} ,
\label{cssol}
\end{equation}
where $C(\alpha_*)$ is an integration constant.  
(We used the fact that $\tbfd(k^2)$ has mass dimension $-2$ to get the $k^2$
dependence in addition to the $\mu$ dependence.)  Thus the naive scaling law 
$\tbfd(k^2)\sim k^{-2}$ 
is changed to $\tbfd(k^2)\sim k^{-2[1-\gamma_\ph(\alpha_*)]}$.
This has applications in the theory of critical phenomena, 
which is beyond the scope of this book.

\vfill\eject

%% file: ch028.tex
\noindent Quantum Field Theory  \hfill   Mark Srednicki

\vskip0.5in

\begin{center}
\large{28: Nonrenormalizable Theories and Effective Field Theory}
\end{center}
\begin{center}
Prerequisite: 27
\end{center}

\vskip0.5in

So far we have been discussing only renormalizable theories.
In this section, we investigate what meaning can be assigned to 
nonrenormalizable theories, following an approach pioneered by 
Ken Wilson.

We will begin by analyzing a renormalizable theory from
a new point of view.   Consider, as an example,
$\ph^4$ theory in four spacetime dimensions:
\begin{equation}
\L = -\half Z_\ph\d^\mu\ph\d_\mu\ph
     -\half Z_m \mph^2\ph^2
     -{\ts{1\over24}}Z_\lam\lamph \ph^4 \;.
\label{lamph4}
\end{equation}
(This example is actually problematic, because $\ph^4$ theory is
{\it trivial}, a technical term we will exlain later.  For now
we proceed with a perturbative analysis.)
We take the renormalizing $Z$ factors to be defined in an on-shell scheme,
and have emphasized this by writing the particle mass as $\mph$
and the coupling constant as $\lamph$.  We define $\lamph$ as the value
of the exact 1PI four-point vertex with zero external four-momenta: 
\begin{equation}
\lamph \equiv \V_4(0,0,0,0)\;.
\label{lamph40000}
\end{equation}
The path integral is given by
\begin{equation}
Z(J) = \int\D\ph\;e^{iS+i\!\int\!J\ph}\;,
\label{zj0}
\end{equation}
where $S=\int\dfx\,\L$ and $\int\!J\ph$ is short for $\int\dfx\,J\ph$.

Our first step in analyzing this theory will be to perform
the Wick rotation (applied to loop integrals in section 14)
directly on the action.  We define a euclidean time
$\tau \equiv it$.  Then we have
\begin{equation}
Z(J) = \int\D\ph\;e^{-S_{\rm E} - \int\!J\ph}\;,
\label{zj02}
\end{equation}
where $S_{\rm E} = \int\dfx\,\L_{\rm E}$,
$\dfx=\dtx\,d\tau$, 
\begin{equation}
\L_{\rm E}  =   \half Z_\ph\d_\mu\ph\d_\mu\ph
     +\half Z_m \mph^2\ph^2
     +{\ts{1\over24}}Z_\lam\lamph \ph^4 \;,
\label{lamph42}
\end{equation}
and 
\begin{equation}
\d_\mu\ph\d_\mu\ph = (\d\ph/\d\tau)^2 + (\nabla\ph)^2 \;.
\label{dphdph}
\end{equation}
Note that each term in $S_{\rm E}$ is always positive (or zero)
for any field configuration $\ph(x)$.  This is the advantage
of working in euclidean space:
\eq{zj02}, the {\it euclidean path integral}, is strongly damped
(rather than rapidly oscillating) at large values of the field
and/or its derivatives, and this makes its convergence properties
more obvious.

Next, we Fourier transform to (euclidean) momentum space via
\begin{equation}
\ph(x) = \int{\dfk\over(2\pi)^4}\,e^{ikx}\,{\widetilde\ph}(k)\;.
\label{ft33}
\end{equation}
The euclidean action becomes
\begin{eqnarray}
S_{\rm E} &=&
{1\over2}\int{\dfk\over(2\pi)^4}\,\tph(-k)\Bigl(Z_\ph k^2+Z_m \mph^2\Bigr)\tph(k)
\nonumber \\
&& {} +{1\over24}Z_\lam\lamph \int
{\dfk_1\over(2\pi)^4}\ldots{\dfk_4\over(2\pi)^4}\,
(2\pi)^4\delta^4(k_1{+}k_2{+}k_3{+}k_4)
\nonumber \\
\noalign{\smallskip}
&& \qquad\qquad\qquad
\times \tph(k_1) \tph(k_2) \tph(k_3) \tph(k_4) \;.
\label{seuc}
\end{eqnarray}
Note that $k^2=\k^2+k_\tau^2\ge0$.  

Now we define a particular energy scale $\Lambda$ as the 
{\it ultraviolet cutoff}.  It should be much larger than the particle mass 
$\mph$, or any other energy scale of practical interest.  Then we
perform the path integral over all ${\widetilde\ph}(k)$ with
$|k|>\Lam$.  We also take 
${\widetilde J}(k)=0$ for $|k|>\Lam$.  Then we find
\begin{equation}
Z(J) = \int\D\ph_{|k|<\Lam}\;e^{-S_{\rm eff}(\ph;\Lam) - \int\!J\ph}\;,
\label{zj03}
\end{equation}
where
\begin{equation}
e^{-S_{\rm eff}(\ph;\Lam)}
= \int\D\ph_{|k|>\Lam}\;e^{-S_{\rm E}(\ph)} \;.
\label{seff}
\end{equation}
$S_{\rm eff}(\ph;\Lam)$ is called the {\it Wilsonian effective action}.
We can write the corresponding lagrangian density as
\begin{eqnarray}
\L_{\rm eff}(\ph;\Lam)  &=& \half Z(\Lam)\d_\mu\ph\d_\mu\ph +\half m^2(\Lam)\ph^2
                             +{\ts{1\over24}}\lam(\Lam)\ph^4 
\nonumber \\
\noalign{\smallskip}
&& {} +\sum_i c_{n,d,i}(\Lam)\O_{n,d,i}\;,
\label{lamph43}
\end{eqnarray}
where the Fourier components of $\ph(x)$ are now cut off at $|k|>\Lam$:
\begin{equation}
\ph(x) = \int_0^\Lam{\dfk\over(2\pi)^4}\,e^{ikx}\,{\widetilde\ph}(k)\;.
\label{ft4}
\end{equation}
The operators $\O_{n,d,i}$ in \eq{lamph43} consist of all terms 
(other than those already written explicitly on the first line)
that we can construct out of an even number $2n$ of $\ph$ fields 
(the number must be even in order to respect 
the $\ph\leftrightarrow -\ph$ symmetry of the original lagrangian)
with an even number $2d$ of derivatives acting on them
(with all the vector indices contracted in pairs).
The index $i$ keeps track of operators with the same values of $n$
and $d$ that are not equivalent after integrations by parts.
The operator $\O_{n,d,i}$ has mass dimension $D=2n+2d\ge 6$.

The coefficients $Z(\Lam)$, $m^2(\Lam)$, $\lam(\Lam)$, and $c_{n,d,i}(\Lam)$
in \eq{lamph43} are all {\it finite} functions of $\Lam$.
This is established by the following argument.
We can differentiate \eq{zj03} with respect to
$J(x)$ to compute correlation functions of the renormalized field
$\ph(x)$, and correlation functions of renormalized fields
are finite.  Using \eq{zj03}, 
we can compute these correlation functions as a series of Feynman diagrams,
with Feynman rules based on $\L_{\rm eff}$.  
These rules include an ultraviolet cutoff $\Lam$ on the loop momenta, 
since the fields with higher momenta have already been {\it integrated out}.
Thus all of the loop integrals in these diagrams are finite.  Therefore
the other parameters that enter the 
diagrams---$Z(\Lam)$, $m^2(\Lam)$, $\lam(\Lam)$, and $c_{n,d,i}(\Lam)$---must 
be finite as well, in order to end up with finite correlation functions.

\begin{figure}
\begin{center}
\epsfig{file=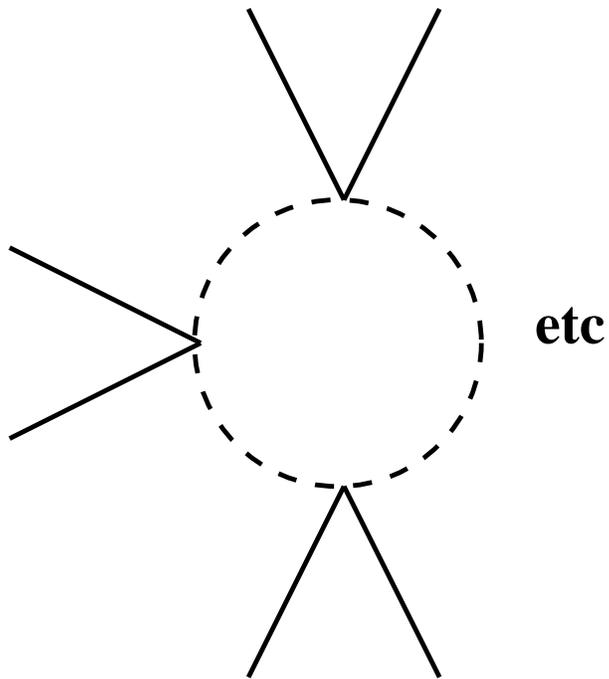}
\end{center}
\caption{A one-loop 1PI diagram with $2n$ external lines.  Each external line
represents a field with $|k|<\Lam$.  The internal (dashed) line represents
a field with $|k|>\Lam$.}
\label{wilson}
\end{figure}

To compute these parameters, 
we can think of \eq{seuc} as the action for two kinds of
fields, those with $|k|<\Lam$ and those with $|k|>\Lam$.  Then we draw
all 1PI diagrams with external lines for $|k|<\Lam$ fields only.
The dominant contribution to $c_{n,d,i}(\Lam)$ 
is then given by a one-loop diagram with $2n$ external lines
(representing $|k|<\Lam$ fields), $n$ vertices, and a $|k|>\Lam$
field circulating in the loop; see \fig{wilson}.  

With $2n$ external lines, there are $(2n)!$ ways of assigning the
external momenta to the lines, but $2^n\times n\times 2$ of these give
the same diagram:  $2^n$ for exchanging the
two external lines that meet at any one vertex; 
$n$ for rotations of the diagram; 
and $2$ for reflection of the diagram.

The simplest case to consider is $\O_{n,0}\equiv\ph^{2n}$.
Since there are no derivatives on the external fields, 
we can set all of the external momenta to zero; then all
$(2n)!/(2^n2n)$ diagrams have the same value.  
With a euclidean action, each internal line contributes a factor of
$1/(k^2+\mph^2)$, and each vertex contributes a factor of 
$-Z_\lam\lamph=-\lamph+O(\lamph^2)$.
The vertex factor associated with the term $c_{n,0}(\Lam)\ph^{2n}$ 
in $\L_{\rm eff}$ is $-(2n)!\,c_{n,0}(\Lam)$.  Thus we have
\begin{eqnarray}
-(2n)!\,c_{n,0}(\Lam) &=& {(-\lamph)^n(2n)!\over2^n 2n}\int_\Lam^\infty
                {\dfk\over(2\pi)^4}\left({1\over k^2+\mph^2}\right)^{\!\!n}
 \nonumber \\
&& \quad {}    +O(\lamph^{n+1})\;.
\label{cn0}
\end{eqnarray}
For $2n\ge6$, the integral converges, and we find
\begin{equation}
c_{n,0}(\Lam) = -{(-\lamph/2)^n\over 32\pi^2 n(n{-}2)}\,{1\over\Lam^{2n-4}}
 +O(\lamph^{n+1})\;.
\label{cn02}
\end{equation}
We have taken $\Lam\gg\mph$, and dropped terms down by powers of 
$\mph/\Lam$.

For $2n=4$, we have to include the tree-level vertex; in this case, we have
\begin{eqnarray}
-\lam(\Lam) &=& -Z_\lam\lamph + {{3\over2}}(-\lamph)^2\int_\Lam^\infty
                {\dfk\over(2\pi)^4}\left({1\over k^2+\mph^2}\right)^{\!\!2}   
\nonumber \\
&& \quad {}  +O(\lamph^{3})\;.
\label{lamLam}
\end{eqnarray}
This integral diverges.  To evaluate it, recall that 
the one-loop contribution to the exact four-point vertex
is given by the {\it same\/} diagram, 
but with fields of {\it all\/} momenta circulating in the loop.  
Thus we have
\begin{eqnarray}
-\V_4(0,0,0,0) &=& -Z_\lam\lamph + {{3\over2}}(-\lamph)^2\int_0^\infty
                {\dfk\over(2\pi)^4}\left({1\over k^2+\mph^2}\right)^{\!\!2}
\nonumber \\
&& \quad {}  +O(\lamph^{3})\;.
\label{v40000}
\end{eqnarray}
Then, using $\V_4(0,0,0,0)=\lamph$ and subtracting \eq{lamLam} from \eq{v40000},
we get
\begin{equation}
- \lamph +\lam(\Lam)= {{3\over2}}(-\lamph)^2\int_0^\Lam
                {\dfk\over(2\pi)^4}\left({1\over k^2+\mph^2}\right)^{\!\!2}
                +O(\lamph^{3})\;.
\label{lamLam2}
\end{equation}
Evaluating the (now finite!) integral and rearranging, we have
\begin{equation}
\lam(\Lam) = \lamph + {{3\over16\pi^2}}\lamph^2 \Bigl[\ln(\Lam/\mph)-\half\Bigr]
             +O(\lamph^3)\;.
\label{lamLam3}
\end{equation}
Note that this result has the problem of a {\it large log\/}; the second term
is smaller than the first only if $\lamph\ln(\Lam/\mph)\ll 1$.  To cure this
problem, we must change the renormalization scheme.  We will take up this issue
shortly, but first let us examine the case of two external lines while continuing
to use the on-shell scheme.

For the case of two external lines, the one-loop diagram has just one
vertex, and by momentum conservation, the loop integral is
completely indepedent of the external momentum.
This implies that the one-loop contribution to $Z(\Lam)$ vanishes, and so
we have
\begin{equation}
Z(\Lam) = 1+O(\lamph^2)\;.
\label{ZLam}
\end{equation}
The one-loop diagram does, however,
give a nonzero contribution to $m^2(\Lam)$; 
after including the tree-level term, we find
\begin{equation}
-m^2(\Lam) = -Z_m \mph^2
+{1\over2}(-\lamph)\int_\Lam^\infty{\dfk\over(2\pi)^4}\,
                                 {1\over k^2+\mph^2}+O(\lamph^2)\;.
\label{m2Lam}
\end{equation}
This integral diverges.  To evaluate it, recall that 
the one-loop contribution to the exact particle mass-squared
is given by the {\it same\/} diagram, 
but with fields of {\it all\/} momenta circulating in the loop.  
Thus we have
\begin{equation}
-\mph^2 = -Z_m\mph^2 + {{1\over2}}(-\lamph)\int_0^\infty
                {\dfk\over(2\pi)^4}\,{1\over k^2+\mph^2} +O(\lamph^2)\;.
\label{mph2}
\end{equation}
Then, subtracting \eq{m2Lam} from \eq{mph2}, we get
\begin{equation}
-\mph^2 +m^2(\Lam)= {{1\over2}}(-\lamph)\int_0^\Lam
                {\dfk\over(2\pi)^4}\,{1\over k^2+\mph^2}
                +O(\lamph^{2})\;.
\label{m2Lam2}
\end{equation}
Evaluating the (now finite!) integral and rearranging, we have
\begin{equation}
m^2(\Lam) = \mph^2-{\lamph\over 16\pi^2}
             \Bigl[\Lam^2-\mph^2\ln(\Lam^2/\mph^2)\Bigr] +O(\lamph^2)\;.
\label{m2Lam3}
\end{equation}
We see that we now have an even worse situation than we did with the large log 
in $\lam(\Lam)$: the correction term is {\it quadratically divergent}.

As already noted, to fix these problems we must change the renormalization scheme.
In the context of an effective action with a specific value of the cutoff $\Lam_0$, 
there is a simple way to do so:  we simply treat $Z(\Lam_0)$, $m^2(\Lam_0)$
$\lam(\Lam_0)$, and $c_{n,d,i}(\Lam_0)$ as the input parameters, and then see
what physics emerges at energy scales well below $\Lam_0$.
We can set $Z(\Lam_0)=1$, with the understanding that the field no longer
has the LSZ normalization (and that we will have to correct the LSZ formula
to account for this).  We will also assume that the parameters $\lam(\Lam_0)$, 
$m^2(\Lam_0)$, and $c_{n,d,i}(\Lam_0)$ are all small when measured in units
of the cutoff:
\begin{eqnarray}
\lam(\Lam_0) &\ll& 1 \;,
\label{lamLam0} \\
\noalign{\medskip}
|m^2(\Lam_0)| &\ll& \Lam^2_0 \;,
\label{m2Lam0} \\
\noalign{\medskip}
c_{n,d,i}(\Lam_0) &\ll& {1\over \Lam^{2n+2d-4}_0} \;.
\label{cndiLam0}
\end{eqnarray}

The proposal to treat the effective action as the fundamental starting point
may not seem very appealing.  For one thing, we now have an infinite number
of parameters to specify, rather than two!  Also, we now have an explicit cutoff
in place, rather than trying to have a theory that works at all energy scales.  

On the other hand, it may well be that quantum field theory does not work at
arbitrarily high energies.  For example, quantum fluctuations in spacetime
itself should become important above the {\it Planck scale}, which is given
by the inverse square root of Newton's constant, and has the numerical value
of $10^{19}\,$GeV (compared to, say, the proton mass, which is 1$\,$GeV).

So, let us leave the cutoff $\Lam_0$ in place for now.  We will then make
a two-pronged analysis.  First, we will see what happens at much lower energies.
Then, we will see what happens if we try to take the limit $\Lam_0\to\infty$.

We begin by examining lower energies.  To make things more tractable, 
we will set
\begin{equation}
c_{n,d,i}(\Lam_0) =0 \;;
\label{cndi00}
\end{equation}
later we will examine the effects of a more general choice.

A nice way to see what happens at lower energies is to integrate out
some more high-energy degrees of freedom.  
Let us, then, perform the functional integral
over Fourier modes $\widetilde\ph(k)$ with $\Lam <|k|<\Lam_0$; we have
\begin{equation}
e^{-S_{\rm eff}(\ph;\Lam)}
= \int\D\ph_{\Lam<|k|<\Lam_0}\;e^{-S_{\rm eff}(\ph;\Lam_0)} \;.
\label{seff2}
\end{equation}
We can do this calculation in perturbation theory, mimicking
the procedure that we used earlier.  We find
\begin{eqnarray}
m^2(\Lam) &=& m^2(\Lam_0)  
+{1\over2}\lam(\Lam_0)\int_\Lam^{\Lam_0}{\dfk\over(2\pi)^4}\,
                                 {1\over k^2+m^2(\Lam_0)} + \ldots \;,
\label{m2Lam4} \\
\noalign{\medskip}
\lam(\Lam) &=& \lam(\Lam_0)-{{3\over2}}\lam^2(\Lam_0)\int_\Lam^{\Lam_0}
                {\dfk\over(2\pi)^4}\left({1\over k^2+m^2(\Lam_0)}\right)^{\!\!2}   
+ \ldots \;,
\label{lamLam4} \\
\noalign{\medskip}
c_{n,0}(\Lam) &=& 
                -{(-1)^n\over2^n 2n} \lam^n(\Lam_0)\int_\Lam^{\Lam_0}
                {\dfk\over(2\pi)^4}\left({1\over k^2+m^2(\Lam_0)}\right)^{\!\!n}
               + \ldots \;,
\label{cnoLam4}
\end{eqnarray}
where the ellipses stand for higher-order corrections.  
For $\Lam$ not too much less than $\Lam_0$
(and, in particular, for $|m^2(\Lam_0)|\ll\Lam^2$), 
we find
\begin{eqnarray}
m^2(\Lam) &=& m^2(\Lam_0)  
+{1\over16\pi^2}\lam(\Lam_0)\Bigl(\Lam_0^2-\Lam^2\Bigr)+ \ldots \;,
\label{m2Lam5} \\
\noalign{\medskip}
\lam(\Lam) &=& \lam(\Lam_0)-{{3\over16\pi^2}}\lam^2(\Lam_0)\ln{\Lam_0\over\Lam}   + \ldots \;,
\label{lamLam5} \\
\noalign{\medskip}
c_{n,0}(\Lam) &=& 
                -{(-1)^n\over32\pi^2 2^n n(n{-}2)} \lam^n(\Lam_0)
                \left({1\over\Lam^{2n-4}}-{1\over\Lam_0^{2n-4}}\right)
                               + \ldots \;.
\label{cn0Lam5}
\end{eqnarray}
We see from this that the corrections to $m^2(\Lam)$, which is
the only coefficient with positive mass dimension, are dominated by contributions from the {\it high\/} end of the integral.  On the other hand, the corrections to $c_{n,d,i}(\Lam)$, coefficients with negative mass dimension, 
are dominated by contributions from the {\it low\/} end of the integral.  
And the corrections to $\lam(\Lam)$, which is dimensionless, come
equally from all portions of the range of integration.

For the $c_{n,d,i}(\Lam)$, this means that their starting values at $\Lam_0$
were not very important, as long as \eq{cndiLam0} is obeyed.  Nonzero
starting values would contribute another term of order $1/\Lam_0^{2n-4}$
to $c_{n,0}(\Lam)$, but all such terms are less important than the one
of order $1/\Lam^{2n-4}$ that comes from doing the integral down to 
$|k|=\Lam$.

\begin{figure}
\begin{center}
\epsfig{file=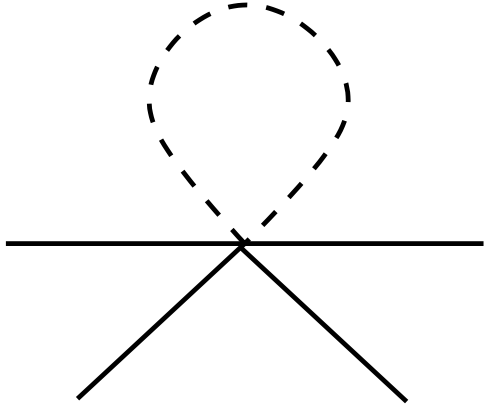}
\end{center}
\caption{A one-loop contribution to the $\ph^4$ vertex for fields
with $|k|<\Lam$.  The internal (dashed) line represents a field with $|k|>\Lam$.}
\label{wilson6}
\end{figure}

Similarly, nonzero values of $c_{n,d,i}(\Lam_0)$ would make subdominant
contributions to $\lam(\Lam)$.  As an example, consider the contribution
of the diagram in \fig{wilson6}.  Ignoring numerical factors,
the vertex factor is $c_{6,0}(\Lam_0)$, and the loop integral is the same
as the one that enters into $m^2(\Lam)$; it yields a factor
of $\Lam_0^2-\Lam^2\sim\Lam_0^2$.  Thus the contribution of this
diagram to $\lam(\Lam)$ is of order $c_{6,0}(\Lam_0)\Lam_0^2$.
This is a pure number that, according to \eq{cndiLam0}, is small.
This contribution is missing the logarithmic enhancement factor
$\ln(\Lam_0/\Lam)$ that we see in \eq{lamLam5}.  

On the other hand, for $m^2(\Lam)$, there are infinitely many contributions
of order $\Lam_0^2$ when $c_{n,d,i}(\Lam_0)\ne 0$.   These must add up
to give $m^2(\Lam)$ a value that is much smaller.  Indeed, we want to
continue the process down to lower and lower values of $\Lam$, with
$m^2(\Lam)$ dropping until it becomes of order $\mph^2$ at
$\Lam\sim\mph$.  For this to happen, there must be very precise cancellations
among all the terms of order $\Lam^2_0$ that contribute to $m^2(\Lam)$.
In some sense, it is more 
``natural'' to have $\mph^2 \sim \lam(\Lam_0)\Lam_0^2$, rather than to 
arrange for these very precise cancellations.  This philosophical issue
is called the {\it fine-tuning problem}, and it generically arises in theories
with spin-zero fields.  

In theories with higher-spin fields only, 
the action typically has more symmetry when these fields
are massless, and this typically prevents divergences that are worse than 
logarithmic.  These theories are said to be {\it technically natural}, 
while theories with spin-zero fields (with physical masses well below the cutoff)
generally are not.  (The only exceptions
are theories where {\it supersymmetry}
relates spin-zero and spin-one-half fields; the spin-zero fields then
inherent the technical naturalness of their spin-one-half partners.)
For now, in $\ph^4$ theory, we will simply accept the necessity of 
fine-tuning in order to have $\mph\ll\Lam$.

Returning to eqs.$\,$({\ref{m2Lam5}--\ref{cn0Lam5}), we can recast
them as differential equations that tell us how these parameters change
with the vaule of the cutoff $\Lam$.  In particular, let us do this for
$\lam(\Lam)$.  We take the derivative of \eq{lamLam5}
with respect to $\Lam$, multiply by $\Lam$, 
and then set $\Lam_0=\Lam$ to get
\begin{equation}
{d\over d\ln\Lam}\, \lam(\Lam) =
{{3\over16\pi^2}}\,\lam^2(\Lam) + \ldots \;.
\label{lamLam6} 
\end{equation}
Notice that the right-hand side of \eq{lamLam5} is apparently
the same as the {\it beta function}
$\beta(\lam)\equiv d\lam/d\ln\mu$ that we calculated in section 26,
where it represented the rate of change in the $\msb$ parameter $\lam$
that was need to compensate for a change in the $\msb$
renormalization scale $\mu$.
\Eq{lamLam5} gives us a new physical interpretation of
the beta function: it is the rate of change in the coefficient of the 
$\ph^4$ term in the effective action as we vary the 
ultraviolet cutoff in that action.

Actually, though, there is a technical detail: it is really 
$Z(\Lam)^{-2}\lam(\Lam)$ that is most closely analogous to the
$\msb$ parameter $\lam$.  This is because,
if we rescaled $\ph$ so that it had a canonical
kinetic term of $\half\d_\mu\ph\d_\mu\ph$, then
$Z(\Lam)^{-2}\lam(\Lam)$ would be the coefficient of the $\ph^4$ term.
Since 
$Z(\Lam)=1+O(\lam^2(\Lam))$, this has no effect at the one-loop level,
but it does matter at two loops.  At three and higher loops, other differences
can arise as well, due to the different underlying definitions of the coupling
$\lam$ in the cutoff scheme and the $\msb$ scheme.

We now have the overall picture of Wilson's approach to quantum
field theory.  First, define a quantum field theory via an action with
an explicit momentum
cutoff in place.  (This can be done in various ways:
for example, we could replace continuous spacetime with a discrete
{\it lattice\/} of points
with {\it lattice spacing\/} $a$; then there is an effective largest
momentum of order $1/a$.  This scheme is very useful as a starting
point for nonperturbative computer calculations.)   Then, lower the
cutoff by integrating out higher-momentum degrees of freedom.
As a result, the coefficients in the effective action will change.
If the field theory is weakly coupled---which in practice means
eqs.$\,$(\ref{lamLam0}--\ref{cndiLam0}) are obeyed---then the
coefficients of the operators with negative mass dimension will
start to take on the values we would have computed for them
in perturbation theory, regardless of their precise initial values.
If we continuously rescale the fields to
have canonical kinetic terms, 
then the dimensionless coupling constant(s) will change according
to their beta functions. Up to at least the two-loop level, these beta
functions are identical to those
that we would find in the $\msb$ scheme.  The final results, 
at an energy scale $E$ well below the initial cutoff $\Lam_0$, 
are the same as we would predict
via renormalized perturbation theory, up to small corrections by powers
of $E/\Lam_0$.

The advantage of the Wilson scheme is that it gives a nonperturbative
definition of the theory which is applicable even if the theory is {\it not\/}
weakly coupled.   With a spacetime lattice providing the cutoff, other
techniques (typically requiring large-scale computer calculations)
can be brought to bear on strongly-coupled theories.

The Wilson scheme also allows us to give physical meaning to nonrenormalizable theories.
Given an action for a nonrenormalizable theory, we can regard it as an
effective action.  We should then impose a momentum cutoff $\Lam_0$, 
where
$\Lam_0$ can be defined by saying that the coefficient of every operator
$\O_i$ with mass dimension $D_i>4$ is given by $c_i/\Lam_0^{D_i-4}$
with $c_i\le 1$.  Then we can use this theory for physics at energies
below $\Lam_0$.  At energies $E$ far below $\Lam_0$, the effective theory
will look like a renormalizable one, 
up to corrections by powers of $E/\Lam_0$.  (This renormalizable theory
might simply be a free-field theory with {\it no\/} interactions, or no theory
at all if there are no particles with physical masses well below $\Lam_0$.)

We now turn to the final issue: can we remove the cutoff completely?

Returning to the example of $\ph^4$ theory, let us suppose that we are
somehow able to compute
 the exact beta function.  Then we can integrate the
renormalization-group equation $d\lam/d\ln\Lam=\beta(\lam)$ from
$\Lam=\mph$ to $\Lam=\Lam_0$ to get
\begin{equation}
\int_{\lam(\mph)}^{\lam(\Lam_0)}{d\lam\over\beta(\lam)} 
= \ln{\Lam_0\over\mph}\;.
\label{intbeta}
\end{equation}
We would like to take the limit $\Lam_0\to\infty$.  
Obviously, the right-hand side of \eq{intbeta}
becomes infinite in this limit, and so the left-hand side must as well.

However, it may not.  
Recall that, for small $\lam$, $\beta(\lam)$ is positive, and it increases
faster than $\lam$.  If this is true for all $\lam$, then the left-hand side
of \eq{intbeta} will approach a fixed, finite value 
as we take the upper limit of integration to infinity.  This 
yields a maximum possible value for the initial cutoff, given by
\begin{equation}
\ln{\Lam_{\rm max}\over\mph}
\equiv\int_{\lam(\mph)}^\infty{d\lam\over\beta(\lam)}\;.
\label{Lammax}
\end{equation}
If we approximate the exact beta function with its leading term, $3\lam^2/16\pi^2$,
and use the leading term in \eq{lamLam3} to get $\lam(\mph)=\lamph$, then we find
\begin{equation}
\Lam_{\rm max}\simeq\mph\,e^{16\pi^2\!/3\lamph}\;.
\label{Lammax2}
\end{equation}
The existence of a maximum possible value for the cutoff 
means that we cannot take the limit as the cutoff
goes to infinity; we {\it must\/} use an effective action with a cutoff
as our starting point.  If we insist on taking the cutoff
to infinity, then the only possible value of $\lamph$ is $\lamph=0$.
Thus, $\ph^4$ theory is {\it trivial\/} in the limit of infinite cutoff:
there are no interactions.  (There is much evidence for this,
but as yet no rigorous proof.  The same is true of quantum electrodynamics,
as was first conjectured by Landau; in this case, $\Lam_{\rm max}$
is known as the location of the {\it Landau pole}.)

However, the cutoff {\it can\/} be removed if the beta function grows no faster
than $\lam$ at large $\lam$; then the left-hand side of \eq{intbeta} would 
diverge as we take the upper limit of integration to infinity.  
Or, $\beta(\lam)$ could drop to zero (and then become negative)
at some finite value $\lam_*$.  Then, if $\lamph<\lam_*$, the
left-hand side of \eq{intbeta} would diverge as the upper limit
of integration approaches $\lam_*$.  In this case, the effective coupling
at higher and higher energies would remain fixed at $\lam_*$,
and $\lam=\lam_*$ is called an {\it ultravioldet fixed point\/} 
of the renormalization group.

If the beta function is negative for $\lam=\lam(\mph)$, the theory is said to
be {\it asymptotically free}, and $\lam(\Lam)$ {\it decreases\/}
as the cutoff is increased. In this case, there is no barrier to
taking the limit $\Lam\to\infty$.  In four spacetimes dimensions,
the only asymptotically free theories are nonabelian gauge theories
with spin-one and spin-one-half fields only.

\vfill\eject

%% file: ch029.tex
\noindent Quantum Field Theory  \hfill   Mark Srednicki

\vskip0.5in

\begin{center}
\large{29: Spontaneous Symmetry Breaking}
\end{center}
\begin{center}
Prerequisite: 3
\end{center}

\vskip0.5in

Consider $\ph^4$ theory, where $\ph$ is a real scalar field with lagrangian
\begin{equation}
\L=-\half\d^\mu \ph\d_\mu \ph -\half m^2\ph^2
- {\ts{1\over24}}\lam\ph^4\;.
\label{ellph428}
\end{equation}
As we discussed in section 23, this theory has a Z$_2$ symmetry:
$\L$ is invariant under $\ph(x)\to-\ph(x)$, and we can define a
unitary operator $Z$ that implements this:
\begin{equation}
Z^{-1}\ph(x)Z = -\ph(x) \;.
\label{zphz28}
\end{equation}
We also have $Z^2=1$, and so $Z^{-1}=Z$.  Since unitarity
implies $Z^{-1}=Z^\dagger$, this makes
$Z$ hermitian as well as unitary.

Now suppose that the parameter $m^2$ is, in spite of its name,
negative rather than positive.  We can write $\L$ in the form
\begin{equation}
\L=-\half\d^\mu \ph\d_\mu \ph -V(\ph) \;,
\label{ellv}
\end{equation}
where the potential is
\begin{eqnarray}
V(\ph) &=& \half m^2\ph^2 + {\ts{1\over24}}\lam\ph^4
\nonumber \\
\noalign{\smallskip}
&=& {\ts{1\over24}}\lam(\ph^2-v^2)^2 - {\ts{1\over24}}\lam v^4 \;.
\label{vofph}
\end{eqnarray}
In the second line, we have defined
\begin{equation}
v \equiv +(6|m^2|/\lam)^{1/2} \;.
\label{vph}
\end{equation}
We can (and will) drop the last, constant, term in \eq{vofph}.  From 
\eq{vofph} it is clear that there are two classical field configurations
that minimize the energy: $\ph(x)=+v$ and $\ph(x)=-v$.
This is in contrast to the usual case of positive $m^2$, for which
the minimum-energy classical field configuration is $\ph(x)=0$.

We can expect that the quantum theory will follow suit.  
For $m^2<0$, there will be two
ground states, $|0{+}\ra$ and $|0{-}\ra$, with
the property that
\begin{eqnarray}
\la 0{+}|\ph(x)|0{+}\ra &=& +v \;,
\nonumber \\
\la 0{-}|\ph(x)|0{-}\ra &=& -v \;,
\label{manyv}
\end{eqnarray}
up to quantum corrections from loop diagrams that we will treat 
in detail in section 30.
These two ground states are exchanged by the operator $Z$,
\begin{equation}
Z|0{+}\ra = |0{-}\ra \;,
\label{zzero}
\end{equation}
and they are orthogonal:   $\la0{+}|0{-}\ra=0$.

This last claim requires some comment.  Consider a similar problem
in quantum mechanics, 
\begin{equation}
H={\ts{1\over2}}p^2 + {\ts{1\over24}}\lam (x^2-v^2)^2 \;.
\label{anharm}
\end{equation}
We could find two approximate ground states in this case,
specified by the approximate wave functions
\begin{equation}
\psi_\pm(x) = \la x|0{\pm}\ra \sim \exp[-\w(x\mp v)^2/2] \;,
\label{psipm}
\end{equation}
where $\w=(\lam v^2/3)^{1/2}$ is the frequency of 
small oscillations about the minimum.  
However, the true ground state
would be a symmetric linear combination of these.
The antisymmetric linear combination would have a slightly
higher energy, due to the effects of quantum tunneling.

We can regard a field theory as an infinite set of oscillators,
one for each point in space,  
each with a hamiltonian like \eq{anharm}, and coupled
together by the $(\nabla\ph)^2$ term in the field-theory
hamiltonian.  There is a tunneling
amplitude for each oscillator, but to turn the field-theoretic
state $|0{+}\ra$ into $|0{-}\ra$, {\it all\/} the oscillators have
to tunnel, and so the tunneling amplitude gets raised to the power of
the number of oscillators, that is, to the power of infinity
(more precisely, to a power that scales like the volume
of space).  Therefore, in the limit of infinite volume,
$\la0{+}|0{-}\ra$ vanishes.

Thus we can pick either $|0{+}\ra$ or $|0{-}\ra$ to 
use as the ground state.  Let us choose $|0{+}\ra$.
Then we can define a shifted field,
\begin{equation}
\rho(x) = \ph(x) - v \;,
\label{phphv}
\end{equation} 
which obeys
$\la 0{+}|\rho(x)|0{+}\ra=0$. 
(We must still worry about loop corrections,
which we will do at the end of this section.)
The potential becomes
\begin{eqnarray}
V(\ph) &=&  {\ts{1\over24}}\lam[(\rho+v)^2-v^2]^2
\nonumber \\
\noalign{\smallskip}
&=& {\ts{1\over6}}\lam v^2\rho^2 
+ {\ts{1\over6}}\lam v\rho^3
+{\ts{1\over24}}\lam \rho^4 \;,
\label{vofph2}
\end{eqnarray}
and so the lagrangian is now
\begin{equation}
\L=-\half  \d^\mu\!\rho\d_\mu \rho 
-{\ts{1\over6}}\lam v^2\rho^2 
-{\ts{1\over6}}\lam v\rho^3
-{\ts{1\over24}}\lam \rho^4\;.
\label{ellshift}
\end{equation}
We see that the coefficient of the $\rho^2$ term
is ${\ts{1\over6}}\lam v^2=|m^2|$.  This coefficient
should be identified as $\half m_\rho^2$, where
$m_\rho$ is the mass of the the corresponding $\rho$ 
particle.  Also, we see that the shifted field now
has a cubic as well as a quartic interaction.

\Eq{ellshift} specifies a perfectly sensible, renormalizable
quantum field theory, but it no longer has 
an obvious Z$_2$ symmetry.  We say that the
Z$_2$ symmetry is {\it hidden\/}, or {\it secret\/},
or (most popular of all) {\it spontaneously broken}.

This leads to a question about renormalization.
If we include renormalizing $Z$ factors in the
original lagrangian, we get
\begin{equation}
\L=-\half Z_\ph \d^\mu \ph\d_\mu \ph -\half  Z_m m^2\ph^2
-{\ts{1\over24}}Z_\lam\lam\ph^4 \;.
\label{ellv2}
\end{equation}
For positive $m^2$, these three $Z$ factors
are sufficient to absorb
infinities for $d\le4$, where the mass dimension of $\lam$
is positive or zero.  On the other hand, looking at the
lagrangian for negative $m^2$ after the shift, \eq{ellshift},
we would seem to need an extra $Z$ factor for the $\rho^3$ term.
Also, once we have a $\rho^3$ term, we would expect
to need to add a $\rho$ term to cancel tadpoles.
So, the question is, are the original three $Z$ factors
sufficient to absorb all the divergences in
the Feynman diagrams derived from \eq{ellv2}?

The answer is yes.  To see why, consider the quantum action
(introduced in section 21) 
\begin{eqnarray}
\Gamma(\ph) &=&
{1\over2}\int{\dfk\over(2\pi)^4}\,\tph(-k)\Bigl(k^2+m^2-\Pi(k^2)\Bigr)\tph(k)
\nonumber \\
&& {} + \sum_{n=3}^\infty {1\over n!} \int
{\dfk_1\over(2\pi)^4}\ldots{\dfk_n\over(2\pi)^4}\,
(2\pi)^4\delta^4(k_1{+}\ldots{+}k_n)
\nonumber \\
&& \qquad\qquad\quad
\times \V_n(k_1,\ldots,k_n)\,\tph(k_1)\ldots\tph(k_n)\;,
\label{qa28}
\end{eqnarray}
computed with $m^2>0$.  The ingredients of $\Gamma(\ph)$---the
self-energy $\Pi(k^2)$ and the exact 1PI vertices $\V_n$---are
all made finite and well-defined (in, say, the $\msb$ renormalization 
scheme) by adjusting the three $Z$ factors in \eq{ellv2}.  
Furthermore, for $m^2>0$,
the quantum action inherits the Z${}_2$ symmetry of the classical action.  
To see this, we note that
$\V_n$ must zero for odd $n$, simply because there
is no way to draw a 1PI diagram with an odd number of external
lines using only a four-point vertex.  Thus $\Gamma(\ph)$ also
has the Z$_2$ symmetry.  

The quantum action also inherits any linear continuous symmetry
of the classical action that is also a symmetry of the integration
measure $\D\ph$; see problem 29.1.
(By {\it linear}, we mean any symmetry whose infinitesimal
transformation $\delta\ph_i(x)$ is linear in $\phi_j(x)$ and its
derivatives.)
This condition of invariance of the integration measure
is almost always met; when it is not, the symmetry is said to be
{\it anomalous}.  We will meet an anomalous symmetry in section 75.  

Once we have computed the quantum action for $m^2>0$,
we can go ahead and consider the case of $m^2<0$.
Recall from section 21 that the
quantum equation of motion in the presence of a source is
$\delta\Gamma/\delta\ph(x)=-J(x)$,
and that the solution of this equation is also the vacuum expectation
value of $\ph(x)$.  
Now set $J(x)=0$, and look for a translationally invariant (that is,
constant) solution $\ph(x)=v$.  If there is more than one such
solution, we want the one(s) with the lowest energy.
This is equivalent to minimizing the quantum potential ${\cal U}(\ph)$,
where 
\begin{equation}
\Gamma(\ph)=\int\dfx\,\left[{}-{\cal U}(\ph)
-\half{\cal Z}(\ph)\d^\mu\ph\d_\mu\ph + \ldots\,\right]\;,
\label{qa29}
\end{equation}
and the ellipses stand for terms with more derivatives.
In a weakly coupled theory, we can expect the loop-corrected potential
${\cal U}(\ph)$ to be qualitatively similar to the
classical potential $V(\ph)$.  Therefore, 
for $m^2< 0$, we expect that there are two minima of ${\cal U}(\ph)$
with equal energy, located at $\ph(x)=\pm v$, where $v=\la0|\ph(x)|0\ra$
is the exact vacuum expectation value of the field.  

Thus we have a description of spontaneous symmetry breaking
in the quantum theory based on the quantum action, 
and the quantum action is made 
finite by adjusting only the three $Z$ factors that appear in the
original, symmetric form of the lagrangian.

In the next section, we will see how this works in explicit calculations.

\vskip0.5in

\begin{center}
Problems
\end{center}

\vskip0.25in

29.1) Consider a classical action $\int\dfx\,\L$ that is invariant under
an infinitesimal transformation of the form 
\begin{equation}
\delta\ph_i(x) = \alpha D_{ij}\ph_j(x)\;,
\label{dphix}
\end{equation}
where $\alpha$ is an infinitesimal parameter, and
$D_{ij}$ is a matrix containing constants and (possibly) derivatives
with respect to $x$.  We assume that the integration measure $\D\ph$ is
also invariant.  Show that $\delta\Gamma(\ph)$ vanishes.

\vfill\eject

%% file: ch030.tex
\noindent Quantum Field Theory  \hfill   Mark Srednicki

\vskip0.5in

\begin{center}
\large{30: Spontaneous Symmetry Breaking and Loop Corrections}
\end{center}
\begin{center}
Prerequisite: 19, 29
\end{center}

\vskip0.5in

Consider $\ph^4$ theory, 
where $\ph$ is a real scalar field with lagrangian
\begin{equation}
\L=-\half Z_\ph \d^\mu \ph\d_\mu \ph -\half Z_m m^2\ph^2
-{\ts{1\over24}}Z_\lam\lam\ph^4 \;.
\label{ellv29}
\end{equation}
In $d=4$ spacetime dimensions, the coupling $\lam$ is 
dimensionless. 

We begin by considering the case $m^2>0$, 
where the Z$_2$ symmetry of $\L$ under $\ph\to-\ph$ is manifest. 
We wish to compute the three renormalizing $Z$ factors.
We work in $d=4-\e$ dimensions, and take $\lam\to\lam \mut^{\e}$
(where $\mut$ has dimensions of mass)
so that $\lam$ remains dimensionless. 

\begin{figure}
\begin{center}
\epsfig{file=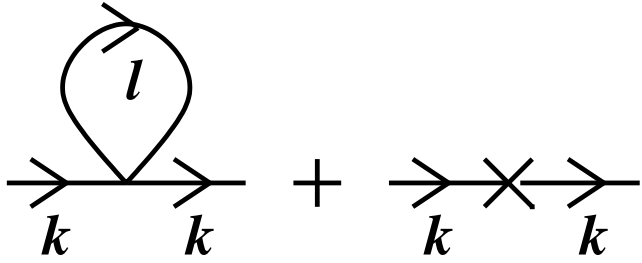}
\end{center}
\caption{$O(\lam)$ corrections to $\Pi(k^2)$.}
\label{prop-phi4}
\end{figure}

The propagator correction $\Pi(k^2)$ is given by the diagrams
of \fig{prop-phi4}, which yield
\begin{equation}
i\Pi(k^2) = {\ts{1\over2}}(-i\lam\mut^\e){\ts{1\over i}}\td(0) 
              -i(Ak^2+Bm^2) \;,
\label{ipiphi4}
\end{equation}
where $A=Z_\ph-1$ and $B=Z_m-1$, and 
\begin{equation}
\td(0) = \int{\ddl\over(2\pi)^d}\,{1\over\ell^2+m^2} \;.
\label{d0phi4}
\end{equation}
Using the usual bag of tricks from section 14, we find
\begin{equation}
\mut^\e\td(0) = {-i\over(4\pi)^2}\left[\,{2\over\e}+1
            +\ln\Bigl(\mu^2/m^2\Bigr)\right]\!m^2 \;,
\label{d1phi4}
\end{equation}
where $\mu^2=4\pi e^{-\gamma}\mut^2$.  Thus
\begin{equation}
\Pi(k^2) = {\lam\over2(4\pi)^2}\left[\,{2\over\e}+1
            +\ln\Bigl(\mu^2/m^2\Bigr)\right]\!m^2
              -Ak^2-Bm^2 \;.
\label{ipiphi42}
\end{equation}
From \eq{ipiphi42} we see that we must have
\begin{eqnarray}
A &=& O(\lam^2) \;,
\label{aphi4} \\
\noalign{\medskip}
B &=& {\lam\over16\pi^2}\left[\,{1\over\e} + \kappa_B\,\right] + O(\lam^2) \;,
\label{bphi4}
\end{eqnarray}
where $\kappa_B$ is a finite constant.  
In the $\msb$ renormalization scheme, we take $\kappa_B=0$, but we
will leave $\kappa_B$ arbitrary for now.

\begin{figure}
\begin{center}
\epsfig{file=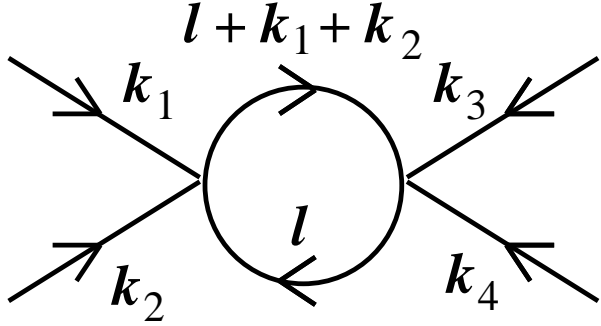}
\end{center}
\caption{The $O(\lam^2)$ correction to $\V_3(k_1,k_2,k_3)$.
Two other diagrams, obtained from this one via
$k_2\leftrightarrow k_3$ and $k_2\leftrightarrow k_4$, also
contribute.}
\label{vert-phi4}
\end{figure}

Next we turn to the vertex correction, given by the diagram of 
\fig{vert-phi4}, plus two others with $k_2\leftrightarrow k_3$ and
$k_2\leftrightarrow k_4$; all momenta are treated as incoming.  We have
\begin{eqnarray}
i\V_4(k_1,k_2,k_3,k_4)
&=& -iZ_\lam \lam 
+ {\ts{1\over2}}(-i\lam)^2\left({\ts{1\over i}}\right)^{\!2} 
\Bigl[iF(-s)+iF(-t)+iF(-u)\Bigr] 
\nonumber \\
&& {} + O(\lam^3)\;.
\label{v4phi4}
\end{eqnarray}
Here we have defined $s=-(k_1+k_2)^2$, $t=-(k_1+k_3)^2$, $u=-(k_1+k_4)^2$,
and
\begin{eqnarray}
iF(k^2) &\equiv& \mut^\e \int{\ddl\over(2\pi)^d}\,
{1\over((\ell{+}k)^2+m^2)(\ell^2+m^2)}
\nonumber \\
\noalign{\medskip}
&=& {i\over16\pi^2}\left[\,{2\over\e}+\int_0^1 dx\,
            \ln\Bigl(\mu^2/D\Bigr)\right] \;,
\label{fphi4}
\end{eqnarray}
where $D=x(1{-}x)k^2+m^2$.
Setting $Z_\lam = 1+C$ in \eq{v4phi4}, we see that we need
\begin{equation}
C = {3\lam\over16\pi^2}
\left[\,{1\over\e} + \kappa_C\,\right] + O(\lam^2) \;,
\label{cphi4}
\end{equation}
where $\kappa_C$ is a finite constant.

We may as well pause to compute the beta function,
$\beta(\lam)=d\lam/d\ln\mu$, where the derivative is taken with the
bare coupling $\lam_0$ held fixed,
and the finite parts of the counterterms set to zero,
in accord with the $\msb$ prescription.
We have
\begin{equation}
\lam_0 = Z_\lam Z_\ph^{-2}\lam\mut^\e \;,
\label{lam0}
\end{equation}
with
\begin{equation}
\ln\Bigl(Z_\lam Z_\ph^{-2}\Bigr) =
{3\lam\over16\pi^2}\,{1\over\e} + O(\lam^2) \;.
\label{loglam0}
\end{equation}
A review of the procedure of section 27 reveals that the first term
in the beta function is given by $\lam$ times the coefficient of $1/\e$ in
\eq{loglam0}.  Therefore,
\begin{equation}
\beta(\lam)={3\lam^2\over16\pi^2} + O(\lam^3) \;.
\label{betaphi4}
\end{equation}
The beta function is positive, which means that the
theory becomes more and more strongly coupled at higher and higher energies.

Now we consider the more interesting case of $m^2<0$, which results
in the spontaneous breakdown of the Z$_2$ symmetry. 

Following the procedure of section 29,
we set $\ph(x)=\rho(x)+v$, where $v=(6|m^2|/\lam)^{1/2}$ minimizes the
potential (without $Z$ factors).  Then the lagrangian becomes
(with $Z$ factors)
\begin{eqnarray}
\L &=& -\half Z_\ph\d^\mu\!\rho\d_\mu \rho
-{\ts{1\over2}}({\ts{3\over4}}Z_\lam {-} {\ts{1\over4}}Z_m)
  m_\rho^2\rho^2
\nonumber \\
\noalign{\smallskip}
&& {} 
+{\ts{1\over2}}(Z_m {-} Z_\lam)(3/\lam\mut^\e)^{1/2}m_\rho^3 \rho
-{\ts{1\over6}}Z_\lam(3\lam\mut^\e)^{1/2}m_\rho\rho^3
-{\ts{1\over24}}Z_\lam\lam\mut^\e\rho^4 \;, \qquad
\label{ellshift29}
\end{eqnarray}
where $m_\rho^2=2|m^2|$.
Now we can compute various one-loop corrections.  

\begin{figure}
\begin{center}
\epsfig{file=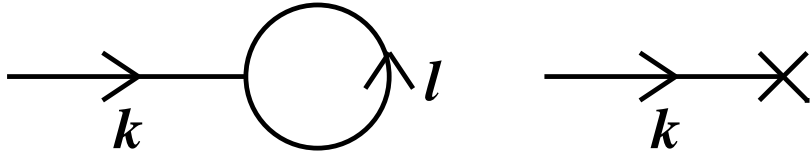}
\end{center}
\caption{The $O(\lam)$ correction to the vacuum expectation
value of the $\rho$ field.}
\label{rho-vev}
\end{figure}

We begin with 
the vacuum expectation value of $\rho$.  The $O(\lam)$ correction
is given by the diagrams of \fig{rho-vev}.  
The three-point vertex factor is $-iZ_\lam g_3$, where
$g_3$ can be read off of \eq{ellshift29}:
\begin{equation}
g_3 = (3\lam\mut^\e)^{1/2}m_\rho \;.
\label{3ptphi4}
\end{equation}
The one-point vertex factor is $iY$, where
$Y$ can also be read off of \eq{ellshift29}:
\begin{equation}
Y = {\ts{1\over2}} (Z_m {-} Z_\lam) (3/\lam\mut^\e)^{1/2}m_\rho^3 \;.
\label{yphi4}
\end{equation}
Following the discussion of section 9, we then find that
\begin{equation}
\la 0|\rho(x)|0\ra =
\left(iY + {\ts{1\over2}}(-iZ_\lam g_3){\ts{1\over i}}\td(0)\right)
\int\dfy\,{\ts{1\over i}}\Delta(x{-}y) \;,
\label{vev29}
\end{equation}
plus higher-order corrections.
Using \eqs{3ptphi4} and (\ref{yphi4}), and \eq{d1phi4} 
with $m^2\to m_\rho^2$,
the factor in large parentheses in \eq{vev29} becomes
\begin{equation}
{i\over2}(3/\lam)^{1/2}m_\rho^3
\left(Z_m{-}Z_\lam + {\lam\over16\pi^2}\left[\,{2\over\e}+1
            +\ln\Bigl(\mu^2/m_\rho^2\Bigr) \right] + O(\lam^2) \right) \;.
\label{vev29a}
\end{equation}
Using $Z_m=1+B$ and $Z_\lam=1+C$, with $B$ and $C$ from
\eqs{bphi4} and (\ref{cphi4}),
the factor in large parentheses in \eq{vev29a} becomes
\begin{equation}
{\lam\over16\pi^2}\left[\,\kappa_B-\kappa_C+1+
\ln\Bigl(\mu^2/m_\rho^2\Bigr)\right] \;.
\label{vev29b}
\end{equation}
All the $1/\e$'s have canceled.  The remaining finite vacuum expectation
value for $\rho(x)$ can now be removed by choosing
\begin{equation}
\kappa_B-\kappa_C = -1 - \ln(\mu^2/m_\rho^2) \;.
\label{bminusc}
\end{equation}
This will also cancel all diagrams with one-loop tadpoles.

\begin{figure}
\begin{center}
\epsfig{file=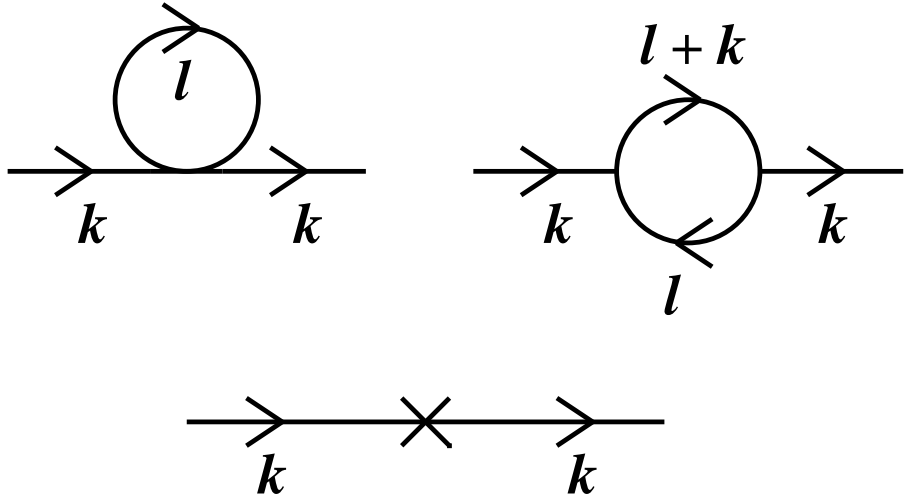}
\end{center}
\caption{$O(\lam)$ corrections to the $\rho$ propagator.}
\label{prop-rho}
\end{figure}

Next we consider the $\rho$ propagator.  The diagrams  
contributing to the $O(\lam)$ correction are shown in \fig{prop-rho}.
The counterterm insertion is $-iX$, where, again reading off of
\eq{ellshift29},
\begin{equation}
X = Ak^2 + ({\ts{3\over4}}C-{\ts{1\over4}}B)m_\rho^2 \;.
\label{xphi4}
\end{equation}
Putting together the results of \eq{ipiphi4} for the first diagram
(with $m^2\to m_\rho^2$),
\eq{fphi4} for the second (ditto), and \eq{xphi4} for the third, we get
\begin{eqnarray}
\Pi(k^2) &=& -{\ts{1\over2}}(\lam\mut^\e){\ts{1\over i}}\td(0) 
          +{\ts{1\over2}}g_3^2 F(k^2) - X +O(\lam^2)
\nonumber \\
\noalign{\medskip}
&=& {\lam\over32\pi^2}\,m_\rho^2
  \left[\,{2\over\e}+1+\ln\Bigl(\mu^2/m_\rho^2\Bigr)\right] 
\nonumber \\
\noalign{\medskip}
&& {} + {\lam\over32\pi^2}\,m_\rho^2
\left[\,{2\over\e}+\int_0^1 dx\, \ln\Bigl(\mu^2/D\Bigr)\right]
\nonumber \\
\noalign{\medskip}
&& {} - Ak^2 - ({\ts{3\over4}}C-{\ts{1\over4}}B)m_\rho^2+O(\lam^2).
\label{pi3phi4}
\end{eqnarray}
Again using \eqs{bphi4} and (\ref{cphi4}) for $B$ and $C$, 
we see that all the $1/\e$'s cancel, and we're left with
\begin{eqnarray}
\Pi(k^2) &=& {\lam\over32\pi^2}\,m_\rho^2
  \left[\,1+\ln\Bigl(\mu^2/m_\rho^2\Bigr)
+\int_0^1 dx\, \ln\Bigl(\mu^2/D\Bigr) + \half(9\kappa_C-\kappa_B)\, \right] 
\nonumber \\
&& {} + O(\lam^2)\;.
\label{pi4phi4}
\end{eqnarray}
We can now choose to work in an OS scheme, where we require
$\Pi(-m_\rho^2)=0$ and $\Pi'(-m_\rho^2)=0$.  
We see that, to this order in $\lam$,
$\Pi(k^2)$ is independent of $k^2$.  Thus, we automatically have 
$\Pi'(-m_\rho^2)=0$, and we can choose $9\kappa_C-\kappa_B$ to fix
$\Pi(-m_\rho^2)=0$.  Together with \eq{bminusc}, this completely
determines $\kappa_B$ and $\kappa_C$ to this order in $\lam$.

\begin{figure}
\begin{center}
\epsfig{file=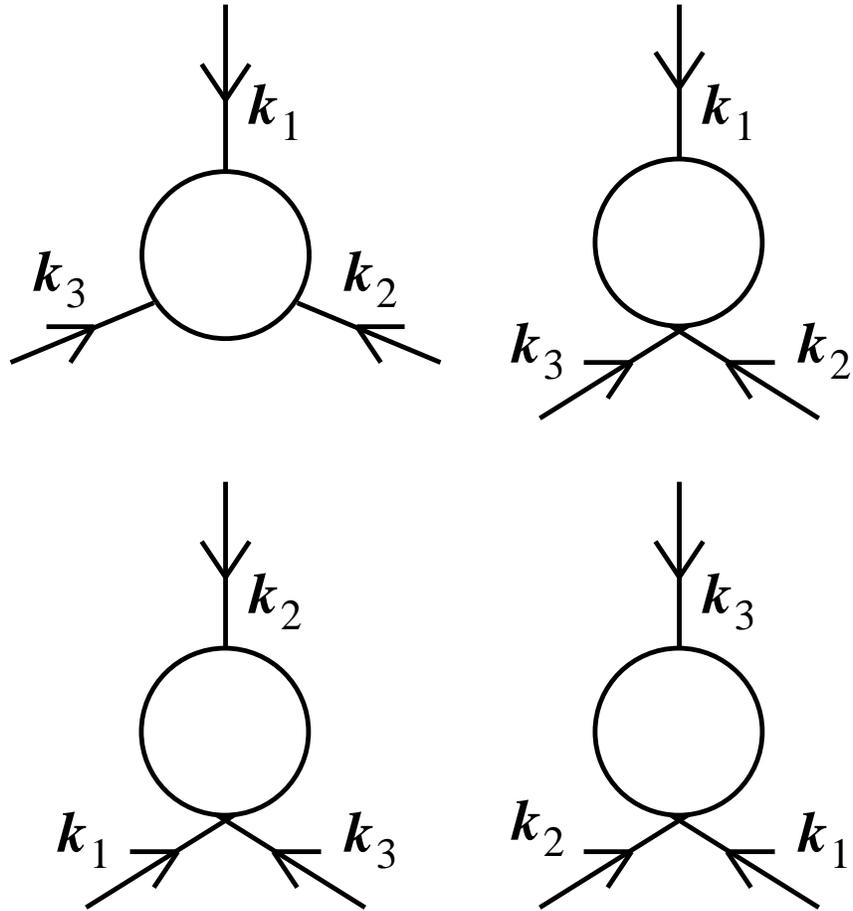}
\end{center}
\caption{$O(\lam)$ corrections to the $\rho$ propagator.}
\label{3vert-rho}
\end{figure}

Next we consider the one-loop correction to the three-point vertex,
given by the diagrams of \fig{3vert-rho}.  
We wish to show that the infinities are canceled by the value of
$Z_\lam=1+C$ that we have already determined.  
The first diagram in \fig{3vert-rho} is finite, and so for our
purposes we can ignore it.  The remaining three, plus the original vertex, 
sum up to give
\begin{eqnarray}
i\V_3(k_1,k_2,k_3)_{\rm div}
&=& -iZ_\lam g_3
+ \half(-i\lam)(-ig_3)\left({\ts{1\over i}}\right)^{\!2} 
\nonumber \\
&& \qquad\qquad\;
{} \times\Bigl[iF(k_1^2)+iF(k_2^2)+iF(k_3^2)\Bigr] 
\nonumber \\
&& {} + O(\lam^{5/2})\;,
\label{v3phi4}
\end{eqnarray}
where the subscript div means that we are keeping only the
divergent part.  Using \eq{fphi4}, we have
\begin{equation}
\V_3(k_1,k_2,k_3)_{\rm div}
= - g_3\left(1+C - {3\lam\over16\pi^2}\,{1\over\e} + O(\lam^2)\right) \;.
\label{v3div}
\end{equation}
{}From \eq{cphi4}, we see that the divergent terms do indeed
cancel to this order in $\lam$.

Finally, we have the correction to the four-point vertex.  In this case,
the divergent diagrams are just those of \fig{prop-phi4}, 
and so the calculation
of the divergent part of $\V_4$ is exactly the same as it is when
$m^2>0$ (but with $m_\rho$ in place of $m$).  
Since we have already done that calculation (it was how
we determined $C$ in the first place), we need not repeat it.

We have thus seen how we can compute the divergent parts of the
counterterms in the simpler case of $m^2>0$, where the Z$_2$ symmetry
is unbroken, and that these counterterms will also serve to cancel
the divergences in the more complicated case of $m^2<0$, where the
Z$_2$ symmetry is spontaneously broken.
This a general rule for renormalizable theories with spontaneous
symmetry breaking, regardless of the nature of the symmetry group.

\vfill\eject

%% file: ch031.tex
\noindent Quantum Field Theory  \hfill   Mark Srednicki

\vskip0.5in

\begin{center}
\large{31: Spontaneous Breakdown of Continuous Symmetries}
\end{center}
\begin{center}
Prerequisite: 29
\end{center}

\vskip0.5in

Consider the theory (introduced in section 22) of a complex
scalar field $\ph$ with
\begin{equation}
\L=-\d^\mu\ph^\dagger\d_\mu\ph-m^2\ph^\dagger\ph
- {\ts{1\over4}}\lam(\ph^\dagger\ph)^2\;.
\label{ellu130}
\end{equation}
This lagrangian is obviously invariant under the U(1) transformation
\begin{equation}
\ph(x)\to e^{-i\alpha}\ph(x)\;, 
\label{u130}
\end{equation}
where $\alpha$ is a real number.
We can also rewrite $\L$ in terms of two real scalar fields by setting
$\ph=(\ph_1+i\ph_2)/\sqrt2$ to get
\begin{equation}
\L=-\half\d^\mu \ph_1\d_\mu \ph_1 -\half\d^\mu \ph_2\d_\mu \ph_2
-\half m^2(\ph_1^2+\ph_2^2)
- {\ts{1\over16}}\lam(\ph_1^2+\ph_2^2)^2\;.
\label{ellab30}
\end{equation}
In terms of $\ph_1$ and $\ph_2$, the U(1) transformation becomes
an SO(2) transformation,
\begin{equation}
\pmatrix{
\ph_1(x) \cr
\noalign{\medskip}
\ph_2(x) \cr}
\to
\pmatrix{ \phantom{-}\cos\alpha & \sin\alpha \cr
\noalign{\medskip}
           {-}\sin\alpha & \cos\alpha \cr}
\pmatrix{
\ph_1(x) \cr
\noalign{\medskip}
\ph_2(x) \cr}.
\label{ab30}
\end{equation}
If we think of $(\ph_1,\ph_2)$ as a two-component vector,
then \eq{ab30} is just a rotation
of this vector in the plane by angle $\alpha$.

Now suppose that $m^2$ is negative.  The minimum of the potential
of \eq{ellu130} is achieved for $\ph(x)=v e^{-i\theta}/\sqrt2$, where
$v=(4|m^2|/\lam)^{1/2}$ and the phase $\theta$ is arbitrary.
(The factor of the square root of two is conventional).
Thus we have a continuous family of minima of the potential,
parameterized by $\theta$.  
Under the U(1) transformation of \eq{u130}, $\theta$ changes to 
$\theta+\alpha$; thus the different minimum-energy field configurations
are all related to each other by the symmetry.

In the quantum theory, we therefore expect to find a continuous family
of ground states, labeled by $\theta$, with the property that
\begin{equation}
\la \theta|\ph(x)|\theta\ra = {\ts{1\over\sqrt2}}ve^{-i\theta} \;.
\label{thphth}
\end{equation}
Also, according to the discussion in section 29, we expect 
$\la\theta'|\theta\ra=0$ for $\theta'\ne\theta$.

Returning to classical language, there is a 
{\it flat direction\/} in field space
that we can move along without changing the energy.  
The physical consequence
of this is the existence of a massless particle called a
{\it Goldstone boson}.

Let us see how this works, first using the SO(2) form of the theory,
\eq{ellab30}.  We will choose the phase $\theta=0$, and write
\begin{eqnarray}
\ph_1(x) &=& v + a(x) \;,
\nonumber \\
\ph_2(x) &=& b(x) \;.
\label{ph1ph2}
\end{eqnarray}
Substituting this into \eq{ellab30}, we find
\begin{eqnarray}
\L &=& -\half\d^\mu a\d_\mu a -\half\d^\mu b\d_\mu b
\nonumber \\
&& {} 
-|m^2|a^2 - \half \lam^{1/2}|m|a(a^2+b^2) 
- {\ts{1\over16}}\lam(a^2+b^2)^2 \;.
\label{ellab301}
\end{eqnarray}
We see from this that the $a$ field has a mass given by
$\half m_a^2 = |m^2|$.
The $b$ field, on the other hand, is massless, and we identify
it as the Goldstone boson.

A different parameterization brings out the role
of the massless field more clearly.  In terms of the complex field
$\ph(x)$, we write
\begin{equation}
\ph(x) = {\ts{1\over\sqrt2}}\Bigl(v+\rho(x)\Bigr)
                            \exp\Bigl(-i\chi(x)/v\Bigr) \;.
\label{phrhochi}
\end{equation}
Substituting this into \eq{ellu130}, we get
\begin{eqnarray}
\L &=& -\half\d^\mu\!\rho\d_\mu\rho 
-\half\Bigl(1+{\rho\over v}\Bigr)^{\!2}\d^\mu\!\chi\d_\mu\chi
\nonumber \\
&& {} 
-|m^2|\rho^2
- \half \lam^{1/2}|m|\rho^3
- {\ts{1\over16}}\lam\rho^4 \;.
\label{ellrhochi}
\end{eqnarray}
We see from this that the $\rho$ field has a mass given by
$\half m_\rho^2 = |m^2|$, and that the $\chi$ field is massless.
These are the same particle masses we found using
the parameterization of \eq{ph1ph2}.  This is not an accident:
the particle masses and scattering amplitudes should be independent
of how we choose to write the fields.

Note that the $\chi$ field does not appear in the potential at all.
Thus it parameterizes the flat direction.   
In terms of the $\rho$ and $\chi$ fields,
the U(1) transformation takes the simple form 
$\chi(x)\to\chi(x)+\alpha$.

Does the masslessness of the $\chi$ field survive loop corrections?
It does.  We will first give
a diagrammatic proof, and then a general argument based on properties
of the quantum action.

Before proceeding to diagrams, recall from section 30 that
(in a renormalizable theory)
we can cancel all divergences by including
renormalizing $Z$ factors in the original, 
symmetric form of the lagrangian 
[in the case at hand, either \eq{ellu130} or (\ref{ellab30})] with $m^2>0$.  
This is important 
because the lagrangian in the form of \eq{ellrhochi} looks nonrenormalizable.
The coefficients of the interaction terms 
$\rho\d^\mu\!\chi\d_\mu\chi$ and $\rho^2\d^\mu\!\chi\d_\mu\chi$ 
are $v^{-1}$ and $v^{-2}$, which have mass dimension $-1$ and $-2$.
Coupling constants with negative mass dimension usually signal
nonrenormalizability, but here we know that
the hidden U(1) symmetry saves us from this disaster.

\begin{figure}
\begin{center}
\epsfig{file=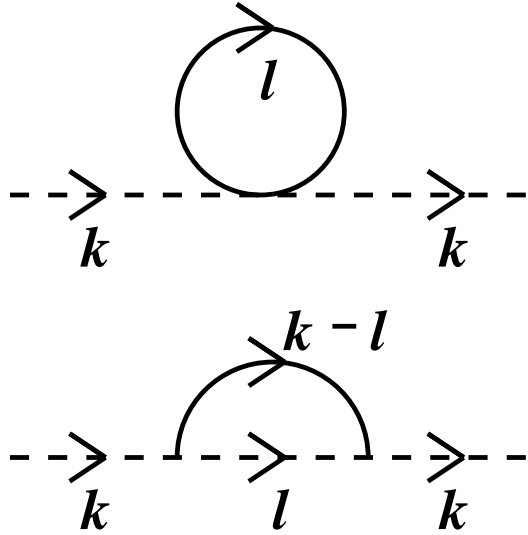}
\end{center}
\caption{$O(\lam)$ corrections to the $\chi$ propagator.}
\label{prop-chi}
\end{figure}

Consider, then, the one-loop corrections to the $\chi$ propagator
shown in \fig{prop-chi}.   The three-point vertex factor is 
$2iv^{-1}k_1\cd k_2$, where $k_1$ and $k_2$ are the two
momenta on the $\chi$ lines (both treated as incoming), and the four-point
vertex factor is $2iv^{-2}k_1\cd k_2$.  The first diagram thus has
a vertex factor of $-2iv^{-2}k^2$, and the loop contributes
a factor of ${1\over i}\td(0)$.  The important point is that
the diagram is proportional to $k^2$; there is no term independent
of $k^2$, which would contribute to a mass term for the $\chi$ field.
The second diagram is proportional to 
$(k\cd\ell)^2/(\ell^2+m_\rho^2)$.
When we integrate over $\ell$, we can use the identity
\begin{equation}
\int \ddl\;\ell^\mu\ell^\nu f(\ell^2) 
= {1\over d}\,g^{\mu\nu}\int\ddl\;\ell^2 f(\ell^2)\;.
\label{symint}
\end{equation}
To get \eq{symint}, we note that, by Lorentz invariance,
the integral on the left-hand side must be proportional to $g^{\mu\nu}$;
we then verify the equality by contracting each side with $g_{\mu\nu}$.  
Therefore,
\begin{equation}
\int \ddl\;{(k\cd\ell)^2 \over \ell^2 + m_\rho^2} \propto k^2 \;.
\label{intklsq}
\end{equation}
Thus we see that the second diagram is also proportional to $k^2$.
It should be clear that this will be true of any diagram we draw,
because of the nature of the vertices.  Thus, the $\chi$ particle
remains exactly massless.

The same conclusion can be reached by considering the quantum action
$\Gamma(\ph)$, which includes all loop corrections.  
According to our discussion in section 29, the
quantum action has the same symmetries as the classical action.
Therefore, in the case at hand,
\begin{equation}
\Gamma(\ph)=\Gamma(e^{-i\alpha}\ph) \;.
\label{Gpha}
\end{equation}
Spontaneous symmetry breaking occurs if the minimum of 
$\Gamma(\ph)$ is at a constant,
nonzero value of $\ph$.  Because of \eq{Gpha},
the phase of this constant is arbitrary.  
Therefore, there must be a flat direction in field
space, corresponding to the phase of $\ph(x)$.  The physical
consequence of this flat direction is a massless particle,
the Goldstone boson.

\vfill\eject

%% file: ch032.tex
\noindent Quantum Field Theory  \hfill   Mark Srednicki

\vskip0.5in

\begin{center}
\large{32: Nonabelian Symmetries}
\end{center}
\begin{center}
Prerequisite: 31
\end{center}

\vskip0.5in

Consider the theory (introduced in section 22) of a two real
scalar fields $\ph_1$ and $\ph_2$ with
\begin{equation}
\L=-\half\d^\mu \ph_1\d_\mu \ph_1 -\half\d^\mu \ph_2\d_\mu \ph_2
-\half m^2(\ph_1^2+\ph_2^2)
- {\ts{1\over16}}\lam(\ph_1^2+\ph_2^2)^2\;.
\label{ellab31}
\end{equation}
We can generalize this to the case of $N$ real scalar fields $\ph_i$ with
\begin{equation}
\L=-\half\d^\mu \ph_i\d_\mu \ph_i
-\half m^2\ph_i\ph_i
- {\ts{1\over16}}\lam(\ph_i\ph_i)^2\;,
\label{elln}
\end{equation}
where a repeated index is summed.  This lagrangian is clearly invariant
under the SO($N$) transformation
\begin{equation}
\ph_i(x) \to R_{ij}\ph_j(x) \;,
\label{phrph}
\end{equation}
where $R$ is an orthogonal matrix with unit determinant: $R^T = R^{-1}$,  
$\det R=+1$.  (This largrangian is also clearly invariant under the
Z${}_2$ symmetry $\ph_i(x)\leftrightarrow -\ph_i(x)$, which enlarges
SO($N$) to O($N$); however, in this section we will be concerned with the
continuous SO($N$) part of the symmetry.)

Next we will need some results from group theory.
Consider an infinitesimal SO($N$) transformation,
\begin{equation}
R_{ij}=\delta_{ij}-\theta_{ij} + O(\theta^2)\;.
\label{rddt}
\end{equation}
Orthogonality of $R$ implies that $\theta$ is real and antisymmetric.
It is convenient to express $\theta$ in terms of a basis set of
hermitian matrices $(T^a)_{ij}$.  The index $a$ runs from
1 to $\half N(N-1)$, the number of linearly independent, hermitian,
antisymmetric, $N\times N$ matrices.  We can, for example, choose each
$T^a$ to have a single nonzero entry $-i$ 
above the main diagonal, and a corresponding $+i$ below the main diagonal.
These matrices obey the normalization condition
\begin{equation}
\tr(T^a T^b)=2\delta^{ab}\;.
\label{trtt}
\end{equation}
In terms of them, we can write $\theta_{jk}=i\theta^a(T^a)_{jk}$,
where $\theta^a$ is a set of $\half N(N-1)$ real, infinitesimal parameters.

The $T^a$'s are the {\it generator matrices\/} of SO($N$).
The product of any two SO($N$) rotations is another SO($N$) rotation.
This implies that the commutator of any two $T^a$'s must be a linear
combination of $T^a$'s:
\begin{equation}
[T^a,T^b]=if^{abc}T^c \;.
\label{ttft}
\end{equation}
The real numerical factors $f^{abc}$ in \eq{ttft} are
the {\it structure coefficients\/} of the group.
If $f^{abc}=0$, the group 
is {\it abelian}.  Otherwise, it is {\it nonabelian}.  Thus, 
U(1) and SO(2) are abelian groups (since they each have only one
generator that obviously must commute with itself), 
and SO($N$) for $N\ge3$ is nonabelian.

If we multiply \eq{ttft} on the right by $T^d$, 
take the trace, and use \eq{trtt},
we find
\begin{equation}
f^{abd}
= -\half i\,\tr\!\left([T^a,T^b]T^d\right) \;.
\label{fabc}
\end{equation}
Using the cyclic property of the trace, we find that
$f^{abd}$ must be completely antisymmetric. 
Taking the complex conjugate of \eq{fabc} 
(and remembering that the $T^a$'s are hermitian matrices),
we find that $f^{abd}$ must be real.

The simplest nonabelian group is SO(3).  In this case,
we can choose $(T^a)_{ij} = -i\e^{a i j}$, where
$\e^{ijk}$ is the completely antisymmetric Levi-Civita symbol,
with $\e^{123}=+1$.  The commutation relations become
\begin{equation}
[T^a,T^b]=i\e^{abc}T^c \;.
\label{ttet}
\end{equation}
That is, the structure coefficients of SO(3) are given by
$f^{abc}=\e^{abc}$.

Let us return to \eq{elln}, and consider the case $m^2<0$.
The minimum of the potential
of \eq{elln} is achieved for $\ph_i(x)=v_i$, where
$v^2=v_i v_i=4|m^2|/\lam$, and the direction in which
the $N$-component vector $\vec v$ points is arbitrary.
In the quantum theory, we interpret $v_i$ as the vacuum
expectation value of the quantum field $\ph_i(x)$.
We can choose our coordinate system so that 
$v_i=v\delta_{iN}$; that is, the vacuum expectation value
lies entirely in the last component.  

Now consider making an infinitesimal SO($N$) transformation,
\begin{eqnarray}
v_i &\to& R_{ij}v_j 
\nonumber \\
&=& v_i - \theta_{ij}v_j
\nonumber \\
&=& v_i - i\theta^a(T^a)_{ij}v_j
\nonumber \\
&=& v\delta_{iN} - i\theta^a(T^a)_{iN}v \;.
\label{vrv}
\end{eqnarray}
For some choices of $\theta^a$, the second term on the right-hand
side of \eq{vrv} vanishes.  This happens if the corresponding
$T^a$ has no nonzero entry in the last column.  Recall that each
$T^a$ has a single $-i$ above the main diagonal
(and a corresponding $+i$ below the main diagonal).
Thus, there are $N{-}1$ $T^a$'s
with a nonzero entry in the last column: those with the $-i$ in the
first row and last column, in the second row and last column, etc,
down to the $N{-}1^{\rm th}$ row and last column.
These $T^a$'s are said to be {\it broken\/} generators:
a generator is broken if $(T^a)_{ij}v_j \ne 0$, and unbroken if
$(T^a)_{ij}v_j = 0$.

An infinitesimal SO($N$) transformation that involves a 
broken generator changes the vacuum expectation value of the field,
but not the energy.  Thus, each
broken generator corresponds to a {\it flat direction\/} in field space.
Each flat direction implies the existence of a
corresponding massless particle.  This is {\it Goldstone's theorem\/}:
there is one massless Goldstone boson for each broken generator.

The unbroken generators, on the other hand, do not change the
vacuum expectation value of the field.  Therefore, after rewriting
the lagrangian in terms of shifted fields (each with zero vacuum expectation
value), there should still be a manifest symmetry corresponding to the
set of unbroken generators.
In the present case, the number of unbroken generators is
\begin{equation}
\half N(N{-}1) - (N{-}1)=\half(N{-}1)(N{-}2) \;.
\label{unbroken}
\end{equation}
This is the number of generators of SO($N{-}1$).  Therefore, we expect
SO($N{-}1$) to be an obvious symmetry of the lagrangian after it is
written in terms of shifted fields.

Let us see how this works in the present case.  We can rewrite \eq{elln} as
\begin{equation}
\L= -\half\d^\mu \ph_i\d_\mu \ph_i - V(\ph)\;,
\label{elln2}
\end{equation}
with
\begin{equation}
V(\ph)={\ts{1\over16}}\lam(\ph_i\ph_i-v^2)^2\;,
\label{vn2}
\end{equation}
where $v=(4|m^2|/\lam)^{1/2}$, and the repeated index $i$ is implicitly
summed from 1 to $N$.  Now let 
$\ph_N(x)=v+\rho(x)$, and plug this into \eq{elln2}.
With the repeated index $i$ now implicitly
summed from 1 to $N{-}1$, we have
\begin{equation}
\L= -\half\d^\mu \ph_i\d_\mu \ph_i 
-\half\d^\mu\!\rho\d_\mu\rho 
- V(\rho,\ph)\;,
\label{elln3}
\end{equation}
where 
\begin{eqnarray}
V(\rho,\ph) &=& {\ts{1\over16}}\lam
[(v{+}\rho)^2 + \ph_i\ph_i-v^2]^2
\nonumber \\
\noalign{\medskip}
&=& {\ts{1\over16}}\lam(2v\rho + \rho^2 + \ph_i\ph_i)^2
\nonumber \\
\noalign{\medskip}
&=& {\ts{1\over4}}\lam v^2\rho^2
+ {\ts{1\over4}}\lam v\rho
(\rho^2 + \ph_i\ph_i)^2
+ {\ts{1\over16}}\lam(\rho^2 + \ph_i\ph_i)^2 \;. \quad
\label{vn3}
\end{eqnarray}
There is indeed a manifest SO($N{-}1$) symmetry in \eqs{elln3} and (\ref{vn3}).
Also, the $N{-}1$ $\ph_i$ fields are massless: 
they are the expected $N{-}1$ Goldstone bosons.

Consider now a theory with $N$ {\it complex\/} scalar fields $\ph_i$,
and a lagrangian
\begin{equation}
\L=-\d^\mu \ph_i^\dagger\d_\mu\ph_i 
-m^2\ph_i^\dagger\ph_i
- {\ts{1\over4}}\lam(\ph^\dagger_i\ph_i)^2\;,
\label{ellsun}
\end{equation}
where a repeated index is summed.  This lagrangian is clearly invariant
under the U($N$) transformation
\begin{equation}
\ph_i(x) \to U_{ij}\ph_j(x) \;,
\label{phuph}
\end{equation}
where $U$ is a unitary matrix: $U^\dagger = U^{-1}$.  
We can write $U_{ij}=e^{-i\theta}{\widetilde U}_{ij}$, where
$\theta$ is a real parameter and
$\det{\widetilde U}_{ij}=+1$;  
${\widetilde U}_{ij}$ is called a {\it special\/} unitary matrix.  
Clearly the product of two 
special unitary matrices is another
special unitary matrix; the $N\times N$
special unitary matrices form the group SU($N$).
The group U($N$) is the {\it direct product\/} of the group
U(1) and the group SU($N$); we write
${\rm U}(N)={\rm U}(1)\times{\rm SU}(N)$.

Consider an infinitesimal SU($N$) transformation,
\begin{equation}
{\widetilde U}_{ij}=\delta_{ij}-i\theta^a(T^a)_{ij} + O(\theta^2)\;,
\label{uddt}
\end{equation}
where $\theta^a$ is a set of real, infinitesimal parameters.
Unitarity of $\widetilde U$ implies that the generator matrices $T^a$
are hermitian, and $\det\widetilde U=+1$ implies that each $T^a$ is 
traceless.  (This follows from the general matrix formula
$\ln\det A = \tr\ln A$.) 
The index $a$ runs from 1 to $N^2{-}1$,
the number of linearly independent, hermitian, traceless, $N\times N$
matrices.  We can choose these matrices to obey the normalization condition
of \eq{trtt}.
For SU(2), the generators can be chosen to be the Pauli matrices; 
the structure coefficients of SU(2) then turn out to be $f^{abc}=2\e^{abc}$,
the same as those of SO(3),
up to an irrelevant overall factor [which could be removed by
changing the numerical factor on right-hand side of \eq{trtt} from 
2 to $\half$].

For SU($N$), we can choose the $T^a$'s in the following way.  First,
there are the SO($N$) generators, with one $-i$ 
above the main diagonal a corresponding $+i$ below;
there are $\half N(N{-}1)$ of these.  
Next, we get another set by putting one $+1$ above the main diagonal
and a corresponding $+1$ below;
there are $\half N(N{-}1)$ of these.  
Finally, there are diagonal matrices with $n$ 1's along the main diagonal,
followed a single entry $-n$, followed by zeros [times an overall normalization
constant to enforce \eq{trtt}]; the are $N{-}1$ of these.
The total is $N^2{-}1$, as required.

We could now return to \eq{ellsun}, consider the case $m^2<0$,
and examine spontaneous breaking of the U($N$) symmetry.
However, the lagrangian of \eq{ellsun} is actually invariant
under a larger symmetry group, namely SO($2N$).  
To see this, write each complex scalar field in terms of two real 
scalar fields,
$\ph_j = (\ph_{j1}+i\ph_{j2})/\sqrt2$.
Then 
\begin{equation}
\ph_j^\dagger\ph_j = \half(\ph_{11}^2 + \ph_{12}^2 + \ldots
+ \ph_{N1}^2 + \ph_{N2}^2)\;.
\label{phiun}
\end{equation}
Thus, we have $2N$ real scalar fields that enter $\L$ symmetrically,
and so the actual symmetry group of \eq{elln2} is SO($2N$), rather than
just the obvious subgroup U($N$).  

We will, however, meet the SU($N$) groups again in Parts II and III,
where they will play a more important role.

\vfill\eject